\documentclass[twocolumn,tighten,times,twocolappendix,numberedappendix]{aastex631}

\usepackage{amsmath}
\usepackage{float}
\usepackage{color}
\usepackage{url}
\usepackage{etoolbox}

\usepackage{graphicx}

\usepackage[flushleft]{threeparttable}

\usepackage{placeins}

\hypersetup{breaklinks=true,colorlinks=true,linkcolor=blue,citecolor=blue,filecolor=blue,urlcolor=blue}
\makeatletter
\patchcmd{\NAT@citex}
  {\@citea\NAT@hyper@{%
     \NAT@nmfmt{\NAT@nm}%
     \hyper@natlinkbreak{\NAT@aysep\NAT@spacechar}{\@citeb\@extra@b@citeb}%
     \NAT@date}}
  {\@citea\NAT@nmfmt{\NAT@nm}%
   \NAT@aysep\NAT@spacechar\NAT@hyper@{\NAT@date}}{}{}

\patchcmd{\NAT@citex}
  {\@citea\NAT@hyper@{%
     \NAT@nmfmt{\NAT@nm}%
     \hyper@natlinkbreak{\NAT@spacechar\NAT@@open\if*#1*\else#1\NAT@spacechar\fi}%
       {\@citeb\@extra@b@citeb}%
     \NAT@date}}
  {\@citea\NAT@nmfmt{\NAT@nm}%
   \NAT@spacechar\NAT@@open\if*#1*\else#1\NAT@spacechar\fi\NAT@hyper@{\NAT@date}}
  {}{}
  
\usepackage{array}
\newcolumntype{C}[1]{>{\centering\let\newline\\\arraybackslash\hspace{0pt}}m{#1}}

\usepackage{xspace}

\def\aj{AJ}
\def\araa{ARA\&A}
\def\apj{ApJ}
\def\apjl{ApJ}
\def\apjs{ApJS}
\def\ao{Appl.~Opt.}
\def\apss{Ap\&SS}
\def\aap{A\&A}

\def\aaps{A\&AS}

\def\mnras{MNRAS}

\def\pasa{PASA}
\def\pasp{PASP}

\def\ssr{Space~Sci.~Rev.}

\def\rmxaa{Rev.~Mex.~Astron.~Astrofis.}

\def\nar{New~Astro.~Rev.}

\def\arcdeg{\hbox{$^\circ$}}

\def\arcsec{\hbox{$^{\prime\prime}$}}

\definecolor{burgundy}{rgb}{0.5, 0.0, 0.13}

\usepackage{xspace}

\newcommand{\orcidicon}{\includegraphics[width=0.26cm]{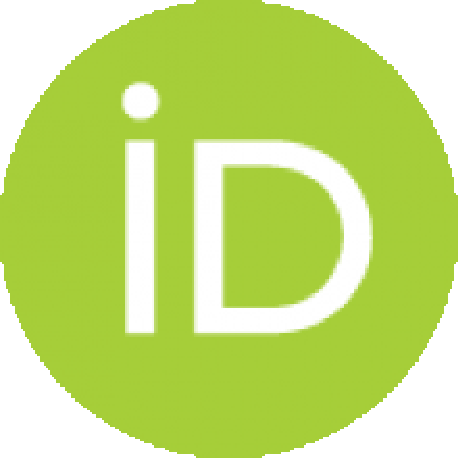}}

\newcommand{\orcidauthor}[1]{\href{https://orcid.org/#1}{\orcidicon}}

\shorttitle{Morpho-kinematic properties of WR planetary nebulae}
\shortauthors{Danehkar}

\makeatletter
\patchcmd{\frontmatter@RRAP@format}{(}{}{}{}
\patchcmd{\frontmatter@RRAP@format}{)}{}{}{}
\renewcommand\Dated@name{}
\makeatother

\begin{document}

\title{Morpho-kinematic properties of Wolf-Rayet planetary nebulae}

\correspondingauthor{A.~Danehkar}
\email{danehkar@umich.edu}

\author[0000-0003-4552-5997]{A.~Danehkar}
\affiliation{Department of Astronomy, University of Michigan, 1085 S. University Ave, Ann Arbor, MI 48109, USA}

\date[ ]{\footnotesize\textit{Received 2021 July 7; revised 2022 February 10; accepted 2022 February 25; published 2022 May 5}}

\begin{abstract}
The majority of planetary nebulae (PNe) show axisymmetric morphologies, whose causes are not well understood. In this work, we present spatially resolved kinematic observations of 14 Galactic PNe surrounding Wolf-Rayet ([WR]) and weak emission-line stars (\textit{wels}) based on the H$\alpha$ and [N~{\sc ii}] emission taken with the Wide Field Spectrograph on the Australian National University 2.3-m telescope. Velocity-resolved channel maps and position--velocity diagrams, together with archival \textit{Hubble Space Telescope} (\textit{HST}) and ground-based images, are employed to construct three-dimensional morpho-kinematic models of 12 objects using the program \textsc{shape}. Our results indicate that these 12 PNe mostly have elliptical morphologies with either open or closed outer ends. The kinematic maps show the on-sky orientations of the interior shells in NGC 6578 and NGC 6629, as well as the compact ($\leq 6$\,arcsec) PNe Pe\,1-1, M\,3-15, M\,1-25, Hen\,2-142, and NGC\,6567, in agreement with the elliptically symmetric morphologies seen in high-resolution \textit{HST} images. Point-symmetric knots in Hb\,4 exhibit deceleration with distance from the central star, which could be due to shock collisions with the ambient medium. The velocity dispersion maps of Pe\,1-1 also disclose the shock interaction between its collimated outflows and the interstellar medium. Collimated bipolar outflows are also visible in the position--velocity diagrams of M\,3-30, M\,1-32, and M\,3-15, which are reconstructed by tenuous prolate ellipsoids extending upward from dense equatorial regions in the kinematic models. The formation of aspherical morphologies and collimated outflows in these PNe could be related to the stellar evolution of hydrogen-deficient [WR] and \textit{wels} nuclei, which require further investigation.
\end{abstract}


\keywords{\href{https://astrothesaurus.org/uat/1249}{Planetary nebulae (1249)};
\href{https://astrothesaurus.org/uat/847}{Interstellar medium (847)};
\href{http://astrothesaurus.org/uat/1607}{Stellar jets (1607)};
\href{http://astrothesaurus.org/uat/1608}{Stellar kinematics (1608)};
\href{https://astrothesaurus.org/uat/1806}{Wolf-Rayet stars(1806)}
\vspace{4pt}
\newline
\textit{Supporting material:} figure sets, interactive figure 
}


\section{Introduction}
\label{wc:sec:introduction}

Planetary nebulae (PNe) are bright hydrogen-rich envelopes expelled by evolved stars with low-to-intermediate progenitor masses (1--8\,M$_{\odot}$) in the asymptotic giant branch (AGB) phase. These envelopes are fully ionized by ultraviolet (UV) radiation from their hot degenerate nuclei in the post-AGB stage. Photoionization produces visible nebulae and also contributes to some thermal effects. Stellar winds from central stars of planetary nebulae (CSPNe) departing from the AGB phase introduce some hydrodynamic effects to these ionized envelopes that successively change their gaseous structures. The line emission produced by the ionized envelope allows us to observe its morpho-kinematic structure. Kinematic observations of PNe not only reveal their morphologies, but also provide insights into mass-loss processes in the AGB phase and during their transition from the AGB to PN, as well as PN evolution in the post-AGB phase \citep[see e.g.][]{Balick1987,Corradi1995,Balick2002,Schonberner2005a,Schonberner2005b,Schonberner2010,Kwok2010}.

Observations of PNe have mostly displayed elliptical and bipolar axisymmetric morphologies \citep[see e.g.][]{Balick1987,Balick1987a,Schwarz1992,Sahai2011,Weidmann2016}, which pose a major challenge to theories describing their formation   \citep[see the review by][]{Balick2002}. Although single stars are expected to produce spherical nebula shells, the interaction with the interstellar medium (ISM) may also result in elliptical shells. However, the theories face considerable difficulties in generating highly axisymmetric morphologies similar to what are seen in observations. According to the interacting stellar wind (ISW) theory  \citep{Kwok1978}, a compressed, dense shell is formed when a slow ($\sim$~$10$~km\,s$^{-1}$), dense wind from the AGB phase is pushed away by a fast ($\sim$~$1000$~km\,s$^{-1}$), tenuous stellar wind from the PN phase. The ISW model extended by \citet{Kahn1985}, known as the generalized interacting stellar wind (GISW) theory, can also produce a highly axisymmetric gas distribution. Nevertheless, the GISW model is unable to fully describe all observed elliptical and bipolar morphologies \citep[][]{Balick2002}. Moreover, the GISW model needs a density contrast to produce an aspherical morphology. Recently, many complex axisymmetric morphologies have been found by high-resolution imaging observations \citep[see e.g.][]{Sahai2011}, which cannot adequately be described by the GISW theory. 

It has been suggested that axisymmetric morphologies are produced by tidal interactions with a binary companion \citep{Soker1992,Soker1994,Soker2006,Nordhaus2006,Nordhaus2007}. This binary companion is usually expected to be a white dwarf that accretes material from an AGB star, though it could also be an exoplanet \citep{DeMarco2011,Hegazi2020,Rapoport2021}. \citet{Paczynski1976} first proposed the role of binarity in shaping PNe through a common-envelope (CE) phase. Alternatively, it has been proposed that a mixture of strong toroidal magnetic fields and rotating stellar winds from a single star  can lead to an equatorial density enhancement and jet-like outflows \citep{Garcia-Segura1997,Garcia-Segura1999,Garcia-Segura2000,Frank2004}. On the other hand, \citet{Soker2006} contended that the magnetic fields measured in PNe cannot play a central role in shaping PNe and that a single star cannot provide enough energy and angular momentum to support complex axisymmetric morphologies. Currently, there is a strong argument that most axisymmetric morphologies of PNe have been shaped through a binary channel \citep{Miszalski2009a,Miszalski2009b,DeMarco2009,Nordhaus2010}. Furthermore, \citet{Miszalski2009b} discovered that nearly 30 percent, possibly as much as 60 percent, of bipolar PNe contain post-CE binaries, implying that the CE phase preferentially shapes axisymmetric PNe. The transformation of the orbital angular momentum of a binary system into the CE unbinds it and shapes an axisymmetric nebula, whose symmetric axis is perpendicular to the orbital plane of the binary system. High-resolution kinematic analyses of some PNe around post-CE binary stars have disclosed alignments between the nebular symmetric axes and binary orbital axes \citep[see e.g.][]{Mitchell2007,Jones2010a,Jones2012,Tyndall2012,Huckvale2013}. 

Small-scale \textit{low-ionization structures} (LISs) inside or outside the main shells have been identified in nearly 10\% of the Galactic PNe \citep{Corradi1996,Gonccalves2001,Gonccalves2009}, which are more visible in low-excited emission like [N\,{\sc ii}] and [S\,{\sc ii}]. They were classified as knots, jets, and jetlike systems by \citet{Gonccalves2001}. \textit{Knots}, either in pairs or isolated, are defined as those LISs with an aspect length-to-width ratio close to 1, while those with an aspect ratio much larger than 1 are classified as filaments. \textit{Jets} are defined as highly collimated filaments that symmetrically appear as a pair on both sides of the central star and move with velocities much larger than the expansion velocity of the main shell. Those filaments with no evidence of velocities higher than the expansion velocity of the main body are called \textit{jetlike}. However, projection effects make it difficult to distinguish between jets and jetlike systems. Highly collimated high-velocity jets or high-velocity pairs of knots constitute around half of the LISs, which are named \textit{fast, low-ionization emission regions} \citep[FLIERs;][]{Balick1993,Balick1994}. FLIERs have radial velocities of 25--200 km\,s$^{-1}$ with respect to the main bodies \citep{Balick1994}. The formation of FLIERs, as well as their associated density and velocity contrasts with the main shells, are not yet well understood. It has been hypothesized that FLIERs are produced by axisymmetric superwind mass-loss through a CE channel, tidal interaction with a low-mass companion, and angular momentum deposition of a binary system \citep{Soker1990,Soker1992}.

\begin{table*}
\caption{Journal of the planetary nebulae observed with the ANU 2.3-m Telescope, including their stellar characteristics.
\label{wc1:tab:obs:journal}
}
\footnotesize
\vspace{-10pt}
\begin{center}
\begin{tabular}{llllllccll}
\hline\hline
\noalign{\smallskip}
Name & PNG & R.A. \& Dec./J2000 & CSPN & $T_{\rm eff}$ -- $\log(L)$  & \multicolumn{1}{c}{$V_{\infty}$}   & $\log \dot{M}$\,$^{\mathrm{a}}$ & $M_{\star}$\,$^{\mathrm{b}}$ & Exp. & Obs. Date\\
     &  (A92)  &   &  (A03)   & (kK) -- (${\rm L}_{\odot}$) & \multicolumn{1}{c}{(km/s)}  & (M${}_{\odot}$/yr)  &(${\rm M}_{\odot}$) & (sec)  \\
\noalign{\smallskip}
\hline
\noalign{\smallskip}
PB 6 	&278.8$+$04.9 & 10\,13\,15.9 $-$50\,19\,59.1 & [WO 1]  	& 103 -- 3.57 (K91) &  2496 (A03) & $-7.15$  &  0.60 & 1200 & 2010 Apr 20  \\
\noalign{\smallskip}
M\,3-30&017.9$-$04.8 & 18\,41\,14.9 $-$15\,33\,43.6 & [WO 1]   	& 49 -- 3.3\,$^{\mathrm{b}}$ (A03) &  2059 (A03) & $-7.50$  &  0.56  & 1200 & 2010 Apr 21 \\
\noalign{\smallskip}
Hb\,4	&003.1$+$02.9 & 17\,41\,52.7 $-$24\,42\,08.0 & [WO 3]   	& 85 -- 3.6 (A03,A02) &  2059 (A03) &  $-7.11$  &  0.60 & 300,1200 & 2010 Apr 21\\
\noalign{\smallskip}
IC\,1297&358.3$-$21.6 & 19\,17\,23.5 $-$39\,36\,46.4 & [WO 3]  	& 91 -- 3.7\,$^{\mathrm{b}}$  (A03) &  2933 (A03) & $-6.98$   &  0.62  & 60,1200 & 2010 Apr 21 \\
\noalign{\smallskip}
Pe\,1-1	&285.4$+$01.5 & 10\,38\,27.6 $-$56\,47\,06.5 & [WO 4]   	& 85 -- 3.3 (A02) &  2870 (A03) & $-7.50$ &  0.55 &  60,1200 & 2010 Apr 21 \\
\noalign{\smallskip}
M\,1-32&011.9$+$04.2 & 17\,56\,20.1 $-$16\,29\,04.6 & [WO 4]{\scriptsize pec}	& 56\,$^{\mathrm{c}}$ --   3.5\,$^{\mathrm{d}}$  &  4867 (A03) & $-7.24$ &  0.60 & 1200 & 2010 Apr 20 \\
\noalign{\smallskip}
M\,3-15&006.8$+$04.1 & 17\,45\,31.7 $-$20\,58\,01.6 & [WC 4]  	& 55 -- 3.6 (A03,A02) &  1872 (A03) & $-7.11$ &  0.59 & 60,1200 & 2010 Apr 20 \\
\noalign{\smallskip}
M\,1-25&004.9$+$04.9 & 17\,38\,30.3 $-$22\,08\,38.8 & [WC 5-6]  	& 56 -- 3.8 (A03,A02) &  1747 (A03) & $-6.85$ & 0.62 &  60,1200 & 2010 Apr 20 \\
\noalign{\smallskip}
Hen\,2-142&327.1$-$02.2 & 15\,59\,57.6 $-$55\,55\,32.9 & [WC 9]   	& 35 --  3.7 (A03,G07) &  884 (A03) &  $-6.98$ &  0.60 & 60,1200 & 2010 Apr 20 \\
\noalign{\smallskip}
K\,2-16&352.9$+$11.4 & 16\,44\,49.1 $-$28\,04\,04.7 & [WC 11]  	& 19 -- 3.3  (A03,A02) &  260 (A03) & $-7.50$ &  0.52 & 1200 & 2010 Apr 20 \\
\noalign{\smallskip}
NGC\,6578&010.8$-$01.8 & 18\,16\,16.5 $-$20\,27\,02.7 & \textit{wels}(T93)  	&  63 -- $<4.03$ (S89) & 1498 (T93) & $<-6.55$ & $<0.68$ & 60,1200 & 2010 Apr 22 \\
\noalign{\smallskip}
NGC\,6567&011.7$-$00.6 & 18\,13\,45.2 $-$19\,04\,34.2 & \textit{wels}(T93)  	& 47 -- 3.62 (G97) & 1747 (T93) & $-7.08$ & 0.60 & 60,1200 & 2010 Apr 22 \\
\noalign{\smallskip}
NGC\,6629&009.4$-$05.0 & 18\,25\,42.5 $-$23\,12\,10.2 & \textit{wels}(T93)  	& 35 -- 3.53 (G97) & 1747 (T93) & $-7.20$ & 0.56 & 60,1200 & 2010 Apr 22 \\
\noalign{\smallskip}
Sa\,3-107&358.0$-$04.6 & 17\,59\,55.0 $-$32\,59\,11.8 & \textit{wels}(D11)  	& 46\,$^{\mathrm{c}}$ -- $< 4.0$\,$^{\mathrm{d}}$ & 874 (D11) & $< -6.60$ & $< 0.70$ & 1200 & 2010 Apr 22 \\
\noalign{\smallskip}
\hline
\end{tabular}
\begin{list}{}{}
\item[$^{\mathrm{a}}$]The mass-loss rates calculated using the formula given by \citet{Nugis2000}. 
\item[$^{\mathrm{b}}$]The evolutionary tracks of the helium-burning model by \citet{Bloecker1995b}.
\item[$^{\mathrm{c}}$]Calculated from the nebular excitation class using the correlation by \citet{Dopita1990,Dopita1991}.
\item[$^{\mathrm{d}}$]The stellar luminosity from the standard bolometric correction method. 
\end{list}
\begin{tablenotes}
\item[1]\textbf{Note.} References are as follows: 
A92 -- \citet{Acker1992}; 
A02 -- \citet{Acker2002}; 
A03 -- \citet{Acker2003}; 
D11 -- \citet{Depew2011}; 
G97 -- \citet{Gorny1997}; 
G07 -- \citet{Gesicki2007}; 
K91 -- \citet{Kaler1991}; 
S89 -- \citet{Shaw1989}; 
T93 -- \citet{Tylenda1993}.
\end{tablenotes}
\end{center}
\end{table*}

Most CSPNe have hydrogen-rich stellar atmospheres, but a substantial fraction ($\lesssim25$\%) of them have been found to possess \textit{hydrogen-deficient} fast expanding atmospheres characterized by high mass-loss rates \citep{Tylenda1993,Leuenhagen1996,Leuenhagen1998,Acker2003}. Their surface abundances exhibit helium, carbon, oxygen, and neon, which are products of the helium-burning phase and a post-helium flash \citep{Werner2006}. The majority of these CSPNe are classified as Wolf-Rayet ([WR]) stars with carbon sequences, resembling massive Wolf–Rayet (WR) stars \citep[][]{vanderHucht1981,vanderHucht2001}, where the square brackets distinguish them from their massive counterparts. Roughly half of them are called early-type ([WCE]) stars, including the spectral classes [WO1]--[WC5] \citep{Koesterke1997,Pena1998}, which have stellar temperatures ranging from 80--150\,kK. Others with stellar temperatures from 20 to 80~kK are named late-type ([WCL]) stars with [WC6--11] \citep{Leuenhagen1996,Leuenhagen1998}. A few CSPNe have narrower and weaker emission lines (C\,{\sc iv} 5805\,{\AA} and C\,{\sc iii} 5695\,{\AA}) that are not identical to those of [WR] stars, the so-called weak emission-line stars \citep[\textit{wels};][]{Tylenda1993}. These \textit{wels} CSPNe are surprisingly higher towards the Galactic bulge and closer to the Galaxy's center, and they appear to have evolved from different stellar populations than [WR] stars \citep{Gorny2004,Gorny2009}. However, medium-resolution spectroscopic studies of 19 stars, that had previously been classified as \textit{wels} using low-resolution spectra, pointed to 12 normal H-rich, 2 probably H-deficient, and 5 unclassified stars \citep{Weidmann2015}.

This work is aimed at the morpho-kinematic properties of PNe surrounding [WR] and \textit{wels} nuclei by means of optical integral field unit (IFU) spectroscopy. This paper is structured as follows.  Section\,\ref{wc1:sec:observations} describes our observational and data reduction processes. In Section\,\ref{wc1:sec:kinematic}, we present our IFU kinematic results. Three-dimensional (3D) morpho-kinematic models and their results are presented in Section\,\ref{wc1:sec:morpho-kinematic}, followed by conclusions and discussion in Section\,\ref{wc1:sec:conclusions}.

\section{Observations}
\label{wc1:sec:observations}

The IFU observations of the PNe studied in this work were collected with the Wide Field Spectrograph \citep[WiFeS;][]{Dopita2007,Dopita2010} at the Siding Spring Observatory in April 2010 under program number 1100147 (PI: Q. A. Parker). The WiFeS is an image-slicing IFU designed for the Australian National University (ANU) 2.3-m telescope that feeds a double-beam spectrograph. It samples 0.5 arcsec along each of twenty five $38$\,arcsec\,$\times$\,$1$\,arcsec slitlets, resulting in a field-of-view of $25\times38$\,arcsec$^2$ with a spatial resolution element of  $1\times0.5$\,arcsec$^2$. The spectroscopic output is projected onto a CCD detector of  $4096 \times 4096$ pixels. Each slitlet is mapped onto 2 pixels on the CCD detector, which leads to a reconstructed point spread function (PSF) with a full width at half-maximum (FWHM) of $\sim2$ arcsec. The spectrograph uses volume phase holographic gratings to provide a spectral resolution of $R \sim 7000$ or $3000$, corresponding to a velocity resolution of either $\sim 21$ or $\sim 60$ km\,s${}^{-1}$, respectively. 

Our targets were observed with a spectral resolution  of $R\sim 7000$ covering the wavelength range of 4415--7070\,{\AA}. As each spectrum is 4096 pixels long in the CCD detector, it provides a linear wavelength dispersion per pixel of $0.36$ {\AA} in the blue arm and $0.45$ {\AA} in the red arm with the $R\sim 7000$ grating. 
The WiFeS spatial resolution with the reconstructed PSF FWHM of 2 arcsec enables us to identify morphologies of PNe with angular diameters larger than 6 arcsec. Although this spatial resolution is not ideal for resolving compact ($\leq 6$\,arcsec) PNe, it can determine the on-sky orientations of morphologies from their extended faint lobes or exterior halos \citep[see][]{Danehkar2015}. All targets were acquired in the classical data accumulation mode. We also collected bias frames, flat-field frames, twilight sky flats, arc lamp exposures, wire frames, and standard stars for bias reduction, flat-fielding, wavelength and spatial calibrations, and flux calibration.

Table~\ref{wc1:tab:obs:journal} presents a journal of the WiFeS observations. The usual PN names, PN in Galactic coordinate (PNG) numbers \citep{Acker1992}, coordinates (RA and Dec.), stellar types of the CSPN, as classified by \citet{Tylenda1993}, \citet{Acker2003} and \citet{Depew2011}, are given in Columns 1--4, respectively. Columns 5--6 provide the effective temperature ($T_{\rm eff}$), stellar luminosity ($L$), and stellar wind terminal velocity ($V_{\infty}$), along with their references, respectively. The mass-loss rates ($\dot{M}$) in Column 7 were calculated using the formula given by \citet{Nugis2000}, with stellar luminosity and typical [WR] chemical composition of $Y=0.43$ and $Z=0.56$. Column 8 presents the stellar mass ($M_{\star}$) estimated from the helium-burning evolutionary models of \citet{Bloecker1995b}. The exposure times used for the observations and the observing date are given in Columns 9 and 10, respectively.

\begin{table*}
\caption{Journal of archival \textit{HST} images of the planetary nebulae in our sample.
\label{wc1:tab:hst}
}
\centering
\footnotesize
\begin{tabular}{lllllclllll}
\hline\hline
\noalign{\smallskip}
Name  & PNG & Instrument  & Aperture & Filters/ & Wavelength & Plate scale  & Exp.       & Obs. Date & \multicolumn{2}{l}{Program} \\
        &    &             &		     &	Gratings & range ({\AA})	& (arcsec/pix) & (sec)  	& 	&  ID   		& PI \\
\noalign{\smallskip}
\hline 
\noalign{\smallskip} 
Hb\,4  		& 003.1$+$02.9 & WFPC2 	& PC			& F658N 	& 6570--6598 & 0.045 & 600 & 1996 Oct 28 & 6347  & K.~Borkowski  \\
\noalign{\smallskip}
Pe\,1-1  	& 285.4$+$01.5 & WFC3 		& UVIS			& F502N 	& 4963--5059 & 0.039 & 600 & 2009 Oct 05 & 11657 & L.~Stanghellini \\
\noalign{\smallskip}
M\,3-15  	& 006.8$+$04.1 & WFPC2 	& PC 			& F656N 	& 6552--6570 & 0.045 & 200 & 2002 Jul 04 & 9356  & A.~Zijlstra  \\
\noalign{\smallskip}
M\,1-25  	& 004.9$+$04.9 & WFPC2  	& PC			& F658N		& 6570--6598 & 0.045 & 480 & 2001 Jun 26 & 8345  & R.~Sahai \\
\noalign{\smallskip}
Hen\,2-142  & 327.1$-$02.2 & WFPC2 	& PC			& F656N		& 6552--6570 & 0.045 & 2040 & 1996 Aug 12 & 6353  & R.~Sahai \\
\noalign{\smallskip}
NGC\,6578   & 010.8$-$01.8 & WFPC2 	& PC			& F658N		& 6570--6598 & 0.045 & 1200 & 2008 Feb 28 & 11122  & B.~Balick \\ 
\noalign{\smallskip}  
NGC\,6567   & 011.7$-$00.6 & NICMOS 	& NIC	        & F108N		& 10797--10836 & 0.043 & 384 & 1998 Aug 14 & 7837  & S.~Pottasch \\
\noalign{\smallskip}
NGC\,6629   & 009.4$-$05.0 & WFPC2 	& PC			& F555W 	& 4789--6025 & 0.045 & 15 & 1995 Aug 16 & 6119  & H.~Bond \\
\noalign{\smallskip}    
\hline
\end{tabular}
\end{table*}

Our sample includes 10 well-known PNe with [WR] central stars chosen from the literature \citep{Crowther1998,Acker2003}. These PNe allow us to identify morphologies at various evolutionary stages with different stellar classes from early- to late-type [WR]. The morpho-kinematic structures of two PNe (Hen\,3-1333 and Hen\,2-113) with [WC\,10] studied by \citet{Danehkar2015} are very compact, since they are too young in comparison to the [WCE] PNe. The CSPN M\,1-32 was classified by \citet{Acker2003} as the \textit{peculiar} [WO\,4] based on the wide FWHM of the C~{\sc iv}-5801/12 doublet, which corresponds to a terminal stellar wind velocity of $V_{\infty}\simeq4900$\,km\,s$^{-1}$. This stellar velocity is higher than $V_{\infty}\simeq2000$--3000\,km\,s$^{-1}$ as typically seen in the usual [WR] CSPNe. The PN Th\,2-A, whose morphological features were disentangled by \citet{Danehkar2015a}, is another object with [WO\,3]${}_{\rm pec}$ classified by \citet{Weidmann2008}. 

The sample also contains 4 PNe with \textit{wels} selected from the literature \citep{Tylenda1993,Depew2011}. A 3D kinematic model of M\,2-42 around another \textit{wels} was also constructed by \citet{Danehkar2016} based on WiFeS observations.
The stellar luminosities listed in Table~\ref{wc1:tab:obs:journal} are mostly from the literature. 
We calculated the stellar luminosity for M\,3-30 and IC\,1297 using \citet{Bloecker1995b}'s evolutionary tracks for helium-burning models and \citet{Pauldrach1988}'s relationship between terminal velocity ($V_{\infty}$), effective temperature ($T_{\rm eff}$), and stellar mass ($M_{\star}$). We also estimated the stellar luminosities according to the standard bolometric correction method for M\,1-32 using $V=17$\,mag \citep{Pena2001}, $D=4796$\,pc \citep{Stanghellini2010}, and $c({\rm H}\beta)=1.35$ \citep{Danehkar2021}; and for Sa\,3-107 using $D=6$\,kpc, $V=16.4$\,mag \citep{Lasker2008}, and $c({\rm H}\beta)=1.62$ \citep{Danehkar2021}.

The data reduction was accomplished using a variety of techniques with the \textsc{iraf} pipeline \textsf{wifes}.\footnote{IRAF is distributed by NOAO, which is operated by AURA, Inc., under contract to the National Science Foundation.} We performed the flat-fielding using the frames taken with exposures from a quartz iodine (QI) lamp. A medium-averaged bias was subtracted from the calibration frames. We carried out the wavelength calibration with Cu--Ar arc exposures taken at the beginning of the night. 
The spatial calibration was conducted by employing wire frames collected from the diffuse illumination of the coronagraphic aperture with a QI lamp. Cosmic rays and bad pixels were removed from the raw data before the sky subtraction. We employed the \textsc{iraf} task \textsf{lacos\_im} \citep[\textsf{LA-Cosmic} package;][]{Dokkum2001} to eliminate cosmic rays from the raw data, while bad pixels and any remaining cosmic rays were manually cleaned off afterward with the \textsc{iraf}/\textsc{stsdas} task \textsf{imedit}. For the purpose of sky subtraction, we chose a suitable sky window from the science data. We calibrated the science data to absolute flux units using observations of the spectrophotometric standard stars EG\,274 and LTT\,3864. The flux-calibrated data were corrected for the atmospheric extinction. We also rebinned the spaxel map by a factor of 2 and then performed a three-point median smoothing to reduce the sampling artifacts created by the image-slicing IFU. 

The high-resolution \textit{Hubble Space Telescope} (\textit{HST}) images of the PNe were retrieved from the Mikulski Archive for Space Telescopes (doi:{\href{https://dx.doi.org/10.17909/t9-pdkg-tg57}{10.17909/t9-pdkg-tg57}), and the 3.5-m ESO New Technology Telescope (NTT) narrow-band images from \citet{Schwarz1992}, which were utilized to 
constrain our morpho-kinematic models. 
Table~\ref{wc1:tab:hst} lists the \textit{HST} images, 
together with their program IDs, instruments, filters/gratings, wavelength bands, spatial resolutions, and exposure times.
The \textsc{iraf} task \textsf{lacos\_im} was utilized to eliminate cosmic rays from the \textit{HST} images. 

\section{Results}
\label{wc1:sec:kinematic}

\subsection{Mass-weighted Mean Expansion and Systemic Velocities}
\label{wc1:sec:kinematic:expansion}

Table~\ref{wc1:tab:kinematics1} lists the kinematic features measured from the integrated emission-line profiles in the spectroscopic observations of our sample. Column 3 presents the systemic velocity ($v_{\rm sys}$) in the frame of the local standard of rest (LSR) based on the H$\alpha$ emission that approximately represents the systemic velocity of the whole nebula, along with the previous value from \citet{Durand1998} in Column 4. We transferred the radial velocity to the LSR frame using the \textsc{iraf}/\textsc{astutil} task \textsf{rvcorrect}. The mass-weighted mean expansion velocities ($v_{\rm HWHM}$) associated with the half width at half maximum (HWHM) of H$\alpha$ $\lambda$6563, [N\,{\sc ii}] $\lambda\lambda$6548,6584 and [S\,{\sc ii}] $\lambda\lambda$6716,6731, and their average velocity are presented in columns 5--8, respectively. The HWHM expansion velocities, $V_{\rm HWHM}=\sqrt{8\,{\rm ln (2)}}\sigma_{\rm true}/2$, were corrected for the instrumental width and the thermal broadening as follows: $\sigma_{\rm true}=\sqrt{{\sigma^2_{\rm obs}-\sigma^2_{\rm ins}-\sigma^2_{\rm th}-\sigma^2_{\rm fs}}}$, where the instrumental width $\sigma_{\rm ins}$ is measured from the $[$O~{\sc i}$]\,\lambda$5577 and $\lambda$6300 night sky lines having a typical value of $\approx 18$\,km\,s$^{-1}$ for the WiFeS ($R\sim7000$), the thermal broadening $\sigma_{\rm th}$ is estimated from the Boltzmann's equation $\sigma_{\rm th}=\sqrt{8.3\,T_e[{\rm kK}]/Z}$~[km\,s$^{-1}$] ($Z$ is the atomic weight of the atom or ion), and the fine structure broadening $\sigma_{\rm fs}$ is typically $\approx3$ km\,s${}^{-1}$ for H$\alpha$ \citep{Clegg1999}. For comparison, Column 9 lists the expansion velocity, together with its reference, from the literature.

\begin{table*}
\caption{Systemic velocities in the LSR frame and mass-weighted mean expansion (HWHM) velocities measured from the integrated emission-line profiles. 
\label{wc1:tab:kinematics1}
}
\centering
\footnotesize
\begin{tabular}{llrcccccl}
\hline\hline
\noalign{\smallskip}
Name 	& PNG & \multicolumn{1}{c}{$v_{\rm sys}$(H$\alpha$)} & $v_{\rm sys}$(km/s) & \multicolumn{4}{c}{$V_{\rm HWHM}$(km/s)} & \multicolumn{1}{c}{$v_{\rm exp}$} \\
\cline{5-8}
\noalign{\smallskip}
       &  & \multicolumn{1}{c}{(km/s)} & {(D98)} & H$\alpha$ &[N\,{\sc ii}] & [S\,{\sc ii}] & Mean &  \multicolumn{1}{c}{(km/s)} \\
\noalign{\smallskip}
\hline
\noalign{\smallskip}
PB\,6 	& 278.8$+$04.9 & ~~~$52.1\pm0.9$  & ~~~$45.9\pm1.3$  & $35.5$ & $32.5$ & $30.6$ & $32.9\pm2.4$  & $38\pm4$ (G09) \\
\noalign{\smallskip}
M\,3-30	& 017.9$-$04.8 & ~~~$79.2\pm0.6$  & ~~~~$67.3\pm14.0$ & $31.6$ & $41.9$ & $35.0$ & $36.2\pm5.1$  & $37.31\pm5$ (M06) \\
\noalign{\smallskip}
Hb\,4\,(shell)	& 003.1$+$02.9 & $-45.9\pm0.2$ & $-48.5\pm1.5$ & $23.2$ & $24.0$ & $22.4$ & $23.2\pm0.8$ & $21.5$ (L97) \\
\noalign{\smallskip}
IC\,1297& 358.3$-$21.6 & ~~~$12.6\pm0.1$      & ~~~$16.6\pm1.5$  & $31.8$  & $31.7$ & $30.1$ & $31.2\pm0.8$  & $32.7\pm1.7$ (A76) \\
\noalign{\smallskip}
Pe\,1-1	& 285.4$+$01.5 & ~~~$7.1\pm0.1$ 	& ~~~~$8.0\pm0.5$   & $21.7$ & $24.3$  & $24.3$ & $23.4\pm1.3$  & $24$  (G96) \\
\noalign{\smallskip}
M\,1-32 & 011.9$+$04.2 & $-73.8\pm1.3$ & $-76.8\pm7.6$ & $33.7$  & $32.0$  & $27.9$ & $31.2\pm2.9$ & $14$: (P01) \\
\noalign{\smallskip}
M\,3-15	& 006.8$+$04.1 & $111.3\pm0.6$ & ~$109.7\pm0.7$ & $23.7$ & $22.3$  & $17.8$ & $21.2\pm3.0$  & $18.2$ (R09) \\
\noalign{\smallskip}
M\,1-25 & 004.9$+$04.9 & ~~~$25.8\pm0.1$ & ~~~$26.3\pm2.0$ & $27.4$  & $28.3$  & $24.2$ & $26.6\pm2.0$ & $23$ (M06)  \\
\noalign{\smallskip}
Hen\,2-142& 327.1$-$02.2 & $-92.5\pm0.2$ & $-94.5\pm0.3$ & $22.1$ & $21.6$ & $17.5$ & $20.4\pm2.3$  & $20$(A02,G07) \\
\noalign{\smallskip}
K\,2-16	& 352.9$+$11.4 & ~~~$23.4\pm0.2$ & ~~~$14.9\pm12.0$ & $31.0$ & $34.1$ & $28.5$ & $31.2\pm2.8$  & $34$ (A02)  \\
\noalign{\smallskip}
NGC\,6578& 010.8$-$01.8 & ~~~$19.3\pm0.1$ & ~~~$17.1\pm1.8$ & $18.7$ & $23.5$ & $23.6$ & $21.9\pm3.0$  & $17.2\pm1.4$ (M06) \\
\noalign{\smallskip}
NGC\,6567& 011.7$-$00.6 & ~~~$136.7\pm0.3$ & ~~~$132.4\pm0.7$ & $24.2$ & $38.9$ & $37.5$ & $33.5\pm8.0$  & $19$ (W89) \\
\noalign{\smallskip}
NGC\,6629& 009.4$-$05.0 & ~~~$25.0\pm0.1$ & ~~~$26.5\pm1.3$ & $16.5$ & $20.8$ & $23.7$ & $20.3\pm3.6$  & $16.3\pm2.5$ (M06)  \\
\noalign{\smallskip}
Sa\,3-107& 358.0$-$04.6 & ~~~$-132.9\pm0.4$ & ~~-- & $17.0$ & $16.9$ & $14.5$ & $16.1\pm1.6$  & -- \\
\noalign{\smallskip}
\hline
\end{tabular}
\begin{list}{}{}
\item[\textbf{Note.}]References are as follows: 
A76 -- \citet{Acker1976}; 
A02 -- \citet{Acker2002};
D98 -- \citet{Durand1998};
G96 -- \citet{Gesicki1996}; 
G07 -- \citet{Gesicki2007}; 
G09 -- \citet{Garcia-Rojas2009}; 
L97 -- \citet{Lopez1997}; 
M06 -- \citet{Medina2006}; 
P01 -- \citet{Pena2001}; 
R09 -- \citet{Richer2009};
W89 -- \citet{Weinberger1989}. 
\end{list}
\end{table*}

The radiation-hydrodynamic simulations conducted by \citet{Schonberner2010} demonstrated that the HWHM velocities of volume-integrated line profiles always underestimate the true expansion velocity, so the HWHM method is suitable for slowly expanding objects, but cannot provide true expansion velocities of extended PNe. For the Magellanic Cloud PNe, \citet{Dopita1985,Dopita1988} assumed that $v_{\rm exp} = 1.82 \, V_{\rm HWHM}$, nearly twice the widespread use of the HWHM velocity. This is associated with the maximum gas velocity behind the outer shock, and it may not be related to the mean expansion velocity of the nebular shell. Here, we assume $v_{\rm exp} = V_{\rm HWHM}$, which may correspond to the mean expansion of a spherical gaseous shell. However, the velocity-resolved channel maps in Section~\ref{wc1:sec:morpho-kinematic} imply that the HWHM velocities of integrated emission-line profiles do not correctly describe the mean expansion velocity of the nebular shell  due to  aspherical morphologies and collimated bipolar outflows extending upward from the main shell.

\subsection{Spatially-resolved Kinematic Maps}
\label{wc1:sec:kinematic:maps}

Figure~\ref{wc1:ifu_map} presents spatially resolved maps of the flux intensity, continuum, radial velocity, and velocity dispersion of the H$\alpha$ and [N\,{\sc ii}] line emission derived from the WiFeS observations of all the PNe in our sample
(the figures for all the objects can be found in the online version of the journal). 
This information is obtained by fitting Gaussian functions to spaxels of the IFU datacube using the IDL library \textsc{mpfit} \citep{Markwardt2009} developed for the nonlinear least-squares minimization fitting. The emission line profile is resolved if its emission line width is wider than the instrumental width ($\sigma_{\rm ins}$, see \S\,\ref{wc1:sec:kinematic:expansion}).
The contour lines in each figure depict the boundaries at $\sim 10$ percent of the mean surface brightness of each object in the H$\alpha$ emission from the SuperCOSMOS H$\alpha$ Sky Survey \citep[SHS;][]{Parker2005}, or in the $R$-band from the SuperCOSMOS Sky Surveys  \citep[SSS;][]{Hambly2001} for M\,1-32, Hen\,2-142, and IC\,1297. This may help us identify the nebular boundaries. 

\setcounter{figure}{0}
\begin{figure*}
\begin{center}
{\footnotesize (c) Hb\,4 H$\alpha$ $\lambda$6563}\\
\includegraphics[width=6.8in]{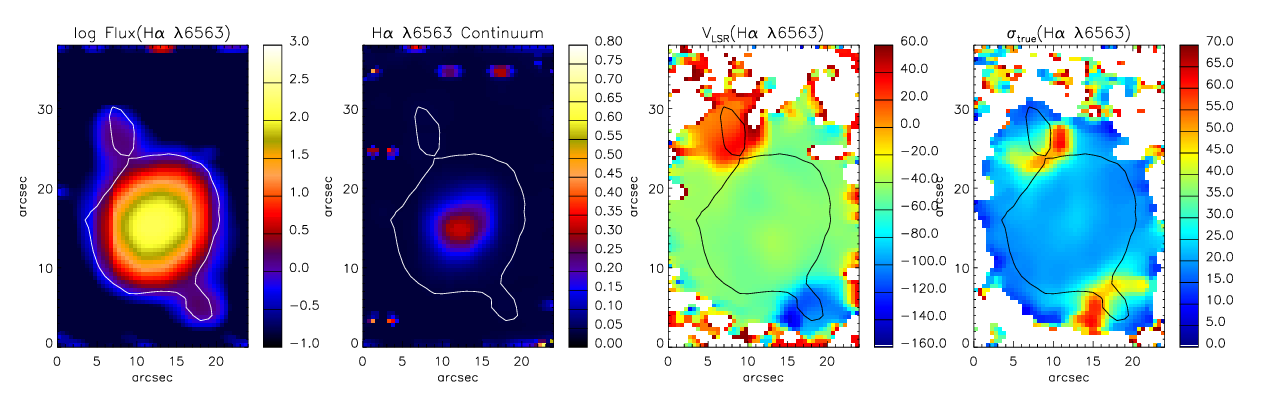}\\
{\footnotesize (c) Hb\,4 [N\,{\sc ii}] $\lambda$6584}\\
\includegraphics[width=6.8in]{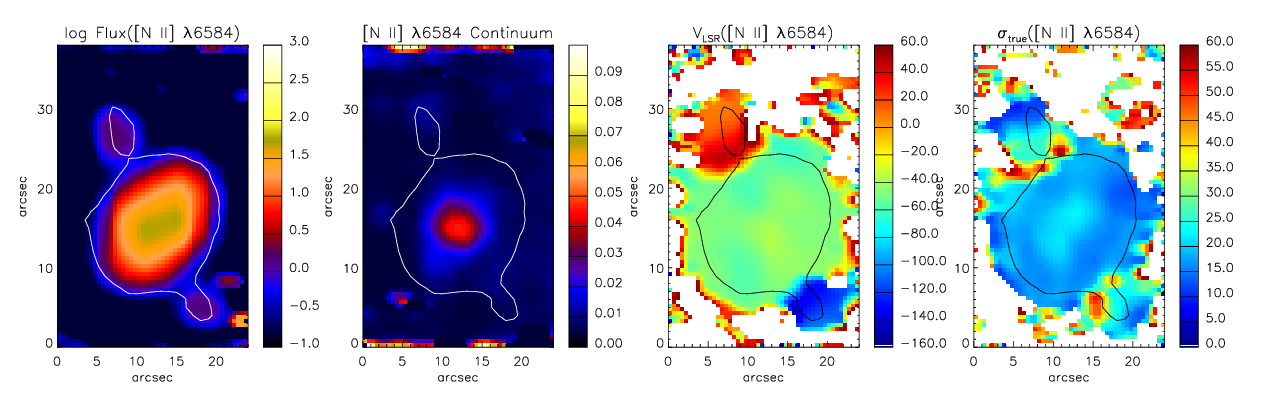}\\
\caption{From left to right, the spatial distribution maps of logarithmic flux intensity, continuum, LSR velocity, and velocity dispersion of the H$\alpha$ $\lambda$6563 and [N\,{\sc ii}] $\lambda$6584 emission lines for: (a) PB\,6, (b) M\,3-30, (c) Hb\,4, (d) IC\,1297, (e) Pe\,1-1, (f) M\,1-32, (g) M\,3-15, (h) M\,1-25, (i) Hen\,2-142, (j) K\,2-16, (k) NGC\,6578, (l) NGC\,6567, (m) NGC\,6629, and (n) Sa\,3-107. Fluxes are in logarithm of $10^{-15}$~erg\,s${}^{-1}$\,cm${}^{-2}$\,spaxel${}^{-1}$ unit, continua in $10^{-15}$~erg\,s${}^{-1}$\,cm${}^{-2}$\,{\AA}$^{-1}$\,spaxel${}^{-1}$, and LSR velocities and velocity dispersion in km\,s${}^{-1}$. The white/black contour in each panel corresponds to $\sim 10$ percent of the mean surface brightness of each object in the H$\alpha$ emission (or $R$-band) retrieved from the SHS (or SSS). North is up and east is toward the left-hand side. The complete figure set (28 images) is available in the online journal.
}
\label{wc1:ifu_map}%
\end{center}

\figsetstart
\figsetnum{1}
\figsettitle{From left to right, the spatial distribution maps of logarithmic flux intensity, continuum, LSR velocity, and velocity dispersion of the H$\alpha$ $\lambda$6563 and [N\,{\sc ii}] $\lambda$6584 emission lines.}

\figsetgrpstart
\figsetgrpnum{1.1}
\figsetgrptitle{(a) PB\,6 H$\alpha$ $\lambda$6563}
\figsetplot{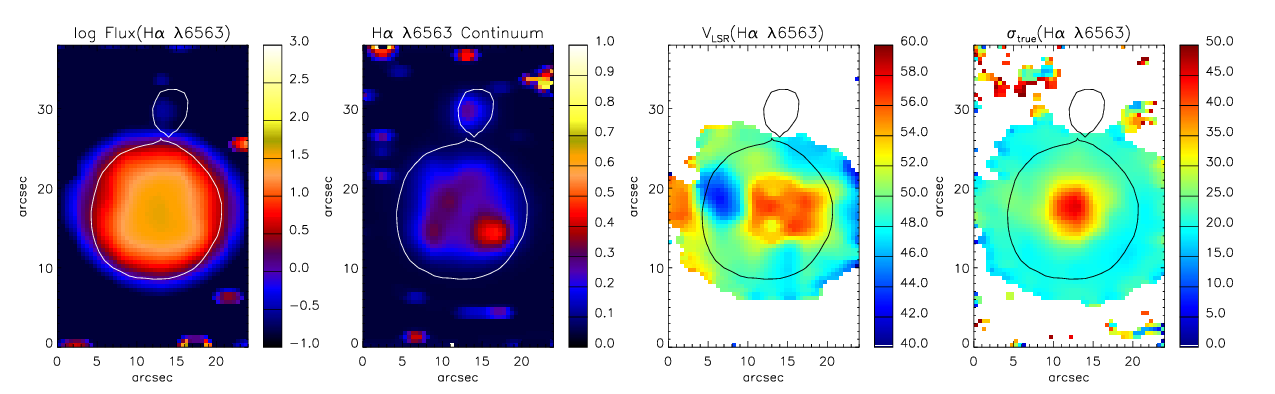}%
\figsetgrpnote{From left to right, the spatial distribution maps of logarithmic 
flux intensity, continuum, LSR velocity, and velocity dispersion of 
the H$\alpha$ $\lambda$6563 emission line for PB\,6.
}
\figsetgrpend

\figsetgrpstart
\figsetgrpnum{1.2}
\figsetgrptitle{(a) PB\,6 [N\,{\sc ii}] $\lambda$6584}
\figsetplot{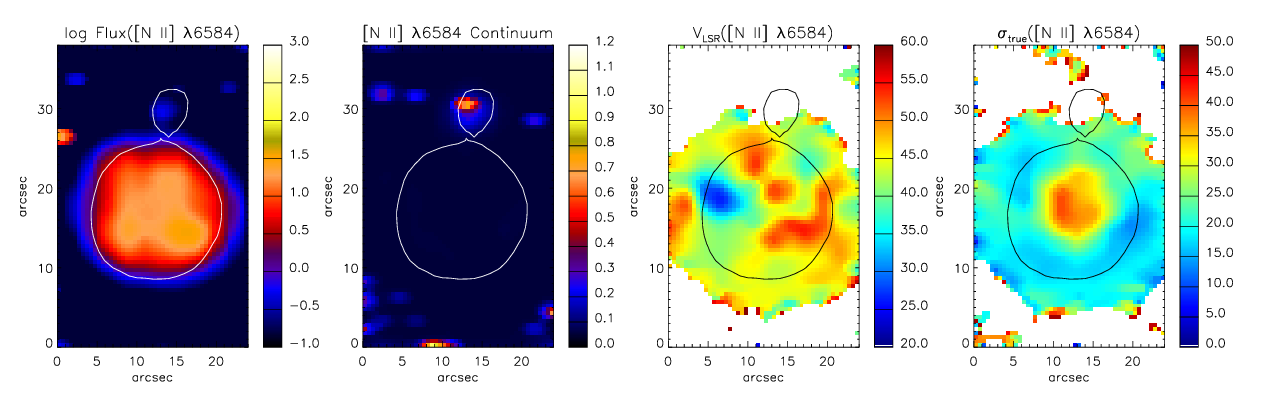}%
\figsetgrpnote{From left to right, the spatial distribution maps of logarithmic 
flux intensity, continuum, LSR velocity, and velocity dispersion of 
the [N\,{\sc ii}] $\lambda$6584 emission line for PB\,6.
}
\figsetgrpend

\figsetgrpstart
\figsetgrpnum{1.3}
\figsetgrptitle{(b) M\,3-30 H$\alpha$ $\lambda$6563}
\figsetplot{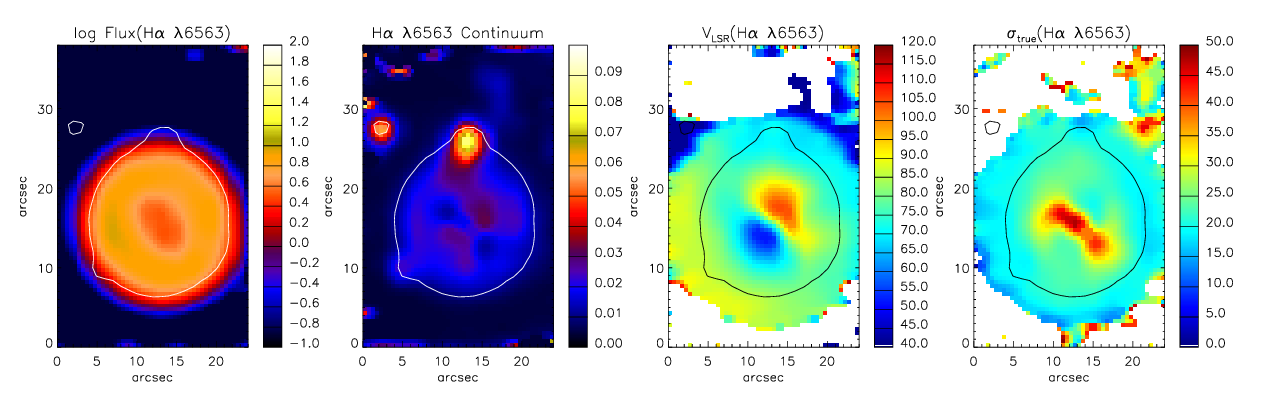}%
\figsetgrpnote{From left to right, the spatial distribution maps of logarithmic 
flux intensity, continuum, LSR velocity, and velocity dispersion of 
the H$\alpha$ $\lambda$6563 emission line for M\,3-30.
}
\figsetgrpend

\figsetgrpstart
\figsetgrpnum{1.4}
\figsetgrptitle{(b) M\,3-30 [N\,{\sc ii}] $\lambda$6584}\
\figsetplot{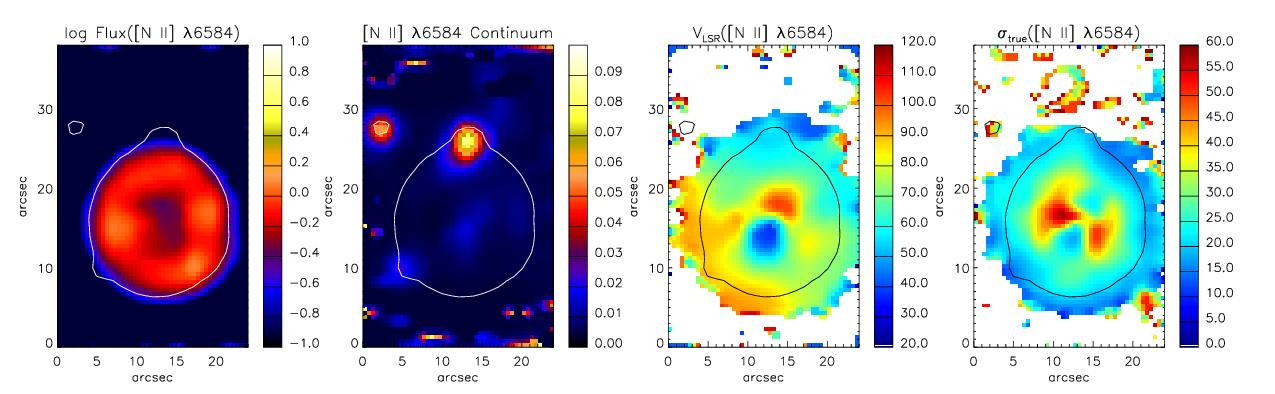}%
\figsetgrpnote{From left to right, the spatial distribution maps of logarithmic 
flux intensity, continuum, LSR velocity, and velocity dispersion of 
the [N\,{\sc ii}] $\lambda$6584 emission line for M\,3-30.
}
\figsetgrpend

\figsetgrpstart
\figsetgrpnum{1.5}
\figsetgrptitle{(c) Hb\,4 H$\alpha$ $\lambda$6563} 
\figsetplot{figure1/fig1_hb4_6563.eps}%
\figsetgrpnote{From left to right, the spatial distribution maps of logarithmic 
flux intensity, continuum, LSR velocity, and velocity dispersion of 
the H$\alpha$ $\lambda$6563 emission line for Hb\,4.
}
\figsetgrpend

\figsetgrpstart
\figsetgrpnum{1.6}
\figsetgrptitle{(c) Hb\,4 [N\,{\sc ii}] $\lambda$6584} 
\figsetplot{figure1/fig1_hb4_6584.eps}%
\figsetgrpnote{From left to right, the spatial distribution maps of logarithmic 
flux intensity, continuum, LSR velocity, and velocity dispersion of 
the [N\,{\sc ii}] $\lambda$6584 emission line for Hb\,4.
}
\figsetgrpend

\figsetgrpstart
\figsetgrpnum{1.7}
\figsetgrptitle{(d) IC\,1297 H$\alpha$ $\lambda$6563}
\figsetplot{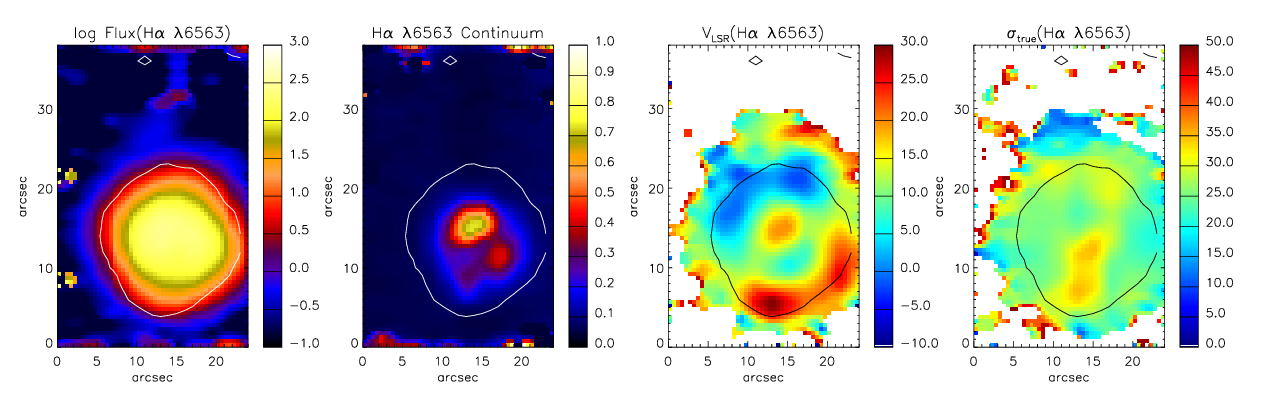}%
\figsetgrpnote{From left to right, the spatial distribution maps of logarithmic 
flux intensity, continuum, LSR velocity, and velocity dispersion of 
the H$\alpha$ $\lambda$6563 emission line for IC\,1297.
}
\figsetgrpend

\figsetgrpstart
\figsetgrpnum{1.8}
\figsetgrptitle{(d) IC\,1297 [N\,{\sc ii}] $\lambda$6584}
\figsetplot{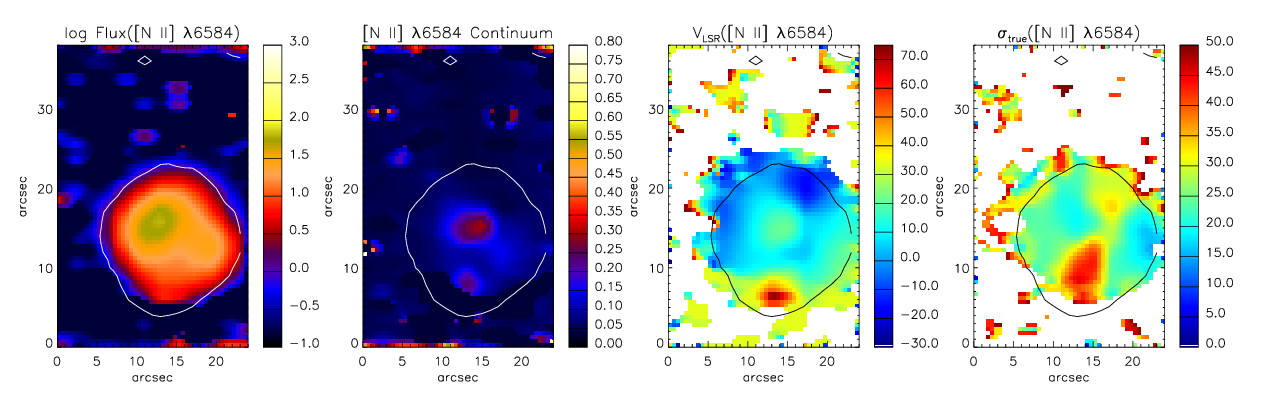}%
\figsetgrpnote{From left to right, the spatial distribution maps of logarithmic 
flux intensity, continuum, LSR velocity, and velocity dispersion of 
the [N\,{\sc ii}] $\lambda$6584 emission line for IC\,1297.
}
\figsetgrpend

\figsetgrpstart
\figsetgrpnum{1.9}
\figsetgrptitle{(e) Pe\,1-1 H$\alpha$ $\lambda$6563}
\figsetplot{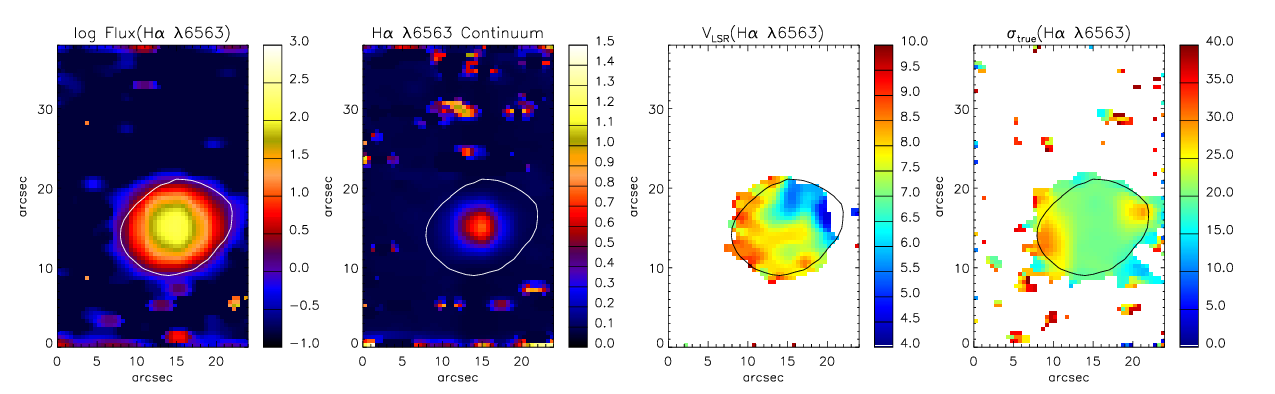}%
\figsetgrpnote{From left to right, the spatial distribution maps of logarithmic 
flux intensity, continuum, LSR velocity, and velocity dispersion of 
the H$\alpha$ $\lambda$6563 emission line for Pe\,1-1.
}
\figsetgrpend

\figsetgrpstart
\figsetgrpnum{1.10}
\figsetgrptitle{(e) Pe\,1-1 [N\,{\sc ii}] $\lambda$6584}
\figsetplot{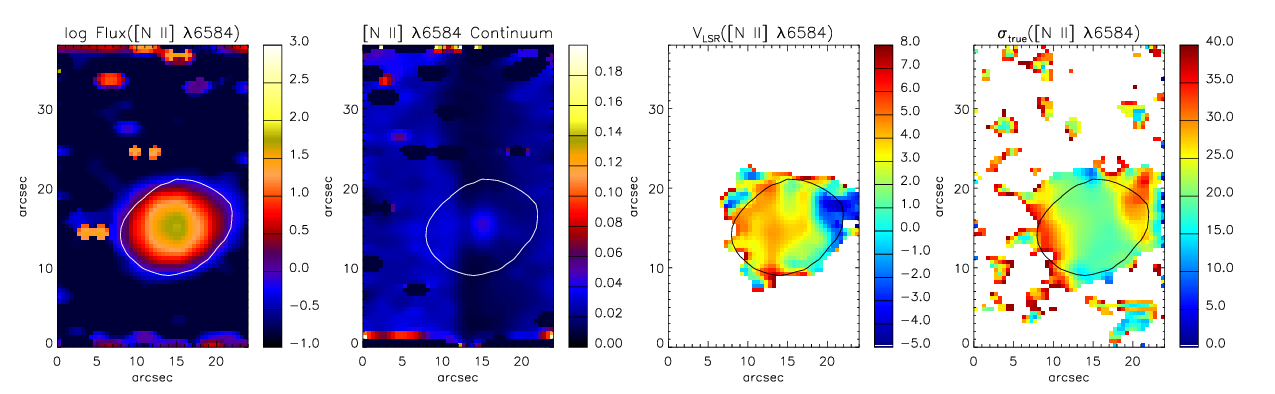}%
\figsetgrpnote{From left to right, the spatial distribution maps of logarithmic 
flux intensity, continuum, LSR velocity, and velocity dispersion of 
the [N\,{\sc ii}] $\lambda$6584 emission line for Pe\,1-1.
}
\figsetgrpend

\figsetgrpstart
\figsetgrpnum{1.11}
\figsetgrptitle{(f) M\,1-32 H$\alpha$ $\lambda$6563} 
\figsetplot{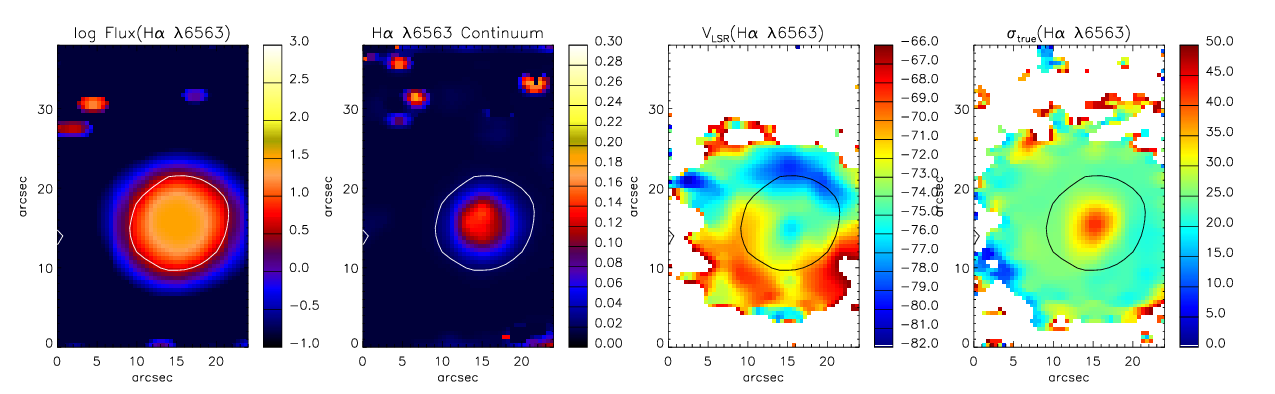}%
\figsetgrpnote{From left to right, the spatial distribution maps of logarithmic 
flux intensity, continuum, LSR velocity, and velocity dispersion of 
the H$\alpha$ $\lambda$6563 emission line for M\,1-32.
}
\figsetgrpend

\figsetgrpstart
\figsetgrpnum{1.12}
\figsetgrptitle{(f) M\,1-32 [N\,{\sc ii}] $\lambda$6584} 
\figsetplot{figure1/fig1_pe1_1_6584.eps}%
\figsetgrpnote{From left to right, the spatial distribution maps of logarithmic 
flux intensity, continuum, LSR velocity, and velocity dispersion of 
the [N\,{\sc ii}] $\lambda$6584 emission line for M\,1-32.
}
\figsetgrpend

\figsetgrpstart
\figsetgrpnum{1.13}
\figsetgrptitle{(g) M\,3-15 H$\alpha$ $\lambda$6563}
\figsetplot{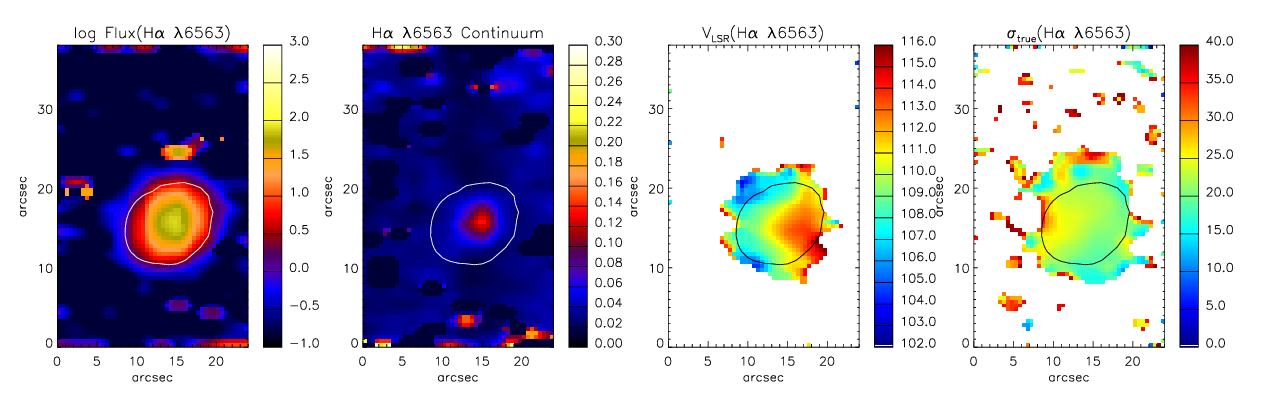}%
\figsetgrpnote{From left to right, the spatial distribution maps of logarithmic 
flux intensity, continuum, LSR velocity, and velocity dispersion of 
the H$\alpha$ $\lambda$6563 emission line for M\,3-15.
}
\figsetgrpend

\figsetgrpstart
\figsetgrpnum{1.14}
\figsetgrptitle{(g) M\,3-15 [N\,{\sc ii}] $\lambda$6584}
\figsetplot{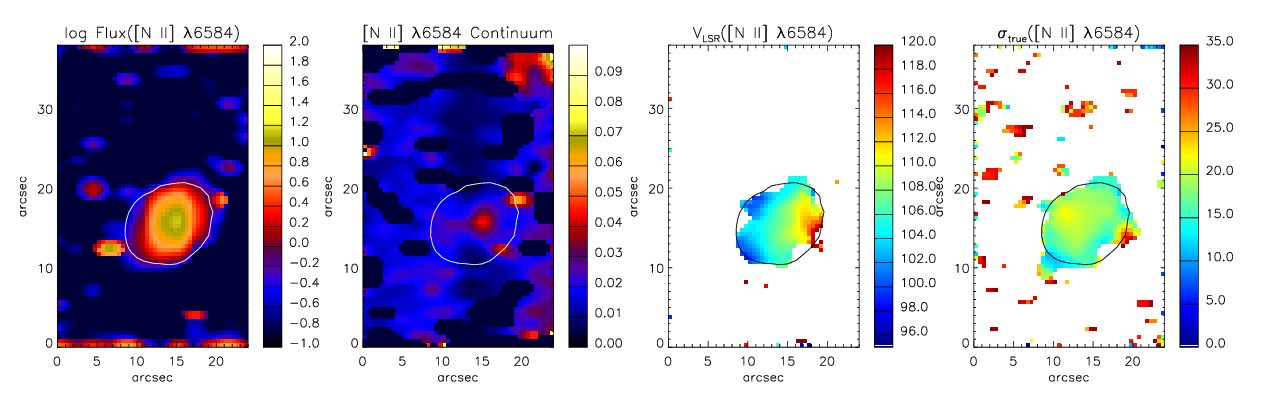}%
\figsetgrpnote{From left to right, the spatial distribution maps of logarithmic 
flux intensity, continuum, LSR velocity, and velocity dispersion of 
the [N\,{\sc ii}] $\lambda$6584 emission line for M\,3-15.
}
\figsetgrpend

\figsetgrpstart
\figsetgrpnum{1.15}
\figsetgrptitle{(h) M\,1-25 H$\alpha$ $\lambda$6563} 
\figsetplot{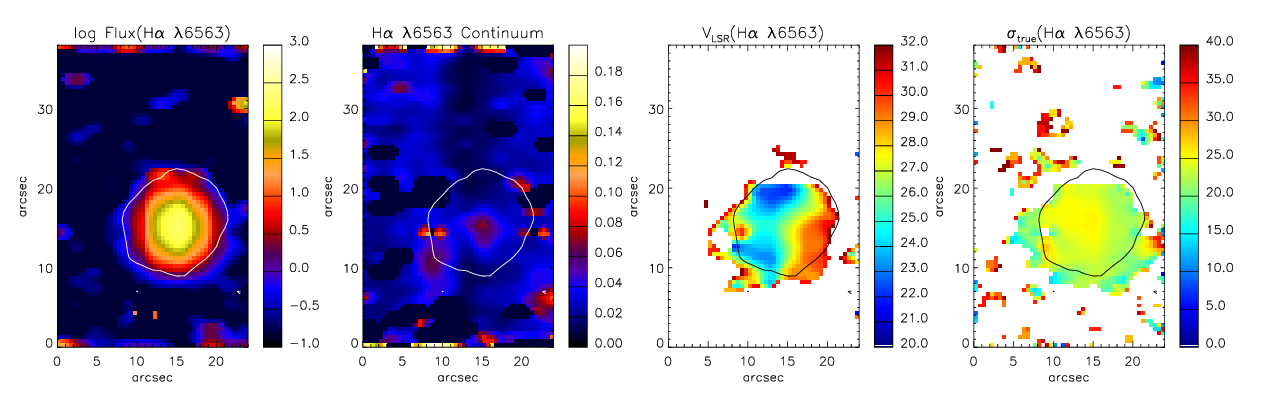}%
\figsetgrpnote{From left to right, the spatial distribution maps of logarithmic 
flux intensity, continuum, LSR velocity, and velocity dispersion of 
the H$\alpha$ $\lambda$6563 emission line for M\,1-25.
}
\figsetgrpend

\figsetgrpstart
\figsetgrpnum{1.16}
\figsetgrptitle{(h) M\,1-25 [N\,{\sc ii}] $\lambda$6584} 
\figsetplot{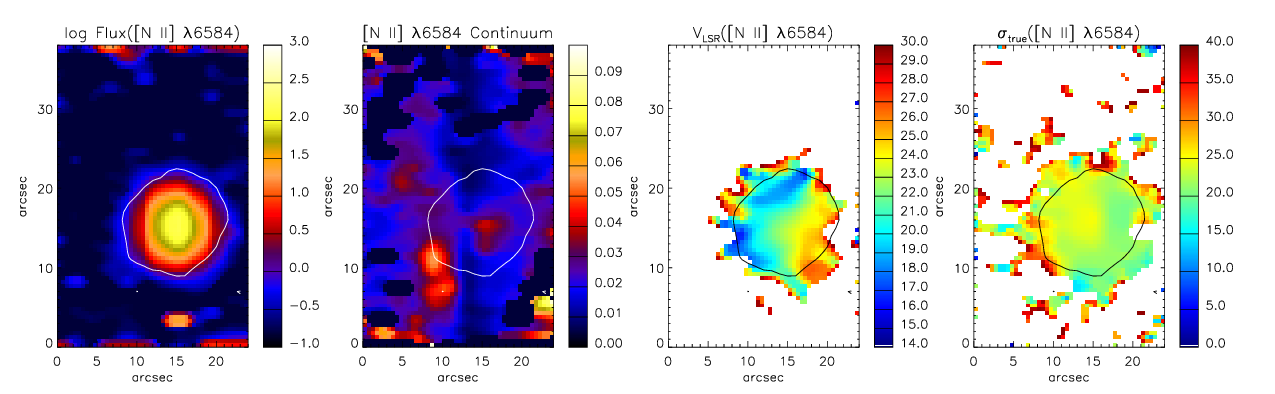}%
\figsetgrpnote{From left to right, the spatial distribution maps of logarithmic 
flux intensity, continuum, LSR velocity, and velocity dispersion of 
the [N\,{\sc ii}] $\lambda$6584 emission line for M\,1-25.
}
\figsetgrpend

\figsetgrpstart
\figsetgrpnum{1.17}
\figsetgrptitle{(i) Hen\,2-142 H$\alpha$ $\lambda$6563}
\figsetplot{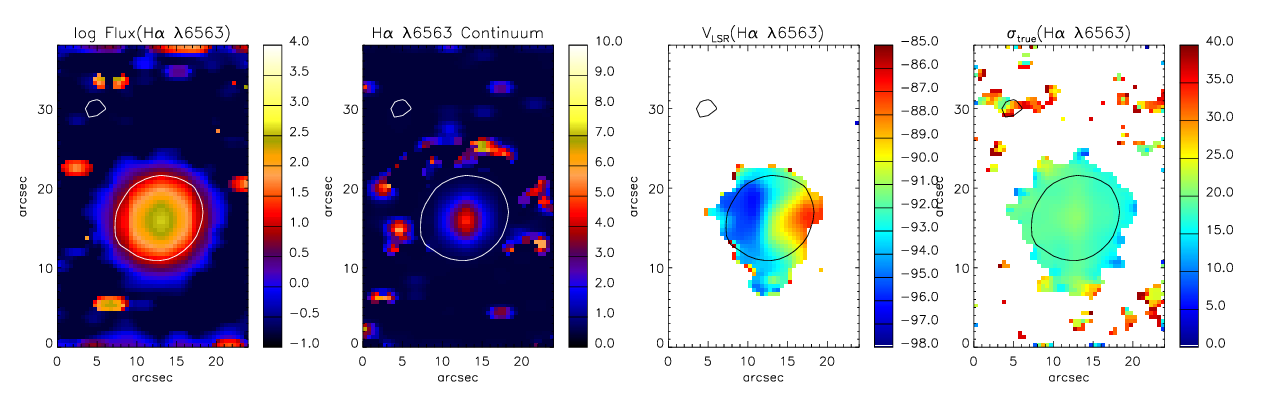}%
\figsetgrpnote{From left to right, the spatial distribution maps of logarithmic 
flux intensity, continuum, LSR velocity, and velocity dispersion of 
the H$\alpha$ $\lambda$6563 emission lines for Hen\,2-142.
}
\figsetgrpend

\figsetgrpstart
\figsetgrpnum{1.18}
\figsetgrptitle{(i) Hen\,2-142 [N\,{\sc ii}] $\lambda$6584}
\figsetplot{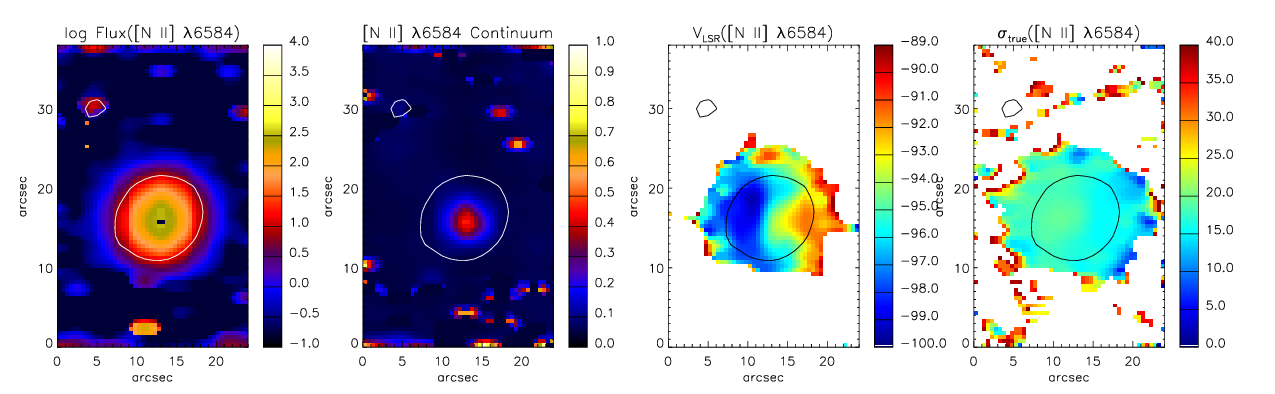}%
\figsetgrpnote{From left to right, the spatial distribution maps of logarithmic 
flux intensity, continuum, LSR velocity, and velocity dispersion of 
the [N\,{\sc ii}] $\lambda$6584 emission lines for Hen\,2-142.
}
\figsetgrpend

\figsetgrpstart
\figsetgrpnum{1.19}
\figsetgrptitle{(j) K\,2-16 H$\alpha$ $\lambda$6563} 
\figsetplot{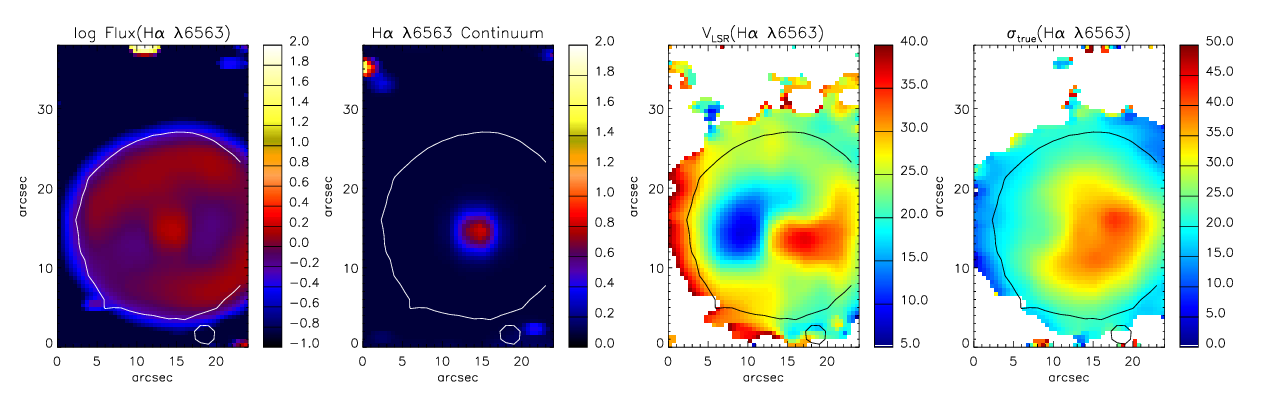}%
\figsetgrpnote{From left to right, the spatial distribution maps of logarithmic 
flux intensity, continuum, LSR velocity, and velocity dispersion of 
the H$\alpha$ $\lambda$6563 emission line for K\,2-16.
}
\figsetgrpend

\figsetgrpstart
\figsetgrpnum{1.20}
\figsetgrptitle{(j) K\,2-16 [N\,{\sc ii}] $\lambda$6584} 
\figsetplot{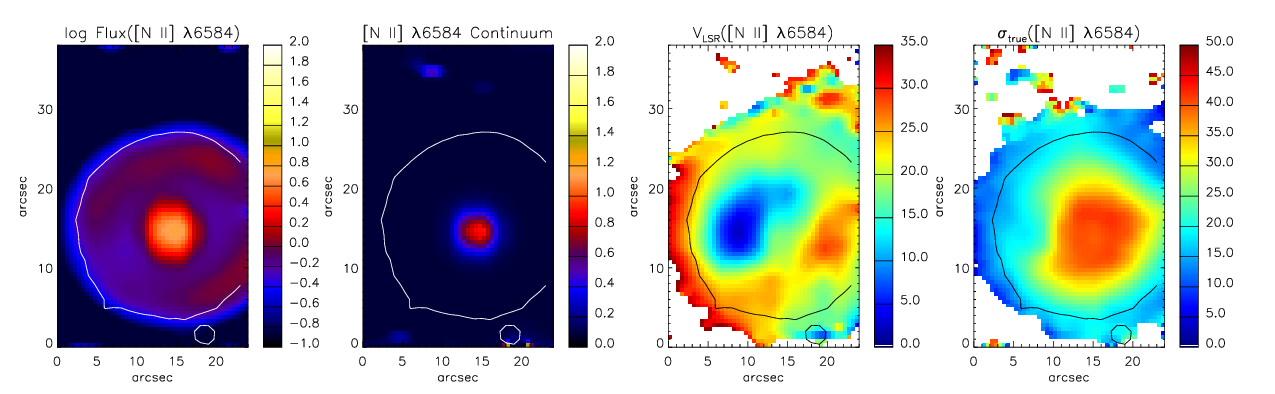}%
\figsetgrpnote{From left to right, the spatial distribution maps of logarithmic 
flux intensity, continuum, LSR velocity, and velocity dispersion of 
the [N\,{\sc ii}] $\lambda$6584 emission line for K\,2-16.
}
\figsetgrpend

\figsetgrpstart
\figsetgrpnum{1.21}
\figsetgrptitle{(k) NGC\,6578 H$\alpha$ $\lambda$6563}
\figsetplot{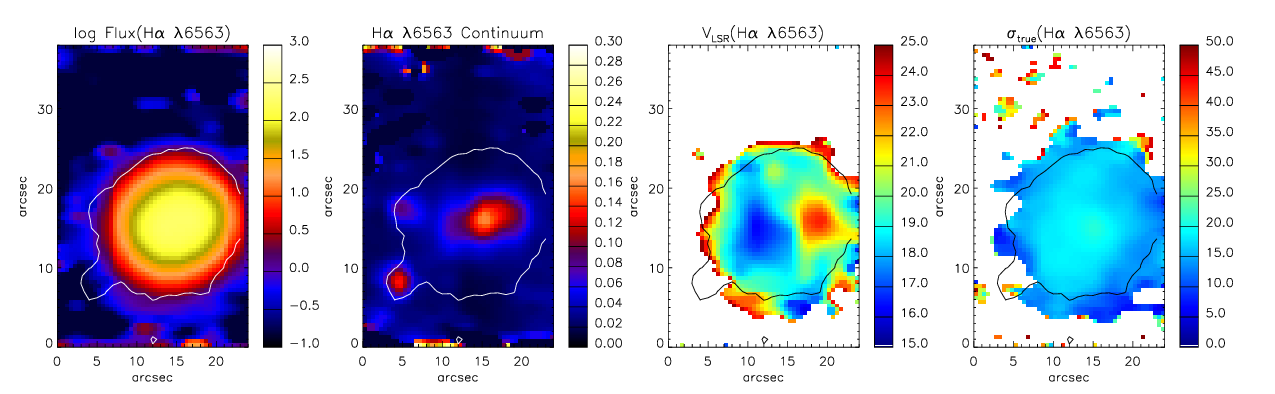}%
\figsetgrpnote{From left to right, the spatial distribution maps of logarithmic 
flux intensity, continuum, LSR velocity, and velocity dispersion of 
the H$\alpha$ $\lambda$6563 emission line for NGC\,6578.
}
\figsetgrpend

\figsetgrpstart
\figsetgrpnum{1.22}
\figsetgrptitle{(k) NGC\,6578 [N\,{\sc ii}] $\lambda$6584}
\figsetplot{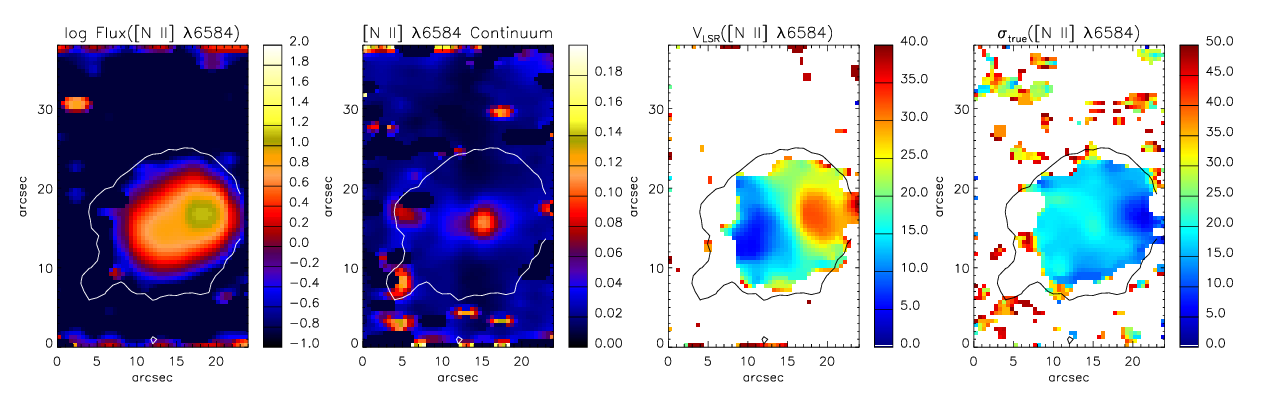}%
\figsetgrpnote{From left to right, the spatial distribution maps of logarithmic 
flux intensity, continuum, LSR velocity, and velocity dispersion of 
the [N\,{\sc ii}] $\lambda$6584 emission line for NGC\,6578.
}
\figsetgrpend

\figsetgrpstart
\figsetgrpnum{1.23}
\figsetgrptitle{(l) NGC\,6567 H$\alpha$ $\lambda$6563}
\figsetplot{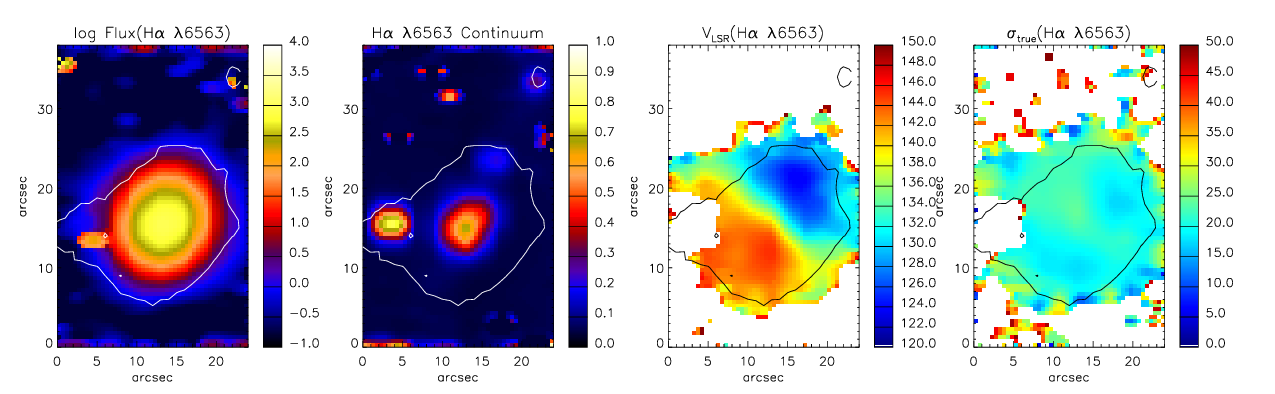}%
\figsetgrpnote{From left to right, the spatial distribution maps of logarithmic 
flux intensity, continuum, LSR velocity, and velocity dispersion of 
the H$\alpha$ $\lambda$6563 emission line for NGC\,6567.
}
\figsetgrpend

\figsetgrpstart
\figsetgrpnum{1.24}
\figsetgrptitle{(l) NGC\,6567 [N\,{\sc ii}] $\lambda$6584}
\figsetplot{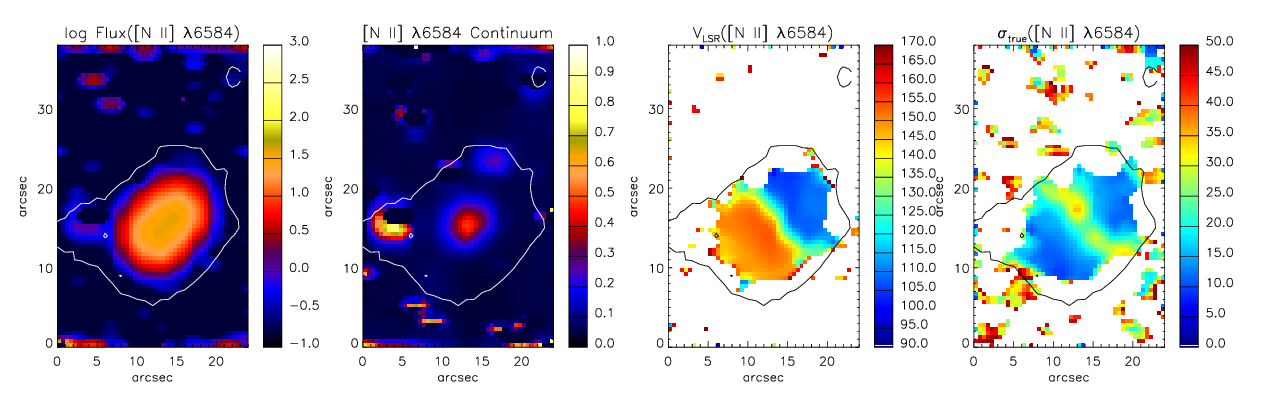}%
\figsetgrpnote{From left to right, the spatial distribution maps of logarithmic 
flux intensity, continuum, LSR velocity, and velocity dispersion of 
the [N\,{\sc ii}] $\lambda$6584 emission line for NGC\,6567.
}
\figsetgrpend

\figsetgrpstart
\figsetgrpnum{1.25}
\figsetgrptitle{(m) NGC\,6629 H$\alpha$ $\lambda$6563}
\figsetplot{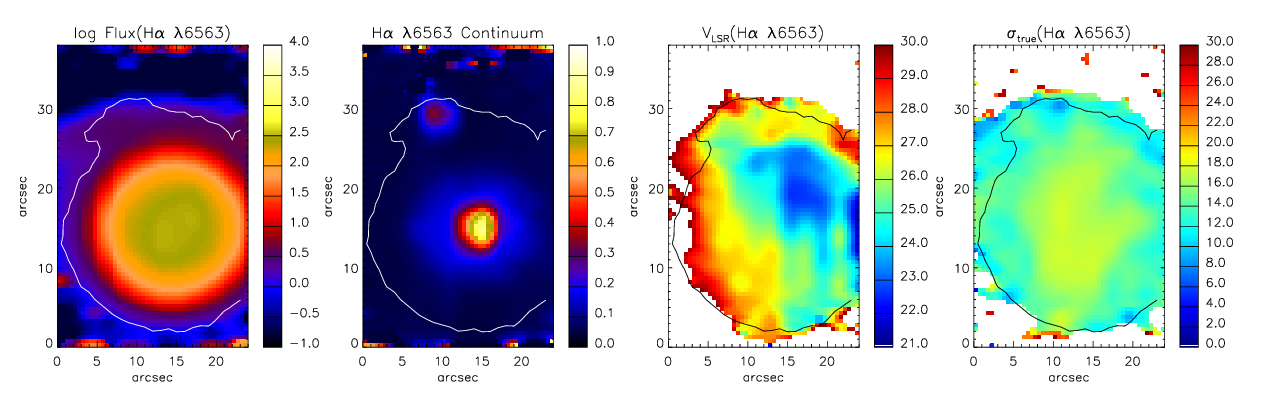}%
\figsetgrpnote{From left to right, the spatial distribution maps of logarithmic 
flux intensity, continuum, LSR velocity, and velocity dispersion of 
the H$\alpha$ $\lambda$6563 emission line for NGC\,6629.
}
\figsetgrpend

\figsetgrpstart
\figsetgrpnum{1.26}
\figsetgrptitle{(m) NGC\,6629 [N\,{\sc ii}] $\lambda$6584}
\figsetplot{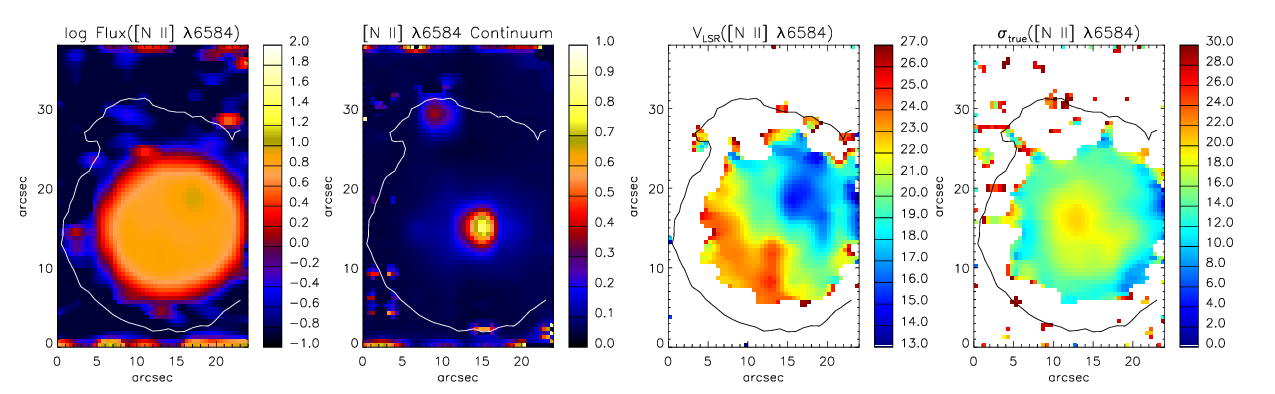}%
\figsetgrpnote{From left to right, the spatial distribution maps of logarithmic 
flux intensity, continuum, LSR velocity, and velocity dispersion of 
the [N\,{\sc ii}] $\lambda$6584 emission line for NGC\,6629.
}
\figsetgrpend

\figsetgrpstart
\figsetgrpnum{1.27}
\figsetgrptitle{(n) Sa\,3-107 H$\alpha$ $\lambda$6563}
\figsetplot{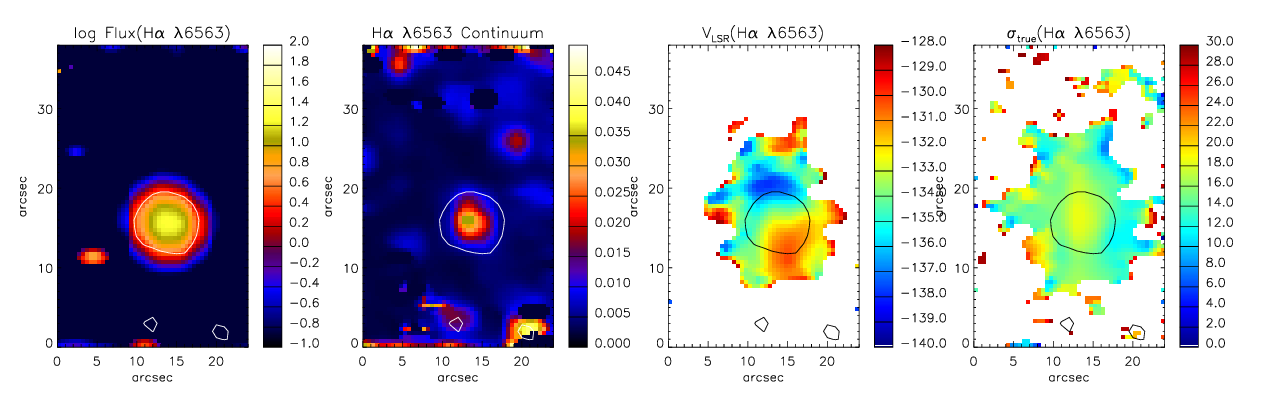}%
\figsetgrpnote{From left to right, the spatial distribution maps of logarithmic 
flux intensity, continuum, LSR velocity, and velocity dispersion of 
the H$\alpha$ $\lambda$6563 emission line for Sa\,3-107.
}
\figsetgrpend

\figsetgrpstart
\figsetgrpnum{1.28}
\figsetgrptitle{(n) Sa\,3-107 [N\,{\sc ii}] $\lambda$6584}
\figsetplot{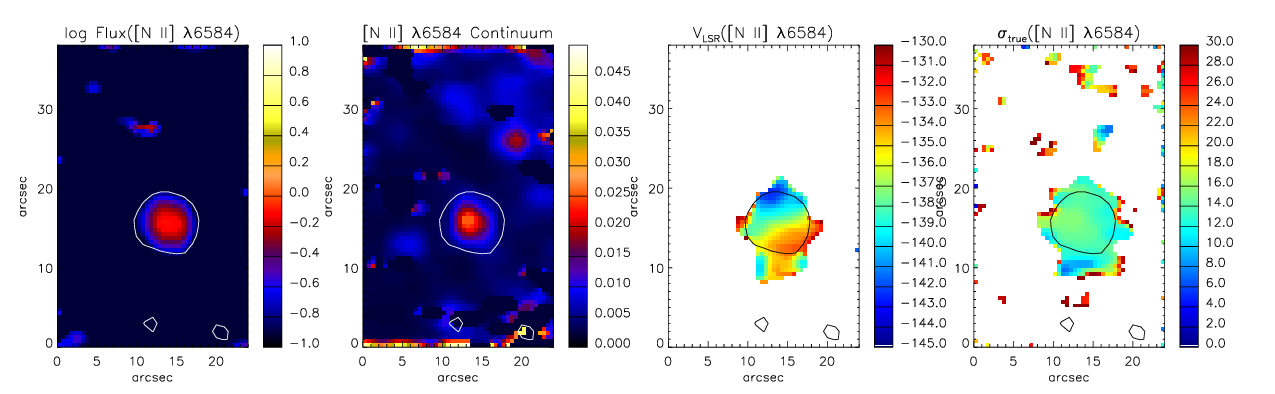}%
\figsetgrpnote{From left to right, the spatial distribution maps of logarithmic 
flux intensity, continuum, LSR velocity, and velocity dispersion of 
the [N\,{\sc ii}] $\lambda$6584 emission line for Sa\,3-107.
}
\figsetgrpend

\figsetend

\end{figure*}

Figure~\ref{wc1:vel:slic} shows the velocity-resolved channels of the H$\alpha$ and [N\,{\sc ii}] flux intensity, on a logarithmic scale, observed in a sequence of 12 or 18 velocity channels with a velocity resolution of $\sim 21$\,km\,s$^{-1}$ relative to the LSR systemic velocity (the corresponding velocity channel maps for all the objects are available in the online journal). The stellar continuum map derived for each line profile was subtracted from the associated flux intensity maps. 
Figure~\ref{wc1:pv:diagram} also presents position--velocity (P--V) diagrams of the H$\alpha$ and [N\,{\sc ii}] line emission of each object extracted from the IFU datacube for two slits passing through the central star that are oriented with the position angles (PA) along and vertical to the symmetric axis of the best-fitting morpho-kinematic model constructed in \S\,\ref{wc1:sec:morpho-kinematic} (the corresponding P--V diagrams for all the PNe are available in the online version of the journal). The velocity axes on the P--V arrays are relative to the systemic velocity in the LSR frame. The stellar continuum of each line profile was also  subtracted from the associated P--V diagrams. In \S\,\ref{wc1:sec:morpho-kinematic}, we use the velocity channels, along with the P--V arrays, to constrain our morpho-kinematic models.

Our IFU velocity maps of PB\,6 taken with a long exposure time ($1200$ sec) reveal that two groups of ionized gas are moving in opposite directions on both sides of the central star. 
The \textit{HST} image of PB\,6, taken with STIS/MIRVIS and a short exposure time of $22$ sec on April 26, 2012 (Program ID: 12600, PI: R. Dufour), also depicts a complex morphology with many multi-scale filamentary structures and several knots, which are analogous to the morphology of NGC\,5189, also around a [WO\,1] star \citep{Crowther1998}. This object could contain low-ionization envelopes similar to those identified in NGC\,5189 \citep{Danehkar2018a}. However, the short exposure time and the MIRVIS long-pass filter (4950--9450 {\AA}) of the \textit{HST} image were inadequate for a detailed morphological analysis of PB\,6. 

The kinematics maps of M\,3-30 likely suggest a pair of bipolar inner knots embedded in the elliptical shell. Previously, \citet{Stanghellini1993} also classified this object as an elliptical morphology with inner knots. However, our morpho-kinematic model (see \S\,\ref{wc1:sec:morpho-kinematic}) constrained by the velocity channel maps and P--V diagrams indicates that they are associated with the outer ends of tenuous collimated bipolar outflows extending from the dense toroidal shell. This toroidal morphology is also visible in the narrow-band H$\alpha$+[N\,{\sc ii}] image taken with the 3.5 m ESO NTT (see Figure\,\ref{wc1:model:shape}). 

The IFU kinematic maps of Hb\,4 illustrate a pair of point-symmetric knots or collimated jets detached from the main shell, with a torus morphology, based on the \textit{HST} imaging observations in Figure\,\ref{wc1:model:shape}.
These point-symmetric structures have extremely low brightness, but high velocity dispersion. 
The H$\alpha$+[N\,{\sc ii}]/[O~{\sc iii}] image collected by \citet{Corradi1996} also depicted the point-symmetric knots. 
Taking the inclination angle of $i=-40^{\circ}$ found by the kinematic model (in \S\,\ref{wc1:sec:morpho-kinematic}), the maximum values of the point-symmetric knots reach $160\pm10$\,km\,s$^{-1}$ with respect to 
the central star, which agrees with the long-slit kinematic analysis \citep{Lopez1997}. 

\begin{figure*}
\begin{center}
{\footnotesize (c) Hb\,4 H$\alpha$ $\lambda$6563}\\ 
\includegraphics[width=6.in]{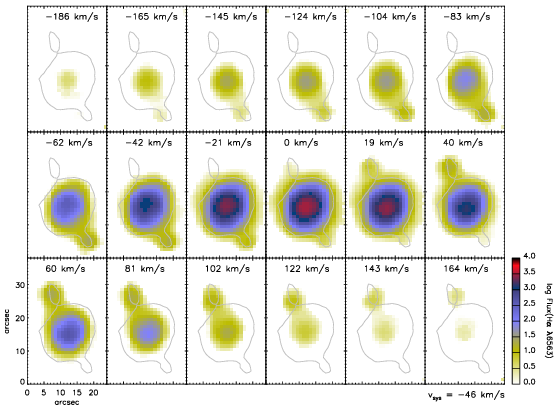}\\
{\footnotesize (c) Hb\,4 Model (H$\alpha$ $\lambda$6563)}\\ 
\includegraphics[width=5.8in]{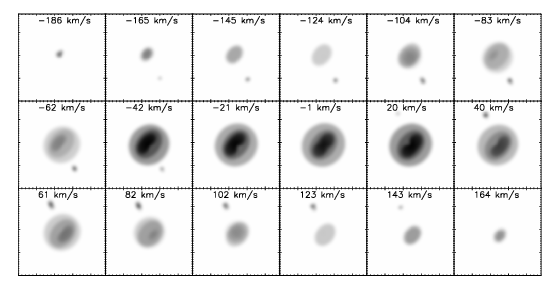}\\
\caption{Velocity slices along the H$\alpha$ $\lambda$6563 and [N\,{\sc ii}] $\lambda$6584 emission-line profiles for: (a) PB\,6, (b) M\,3-30, (c) Hb\,4, (d) IC\,1297, (e) Pe\,1-1, (f) M\,1-32, (g) M\,3-15, (h) M\,1-25, (i) Hen\,2-142, (j) K\,2-16, (k) NGC\,6578, (l) NGC\,6567, (m) NGC\,6629, and (n) Sa\,3-107, followed by the associated synthetic velocity-resolved channel maps obtained for all the PNe, except for PB\,6 and Sa\,3-107, produced by the best-fitting morpho-kinematic models with the parameters given in Table~\ref{wc1:tab:shapemodel}. Each observed slice has a $\sim 21$ km\,s${}^{-1}$ width, whose central velocity is given in km\,s${}^{-1}$ unit at the top of the panel. The LSR systemic velocity ($v_{\rm sys}$) of each object is given in km\,s${}^{-1}$ unit in the right bottom corner of each observed velocity channel map. The flux color  in each observed slice is in logarithm of $10^{-15}$~erg\,s${}^{-1}$\,cm${}^{-2}$\,spaxel${}^{-1}$ unit. The gray contour in each panel corresponds to $\sim 10$ percent of the mean surface brightness of each object in the H$\alpha$ emission (or $R$-band) retrieved from the SHS (or SSS). North is up and east is toward the left-hand side. The complete figure set (52 images) is available in the online journal.
}
\label{wc1:vel:slic}%
\end{center}

\figsetstart
\figsetnum{2}
\figsettitle{Velocity slices along the H$\alpha$ $\lambda$6563 and [N\,{\sc ii}] $\lambda$6584 emission-line profiles.}

\figsetgrpstart
\figsetgrpnum{2.1}
\figsetgrptitle{(a) PB\,6 H$\alpha$ $\lambda$6563}
\figsetplot{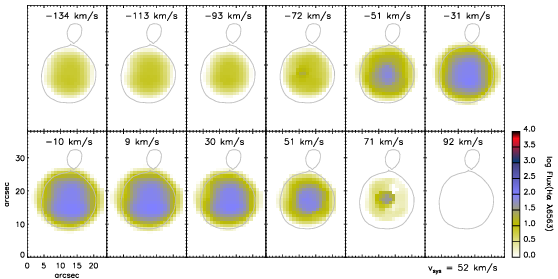}
\figsetgrpnote{Velocity slices along the H$\alpha$ $\lambda$6563 emission for PB\,6.}
\figsetgrpend

\figsetgrpstart
\figsetgrpnum{2.2}
\figsetgrptitle{(a) PB\,6 [N\,{\sc ii}] $\lambda$6584}
\figsetplot{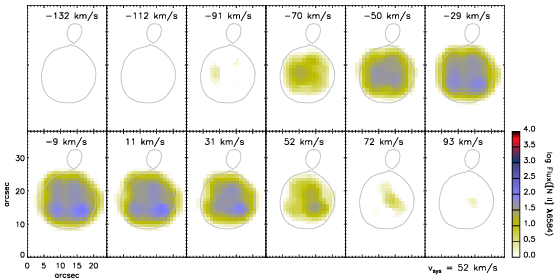}
\figsetgrpnote{Velocity slices along the [N\,{\sc ii}] $\lambda$6584 emission for PB\,6.}
\figsetgrpend

\figsetgrpstart
\figsetgrpnum{2.3}
\figsetgrptitle{(b) M\,3-30 H$\alpha$ $\lambda$6563}
\figsetplot{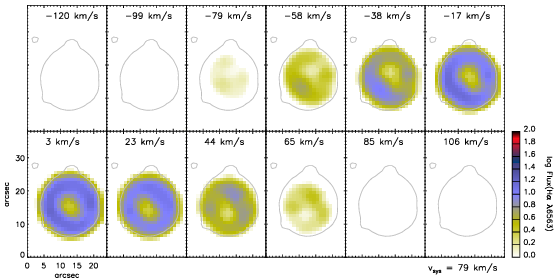}
\figsetgrpnote{Velocity slices along the H$\alpha$ $\lambda$6563 emission for M\,3-30.}
\figsetgrpend

\figsetgrpstart
\figsetgrpnum{2.4}
\figsetgrptitle{(b) M\,3-30 Model (H$\alpha$ $\lambda$6563)}
\figsetplot{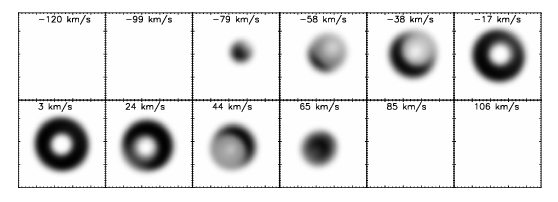}
\figsetgrpnote{Synthetic velocity-resolved channel maps from the morpho-kinematic model of M\,3-30 for the velocity slices along H$\alpha$ $\lambda$6563.}
\figsetgrpend

\figsetgrpstart
\figsetgrpnum{2.5}
\figsetgrptitle{(b) M\,3-30 [N\,{\sc ii}] $\lambda$6584}
\figsetplot{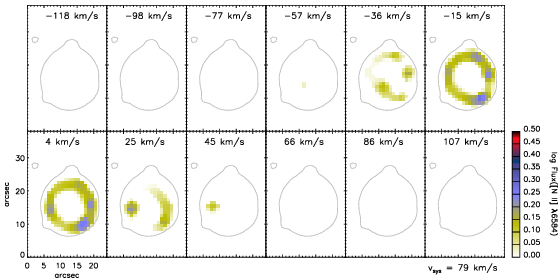}
\figsetgrpnote{Velocity slices along the [N\,{\sc ii}] $\lambda$6584 emission for M\,3-30.}
\figsetgrpend

\figsetgrpstart
\figsetgrpnum{2.6}
\figsetgrptitle{(b) M\,3-30 Model ([N\,{\sc ii}] $\lambda$6584)}
\figsetplot{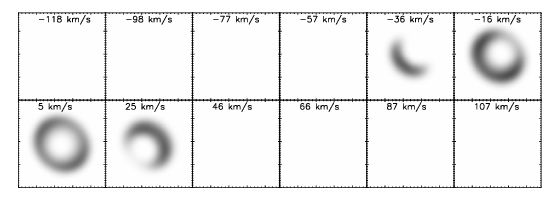}
\figsetgrpnote{Synthetic velocity-resolved channel maps from the morpho-kinematic model of M\,3-30 for the velocity slices along [N\,{\sc ii}] $\lambda$6584.}
\figsetgrpend

\figsetgrpstart
\figsetgrpnum{2.7}
\figsetgrptitle{(c) Hb\,4 H$\alpha$ $\lambda$6563}
\figsetplot{figure2/fig2_hb4_6563_vmap.eps}
\figsetgrpnote{Velocity slices along the H$\alpha$ $\lambda$6563 emission for Hb\,4.}
\figsetgrpend

\figsetgrpstart
\figsetgrpnum{2.8}
\figsetgrptitle{(c) Hb\,4 Model (H$\alpha$ $\lambda$6563)}
\figsetplot{figure2/fig2_hb4_shape_vmap_ha.eps}
\figsetgrpnote{Synthetic velocity-resolved channel maps from the morpho-kinematic model of Hb\,4 for the velocity slices along H$\alpha$ $\lambda$6563.}
\figsetgrpend

\figsetgrpstart
\figsetgrpnum{2.9}
\figsetgrptitle{(c) Hb\,4 [N\,{\sc ii}] $\lambda$6584}
\figsetplot{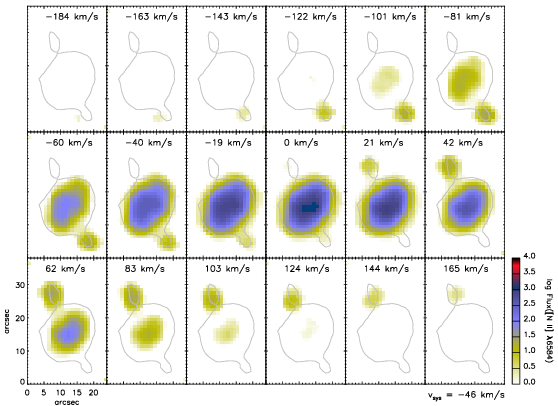}
\figsetgrpnote{Velocity slices along the [N\,{\sc ii}] $\lambda$6584 emission for Hb\,4.}
\figsetgrpend

\figsetgrpstart
\figsetgrpnum{2.10}
\figsetgrptitle{(c) Hb\,4 Model ([N\,{\sc ii}] $\lambda$6584)}
\figsetplot{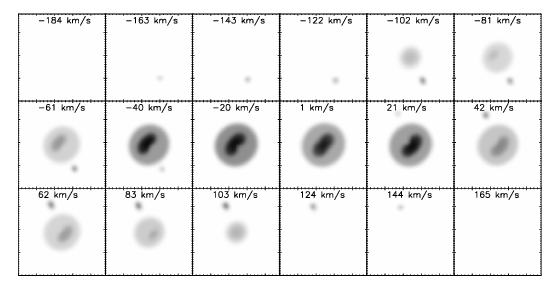}
\figsetgrpnote{Synthetic velocity-resolved channel maps from the morpho-kinematic model of Hb\,4 for the velocity slices along [N\,{\sc ii}] $\lambda$6584.}
\figsetgrpend

\figsetgrpstart
\figsetgrpnum{2.11}
\figsetgrptitle{(d) IC\,1297 H$\alpha$ $\lambda$6563}
\figsetplot{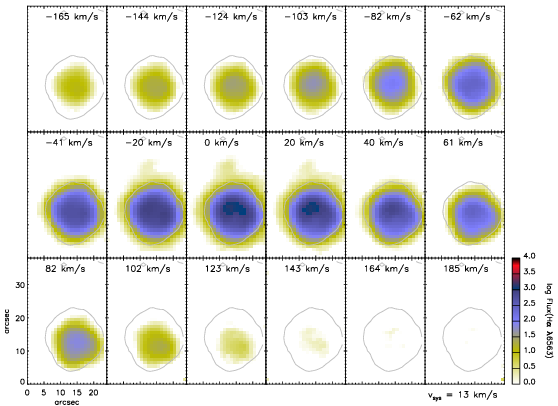}
\figsetgrpnote{Velocity slices along the H$\alpha$ $\lambda$6563 emission for IC\,1297.}
\figsetgrpend

\figsetgrpstart
\figsetgrpnum{2.12}
\figsetgrptitle{(d) IC\,1297 Model (H$\alpha$ $\lambda$6563)}
\figsetplot{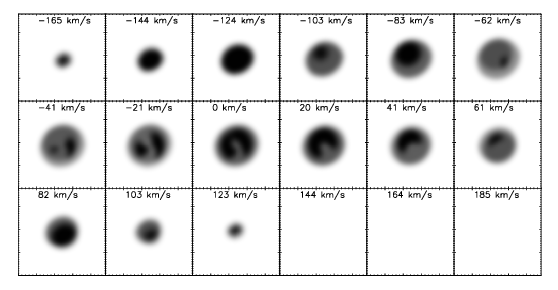}
\figsetgrpnote{Synthetic velocity-resolved channel maps from the morpho-kinematic model of IC\,1297 for the velocity slices along H$\alpha$ $\lambda$6563.}
\figsetgrpend

\figsetgrpstart
\figsetgrpnum{2.13}
\figsetgrptitle{(d) IC\,1297 [N\,{\sc ii}] $\lambda$6584}
\figsetplot{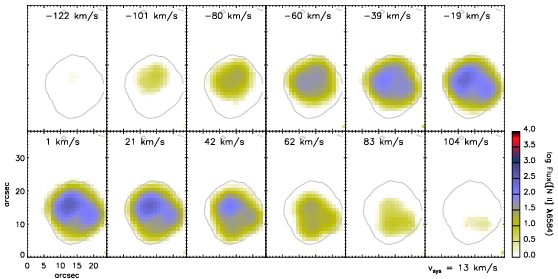}
\figsetgrpnote{Velocity slices along the [N\,{\sc ii}] $\lambda$6584 emission for IC\,1297.}
\figsetgrpend

\figsetgrpstart
\figsetgrpnum{2.14}
\figsetgrptitle{(d) IC\,1297 Model ([N\,{\sc ii}] $\lambda$6584)}
\figsetplot{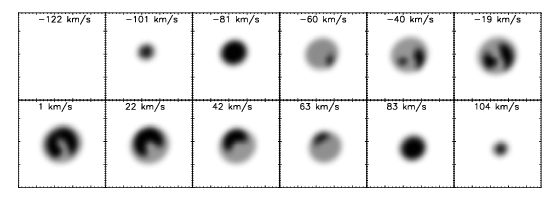}
\figsetgrpnote{Synthetic velocity-resolved channel maps from the morpho-kinematic model of IC\,1297 for the velocity slices along [N\,{\sc ii}] $\lambda$6584.}
\figsetgrpend

\figsetgrpstart
\figsetgrpnum{2.15}
\figsetgrptitle{(e) Pe\,1-1 H$\alpha$ $\lambda$6563}
\figsetplot{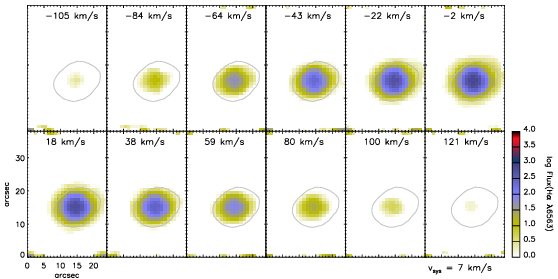}
\figsetgrpnote{Velocity slices along the H$\alpha$ $\lambda$6563 emission for Pe\,1-1.}
\figsetgrpend

\figsetgrpstart
\figsetgrpnum{2.16}
\figsetgrptitle{(e) Pe\,1-1 Model (H$\alpha$ $\lambda$6563)}
\figsetplot{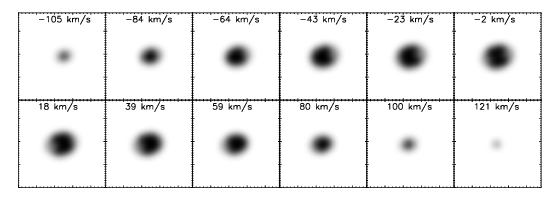}
\figsetgrpnote{Synthetic velocity-resolved channel maps from the morpho-kinematic model of Pe\,1-1 for the velocity slices along H$\alpha$ $\lambda$6563.}
\figsetgrpend

\figsetgrpstart
\figsetgrpnum{2.17}
\figsetgrptitle{(e) Pe\,1-1 [N\,{\sc ii}] $\lambda$6584}
\figsetplot{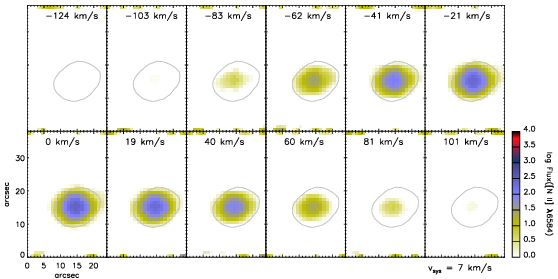}
\figsetgrpnote{Velocity slices along the [N\,{\sc ii}] $\lambda$6584 emission for Pe\,1-1.}
\figsetgrpend

\figsetgrpstart
\figsetgrpnum{2.18}
\figsetgrptitle{(e) Pe\,1-1 Model ([N\,{\sc ii}] $\lambda$6584)}
\figsetplot{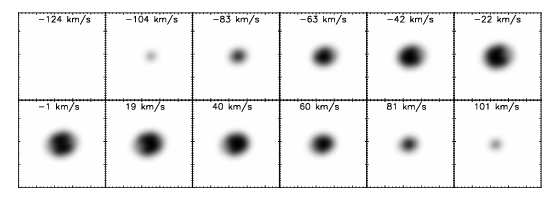}
\figsetgrpnote{Synthetic velocity-resolved channel maps from the morpho-kinematic model of Pe\,1-1 for the velocity slices along [N\,{\sc ii}] $\lambda$6584.}
\figsetgrpend

\figsetgrpstart
\figsetgrpnum{2.19}
\figsetgrptitle{(f) M\,1-32 H$\alpha$ $\lambda$6563}
\figsetplot{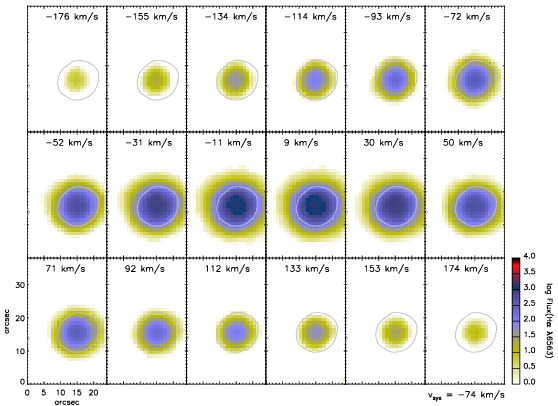}
\figsetgrpnote{Velocity slices along the H$\alpha$ $\lambda$6563 emission for M\,1-32.}
\figsetgrpend

\figsetgrpstart
\figsetgrpnum{2.20}
\figsetgrptitle{(f) M\,1-32 Model (H$\alpha$ $\lambda$6563)}
\figsetplot{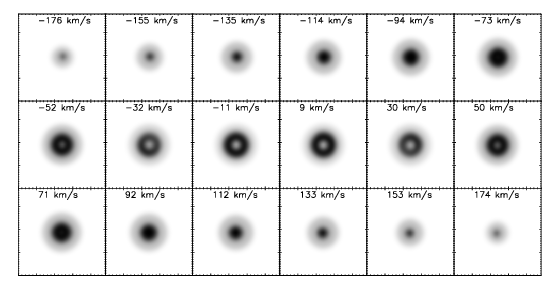}
\figsetgrpnote{Synthetic velocity-resolved channel maps from the morpho-kinematic model of M\,1-32 for the velocity slices along H$\alpha$ $\lambda$6563.}
\figsetgrpend

\figsetgrpstart
\figsetgrpnum{2.21}
\figsetgrptitle{(f) M\,1-32 [N\,{\sc ii}] $\lambda$6584}
\figsetplot{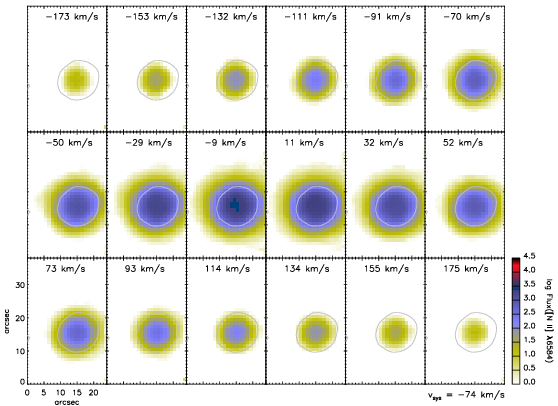}
\figsetgrpnote{Velocity slices along the [N\,{\sc ii}] $\lambda$6584 emission for M\,1-32.}
\figsetgrpend

\figsetgrpstart
\figsetgrpnum{2.22}
\figsetgrptitle{(f) M\,1-32 Model ([N\,{\sc ii}] $\lambda$6584)}
\figsetplot{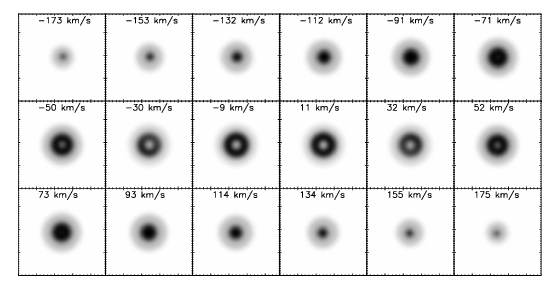}
\figsetgrpnote{Synthetic velocity-resolved channel maps from the morpho-kinematic model of M\,1-32 for the velocity slices along [N\,{\sc ii}] $\lambda$6584.}
\figsetgrpend

\figsetgrpstart
\figsetgrpnum{2.23}
\figsetgrptitle{(g) M\,3-15 H$\alpha$ $\lambda$6563}
\figsetplot{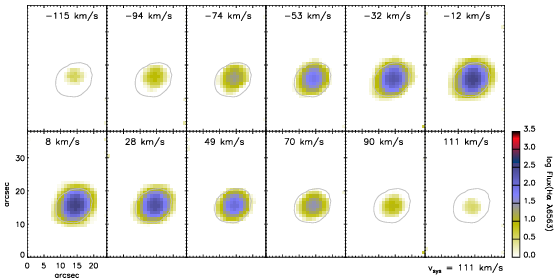}
\figsetgrpnote{Velocity slices along the H$\alpha$ $\lambda$6563 emission for M\,3-15.}
\figsetgrpend

\figsetgrpstart
\figsetgrpnum{2.24}
\figsetgrptitle{(g) M\,3-15 Model (H$\alpha$ $\lambda$6563)}
\figsetplot{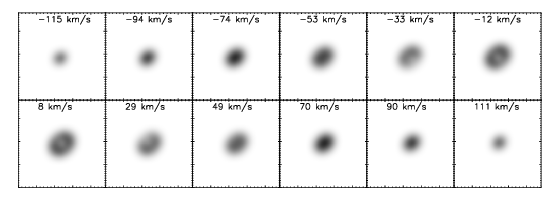}
\figsetgrpnote{Synthetic velocity-resolved channel maps from the morpho-kinematic model of M\,3-15 for the velocity slices along H$\alpha$ $\lambda$6563.}
\figsetgrpend

\figsetgrpstart
\figsetgrpnum{2.25}
\figsetgrptitle{(g) M\,3-15 [N\,{\sc ii}] $\lambda$6584}
\figsetplot{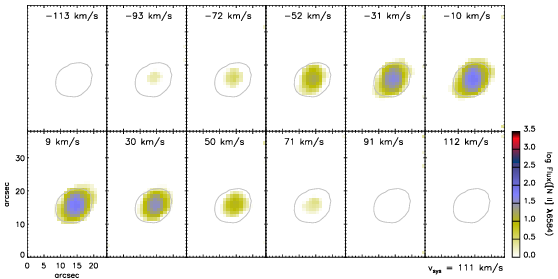}
\figsetgrpnote{Velocity slices along the [N\,{\sc ii}] $\lambda$6584 emission for M\,3-15.}
\figsetgrpend

\figsetgrpstart
\figsetgrpnum{2.26}
\figsetgrptitle{(g) M\,3-15 Model ([N\,{\sc ii}] $\lambda$6584)}
\figsetplot{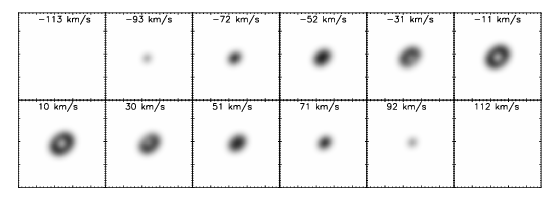}
\figsetgrpnote{Synthetic velocity-resolved channel maps from the morpho-kinematic model of M\,3-15 for the velocity slices along [N\,{\sc ii}] $\lambda$6584.}
\figsetgrpend

\figsetgrpstart
\figsetgrpnum{2.27}
\figsetgrptitle{(h) M\,1-25 H$\alpha$ $\lambda$6563} 
\figsetplot{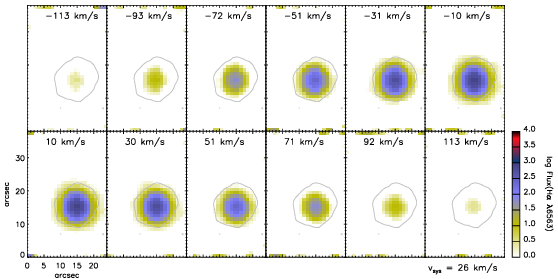}
\figsetgrpnote{Velocity slices along the H$\alpha$ $\lambda$6563 emission for M\,1-25.}
\figsetgrpend

\figsetgrpstart
\figsetgrpnum{2.28}
\figsetgrptitle{(h) M\,1-25 Model (H$\alpha$ $\lambda$6563)} 
\figsetplot{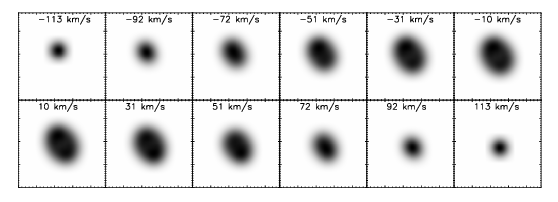}
\figsetgrpnote{Synthetic velocity-resolved channel maps from the morpho-kinematic model of M\,1-25 for the velocity slices along H$\alpha$ $\lambda$6563.}
\figsetgrpend

\figsetgrpstart
\figsetgrpnum{2.29}
\figsetgrptitle{(h) M\,1-25 [N\,{\sc ii}] $\lambda$6584}
\figsetplot{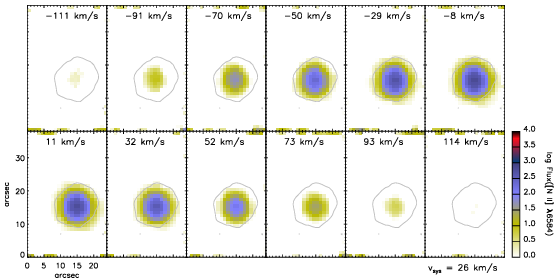}
\figsetgrpnote{Velocity slices along the [N\,{\sc ii}] $\lambda$6584 emission for M\,1-25.}
\figsetgrpend

\figsetgrpstart
\figsetgrpnum{2.30}
\figsetgrptitle{(h) M\,1-25 Model ([N\,{\sc ii}] $\lambda$6584)} 
\figsetplot{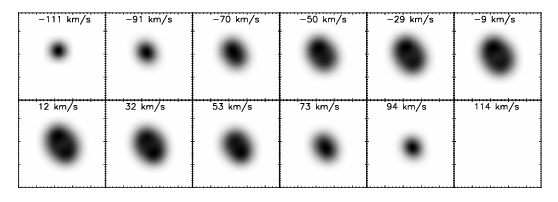}
\figsetgrpnote{Synthetic velocity-resolved channel maps from the morpho-kinematic model of M\,1-25 for the velocity slices along [N\,{\sc ii}] $\lambda$6584.}
\figsetgrpend

\figsetgrpstart
\figsetgrpnum{2.31}
\figsetgrptitle{(i) Hen\,2-142 H$\alpha$ $\lambda$6563}
\figsetplot{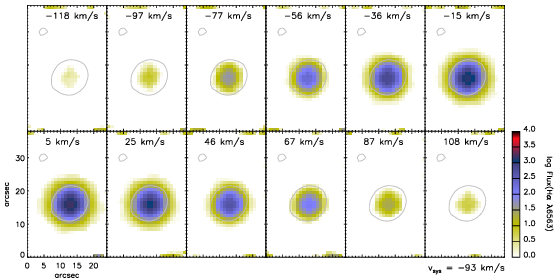}
\figsetgrpnote{Velocity slices along the H$\alpha$ $\lambda$6563 emission for Hen\,2-142.}
\figsetgrpend

\figsetgrpstart
\figsetgrpnum{2.32}
\figsetgrptitle{(i) Hen\,2-142 Model (H$\alpha$ $\lambda$6563)}
\figsetplot{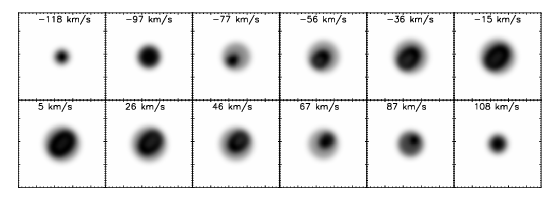}
\figsetgrpnote{Synthetic velocity-resolved channel maps from the morpho-kinematic model of Hen\,2-142 for the velocity slices along H$\alpha$ $\lambda$6563.}
\figsetgrpend

\figsetgrpstart
\figsetgrpnum{2.33}
\figsetgrptitle{(i) Hen\,2-142 [N\,{\sc ii}] $\lambda$6584}
\figsetplot{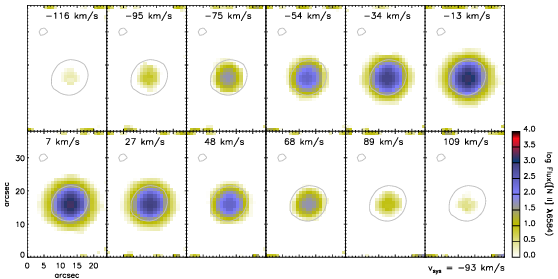}
\figsetgrpnote{Velocity slices along the [N\,{\sc ii}] $\lambda$6584 emission for Hen\,2-142.}
\figsetgrpend

\figsetgrpstart
\figsetgrpnum{2.34}
\figsetgrptitle{(i) Hen\,2-142 Model ([N\,{\sc ii}] $\lambda$6584)}
\figsetplot{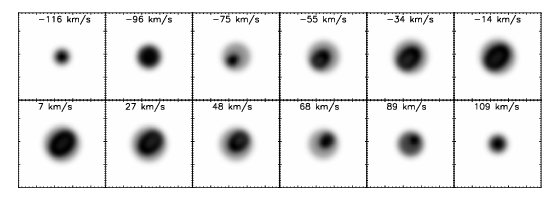}
\figsetgrpnote{Synthetic velocity-resolved channel maps from the morpho-kinematic model of Hen\,2-142 for the velocity slices along [N\,{\sc ii}] $\lambda$6584.}
\figsetgrpend

\figsetgrpstart
\figsetgrpnum{2.35}
\figsetgrptitle{(j) K\,2-16 H$\alpha$ $\lambda$6563} 
\figsetplot{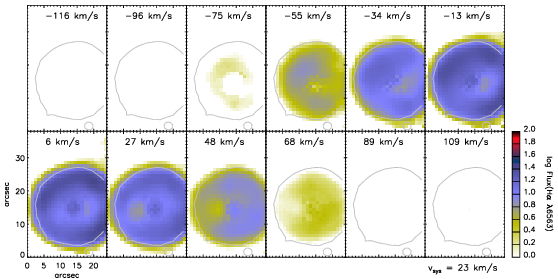}
\figsetgrpnote{Velocity slices along the H$\alpha$ $\lambda$6563 emission for K\,2-16.}
\figsetgrpend

\figsetgrpstart
\figsetgrpnum{2.36}
\figsetgrptitle{(j) K\,2-16 Model (H$\alpha$ $\lambda$6563)}
\figsetplot{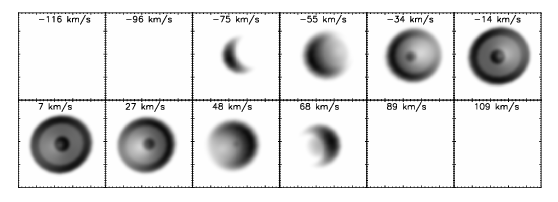}
\figsetgrpnote{Synthetic velocity-resolved channel maps from the morpho-kinematic model of K\,2-16 for the velocity slices along H$\alpha$ $\lambda$6563.}
\figsetgrpend

\figsetgrpstart
\figsetgrpnum{2.37}
\figsetgrptitle{(j) K\,2-16 [N\,{\sc ii}] $\lambda$6584} 
\figsetplot{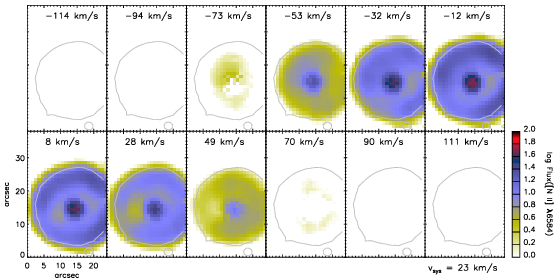}
\figsetgrpnote{Velocity slices along the [N\,{\sc ii}] $\lambda$6584 emission for K\,2-16.}
\figsetgrpend

\figsetgrpstart
\figsetgrpnum{2.38}
\figsetgrptitle{(j) K\,2-16 Model ([N\,{\sc ii}] $\lambda$6584)}
\figsetplot{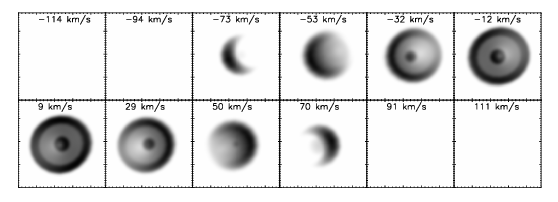}
\figsetgrpnote{Synthetic velocity-resolved channel maps from the morpho-kinematic model of K\,2-16 for the velocity slices along [N\,{\sc ii}] $\lambda$6584.}
\figsetgrpend

\figsetgrpstart
\figsetgrpnum{2.39}
\figsetgrptitle{(k) NGC\,6578 H$\alpha$ $\lambda$6563}
\figsetplot{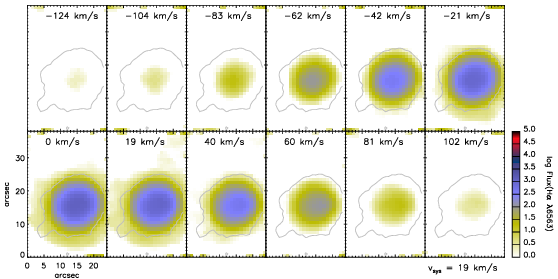}
\figsetgrpnote{Velocity slices along the H$\alpha$ $\lambda$6563 emission for NGC\,6578.}
\figsetgrpend

\figsetgrpstart
\figsetgrpnum{2.40}
\figsetgrptitle{(k) NGC\,6578 Model (H$\alpha$ $\lambda$6563)}
\figsetplot{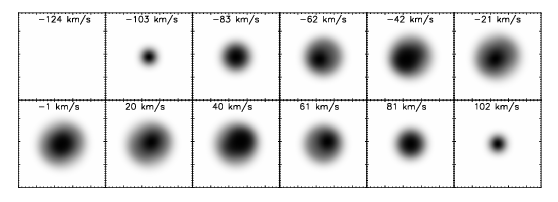}
\figsetgrpnote{Synthetic velocity-resolved channel maps from the morpho-kinematic model of NGC\,6578 for the velocity slices along H$\alpha$ $\lambda$6563.}
\figsetgrpend

\figsetgrpstart
\figsetgrpnum{2.41}
\figsetgrptitle{(k) NGC\,6578 [N\,{\sc ii}] $\lambda$6584}
\figsetplot{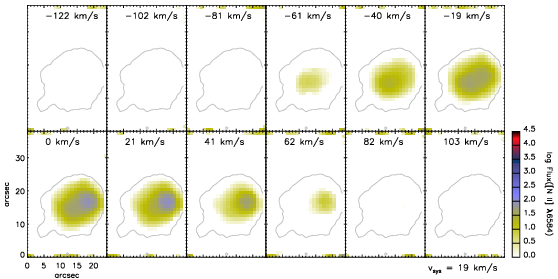}
\figsetgrpnote{Velocity slices along the [N\,{\sc ii}] $\lambda$6584 emission for NGC\,6578.}
\figsetgrpend

\figsetgrpstart
\figsetgrpnum{2.42}
\figsetgrptitle{(k) NGC\,6578 Model ([N\,{\sc ii}] $\lambda$6584)}
\figsetplot{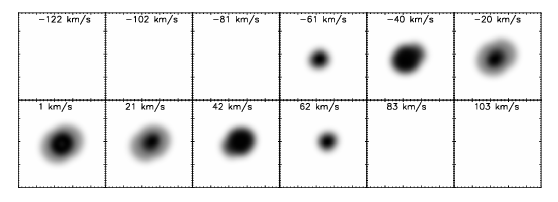}
\figsetgrpnote{Synthetic velocity-resolved channel maps from the morpho-kinematic model of NGC\,6578 for the velocity slices along [N\,{\sc ii}] $\lambda$6584.}
\figsetgrpend

\figsetgrpstart
\figsetgrpnum{2.43}
\figsetgrptitle{(l) NGC\,6567 H$\alpha$ $\lambda$6563}
\figsetplot{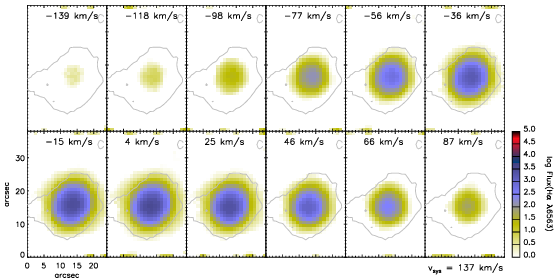}
\figsetgrpnote{Velocity slices along the H$\alpha$ $\lambda$6563 emission for NGC\,6567.}
\figsetgrpend

\figsetgrpstart
\figsetgrpnum{2.44}
\figsetgrptitle{(l) NGC\,6567 Model (H$\alpha$ $\lambda$6563)}
\figsetplot{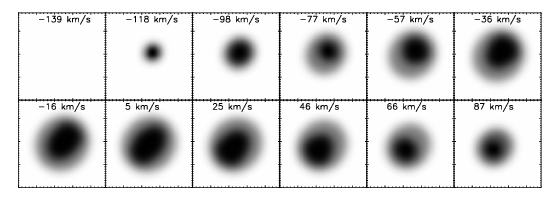}
\figsetgrpnote{Synthetic velocity-resolved channel maps from the morpho-kinematic model of NGC\,6567 for the velocity slices along H$\alpha$ $\lambda$6563.}
\figsetgrpend

\figsetgrpstart
\figsetgrpnum{2.45}
\figsetgrptitle{(l) NGC\,6567 [N\,{\sc ii}] $\lambda$6584}
\figsetplot{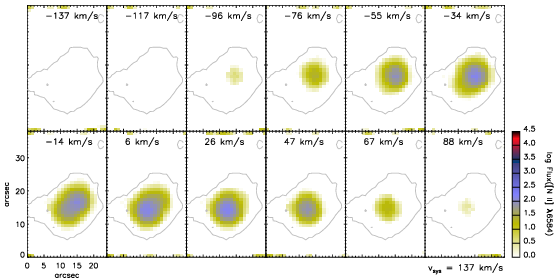}
\figsetgrpnote{Velocity slices along the [N\,{\sc ii}] $\lambda$6584 emission for NGC\,6567.}
\figsetgrpend

\figsetgrpstart
\figsetgrpnum{2.46}
\figsetgrptitle{(l) NGC\,6567 Model ([N\,{\sc ii}] $\lambda$6584)}
\figsetplot{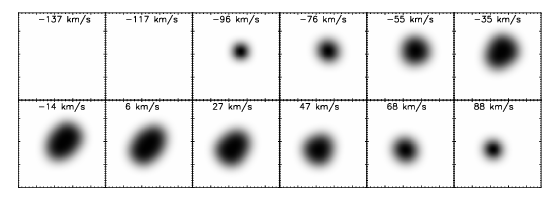}
\figsetgrpnote{Synthetic velocity-resolved channel maps from the morpho-kinematic model of NGC\,6567 for the velocity slices along [N\,{\sc ii}] $\lambda$6584.}
\figsetgrpend

\figsetgrpstart
\figsetgrpnum{2.47}
\figsetgrptitle{(m) NGC\,6629 H$\alpha$ $\lambda$6563}
\figsetplot{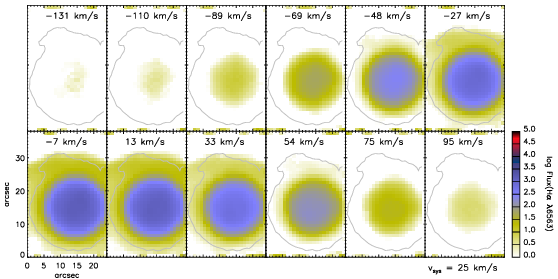}
\figsetgrpnote{Velocity slices along the H$\alpha$ $\lambda$6563 emission for NGC\,6629.}
\figsetgrpend

\figsetgrpstart
\figsetgrpnum{2.48}
\figsetgrptitle{(m) NGC\,6629 Model (H$\alpha$ $\lambda$6563)}
\figsetplot{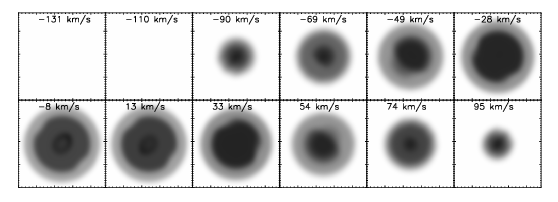}
\figsetgrpnote{Synthetic velocity-resolved channel maps from the morpho-kinematic model of NGC\,6629 for the velocity slices along H$\alpha$ $\lambda$6563.}
\figsetgrpend

\figsetgrpstart
\figsetgrpnum{2.49}
\figsetgrptitle{(m) NGC\,6629 [N\,{\sc ii}] $\lambda$6584}
\figsetplot{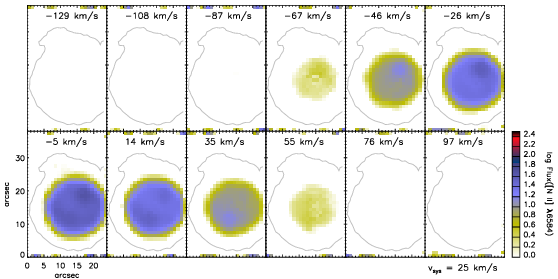}
\figsetgrpnote{Velocity slices along the [N\,{\sc ii}] $\lambda$6584 emission for NGC\,6629.}
\figsetgrpend

\figsetgrpstart
\figsetgrpnum{2.50}
\figsetgrptitle{(m) NGC\,6629 Model ([N\,{\sc ii}] $\lambda$6584)}
\figsetplot{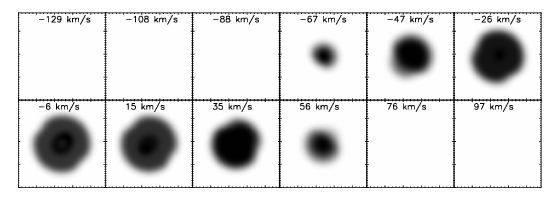}
\figsetgrpnote{Synthetic velocity-resolved channel maps from the morpho-kinematic model of NGC\,6629 for the velocity slices along [N\,{\sc ii}] $\lambda$6584.}
\figsetgrpend

\figsetgrpstart
\figsetgrpnum{2.51}
\figsetgrptitle{(n) Sa\,3-107 H$\alpha$ $\lambda$6563}
\figsetplot{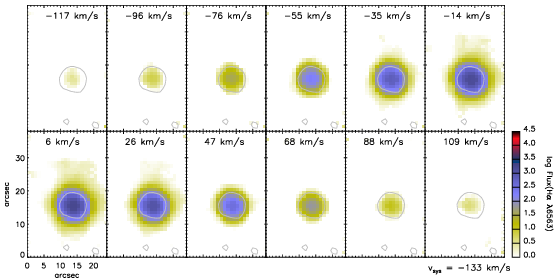}
\figsetgrpnote{Velocity slices along the H$\alpha$ $\lambda$6563 emission for Sa\,3-107.}
\figsetgrpend

\figsetgrpstart
\figsetgrpnum{2.52}
\figsetgrptitle{(n) Sa\,3-107 [N\,{\sc ii}] $\lambda$6584}
\figsetplot{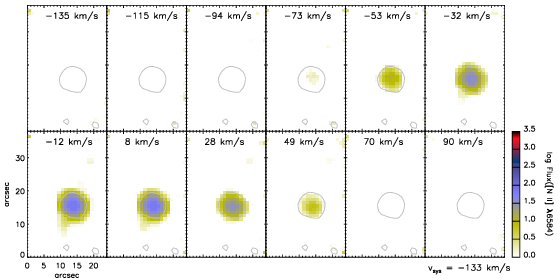}
\figsetgrpnote{Velocity slices along the [N\,{\sc ii}] $\lambda$6584 emissions for Sa\,3-107.}
\figsetgrpend

\figsetend

\end{figure*}

\setcounter{figure}{1}
\begin{figure*}
\begin{center}
{\footnotesize (c) Hb\,4 [N\,{\sc ii}] $\lambda$6584}\\ 
\includegraphics[width=6.in]{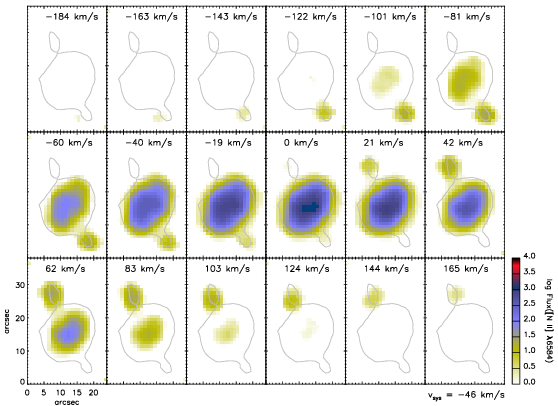}\\
{\footnotesize (c) Hb\,4 Model ([N\,{\sc ii}] $\lambda$6584)}\\ 
\includegraphics[width=5.8in]{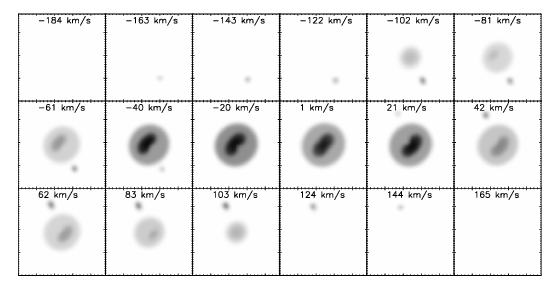}\\
\caption{\textit{-- continued}}
\end{center}
\end{figure*}

The IFU observations of IC\,1297 likely imply an elliptical morphology, which is also supported by the H$\alpha$+[N\,{\sc ii}] image \citep{Schwarz1992}. However, \citet{Zuckerman1986} described it as a broken ring, and \citet{Stanghellini1993} classified it under the irregular nebulae. Furthermore, \citet{Corradi1996} proposed the presence of a single isolated knot, which was later dismissed by \citet{Gonccalves2001}, and turned out to be a field star. The H$\alpha$+[N\,{\sc ii}] and [O\,{\sc iii}] images taken by \citet{Corradi1996} again pointed to a broken ring-shaped shell.

The velocity dispersion maps of Pe\,1-1 depict two regions with high values of $\sigma_{\rm true}=35\pm10$\,km\,s$^{-1}$ on both sides, which are analogous to those in Hb\,4. Its \textit{HST} image in Figure\,\ref{wc1:model:shape} shows a barrel morphology having two collimated outflows in the regions with high velocity dispersion in the IFU maps. The on-sky orientation of the bipolar outflows determined from the IFU velocity maps is consistent with what is seen in the high-resolution \textit{HST} image. 

It can be seen that the IFU velocity and dispersion maps of M\,1-32 are akin to what is found for Th\,2-A, which is also viewed almost pole-on \citep{Danehkar2015a}. Similarly, the P--V arrays (Figure~\ref{wc1:pv:diagram}) imply that this object also has collimated bipolar outflows with very low inclination ($i = 4^{\circ}$; in \S\,\ref{wc1:sec:morpho-kinematic}) relative to the line of sight. \citet{Pena2001} discovered line profile wings with speeds of up to $100$\,km\,s$^{-1}$  relative to the central star, implying high-velocity bipolar or multipolar ejection. The narrow-band H$\alpha$+[N\,{\sc ii}] image (Figure\,\ref{wc1:model:shape}) also depicts a compact torus morphology. 

The WiFeS spatial resolution is inadequate to distinguish the morphological features of the compact ($\leq 6$\,arcsec) PNe M\,3-15, M\,1-25, Hen\,2-142, and NGC\,6567, as well as the interior shells in NGC\,6578 and NGC\,6629. The \textit{HST} observations (Figure~\ref{wc1:model:shape}) exhibit elliptical morphologies for M\,1-25, Hen\,2-142, NGC\,6578, NGC\,6567, and NGC\,6629, and a ring-like morphology for M\,3-15. The overall orientations of their aspherical gas distributions on the sky plane seen in IFU velocity maps are similar to those of their axisymmetric morphologies, with detailed features in the high-angular resolution \textit{HST} images. The \textit{Hubble} imaging observations of NGC\,6578 and NGC\,6629 also indicate that their compact elliptical shells are surrounded by tenuous exterior halos. 

The velocity IFU maps of K\,2-16 also point to an elliptical nebula with upward bipolar extension. The narrow H$\alpha$+[N\,{\sc ii}] image of K\,2-16 \citep{Schwarz1992} exhibits a faint, circular nebula with a diameter of $23\arcsec$. \citet{Kohoutek1977} also suggested that it could be a large and old PN. 

The velocity maps of Sa\,3-107 likely hint at an axisymmetric morphology, though the detailed structures of this compact object cannot be resolved by WiFeS. Similarly, WiFeS observations of the compact PN PB\,8 around a [WN/WC]-hybrid star pointed to an aspherical morphology but without any precise details \citep{Danehkar2018}. In particular, tenuous lobes extending from a compact nebula can spatially be mapped through IFU spectroscopy, which may help us determine the on-sky orientation of its axisymmetric morphology \citep[e.g.][]{Danehkar2015}. Thus, our WiFeS observations of Sa\,3-107 could expose the on-sky orientation of its elliptical or bipolar morphology. 


\section{Morpho-kinematic modeling}
\label{wc1:sec:morpho-kinematic}

To construct 3D morphological models based on velocity-resolved channel maps and P--V diagrams, we have used the interactive kinematic modeling program \textsc{shape} v5.0 \citep[][]{Steffen2006,Steffen2011}, which applies molded polygon meshes to 3D geometrical models of gaseous nebulae. It assembles a morphological model on a grid of volumetric cells and utilizes a ray-casting process to transfer radiation fields to these cells. The program produces outputs that can be directly compared with kinematic observations, namely the P--V diagram and velocity channel maps. P--V diagrams have been used for long-slit observations of many PNe \citep[e.g.][]{Garcia-Diaz2009,Jones2010a,Lopez2012,Clark2013}, whereas velocity channels have recently been employed to interpret kinematic IFU maps of some objects \citep[][]{Danehkar2015a,Danehkar2016}. It also produces 2D rendered images of ionized nebulae that can be matched with high-resolution images. In particular, \textsc{shape} can also export its 3D mesh model to the standard tessellation language (STL) file format, which can be converted to the extensible 3D graphics (X3D) format by mesh processing programs such as MeshLab \citep{Cignoni2008} for inclusion in publications. This program has been used to create 3D morpho-kinematic models of many PNe, including NGC~6337 \citep{Garcia-Diaz2009}, Abell~41 \citep{Jones2010a}, Hb~5 \citep{Lopez2012}, HaTr~4 \citep{Tyndall2012}, NGC~7026 \citep{Clark2013}, Abell~65 \citep{Huckvale2013}, Th 2-A \citep{Danehkar2015a}, M2-42 \citep{Danehkar2016}, and Abell 14 \citep{Akras2016}.

For this work, the velocity field is defined by a homologous expansion \citep{Steffen2009} that increases linearly with distance from the nebular center. However, to reproduce the point-symmetric knots in the P--V diagrams of Hb 4, we use a decelerating velocity law that linearly decreases with distance from the nebular center.
To model each object, we carefully choose a 3D geometry model and adjust various geometric parameters, along with the inclination and origination angles, in order to reproduce those kinematic features seen in the P--V arrays and velocity channels of each object, while the 2D rendered images are also expected to resemble the associated archival image. 
The inclination ($i$) relative to the line of sight, the position angle (PA) of the on-sky orientation, and geometric parameters are modified in an iterative process until the model velocity maps and P--V diagram closely resemble the kinematic observations. 
We were able to construct 3D morpho-kinematic models of the 12 PNe. 
For the compact PN Sa\,3-107, we do not have any available high-resolution images. 
The \textit{HST}/MIRVIS image of PB\,6 
is not suitable for morpho-kinematic modeling, but it may suggest a complex morphology with multi-scale features similar to NGC\,5189 \citep{Danehkar2018a}. To reproduce the WiFeS observations, we configured the seeing parameters in \textsc{shape} according to the WiFeS spatial resolution that resulted in the convoluted P--V and channel outputs, while 2D rendered images were made without the convolution processing for comparison with the high-resolution archival images.

Figure~\ref{wc1:model:shape} depicts the final \textsc{shape} mesh models of the 12 objects viewed 
from the top ($i=0^{\circ}$) and front ($i=90^{\circ}$), as well as the best-fitting inclinations ($i$) and orientations (PA; see Table~\ref{wc1:tab:shapemodel}) before rendering and the rendered images, respectively 
(for all the objects available in the online journal). 
Additionally, an interactive X3D viewer of each \textsc{shape} mesh model is provided in Figure~\ref{wc1:figures:shape:models} for the online journal.\footnote{The files of the interactive figure (14 three-dimensional models), including those for 
Th\,2-A \citep{Danehkar2015a} and M\,2-42 \citep{Danehkar2016}, are available in the X3D file format and archived on Zenodo (doi:\href{https://doi.org/10.5281/zenodo.5393974}{10.5281/zenodo.5393974}).}

\begin{figure}
\begin{center}
{\footnotesize (c) Hb\,4 H$\alpha$ $\lambda$6563}\\ 
\includegraphics[width=3.45in]{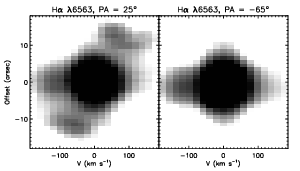}\\
{\footnotesize (c) Hb\,4 Model (H$\alpha$ $\lambda$6563)} \\ 
\includegraphics[width=3.45in]{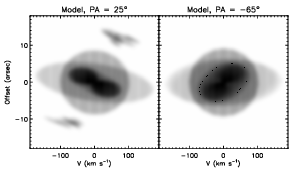}\\
\caption{P--V arrays in the H$\alpha$ $\lambda$6563 and [N\,{\sc ii}] $\lambda$6584 emissions for: (a) PB\,6, (b) M\,3-30, (c) Hb\,4, (d) IC\,1297, (e) Pe\,1-1, (f) M\,1-32, (g) M\,3-15, (h) M\,1-25, (i) Hen\,2-142, (j) K\,2-16, (k) NGC\,6578, (l) NGC\,6567, (m) NGC\,6629, and (n) Sa\,3-107, followed by the associated synthetic P--V diagrams obtained from the best-fitting morpho-kinematic model of all the PNe, except for PB\,6 and Sa\,3-107, with the parameters given in Table~\ref{wc1:tab:shapemodel}. Slits oriented with the position angles (PA) along and vertical to the symmetric axis of the morpho-kinematic model passing through the central star, respectively. The velocity in each observed P--V arrays is with respect to the systemic velocity of the object, given in km\,s${}^{-1}$ unit. The angular offset at 0 arcsec is the nebular center.
The complete figure set (52 images) is available in the online journal.
}
\label{wc1:pv:diagram}%
\end{center}

\figsetstart
\figsetnum{3}
\figsettitle{P--V arrays in the H$\alpha$ $\lambda$6563 and [N\,{\sc ii}] $\lambda$6584 emissions.
}

\figsetgrpstart
\figsetgrpnum{3.1}
\figsetgrptitle{(a) PB\,6 H$\alpha$ $\lambda$6563}
\figsetplot{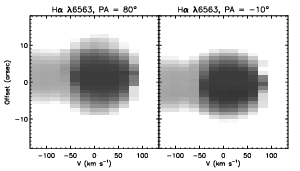}

\figsetgrpnote{P--V array in the H$\alpha$ $\lambda$6563 emission for PB\,6.
}
\figsetgrpend

\figsetgrpstart
\figsetgrpnum{3.2}
\figsetgrptitle{(a) PB\,6 [N\,{\sc ii}] $\lambda$6584}
\figsetplot{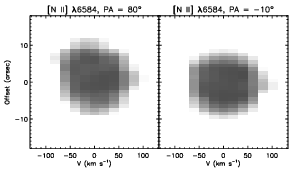}
\figsetgrpnote{P--V array in the [N\,{\sc ii}] $\lambda$6584 emission for PB\,6.
}
\figsetgrpend

\figsetgrpstart
\figsetgrpnum{3.3}
\figsetgrptitle{(b) M\,3-30 H$\alpha$ $\lambda$6563}
\figsetplot{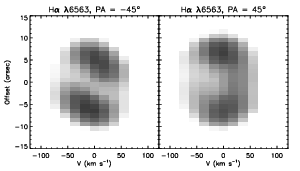}
\figsetgrpnote{P--V array in the H$\alpha$ $\lambda$6563 emission for M\,3-30.
}
\figsetgrpend

\figsetgrpstart
\figsetgrpnum{3.4}
\figsetgrptitle{(b) M\,3-30 Model (H$\alpha$ $\lambda$6563)}
\figsetplot{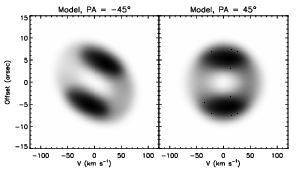}
\figsetgrpnote{Synthetic P--V diagram from the morpho-kinematic model of M\,3-30 for the P--V array in H$\alpha$ $\lambda$6563.
}
\figsetgrpend

\figsetgrpstart
\figsetgrpnum{3.5}
\figsetgrptitle{(b) M\,3-30 [N\,{\sc ii}] $\lambda$6584}
\figsetplot{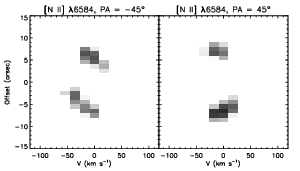}
\figsetgrpnote{P--V array in the [N\,{\sc ii}] $\lambda$6584 emission for M\,3-30.
}
\figsetgrpend

\figsetgrpstart
\figsetgrpnum{3.6}
\figsetgrptitle{(b) M\,3-30 Model ([N\,{\sc ii}] $\lambda$6584)}
\figsetplot{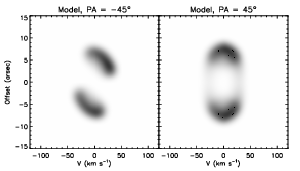}
\figsetgrpnote{Synthetic P--V diagram from the morpho-kinematic model of M\,3-30 for the P--V array in [N\,{\sc ii}] $\lambda$6584.
}
\figsetgrpend

\figsetgrpstart
\figsetgrpnum{3.7}
\figsetgrptitle{(c) Hb\,4 H$\alpha$ $\lambda$6563} 
\figsetplot{figure3/fig3_hb4_6563_pv.eps}
\figsetgrpnote{P--V array in the H$\alpha$ $\lambda$6563 emission for Hb\,4.
}
\figsetgrpend

\figsetgrpstart
\figsetgrpnum{3.8}
\figsetgrptitle{(c) Hb\,4 Model (H$\alpha$ $\lambda$6563)} 
\figsetplot{figure3/fig3_hb4_shape_pv_ha.eps}
\figsetgrpnote{Synthetic P--V diagram from the morpho-kinematic model of Hb\,4 for the P--V array in H$\alpha$ $\lambda$6563.
}
\figsetgrpend

\figsetgrpstart
\figsetgrpnum{3.9}
\figsetgrptitle{(c) Hb\,4 [N\,{\sc ii}] $\lambda$6584} 
\figsetplot{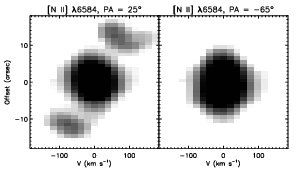}
\figsetgrpnote{P--V array in the [N\,{\sc ii}] $\lambda$6584 emission for Hb\,4.
}
\figsetgrpend

\figsetgrpstart
\figsetgrpnum{3.10}
\figsetgrptitle{(c) Hb\,4 Model ([N\,{\sc ii}] $\lambda$6584)} 
\figsetplot{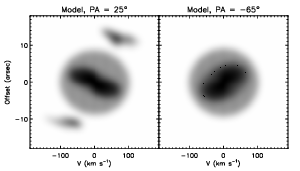}
\figsetgrpnote{Synthetic P--V diagram from the morpho-kinematic model of Hb\,4 for the P--V array in [N\,{\sc ii}] $\lambda$6584.
}
\figsetgrpend

\figsetgrpstart
\figsetgrpnum{3.11}
\figsetgrptitle{(d) IC\,1297 H$\alpha$ $\lambda$6563}
\figsetplot{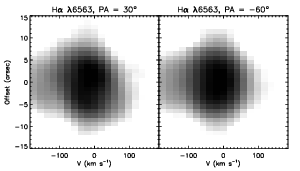}
\figsetgrpnote{P--V array in the H$\alpha$ $\lambda$6563 emission for IC\,1297.
}
\figsetgrpend

\figsetgrpstart
\figsetgrpnum{3.12}
\figsetgrptitle{(d) IC\,1297 Model (H$\alpha$ $\lambda$6563)}
\figsetplot{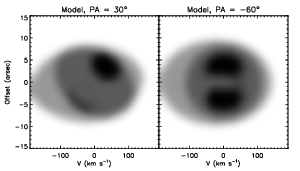}
\figsetgrpnote{Synthetic P--V diagram from the morpho-kinematic model of IC\,1297 for the P--V array in H$\alpha$ $\lambda$6563.
}
\figsetgrpend

\figsetgrpstart
\figsetgrpnum{3.13}
\figsetgrptitle{(d) IC\,1297 [N\,{\sc ii}] $\lambda$6584}
\figsetplot{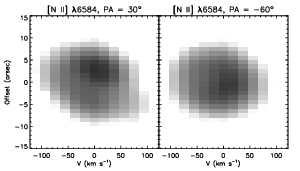}
\figsetgrpnote{P--V array in the [N\,{\sc ii}] $\lambda$6584 emission for IC\,1297.
}
\figsetgrpend

\figsetgrpstart
\figsetgrpnum{3.14}
\figsetgrptitle{(d) IC\,1297 Model ([N\,{\sc ii}] $\lambda$6584)}
\figsetplot{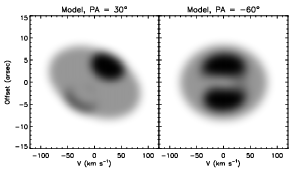}
\figsetgrpnote{Synthetic P--V diagram from the morpho-kinematic model of IC\,1297 for the P--V array in [N\,{\sc ii}] $\lambda$6584.
}
\figsetgrpend

\figsetgrpstart
\figsetgrpnum{3.15}
\figsetgrptitle{(e) Pe\,1-1 H$\alpha$ $\lambda$6563}
\figsetplot{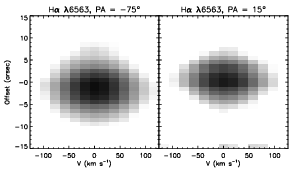}
\figsetgrpnote{P--V array in the H$\alpha$ $\lambda$6563 emission for Pe\,1-1.
}
\figsetgrpend

\figsetgrpstart
\figsetgrpnum{3.16}
\figsetgrptitle{(e) Pe\,1-1 Model (H$\alpha$ $\lambda$6563)}
\figsetplot{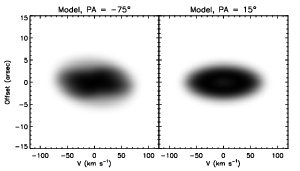}
\figsetgrpnote{Synthetic P--V diagram from the morpho-kinematic model of Pe\,1-1 for the P--V array in H$\alpha$ $\lambda$6563.
}
\figsetgrpend

\figsetgrpstart
\figsetgrpnum{3.17}
\figsetgrptitle{(e) Pe\,1-1 [N\,{\sc ii}] $\lambda$6584}
\figsetplot{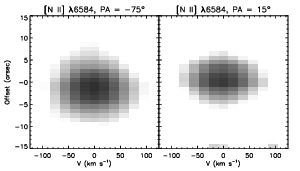}
\figsetgrpnote{P--V array in the [N\,{\sc ii}] $\lambda$6584 emission for Pe\,1-1.
}
\figsetgrpend

\figsetgrpstart
\figsetgrpnum{3.18}
\figsetgrptitle{(e) Pe\,1-1 Model ([N\,{\sc ii}] $\lambda$6584)}
\figsetplot{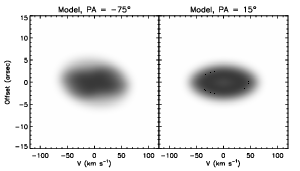}
\figsetgrpnote{Synthetic P--V diagram from the morpho-kinematic model of Pe\,1-1 for the P--V array in [N\,{\sc ii}] $\lambda$6584.
}
\figsetgrpend

\figsetgrpstart
\figsetgrpnum{3.19}
\figsetgrptitle{(f) M\,1-32 H$\alpha$ $\lambda$6563} 
\figsetplot{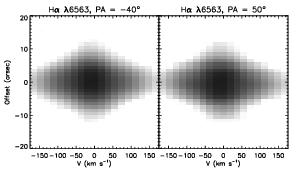}
\figsetgrpnote{P--V array in the H$\alpha$ $\lambda$6563 emission for M\,1-32.
}
\figsetgrpend

\figsetgrpstart
\figsetgrpnum{3.20}
\figsetgrptitle{(f) M\,1-32 Model (H$\alpha$ $\lambda$6563)} 
\figsetplot{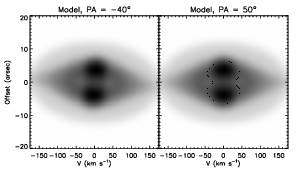}
\figsetgrpnote{Synthetic P--V diagram from the morpho-kinematic model of M\,1-32 for the P--V array in H$\alpha$ $\lambda$6563.
}
\figsetgrpend

\figsetgrpstart
\figsetgrpnum{3.21}
\figsetgrptitle{(f) M\,1-32 [N\,{\sc ii}] $\lambda$6584} 
\figsetplot{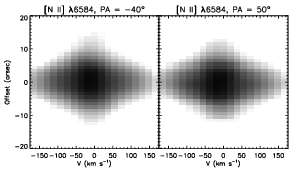}
\figsetgrpnote{P--V array in the [N\,{\sc ii}] $\lambda$6584 emission for M\,1-32.
}
\figsetgrpend

\figsetgrpstart
\figsetgrpnum{3.22}
\figsetgrptitle{(f) M\,1-32 Model ([N\,{\sc ii}] $\lambda$6584)} 
\figsetplot{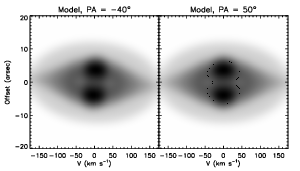}
\figsetgrpnote{Synthetic P--V diagram from the morpho-kinematic model of M\,1-32 for the P--V array in [N\,{\sc ii}] $\lambda$6584.
}
\figsetgrpend

\figsetgrpstart
\figsetgrpnum{3.23}
\figsetgrptitle{(g) M\,3-15 H$\alpha$ $\lambda$6563}
\figsetplot{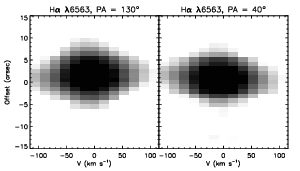}
\figsetgrpnote{P--V array in the H$\alpha$ $\lambda$6563 emission for M\,3-15.
}
\figsetgrpend

\figsetgrpstart
\figsetgrpnum{3.24}
\figsetgrptitle{(g) M\,3-15 Model (H$\alpha$ $\lambda$6563)}
\figsetplot{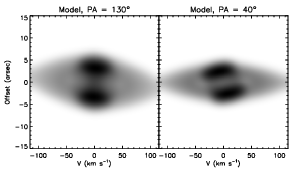}
\figsetgrpnote{Synthetic P--V diagram from the morpho-kinematic model of M\,3-15 for the P--V array in H$\alpha$ $\lambda$6563.
}
\figsetgrpend

\figsetgrpstart
\figsetgrpnum{3.25}
\figsetgrptitle{(g) M\,3-15 [N\,{\sc ii}] $\lambda$6584}
\figsetplot{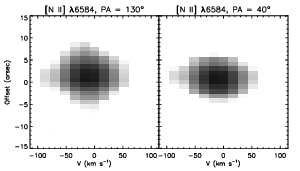}
\figsetgrpnote{P--V array in the [N\,{\sc ii}] $\lambda$6584 emission for M\,3-15.
}
\figsetgrpend

\figsetgrpstart
\figsetgrpnum{3.26}
\figsetgrptitle{(g) M\,3-15 Model ([N\,{\sc ii}] $\lambda$6584)}
\figsetplot{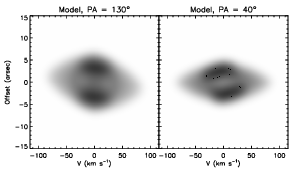}
\figsetgrpnote{Synthetic P--V diagram from the morpho-kinematic model of M\,3-15 for the P--V array in [N\,{\sc ii}] $\lambda$6584.
}
\figsetgrpend

\figsetgrpstart
\figsetgrpnum{3.27}
\figsetgrptitle{(h) M\,1-25 H$\alpha$ $\lambda$6563} 
\figsetplot{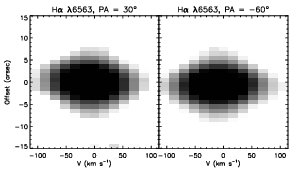}
\figsetgrpnote{P--V array in the H$\alpha$ $\lambda$6563 emission for M\,1-25.
}
\figsetgrpend

\figsetgrpstart
\figsetgrpnum{3.28}
\figsetgrptitle{(h) M\,1-25 Model (H$\alpha$ $\lambda$6563)} 
\figsetplot{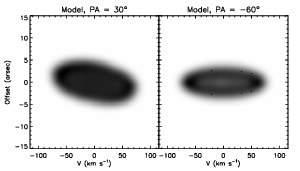}
\figsetgrpnote{Synthetic P--V diagram from the morpho-kinematic model of M\,1-25 for the P--V array in H$\alpha$ $\lambda$6563.
}
\figsetgrpend

\figsetgrpstart
\figsetgrpnum{3.29}
\figsetgrptitle{(h) M\,1-25 [N\,{\sc ii}] $\lambda$6584} 
\figsetplot{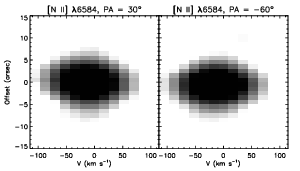}
\figsetgrpnote{P--V array in the [N\,{\sc ii}] $\lambda$6584 emission for M\,1-25.
}
\figsetgrpend

\figsetgrpstart
\figsetgrpnum{3.30}
\figsetgrptitle{(h) M\,1-25 Model ([N\,{\sc ii}] $\lambda$6584)} 
\figsetplot{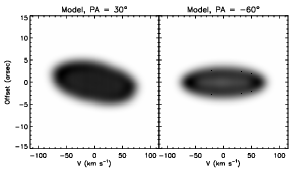}
\figsetgrpnote{Synthetic P--V diagram from the morpho-kinematic model of M\,1-25 for the P--V array in [N\,{\sc ii}] $\lambda$6584.
}
\figsetgrpend

\figsetgrpstart
\figsetgrpnum{3.31}
\figsetgrptitle{(i) Hen\,2-142 H$\alpha$ $\lambda$6563}
\figsetplot{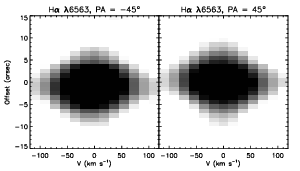}
\figsetgrpnote{P--V array in the H$\alpha$ $\lambda$6563 emission for Hen\,2-142.
}
\figsetgrpend

\figsetgrpstart
\figsetgrpnum{3.32}
\figsetgrptitle{(i) Hen\,2-142 Model (H$\alpha$ $\lambda$6563)}
\figsetplot{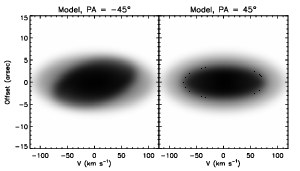}
\figsetgrpnote{Synthetic P--V diagram from the morpho-kinematic model of Hen\,2-142 for the P--V array in H$\alpha$ $\lambda$6563.
}
\figsetgrpend

\figsetgrpstart
\figsetgrpnum{3.33}
\figsetgrptitle{(i) Hen\,2-142 [N\,{\sc ii}] $\lambda$6584}
\figsetplot{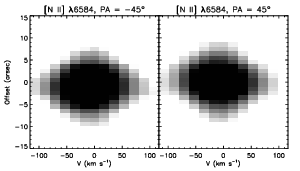}
\figsetgrpnote{P--V array in the [N\,{\sc ii}] $\lambda$6584 emission for Hen\,2-142.
}
\figsetgrpend

\figsetgrpstart
\figsetgrpnum{3.34}
\figsetgrptitle{(i) Hen\,2-142 Model ([N\,{\sc ii}] $\lambda$6584)}
\figsetplot{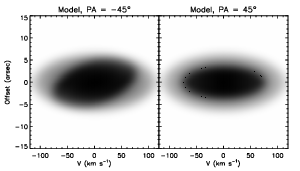}
\figsetgrpnote{Synthetic P--V diagram from the morpho-kinematic model of Hen\,2-142 for the P--V array in [N\,{\sc ii}] $\lambda$6584.
}
\figsetgrpend

\figsetgrpstart
\figsetgrpnum{3.35}
\figsetgrptitle{(j) K\,2-16 H$\alpha$ $\lambda$6563} 
\figsetplot{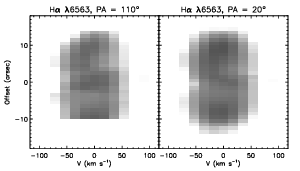}
\figsetgrpnote{P--V array in the H$\alpha$ $\lambda$6563 emission for K\,2-16.
}
\figsetgrpend

\figsetgrpstart
\figsetgrpnum{3.36}
\figsetgrptitle{(j) K\,2-16 Model (H$\alpha$ $\lambda$6563)} 
\figsetplot{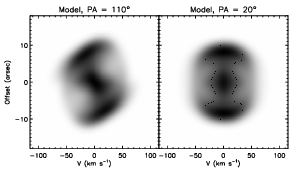}
\figsetgrpnote{Synthetic P--V diagram from the morpho-kinematic model of K\,2-16 for the P--V array in H$\alpha$ $\lambda$6563.
}
\figsetgrpend

\figsetgrpstart
\figsetgrpnum{3.37}
\figsetgrptitle{(j) K\,2-16 [N\,{\sc ii}] $\lambda$6584} 
\figsetplot{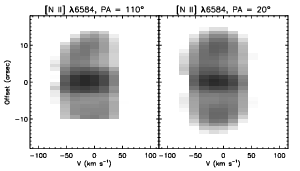}
\figsetgrpnote{P--V array in the [N\,{\sc ii}] $\lambda$6584 emission for K\,2-16.
}
\figsetgrpend

\figsetgrpstart
\figsetgrpnum{3.38}
\figsetgrptitle{(j) K\,2-16 Model ([N\,{\sc ii}] $\lambda$6584)} 
\figsetplot{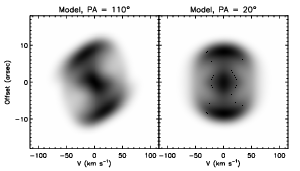}
\figsetgrpnote{Synthetic P--V diagram from the morpho-kinematic model of K\,2-16 for the P--V array in [N\,{\sc ii}] $\lambda$6584.
}
\figsetgrpend

\figsetgrpstart
\figsetgrpnum{3.39}
\figsetgrptitle{(k) NGC\,6578 H$\alpha$ $\lambda$6563}
\figsetplot{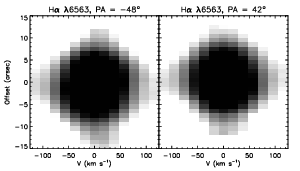}
\figsetgrpnote{P--V array in the H$\alpha$ $\lambda$6563 emission for NGC\,6578.
}
\figsetgrpend

\figsetgrpstart
\figsetgrpnum{3.40}
\figsetgrptitle{(k) NGC\,6578 Model (H$\alpha$ $\lambda$6563)}
\figsetplot{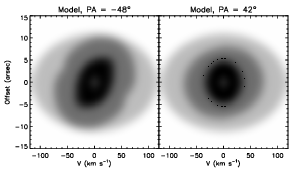}
\figsetgrpnote{Synthetic P--V diagram from the morpho-kinematic model of NGC\,6578 for the P--V array in H$\alpha$ $\lambda$6563.
}
\figsetgrpend

\figsetgrpstart
\figsetgrpnum{3.41}
\figsetgrptitle{(k) NGC\,6578 [N\,{\sc ii}] $\lambda$6584}
\figsetplot{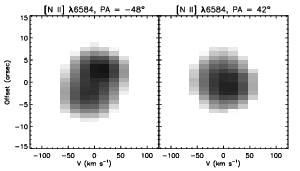}
\figsetgrpnote{P--V array in the [N\,{\sc ii}] $\lambda$6584 emission for NGC\,6578.
}
\figsetgrpend

\figsetgrpstart
\figsetgrpnum{3.42}
\figsetgrptitle{(k) NGC\,6578 Model ([N\,{\sc ii}] $\lambda$6584)}
\figsetplot{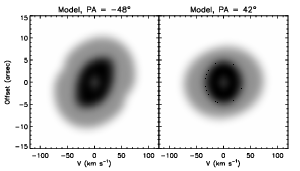}
\figsetgrpnote{Synthetic P--V diagram from the morpho-kinematic model of NGC\,6578 for the P--V array in [N\,{\sc ii}] $\lambda$6584.
}
\figsetgrpend

\figsetgrpstart
\figsetgrpnum{3.43}
\figsetgrptitle{(l) NGC\,6567 H$\alpha$ $\lambda$6563}
\figsetplot{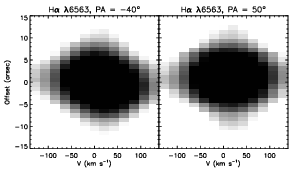}
\figsetgrpnote{P--V array in the H$\alpha$ $\lambda$6563 emission for NGC\,6567.
}
\figsetgrpend

\figsetgrpstart
\figsetgrpnum{3.44}
\figsetgrptitle{(l) NGC\,6567 Model (H$\alpha$ $\lambda$6563)}
\figsetplot{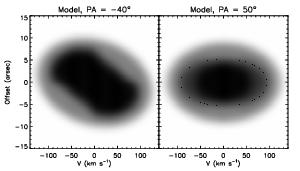}
\figsetgrpnote{Synthetic P--V diagram from the morpho-kinematic model of NGC\,6567 for the P--V array in H$\alpha$ $\lambda$6563.
}
\figsetgrpend

\figsetgrpstart
\figsetgrpnum{3.45}
\figsetgrptitle{(l) NGC\,6567 [N\,{\sc ii}] $\lambda$6584}
\figsetplot{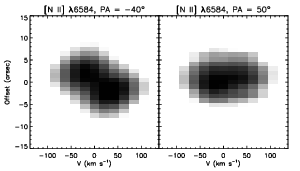}
\figsetgrpnote{P--V array in the [N\,{\sc ii}] $\lambda$6584 emission for NGC\,6567.
}
\figsetgrpend

\figsetgrpstart
\figsetgrpnum{3.46}
\figsetgrptitle{(l) NGC\,6567 Model ([N\,{\sc ii}] $\lambda$6584)}
\figsetplot{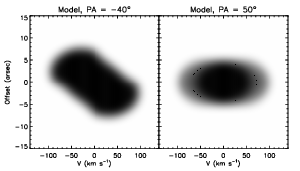}
\figsetgrpnote{Synthetic P--V diagram from the morpho-kinematic model of NGC\,6567 for the P--V array in [N\,{\sc ii}] $\lambda$6584.
}
\figsetgrpend

\figsetgrpstart
\figsetgrpnum{3.47}
\figsetgrptitle{(m) NGC\,6629 H$\alpha$ $\lambda$6563}
\figsetplot{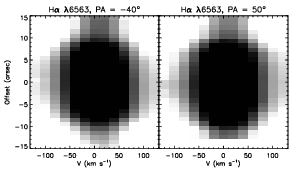}
\figsetgrpnote{P--V array in the H$\alpha$ $\lambda$6563 emission for NGC\,6629.
}
\figsetgrpend

\figsetgrpstart
\figsetgrpnum{3.48}
\figsetgrptitle{(m) NGC\,6629 Model (H$\alpha$ $\lambda$6563)}
\figsetplot{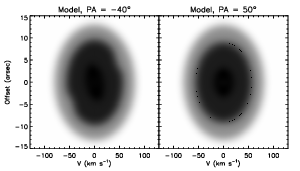}
\figsetgrpnote{Synthetic P--V diagram from the morpho-kinematic model of NGC\,6629 for the P--V array in H$\alpha$ $\lambda$6563.
}
\figsetgrpend

\figsetgrpstart
\figsetgrpnum{3.49}
\figsetgrptitle{(m) NGC\,6629 [N\,{\sc ii}] $\lambda$6584}
\figsetplot{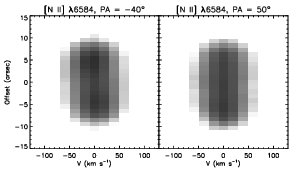}
\figsetgrpnote{P--V array in the [N\,{\sc ii}] $\lambda$6584 emission for NGC\,6629.
}
\figsetgrpend

\figsetgrpstart
\figsetgrpnum{3.50}
\figsetgrptitle{(m) NGC\,6629 Model ([N\,{\sc ii}] $\lambda$6584)}
\figsetplot{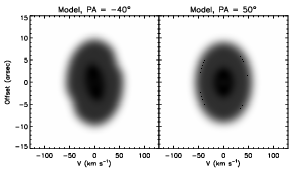}
\figsetgrpnote{Synthetic P--V diagram from the morpho-kinematic model of NGC\,6629 for the P--V array in [N\,{\sc ii}] $\lambda$6584.
}
\figsetgrpend

\figsetgrpstart
\figsetgrpnum{3.51}
\figsetgrptitle{(n) Sa\,3-107 H$\alpha$ $\lambda$6563}
\figsetplot{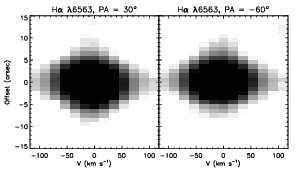}
\figsetgrpnote{P--V array in the H$\alpha$ $\lambda$6563 emission for Sa\,3-107.
}
\figsetgrpend

\figsetgrpstart
\figsetgrpnum{3.52}
\figsetgrptitle{(n) Sa\,3-107 [N\,{\sc ii}] $\lambda$6584}
\figsetplot{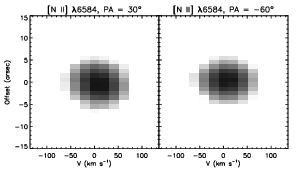}
\figsetgrpnote{P--V array in the [N\,{\sc ii}] $\lambda$6584 emission for Sa\,3-107.
}
\figsetgrpend

\figsetend

\end{figure}

\setcounter{figure}{2}
\begin{figure}
\begin{center}
{\footnotesize (c) Hb\,4 [N\,{\sc ii}] $\lambda$6584}\\ 
\includegraphics[width=3.45in]{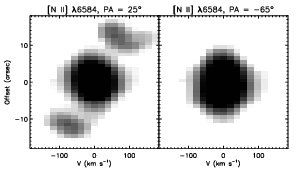}\\
{\footnotesize (c) Hb\,4 Model ([N\,{\sc ii}] $\lambda$6584)} \\ 
\includegraphics[width=3.45in]{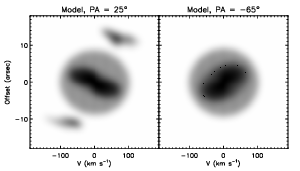}\\
\caption{\textit{-- continued}}
\end{center}
\end{figure}


\begin{figure*}
\begin{center}
{\footnotesize (a) M\,3-30}\\ 
\includegraphics[width=7.in]{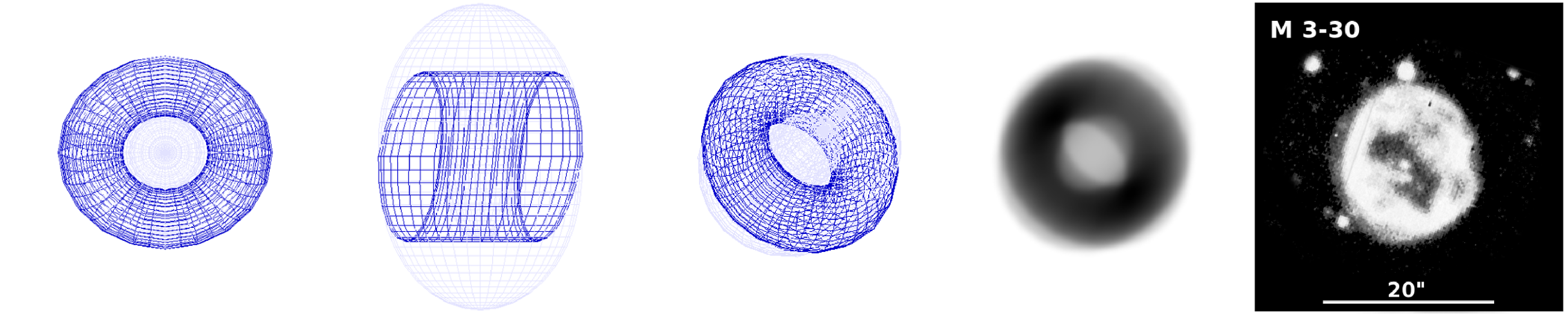}\\
{\footnotesize (b) Hb\,4}\\ 
\includegraphics[width=7.in]{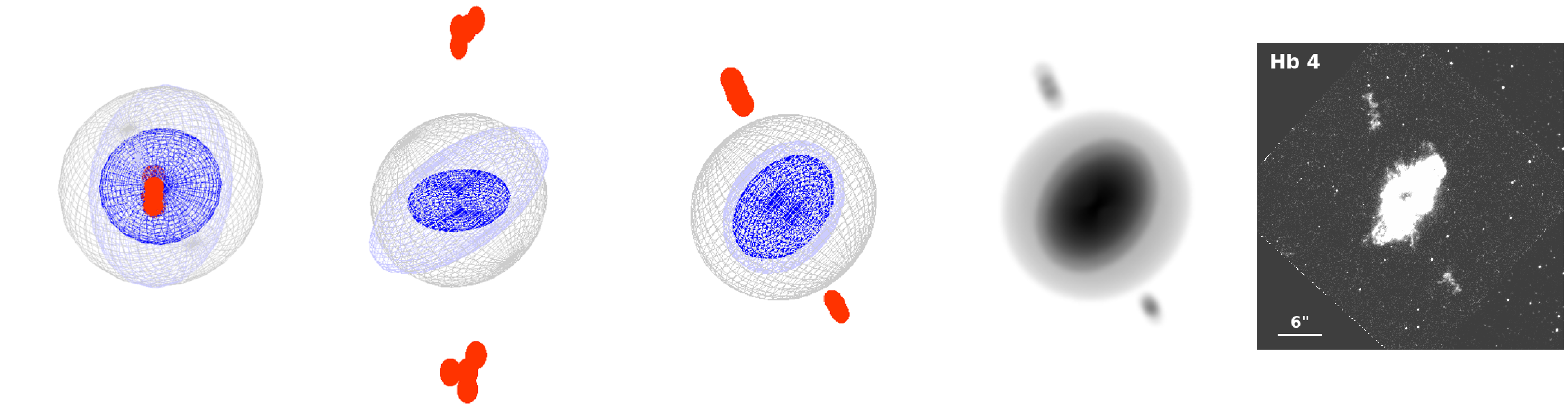}
\caption{\textsc{shape} wireframe models of (a) M\,3-30, (b) Hb\,4, (c) IC\,1297, (d) Pe\,1-1, (e) M\,1-32, (f) M\,3-15, (g) M\,1-25, (h) Hen\,2-142, (i) K\,2-16, (j) NGC\,6578, (k) NGC\,6567, and (l) NGC\,6629, in the top view ($i=0^{\circ}$; first column), the front view ($i=90^{\circ}$; second column), and the best-fitting inclination and orientation (third column; see Figure~\ref{wc1:figures:shape:models} for an interactive viewer in the online journal), followed by the rendered image (fourth column) and the associated archival imaging observation (fifth column), respectively. The parameters of the best-fitting models are listed in Table~\ref{wc1:tab:shapemodel}. The archival data include the \textit{HST} images of Hb\,4, Pe\,1-1, M\,3-15, M\,1-25, Hen\,2-142, NGC\,6578, NGC\,6567 and NGC\,6629, with the instruments and filters/gratings listed in Table~\ref{wc1:tab:hst}; and the narrow-band H$\alpha$+[N\,{\sc ii}] filter images of M\,3-30, IC\,1297, M\,1-32 and K\,2-16 taken with the 3.5-m ESO NTT from \citet{Schwarz1992}, with the image scales shown by solid lines. North is up and east is toward the left-hand side in each archival image. The complete figure set (12 images) is available in the online journal.
}
\label{wc1:model:shape}%
\end{center}

\figsetstart
\figsetnum{4}
\figsettitle{\textsc{shape} wireframe models in the top view ($i=0^{\circ}$), the front view ($i=90^{\circ}$), and the best-fitting inclination and orientation, followed by the rendered images and the associated archival imaging observations.}

\figsetgrpstart
\figsetgrpnum{4.1}
\figsetgrptitle{(a) M\,3-30}
\figsetplot{figure4/fig4_m3_30_shape_schwarz.eps}
\figsetgrpnote{\textsc{shape} mesh model for M\,3-30.
}
\figsetgrpend

\figsetgrpstart
\figsetgrpnum{4.2}
\figsetgrptitle{(b) Hb\,4} 
\figsetplot{figure4/fig4_hb4_shape_hst.eps}
\figsetgrpnote{\textsc{shape} mesh model for Hb\,4.
}
\figsetgrpend

\figsetgrpstart
\figsetgrpnum{4.3}
\figsetgrptitle{(c) IC\,1297}
\figsetplot{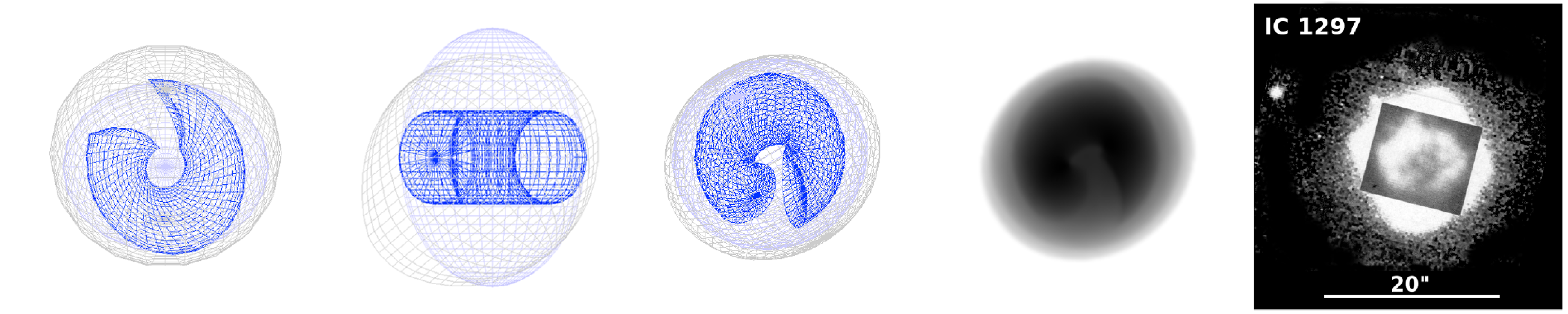}
\figsetgrpnote{\textsc{shape} mesh model for IC\,1297.
}
\figsetgrpend

\figsetgrpstart
\figsetgrpnum{4.4}
\figsetgrptitle{(d) Pe\,1-1}
\figsetplot{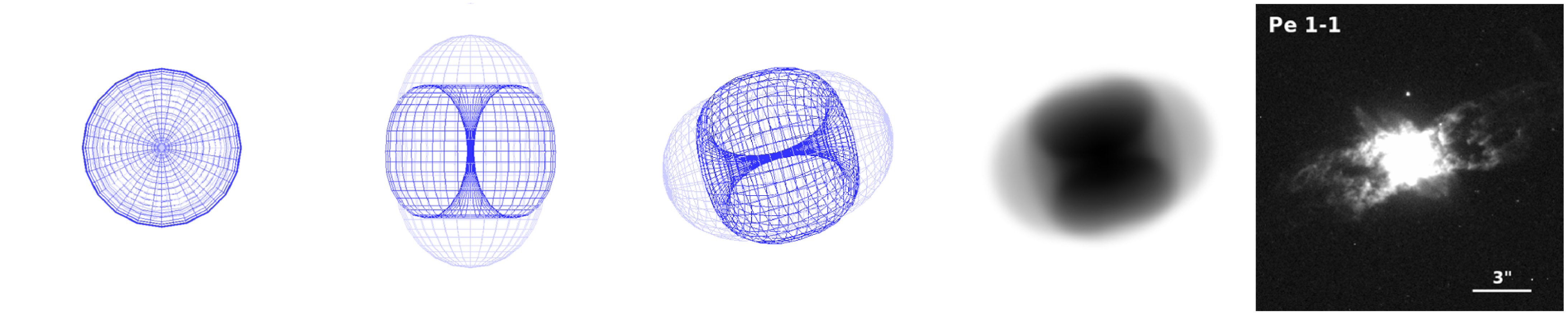}
\figsetgrpnote{\textsc{shape} mesh model for Pe\,1-1.
}
\figsetgrpend

\figsetgrpstart
\figsetgrpnum{4.5}
\figsetgrptitle{(e) M\,1-32} 
\figsetplot{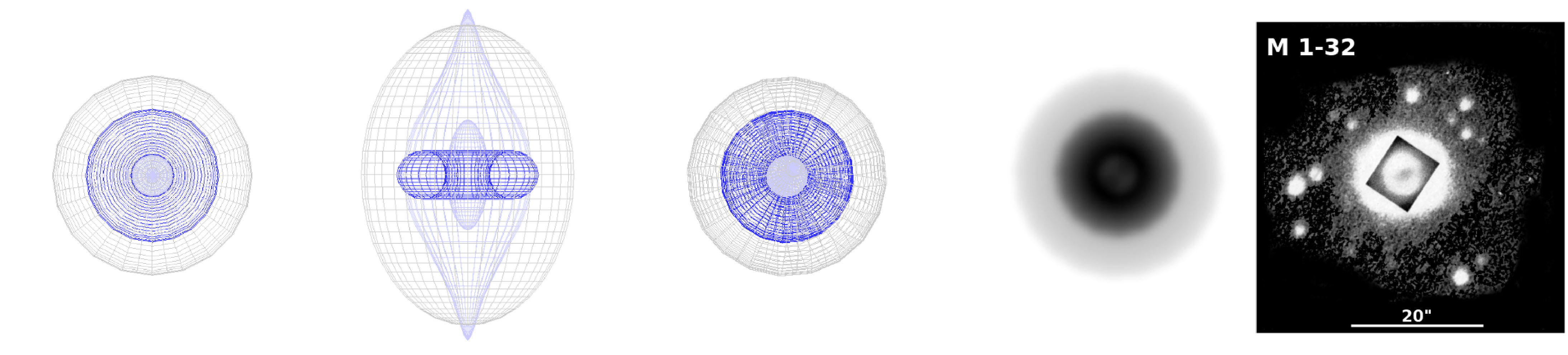}
\figsetgrpnote{\textsc{shape} mesh model for M\,1-32.
}
\figsetgrpend

\figsetgrpstart
\figsetgrpnum{4.6}
\figsetgrptitle{(f) M\,3-15}
\figsetplot{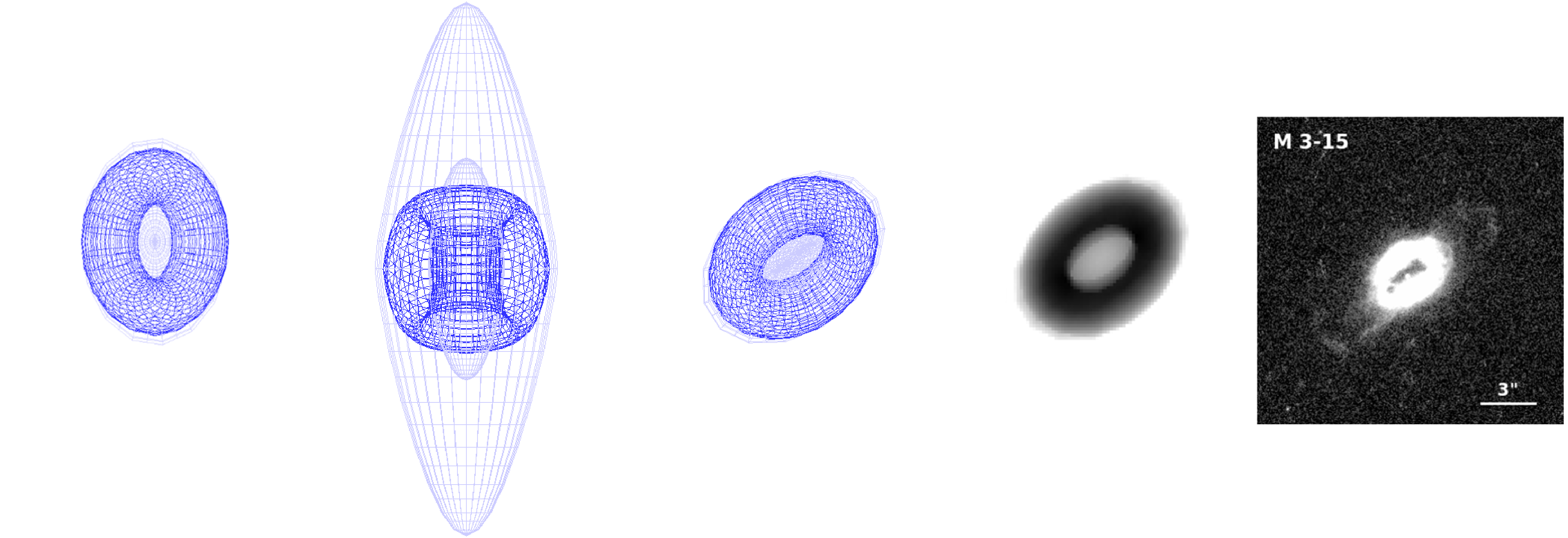}
\figsetgrpnote{\textsc{shape} mesh model for M\,3-15.
}
\figsetgrpend

\figsetgrpstart
\figsetgrpnum{4.7}
\figsetgrptitle{(g) M\,1-25} 
\figsetplot{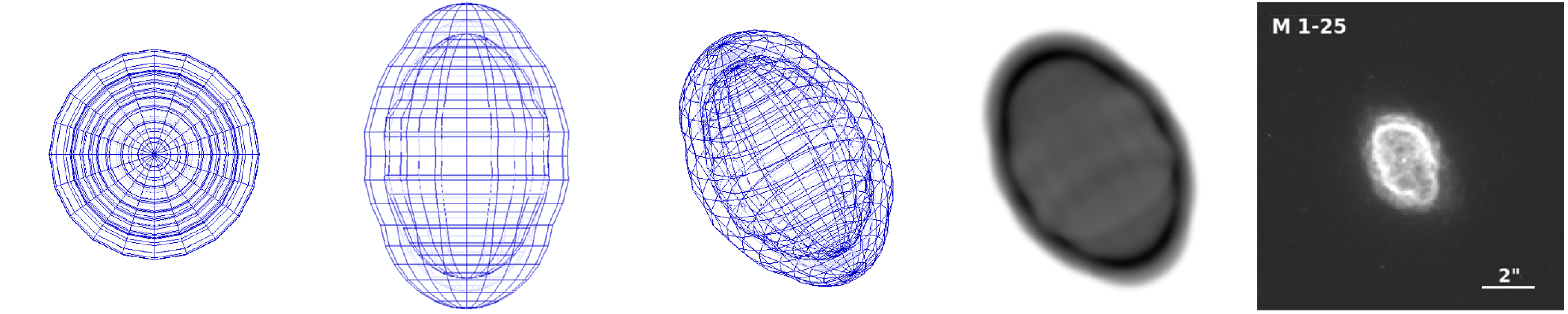}
\figsetgrpnote{\textsc{shape} mesh model for M\,1-25.
}
\figsetgrpend

\figsetgrpstart
\figsetgrpnum{4.8}
\figsetgrptitle{(h) Hen\,2-142}
\figsetplot{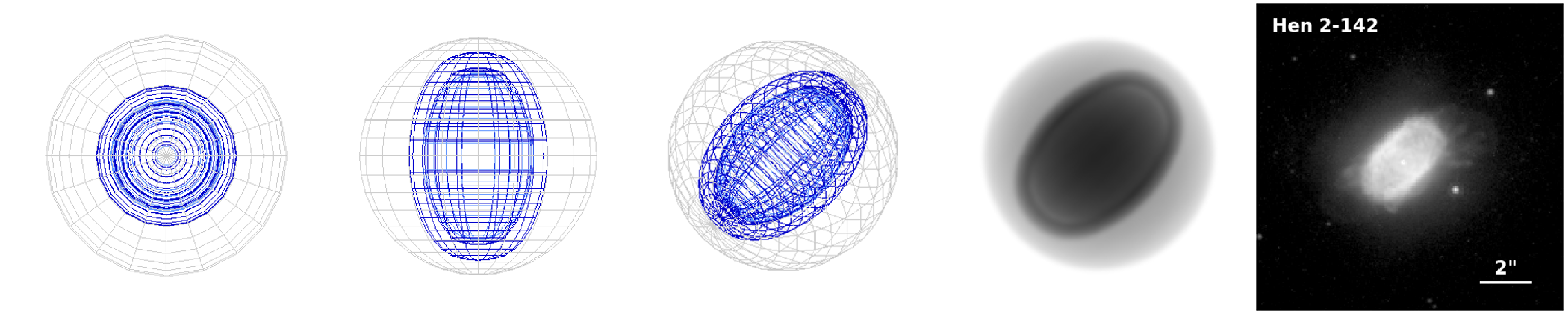}
\figsetgrpnote{\textsc{shape} mesh model for Hen\,2-142.
}
\figsetgrpend

\figsetgrpstart
\figsetgrpnum{4.9}
\figsetgrptitle{(i) K\,2-16} 
\figsetplot{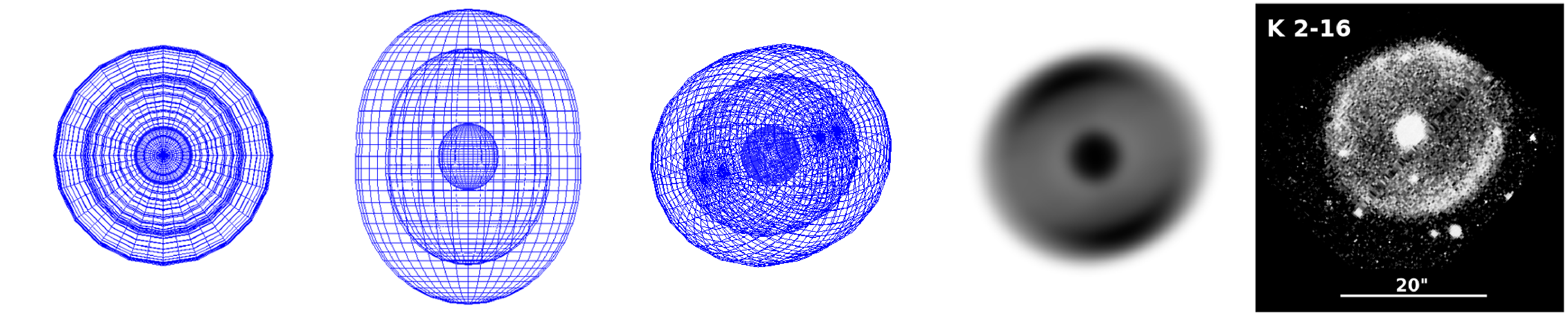}
\figsetgrpnote{\textsc{shape} mesh model for K\,2-16.
}
\figsetgrpend

\figsetgrpstart
\figsetgrpnum{4.10}
\figsetgrptitle{(j) NGC\,6578}
\figsetplot{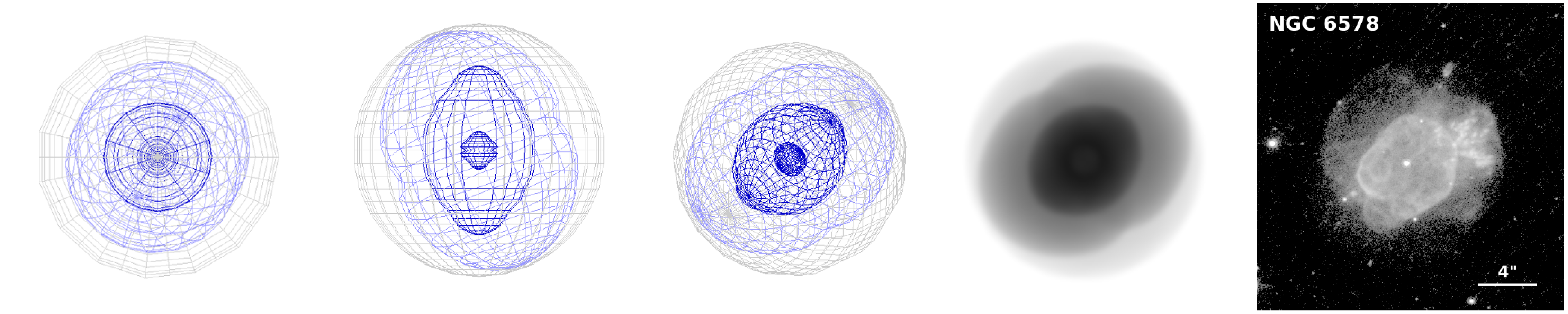}
\figsetgrpnote{\textsc{shape} mesh model for NGC\,6578.
}
\figsetgrpend

\figsetgrpstart
\figsetgrpnum{4.11}
\figsetgrptitle{(k) NGC\,6567}
\figsetplot{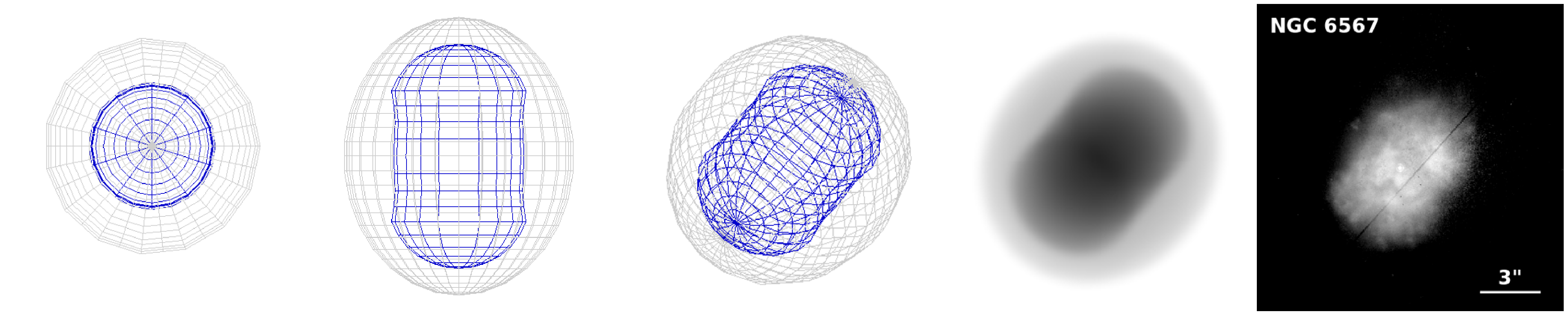}
\figsetgrpnote{\textsc{shape} mesh model for NGC\,6567.
}
\figsetgrpend

\figsetgrpstart
\figsetgrpnum{4.12}
\figsetgrptitle{(l) NGC\,6629}
\figsetplot{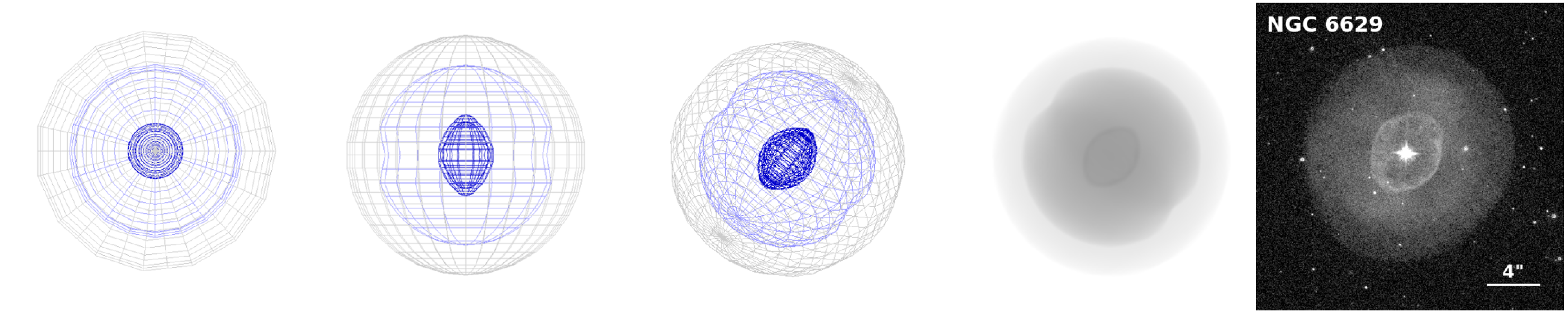}
\figsetgrpnote{\textsc{shape} mesh model for NGC\,6629.
}
\figsetgrpend

\figsetend

\end{figure*}

In Figure~\ref{wc1:vel:slic}, we also present the synthetic velocity-resolved channel maps created by the best-fitting kinematic models aimed at reproducing the velocity channel maps of the H$\alpha$ and [N\,{\sc ii}] emission shown in the same figure. The synthetic P--V diagrams made by the \textsc{shape} models are also presented in Figure~\ref{wc1:pv:diagram} to match the P--V arrays of H$\alpha$ and [N\,{\sc ii}] (also shown in this figure) extracted from the IFU datacube for two slits oriented with the PA along and vertical to the symmetric axis passing through the center star of each object.

Table~\ref{wc1:tab:shapemodel} lists the key parameters of the best-fitting morpho-kinematic models. Columns 3--5 provide the position angle (PA) of the nebular symmetric axis projected onto the sky plane in the equatorial coordinate system, the Galactic position in the Galactic coordinate system, and the inclination ($i$) of the symmetric axis with respect to the line of sight, respectively. Column 6 presents the linear velocity factor ($k_{\rm shell}$; km\,s$^{-1}$\,arcsec$^{-1}$) describing the homologous expansion law  $v$(km\,s$^{-1}$)\,$=k_{\rm shell} \cdot r$(arcsec) of the primary shell. Columns 7--8 give the mass-weighted average expansion velocity of the primary shell ($v_{\rm exp}$) adopted according to the integrated flux HWHM measurements, and the polar expanding velocities of collimated outflows ($v_{\rm out}$) with respect to the central star estimated from the P--V diagrams and velocity channels in H$\alpha$ and [N~{\sc ii}], respectively. For comparison, we also include the kinematic parameters of the PNe Th\,2-A and M\,2-42, which were determined by \citet{Danehkar2015a} and \citet{Danehkar2016}, respectively. The \textsc{shape} models of the 12 PNe studied here, besides Th\,2-A and M\,2-42 previously modeled, before rendering together with the rendered results at their best-fitting inclinations and orientations, are summarized in Figure~\ref{wc1:figures:shape:models}, with fully interactive 3D mesh models in the X3DOM framework \citep{Behr2009} available in the online version of this article.

\begin{figure*}
\begin{center}
\begin{interactive}{js}{figure5.tar.gz}
(a) 3D mesh models\\
\includegraphics[width=5.5in]{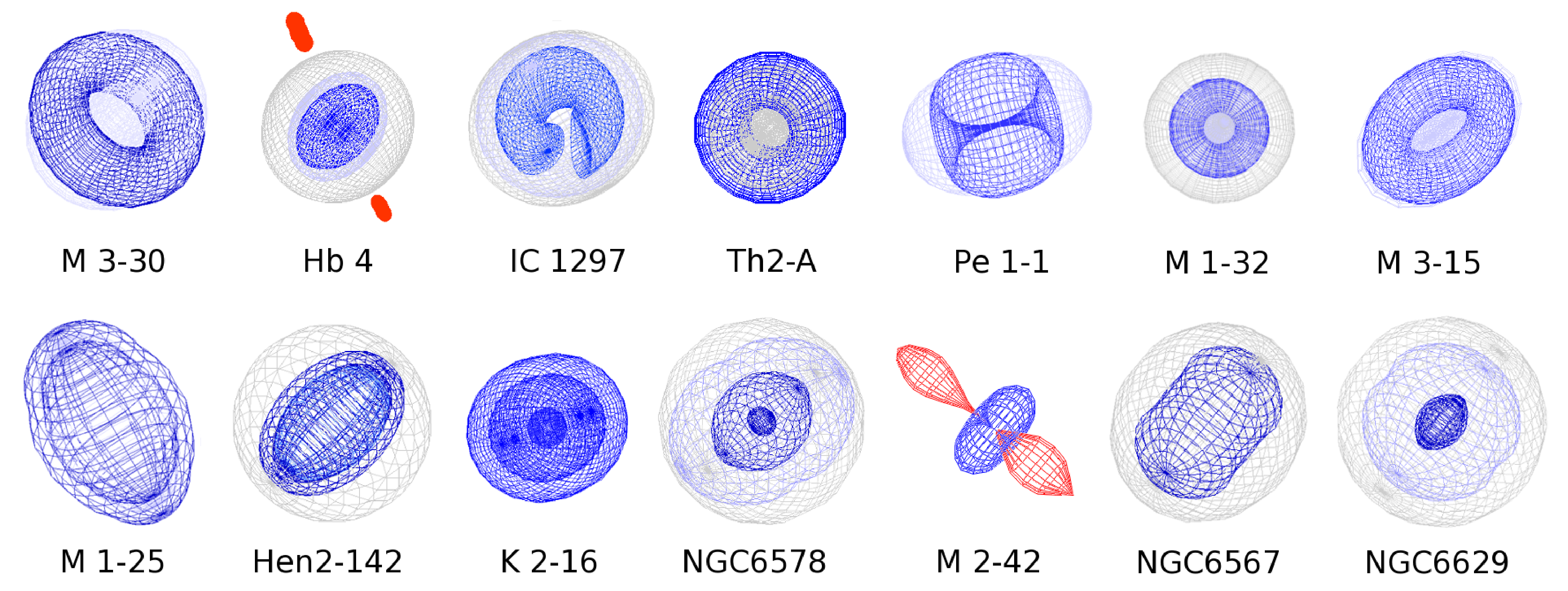}\\~\\
(b) Rendered mesh models\\
\includegraphics[width=5.5in]{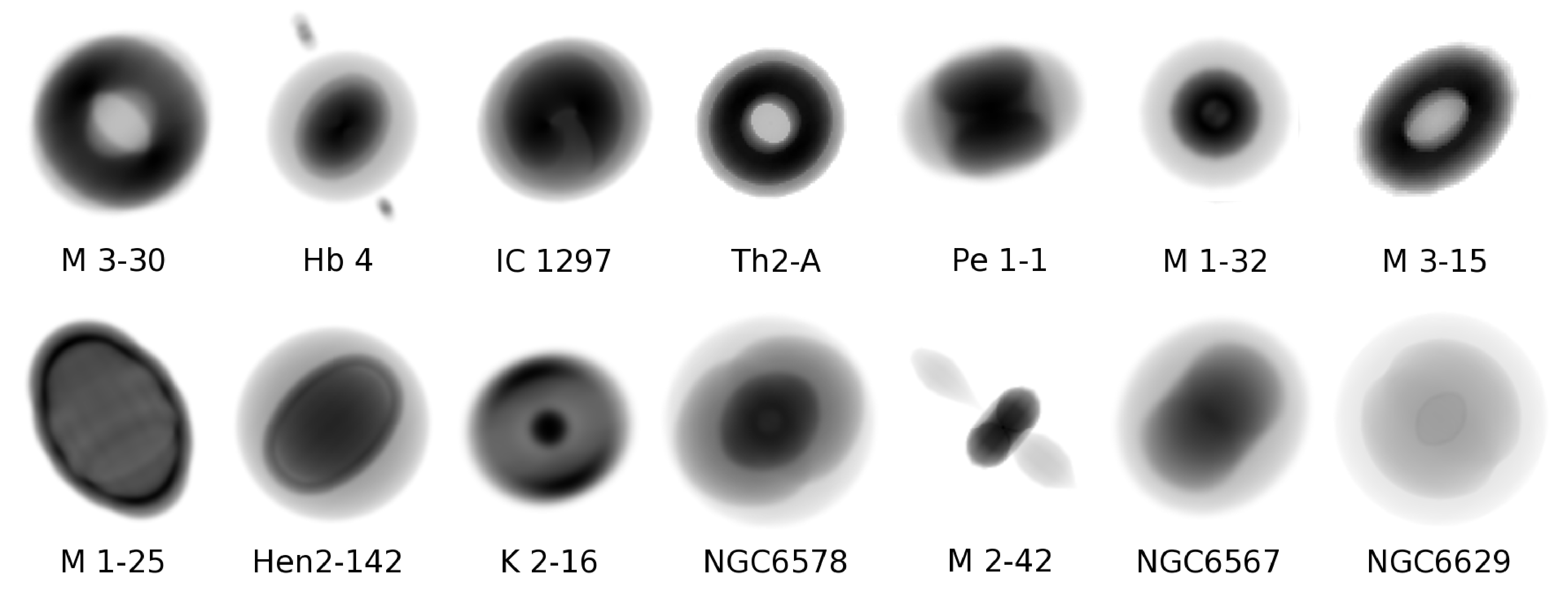}%
\end{interactive}
\caption{Three-dimensional mesh models of the PNe M\,3-30, Hb\,4, IC\,1297, 
Th\,2-A \citep{Danehkar2015a}, Pe\,1-1, M\,1-32, M\,3-15, M\,1-25, Hen\,2-142, K\,2-16, MGC\,6578, M\,2-42 \citep{Danehkar2016}, NGC\,6567 and NGC\,6629, (a) before and (b) after rendering at the best-fitting inclinations and orientations.
The 3D mesh models (14 models in the X3D file format) are available in the interactive figure in the online journal.
}%
\label{wc1:figures:shape:models}%
\end{center}
\end{figure*}

The primary and secondary morphological features determined from our morpho-kinematic models are also summarized in Columns 10 and 11 of Table~\ref{wc1:tab:shapemodel}, respectively. Following \citet{Sahai2011}, the classification codes include four primary classes: bipolar (\textit{B}), elliptical (\textit{E}), multipolar (\textit{M}), and irregular (\textit{I}). The \textit{B} class is defined as objects with two primary, diametrically opposed lobes, centered on their expected location. The \textit{E} class represents objects that are elliptical along a specific axis, and it also comprises ring-shaped or torus morphologies. The \textit{M} class defines objects showing more than one primary lobe pair whose axes are not aligned, such as NGC 5189 \citep[see][]{Danehkar2018a}. The \textit{I} class is defined by objects do not display any geometrical symmetry. The structure may be open or closed at its outer ends, denoted by `\textit{o}' or `\textit{c}', respectively. 
Other morphological features described by secondary descriptors include point symmetry (\textit{ps}), two or more pairs of diametrically opposed lobes (\textit{ps(m)}), diametrically opposed ansae (\textit{ps(an)}), the overall shape of the lobes (\textit{ps(s)}), rings projected on lobes (\textit{rg}), and the central region with a toroidal structure (\textit{t}). 

The \textsc{shape} model of M\,3-30 was constructed using a thick toroidal shell together with a thin prolate ellipsoid describing collimated outflows. For the thin prolate ellipsoid, we employed an inhomogeneous density model defined by a sinusoidal function in the spherical coordinate system, where the density profile decreases as it moves from the polar angle $\phi=90^{\circ}$ (equator) toward $0^{\circ}$ and $180^{\circ}$ (polar) with a slightly higher density reduction in the outflow moving toward us. 
For the PN BD\,+30$^{\circ}$3639 around a [WC9] star, which has P--V diagrams similar to M\,3-30, \citet{Akras2012a} similarly adopted a prolate ellipsoid with thinner and faster polar regions. 
To reproduce the [N~{\sc ii}] observations, a different density law is applied to the toroidal shell, which has a power-law radial distribution with a power-law index of 3, similar to the [N~{\sc ii}] model built for Abell 48 \citep{Danehkar2022}. 
Moreover, the prolate ellipsoid is deactivated to generate the [N~{\sc ii}] synthetic results.
To improve the velocity channels, the thick toroidal shell was also modified by the shear operator in  \textsc{shape}. A sheared toroidal model was built in a similar way by \citet{Jones2010a} to reproduce the P--V arrays of the PN Abell 41. A sheared elliptical shell could be a result of the PN-ISM interaction that occurs as the PN passes though the ISM.  The rendered \textsc{shape} representation of M\,3-30 is akin to the narrow band  H$\alpha$+[N\,{\sc ii}] image from \citet{Schwarz1992}. The toroidal shell seen in the image is similar to the PN Abell 48 \citep{Danehkar2014}, while M\,3-30 also contains collimated outflows. The maximum velocity of the collimated outflows in the P--V diagram of the H$\alpha$ emission corresponds to the outflow velocity of $v_{\rm out}\approx 80$\,km\,s$^{-1}$ at the inclination of $25^{\circ}$ found by our model. The waist expansion velocity of the equatorial shell is about $v_{\rm exp}=36\pm 10$\,km\,s$^{-1}$ similar to $v_{\rm HWHM}= 36 \pm 5$ \,km\,s$^{-1}$ estimated from the integrated emission-line profiles. 

The \textit{HST} observation of Hb\,4 on a linear scale in Figure~\ref{wc1:model:shape} shows that the main shell has a deformed ring-like structure, probably owing to the interaction with the ISM. Moreover, the \textit{HST} observation shows that the point-symmetric knots or filaments are not vertically aligned, which again hints at the ISM interaction. Our IFU observations on a logarithmic scale indicate that the ring-shaped dense shell is also surrounded by a tenuous exterior halo. Accordingly, we used a \textsc{shape} model that consisted of an equatorial dense torus, a tenuous oblate spheroid, and two sets of four point-symmetric knots to reproduce the [N~{\sc ii}] observations. Recently, \citet{Derlopa2019} reconstructed the point-symmetric structures of Hb\,4 using a string of four knots for each of them. Our P--V diagrams imply that the expansion velocity of the dense shell reaches $80\pm 20$\,km\,s$^{-1}$ at the inclination of $40^{\circ}$ found by the morpho-kinematic model, while the mass-weighted average HWHM expansion velocity of $v_{\rm HWHM}= 23 \pm 1$ \,km\,s$^{-1}$ is estimated from the line profiles integrated over an aperture covering the main shell. The deprojected velocities of the bipolar collimated outflows are found to be in the range of $v_{\rm out}= 30$--$160$ \,km\,s$^{-1}$ at $i=40^{\circ}\pm 3^{\circ}$. To match better the H$\alpha$ P--V arrays, we also include a thin oblate ellipsoid around the dense toroidal shell, but it is deactivated in the [N~{\sc ii}] model.

The P--V diagrams reveal the deceleration of the point-symmetric knots in Hb\,4, which were also identified by \citet{Derlopa2019}. This deceleration is not the homologous expansion of typical PNe. We should note that hydrodynamic simulations of bipolar jets and ring-shaped nebulae \citep[see e.g.][]{GarciaSegura2001,Lee2004} and nebular shells \citep{Schonberner2005a} indicate that the radial expansion velocity increases with the distance from the central star. Decelerating bow shocks have been discovered in the Herbig-Haro (HH) object HH34 \citep{Devine1997} and post-AGB water-fountain star IRAS\,18113--2503 \citep{Orosz2019}, which could be due to shock collisions and interactions with ambient media. Point-symmetric knots and LISs in other PNe also demonstrate some signs of shock excitation \citep{Akras2016a} and molecular H$_{2}$ emission \citep{Akras2017,Akras2020} because of the interaction with the ISM or their exterior nebula halos. To model the decelerating features of the point-symmetric structures, we employed two sets of four knots whose velocities linearly decrease with distance from the nebular center. This allowed us to reproduce the knot features in the P--V arrays and velocity channels. Although the knots adopted in our \textsc{shape} model can reproduce our IFU observations, having a reconstructed PSF FWHM of $\sim2$\,arcsec, we caution that the knots could be smaller, according to the high-resolution \textit{HST} observations. The velocity dispersion maps of Hb\,4 also depict higher values in the locations of these knots, which again suggest shock collisions with the surrounding medium.

\begin{table*}
\caption{Key parameters of the best-fitting morpho-kinematic models and morphological classification.
\label{wc1:tab:shapemodel}
}
\centering
\footnotesize
\begin{tabular}{llrrrccllll}
\hline\hline
\noalign{\smallskip}
Name & PNG & \multicolumn{1}{c}{PA} &  \multicolumn{1}{c}{GPA} & \multicolumn{1}{c}{$i$} & $k_{\rm shell}$ & \multicolumn{1}{c}{$v_{\rm exp}$}  & \multicolumn{1}{c}{$v_{\rm out}$}  & \multicolumn{1}{c}{$v_{\rm out}$} & \multicolumn{2}{c}{Morphology}\\
\cline{10-11}
& &\multicolumn{1}{c}{($^{\circ}$)} & \multicolumn{1}{c}{($^{\circ}$)}		& \multicolumn{1}{c}{($^{\circ}$)} &  & \multicolumn{1}{c}{(km/s)}  & \multicolumn{1}{c}{H$\alpha$\,(km/s)} & \multicolumn{1}{c}{[N~{\sc ii}]\,(km/s)}  & \multicolumn{1}{c}{P} & \multicolumn{1}{c}{S} \\
\noalign{\smallskip}
\hline
\noalign{\smallskip}
PB\,6 	& 278.8$+$04.9 &  $80 \pm 7$  &  $45.3 \pm 7$ & \multicolumn{1}{c}{--}    & -- &  $33$    & $\gtrsim 80$\,$^{\mathrm{b}}$  & $\gtrsim 70$\,$^{\mathrm{b}}$  & \textit{M}/\textit{I}?  &   \textit{ps(m)}?\\
\noalign{\smallskip}
M\,3-30	& 017.9$-$04.8 &  $-45 \pm 4$ & $18.2 \pm 4$ & $25 \pm 5$ & 3.1 & $36$ & $80\pm20$ & $45\pm20$  & \textit{E,o} & \textit{rg/t,ps(s)}  \\
\noalign{\smallskip}
Hb\,4	& 003.1$+$02.9 &  $25 \pm 3$ & $83.3 \pm 3$ & $-40 \pm 3$ & 13.2 & $23$ & $30\cdots160$ & $30\cdots160$  & \textit{E,o} & \textit{rg/t,ps(an)}  \\
\noalign{\smallskip}
IC\,1297& 358.3$-$21.6 &  $30 \pm 5$ & $102.0 \pm 5$ & $-145 \pm 6$ & 12.8 & $31$ & $110\pm10$  & $100\pm10$  & \textit{E/I,o} &  {} \\
\noalign{\smallskip}
Th\,2-A\,$^{\mathrm{a}}$	&306.4$-$00.6&  $-45 \pm 5$ & $-38.1 \pm 5$ & $-10 \pm 5$ & 4.0 & $40$ & $90 \pm 20$  & --  & \textit{E,o} & \textit{rg/t,ps(s)}  \\
\noalign{\smallskip}
Pe\,1-1	& 285.4$+$01.5 &  $-75 \pm 4$ & $-104.2 \pm 4$ & $75 \pm 4$ & 51.0 & $23$ & $200\pm 60$ & $200\pm 60$  & \textit{E,o} & \textit{t,ps(s)}   \\
\noalign{\smallskip}
M\,1-32 & 011.9$+$04.2 &  $-40 \pm 6$ & $20.1 \pm 6$ & $4 \pm 3$ & 14.3 & $31$ & $170 \pm 20$ & $170 \pm 20$  & \textit{E,o}  & \textit{t,ps(s)} \\
\noalign{\smallskip}
M\,3-15	& 006.8$+$04.1 &  $130 \pm 4$ & $188.8 \pm 4$ & $4 \pm 4$ & 17.0 & $21$ & $110 \pm 20$ & $70 \pm 20$  & \textit{E,o} & \textit{rg/t,ps(s)}  \\
\noalign{\smallskip}
M\,1-25 & 004.9$+$04.9 &  $30 \pm 5$ & $88.0 \pm 5$ & $65 \pm 3$ & 65.0 & $27$ & $190\pm 50$  & $190\pm 50$  & \textit{E,c}  & { }  \\
\noalign{\smallskip}
Hen\,2-142& 327.1$-$02.2 &  $-45 \pm 3$ & $-4.3 \pm 3$ & $-60 \pm 3$ & 45.0 & $20$ & $200\pm 40$ & $200\pm 40$  & \textit{E,c}  & { }  \\
\noalign{\smallskip}
K\,2-16	& 352.9$+$11.4 &  $110 \pm 6$ & $160.6 \pm 6$ & $-30 \pm 5$ & 5.7 & $31$ & $70\pm10$ & $70\pm10$  & \textit{E,c/o} & \textit{ }  \\
\noalign{\smallskip}
NGC\,6578 	& 010.8$-$01.8 &  $-48 \pm 6$ & $13.6 \pm 6$ & $-45 \pm 5$ & 12.4 & $22$ & $110\pm20$ & $80\pm20$  & \textit{E,c} & \\
\noalign{\smallskip}
M\,2-42\,$^{\mathrm{a}}$ 	& 008.2$-$04.8	&  $50 \pm 5$ & $112.4 \pm 5$ & $-82 \pm 4$ & 27.0 & $20$ & --  & $120 \pm 40$  & \textit{E,o} & \textit{rg/t,ps(an)}  \\
\noalign{\smallskip}
NGC\,6567 	& 011.7$-$00.6 &  $-40 \pm 3$ & $21.4 \pm 3$ & $46 \pm 3$ & 29.0 & $34$ & $190\pm30$ & $100\pm30$  & \textit{E,c} & \\
\noalign{\smallskip}
NGC\,6629 	& 009.4$-$05.0 &  $-40 \pm 3$ & $22.6 \pm 3$ & $60 \pm 4$ & 8.7 & $20$ & $160\pm40$ & $100\pm40$  & \textit{E,c} & { }  \\
\noalign{\smallskip}
Sa\,3-107 & 358.0$-$04.6 &  $30 \pm 8$ & $90.5 \pm 8$ & \multicolumn{1}{c}{--}  & -- & $16$ & $\gtrsim 100$\,$^{\mathrm{b}}$ & $\gtrsim 50$\,$^{\mathrm{b}}$  & {\textit{B}/\textit{E}?} & { } \\
\noalign{\smallskip}
\hline
\end{tabular}
\begin{list}{}{}
\item[$^{\mathrm{a}}$]The results of Th\,2-A and M\,2-42 are from \citet{Danehkar2015a} and \citet{Danehkar2016}, respectively.
\item[$^{\mathrm{b}}$]The deprojected outflow velocities are not available for PB\,6 and Sa\,3-107.
\item[\textbf{Note.}]The morphological classification codes in Columns 10 and 11 are based on \citet{Sahai2011}. The symbol `?' means that further observation is required to confirm the structure. 
The deprojected outflow velocities ($v_{\rm out}$) of the collimated outflows or polar expansions are with respect to the central stars. The mass-weighted average expansion velocities ($v_{\rm exp}$) are from the HWHM measurements of the integrated emission-line profiles (see Table~\ref{wc1:tab:kinematics1}). The linear factor $k_{\rm shell}$(km\,s$^{-1}$\,arcsec$^{-1}$) of the homologous velocity law, $v$(km\,s$^{-1}$)\,$=k_{\rm shell} \cdot r$(arcsec), corresponds to the primary shell in each model.
\end{list}
\end{table*}

Based on the narrow-band H$\alpha$+[N\,{\sc ii}] image of IC\,1297 \citep{Schwarz1992,Corradi1996}, a broken thick elliptic torus and a thin prolate spheroid are used to assemble a 3D model for this object. Additionally, another shifted, rotated tenuous prolate spheroid is included in the H$\alpha$ kinematic model to recreate the high H$\alpha$ expansion toward us projected mostly in the southern part. The velocity dispersion maps exhibit higher values in the southern part of the nebula, where a part of the torus is missing. High dispersion could be a sign of the ISM interaction or mass-accretion to a companion. Although the radial velocity map of IC\,1297 is similar to that of M\,3-30, the velocity channel slices and P--V arrays do not suggest the same kinematic structure. The H$\alpha$+[N\,{\sc ii}] image depicts a broken ring-shaped nebula with the same on-sky orientations seen in the WiFeS velocity maps. A maximum polar expansion velocity of $v_{\rm out}=100\pm10$\,km\,s$^{-1}$ is determined from the [N\,{\sc ii}] P--V diagrams, while a mass-weighted expansion velocity of $v_{\rm exp}=31\pm1$\,km\,s$^{-1}$ is pointed out by the HWHM measurements.

The \textit{HST} image of Pe\,1-1 (Figure~\ref{wc1:model:shape}) allowed us to mold a 3D morphological model for this object. Moreover, the IFU velocity dispersion maps in Figure~\ref{wc1:ifu_map} manifest two regions with possible shock collisions on both sides of the nebula, similar to the point-symmetric knots seen in the dispersion maps of Hb\,4. The high values of the velocity dispersion in Pe\,1-1 could be due to the interaction of collimated outflows with the ambient medium. We therefore included a tenuous prolate ellipsoid associated with bipolar outflows protruding from both sides of a dense torus. 

To model M\,1-32, we exploited an equatorial thick torus resembling its H$\alpha$+[N\,{\sc ii}] image (see Figure~\ref{wc1:model:shape}), and a faint bi-conical shape reproducing collimated bipolar outflows reaching $v_{\rm out}=170\pm20$\,km\,s$^{-1}$ relative to the central star in the P--V arrays (see Figure~\ref{wc1:pv:diagram}). Additionally, a tenuous exterior halo was included to match the velocity channels, resulting in a model similar to \citet{Akras2012}. The velocity dispersion maps depict high values in the central region analogous to that of Th\,2-A, which is similarly viewed almost pole-on \citep{Danehkar2015a}. Those high values in the dispersion maps are produced by high-velocity collimated bipolar outflows with a very low inclination ($i=4^{\circ}$). \citet{Akras2012} also estimated $v_{\rm out}= 200$\,km\,s$^{-1}$ relative to the central star.
The HWHM measurements of the integrated emission-line profiles also yield a mean expansion velocity of $31\pm 3$\,km\,s$^{-1}$. 

The \textsc{shape} model of M\,3-15 was fabricated using a stretched, sheared torus based on the \textit{HST} observation (Figure~\ref{wc1:model:shape}), and a thin prolate ellipsoid associated with collimated bipolar outflows with $v_{\rm out}=
110$ and $70$\,km\,s$^{-1}$ according to the P--V diagrams (Figure~\ref{wc1:pv:diagram}) for H$\alpha$ and [N\,{\sc ii}], respectively. 
Two dissimilar linear velocity factors are employed in the velocity law of the prolate ellipsoid to generate the synthetic H$\alpha$ and [N\,{\sc ii}] results. 
The main shell in this PN could be deformed similar to M\,3-30 and Abell 41 \citep{Jones2010a} owing to the ISM interaction. However, we caution that our WiFeS spatial resolution is insufficient for properly constraining the ring-shaped shell of this compact ($\leq 6$\,arcsec) PN, so our 3D model is one of several possible solutions that can be achieved by modifying the main shell and its orientation. \citet{Akras2012} also found outflow velocities of $100$ \,km\,s$^{-1}$ in a long-slit observation of M\,3-15. Our inclination of $4^{\circ}$ with respect to the line of sight is also in agreement with \citet{Akras2012}. The \textit{HST} image of this PN looks very similar to the PN Vy\,1-2 with [WR] or \textit{wels}, which has a similar morphology seen almost pole-on with an inclination angle of $10^{\circ}$ \citep{Akras2015}.

\begin{table*}
\caption{Nebular evolutionary ages obtained from distances, sizes, and expansion velocities.
\label{wc1:tab:age}
}
\centering
\footnotesize
\begin{tabular}{lllccccccccccccc}
\hline\hline
\noalign{\smallskip}
Name & PNG & Optical Angular & Distance & $r_{\rm opt}$ & $r_{\rm out}$ & $r_{\rm out}$ && \multicolumn{2}{c}{$\tau_{\rm mean}$(yr)} && \multicolumn{2}{c}{$\tau_{\rm out}$\,H$\alpha$\,(yr)} && \multicolumn{2}{c}{$\tau_{\rm out}$\,[N~{\sc ii}]\,(yr)} \\
\cline{9-10}\cline{12-13}\cline{15-16}
    &     &    Dimensions (arcsec$^2$)      & (pc)     & (pc) & H$\alpha$(pc) & [N~{\sc ii}](pc) && D96 & G00 && D96 & G00 && D96 & G00 \\
\noalign{\smallskip}
\hline
\noalign{\smallskip}
M\,3-30 & 017.9$-$04.8 & $19.2\times18.5$\,(T03) & 4536\,(S10) & 0.207 & 0.311 & 0.207 && 8440 & 7490 &&5700	&5050	&&6750	&5990\\
\noalign{\smallskip}
Hb\,4 & 003.1$+$02.9 & $11.4\times7.4$\,(T03) &5096\,(S10) & 0.113 & 0.415 & 0.415 && 7240 & 6420 &&3810	&3380	&&3810	&3380\\
\noalign{\smallskip}
IC\,1297 & 358.3$-$21.6 & $10.9\times9.9$\,(T03) & 4100\,(T98) & 0.103 & 0.145 & 0.145 && 4890 & 4330 &&1930	&1710	&&2120	&1880\\
\noalign{\smallskip}
Th\,2-A & 306.4$-$00.6 & $27.7\times25.2$\,(T03) & 2500\,(S10) & 0.160 & 0.304 & -- && 5870 & 5210 &&4960	&4400	&&	-- & -- \\
\noalign{\smallskip}
Pe\,1-1 & 285.4$+$01.5 & $3.0\times3.0$\,(A92) & 6569\,(S10) & 0.048 & 0.067 & 0.067 && 3050 & 2700 &&490	&440	&&490	&440\\
\noalign{\smallskip}
M\,1-32 & 011.9$+$04.2 & $9.4\times8.3$\,(T03) & 4796\,(S10) & 0.103 & 0.257 & 0.257 && 4860 & 4310 &&2220	&1960	&&2220	&1960\\
\noalign{\smallskip}
M\,3-15 & 006.8$+$04.1 & $4.2\times4.2$\,(A92) & 6825\,(S10) & 0.069 & 0.222 & 0.118 && 4850 & 4300 &&2970	&2630	&&2480	&2200\\
\noalign{\smallskip}
M\,1-25 & 004.9$+$04.9 & $4.6\times4.6$\,(A92) & 6682\,(S10) & 0.075 & 0.112 & 0.112 && 4050 & 3590 &&860	&770	&&860	&770\\
\noalign{\smallskip}
Hen\,2-142 & 327.1$-$02.2 & $4.4\times3.5$\,(T03) & 4620\,(Z95) & 0.044 & 0.066 & 0.066 && 3220 & 2860 &&480	&430	&&480	&430\\
\noalign{\smallskip}
K\,2-16 & 352.9$+$11.4 & $26.6\times24.3$\,(T03) & 2200\,(T98) & 0.136 & 0.176 & 0.176 && 6420 & 5690 &&3690	&3280	&&3690	&3280\\
\noalign{\smallskip}
NGC\,6578 & 010.8$-$01.8 & $12.1\times11.8$\,(T03) & 3680\,(S10) & 0.107 & 0.149 & 0.149 && 7110 & 6300 &&1990	&1760	&&2740	&2430\\
\noalign{\smallskip}
M\,2-42 & 008.2$-$04.8 & $4.0\times4.0$\,(S08) & 7400\,(D16) & 0.072 & -- & 0.151 && 5260 & 4670 &&	-- & -- &&1840	&1630\\
\noalign{\smallskip}
NGC\,6567 & 011.7$-$00.6 & $8.1\times6.4$\,(T03) & 3652\,(S10) & 0.064 & 0.127 & 0.105 && 2750 & 2440 &&980	&870	&&1540	&1370\\
\noalign{\smallskip}
NGC\,6629 & 009.4$-$05.0 & $16.6\times15.5$\,(T03) & 2399\,(S10) & 0.093 & 0.149 & 0.117 && 8550 & 7580 &&1370	&1210	&&1710	&1520\\
\noalign{\smallskip}
\hline
\end{tabular}
\begin{list}{}{}
\item[\textbf{Note.}]
The actual evolutionary ages estimated from $v_{\rm exp}$ and $v_{\rm out}$ as listed in Table~\ref{wc1:tab:shapemodel} using the $1.5 \times$ dynamical age derived by \citet[][D96]{Dopita1996} and the initial expansion velocity of $v_{\rm init} = 0.5 v$ assumed by \citet[][G00]{Gesicki2000}, discussed in text.  A distance to M\,2-42 was derived by \citet{Danehkar2016} using the H$\alpha$ surface brightness-radius relation of \citet{Frew2016}.
References for diameters and distances are as follows: 
A92 -- \citet{Acker1992}; 
D16 -- \citet{Danehkar2016}; 
S08 -- \citet{Stanghellini2008}; 
S10 -- \citet{Stanghellini2010};
T03 -- \citet{Tylenda2003};
T98 -- \citet{Tajitsu1998};
Z95 -- \citet{Zhang1995}.
\end{list}
\end{table*}

K\,2-16's \textsc{shape} model is made up of an elliptical shell and a thinner interior prolate spheroid with nonuniform density laws (defined with $\sin \phi$) decreasing from the equator ($\phi=90^{\circ}$) toward the polar direction ($\phi=0^{\circ}$ and $180^{\circ}$). A small interior sphere was also included to better match the P--V arrays. This model reproduces the P--V diagrams in the H$\alpha$ emission (see Figure~\ref{wc1:pv:diagram}), while the lower part of the P--V diagram, extracted from the slit over the symmetric axis, was not covered by the WiFeS footprint. Taking the inclination of $-30^{\circ}$ used by the 3D model, we derived a maximum polar expansion velocity of $v_{\rm out}=70\pm10$\,km\,s$^{-1}$. The HWHM measurements also point to a mean expansion velocity of $v_{\rm exp}= 31\pm3$\,km\,s$^{-1}$.

The high-resolution \textit{HST} images of M\,1-25, Hen\,2-142, NGC\,6578, NGC\,6567, and NGC\,6629 disclose their elliptically symmetric morphologies. Although the spatial resolution of our IFU observations did not provide much detail of their elliptical morphologies, the IFU velocity maps display their on-sky orientations similar to what are seen in the \textit{HST} images. Their 3D morpho-kinematic models (Figure~\ref{wc1:model:shape}) were constructed with modified prolate ellipsoids in a way where the rendered images resemble their inherent shapes, as seen in the \textit{HST} images. We also included tenuous exterior halos in Hen\,2-142, NGC\,6578, and NGC\,6629, which are visible in the \textit{HST} observations. 
An exterior halo was utilized only for the H$\alpha$ kinematic model of NGC 6567. 
Similarly, second larger exterior halos were added to the H$\alpha$ models of NGC 6578 and NGC 6629, which are able to 
reproduce the H$\alpha$ kinematic observations.
The orientations and inclinations of these models were constrained by the P--V arrays (Figure~\ref{wc1:pv:diagram}) and velocity-resolved channels (Figure~\ref{wc1:vel:slic}). 


\subsection{Nebular Evolutionary Ages}
\label{wc1:sec:kinematic:maps}

The expansion velocity $v$ and radius $r$ can be used to estimate the dynamical age 
$t = r/v$ of each nebular component.
However, a semi-empirical analysis by \citet[][D96]{Dopita1996} suggested that
the actual evolutionary age should be 1.5 times greater than 
the dynamical age, i.e. $\tau = 1.5 ( r/v)$.
Moreover, \citet[][G00]{Gesicki2000} assumed an AGB velocity of $v_{\rm init} = 0.5 v$ as the initial value, resulting in a mean explanation velocity of $(v+v_{\rm init})/2$ and a true dynamical age of 
$\tau = 1.33 (r/v)$.

The nebular evolutionary ages are presented in Table~\ref{wc1:tab:age}, including the optical angular dimensions $a\times b$ (Column 3),
the distance (Column 4), the optical radius $r_{\rm opt}= \sqrt{a b}/2$ (Column 5) associated with the optical angular dimensions at the distance $D$,
and the radial distances $r_{\rm out}$ of the collimated outflow or polar expansion from the central star based on the H$\alpha$ and [N~{\sc ii}] kinematic models (Columns 6 and 7).
The mean actual dynamical ages $\tau_{\rm mean}$ of the main structure given in Columns 8 and 9 were calculated using the optical radius $r_{\rm opt}$ and the mass-weighted average HWHM expansion $v_{\rm exp}$ from Table~\ref{wc1:tab:shapemodel} under the different assumptions, namely $\tau_{\rm mean} = 1.5\, r_{\rm opt}/v_{\rm exp}$ (D96) and $
\tau_{\rm mean} = 1.33\, r_{\rm opt}/v_{\rm exp}$ (G00). 
Similarly, the actual dynamical ages $\tau_{\rm out}$ of the bipolar outflow or polar expansions (Columns 10--13) were determined from the radial distance $r_{\rm out}$ and the deprojected outflow velocities $v_{\rm out}$ (listed in Table~\ref{wc1:tab:shapemodel}) for the H$\alpha$ and [N~{\sc ii}] models. 

We note that the optical radius $r_{\rm opt}$ based on the optical angular dimensions (typically measured at 10-20\% nebular brightness in the literature) may not be the same as the shell radius used in the \textsc{shape} model, which was constructed in a way to replicate the observed P--V arrays and velocity channels. For example, the outer radius of 6 arcsec is employed for the primary shell in the 3D model of Hb\,4, corresponding to $r_{\rm out}=0.148$\,pc, whereas $r_{\rm opt}=0.113$\,pc according to the optical angular dimensions. This is consistent with a radius of 0.8 times the outer radius assumed by \citet{Gesicki2000} in order to account for the velocity gradient in the age calculation using a mass-weighted average velocity. The adopted optical radius and mass-weighted HWHM expansion velocity estimated from the integrated H$\alpha$ emission might yield the average age of the main nebula, while the collimated outflows and point-symmetric structure could be younger than the primary shell.

The true dynamical ages of the point-symmetric structures in Hb\,4 were calculated using the maximum deprojected outflow velocity of $v_{\rm out}=160$\,km\,s$^{-1}$ at the radial distance $r_{\rm out}=0.415$\,pc (16.8 arcsec) being projected onto 10.8 arcsec from the central star where the nearest knots are located. We obtain the mean evolutionary age of $\tau_{\rm mean} = 6420$\,yr (G00) at $D=5096$\,pc \citep{Stanghellini2010}, which is lower than the kinematic age of 3190\,yr estimated by \citet{Derlopa2019} at $D=2.88$\,kpc. Moreover, \citet{Derlopa2019} derived a kinematic age of around 890\,yr for the fastest knots ($v_{\rm out}=160$\,km\,s$^{-1}$), whereas we get a true dynamical age of $\tau_{\rm out} = 3380$\,yr (G00).  These discrepancies can be explained by dissimilar distances and the AGB velocity $v_{\rm init} = 0.5 v_{\rm out}$ assumed in our dynamical age calculation. Although the point-symmetric knots in Hb\,4 may not be as young as what was found by \citet{Derlopa2019}, they could still be around 3000\,yr younger than the primary shell. 

It can be seen that the mean ages $\tau_{\rm mean}$  estimated from the mass-weighted average expansion velocities are not always the same as the outflow ages $\tau_{\rm out}$ estimated from the outflow or polar expansion velocities. 
Although these values are close in the PN\,Th 2-A, the outflow ages $\tau_{\rm out}$ are 1500 to 7200 years younger than the mean ages $\tau_{\rm mean}$ in other objects. In particular, the dynamical ages of the collimated outflows are about 500\,yr in Pe\,1-1 and Hen\,2-142, which are lower than the mean ages of $\gtrsim 2700$\,yr. Furthermore, the dynamical ages of the polar H$\alpha$ expansion were found to be slightly below 1000\,yr in M\,1-25 and NGC\,6567. This could be an indication of some recent outbursts from the post-AGB stars. However, the mean age of a nebula may be overestimated by using a mass-weighted average velocity, so the actual age could be lower.

\section{Conclusions and Discussion}
\label{wc1:sec:conclusions}

We have studied the morpho-kinematic features of a sample of 14 Galactic PNe surrounding [WR]-type and \textit{wels} nuclei, using spatially resolved kinematic maps and position--velocity profiles attained from WiFeS integral field observations. The kinematic IFU data, together with archival high-resolution images, have been deployed to carry out 3D morpho-kinematic modeling of 12 PNe. Our results imply that these 12 objects mostly have elliptical axisymmetric morphologies, with either open or closed outer ends. Collimated outflows are found to be present in the PNe Hb\,4, Pe\,1-1, M\,3-30, M\,1-32, and M\,3-15, while point-symmetric knots or jets are detached in Hb\,4. Axisymmetric morphologies and collimated outflows have been recognized in other PNe around [WR] stars \citep{Akras2012}. Some elliptical PNe with FLIERs were also observed to have ``diffuse X-ray'' sources \citep{Kastner2012}. Other aspherical PNe around [WR] stars, such as BD\,+30$^{\circ}$3639 \citep{Arnaud1996}, NGC\,40 \citep{Montez2005}, and NGC\,5315 \citep{Kastner2008}, have been reported to contain diffuse X-ray sources and hot bubbles. The soft X-ray emission could be associated with collimated jets and wind-wind shocks \citep{Kastner2012}, whereas the hard X-ray emission may be an indication of the mass accretion into a binary companion. 

The PNe in our sample have elliptically symmetric morphologies, apart from PB\,6, IC\,1297, and Sa\,3-107, which could imply a possible link between the nebulae and their WR-type nuclei. PB\,6 around a [WO\,1] star likely has a complex morphology containing multi-scale structures similar to NGC\,5189 with a [WO\,1] star \citep{Danehkar2018a}. IC\,1297 also has an elliptical broken ring, while the missing ring part exhibits high velocity dispersion, which could be due to either the ISM interaction or mass-accretion into a companion. Although we do not have any imaging observations of Sa\,3-107, its IFU velocity maps suggest an axisymmetric morphology. The theoretical evolutionary models by \citet{Schonberner2005a,Schonberner2005b} predict that the expansion velocity increases as the nebula evolves and the central star becomes hotter. However, we cannot distinguish any link between the nebular kinematics and stellar characteristics listed in Table~\ref{wc1:tab:obs:journal}. Some PNe with hot stars such as Pe\,1-1 and Hb\,4 have low mass-weighted average expansion velocities, while some PNe around cool stars such as K\,2-16 demonstrate higher values. This tendency does not seem to be explained by single star evolution, since the mean nebular expansion velocities are not linked to their stellar parameters. Moreover, collimated bipolar outflows were found in some objects, Hb\,4, M\,3-30, M\,1-32, and M 3-15, reaching outflow velocities in the rang of 60--200\,km\,s$^{-1}$ with respect to the central stars, which cannot be expected from the GISW model. 

Recent hydrodynamic simulations demonstrate that binary stellar evolution can shape axisymmetric morphologies \citep[e.g.][]{Akashi2015,Chen2016,Akashi2016,Garcia-Segura2018,Garcia-Segura2021,Zou2020,Castellanos-Ramirez2021}. It is worthwhile mentioning that magnetohydrodynamic simulations of the envelope of a single AGB star could not support bipolar PNe at physically possible rotating speeds \citep{Garcia-Segura2014}. However, 3D hydrodynamic simulations of a binary system reproduced the wide-bipolar lobes of an AGB Mira-like variable that may be in transition to a PN \citep{Chen2016}. Recently, 2D hydrodynamic simulations of a system that has undergone a CE phase by \citet{Garcia-Segura2018} were able to generate elliptical shapes similar to M\,1-25, Hen\,2-142, and NGC\,6578, as well as ring-shaped elliptical morphologies with a pair of external bipolar point-symmetric structures, similar to Hb\,4 and M\,2-42. Additionally, 3D  hydrodynamic CE simulations illustrate how the interaction between CE ejecta and spherical wind from the post-AGB star results in the formation of highly collimated outflows \citep{Zou2020}. The implication of triple systems was also investigated by \citet{Akashi2017} through 3D hydrodynamic simulations that generate highly asymmetrical filamentary structures similar to those seen in PB\,6 and NGC\,5189.

As confirmed by \textit{HST} imaging observations, our WiFeS integral field observations reveal the on-sky orientations of the elliptical morphologies of the interior shells in the PNe NGC\,6578 and NGC\,6629, as well as the compact ($\leq 6$\,arcsec) PNe Pe\,1-1, M\,3-15, M\,1-25, Hen\,2-142, and NGC\,6567. Although the WiFeS spatial resolution is inadequate for unveiling morphological details of compact objects with angular diameters of less than 6 arcsec, it can help us unmask the on-sky orientations of overall morphologies, as previously shown by \citet{Danehkar2015}. This study demonstrates that the WiFeS has the capability to conduct an extensive IFU survey of numerous objects where the precise morphological details are unimportant.

Future high-resolution imaging and kinematic observations of other PNe around [WR] and \textit{wels} nuclei will definitely shed light on the mechanism that shapes the axisymmetric morphologies in these objects. In-depth spectroscopy and lightcurve monitoring of their central stars will help us gain a better understanding of them. The presence of binary companions should be investigated as they can be responsible for axisymmetric morphologies. Nevertheless, an aspherical morphology can also be formed by an exoplanet whose existence is extremely difficult to verify. Further studies of [WR] CSPNe will unravel the puzzle of elliptical and bipolar morphologies in PNe. 

\begin{acknowledgments}

This work is partially based on a dissertation by the author \citep{Danehkar2014b}, supported by an international Macquarie University Research Excellence Scholarship (iMQRES) and a Sigma Xi Grant-in-Aid of Research (GIAR). The author wishes to thank the referee for useful suggestions and comments, Quentin~Parker for supporting the 2010 ANU observations, Milorad Stupar for his guidance on \textsc{iraf} data reduction, Wolfgang Steffen for helpful discussions and guidance on \textsc{shape}, Nico Koning for upgrading \textsc{shape} and support, David~Frew for helping with the observing proposal, Kyle~DePew for carrying out the ANU observations in 2010, and the ANU telescope time allocation committee and the staff at the ANU Siding Spring Observatory for their support. Based on observations made with the NASA/ESA \textit{Hubble Space Telescope}, obtained from the Data Archive at the Space Telescope Science Institute, which is operated by the Association of Universities for Research in Astronomy, Inc., under NASA contract NAS 5-26555. 

\end{acknowledgments}



\software{IDL Astronomy User's Library \citep{Landsman1993}, IDL Coyote Library \citep{Fanning2011}.}
   
\facilities{ATT (WiFeS), \textit{HST}}   


\begin{appendix}


\section{Supplementary Data}
\label{appendix:a}

\noindent The following figure sets and interactive figures are available for the electronic edition of this article:
\newline
\textbf{Fig. Set~\ref{wc1:ifu_map}.} The spatial distribution maps of logarithmic flux intensity, continuum, LSR velocity, and velocity dispersion of the H$\alpha$ $\lambda$6563 and [N\,{\sc ii}] $\lambda$6584 emission lines for: (a) PB\,6, (b) M\,3-30, (c) Hb\,4, (d) IC\,1297, (e) Pe\,1-1, (f) M\,1-32, (g) M\,3-15, (h) M\,1-25, (i) Hen\,2-142, (j) K\,2-16, (k) NGC\,6578, (l) NGC\,6567, (m) NGC\,6629, and (n) Sa\,3-107.
\newline
\textbf{Fig. Set~\ref{wc1:vel:slic}.} Velocity slices along the H$\alpha$ $\lambda$6563 and [N\,{\sc ii}] $\lambda$6584 emissions for: (a) PB\,6, (b) M\,3-30, (c) Hb\,4, (d) IC\,1297, (e) Pe\,1-1, (f) M\,1-32, (g) M\,3-15, (h) M\,1-25, (i) Hen\,2-142, (j) K\,2-16, (k) NGC\,6578, (l) NGC\,6567, (m) NGC\,6629, and (n) Sa\,3-107, followed by the associated synthetic velocity-resolved channel maps from the morpho-kinematic model of all the PNe, except for PB\,6 and Sa\,3-107.
\newline
\textbf{Fig. Set~\ref{wc1:pv:diagram}.} P--V arrays in the H$\alpha$ $\lambda$6563 and [N\,{\sc ii}] $\lambda$6584 emissions for: (a) PB\,6, (b) M\,3-30, (c) Hb\,4, (d) IC\,1297, (e) Pe\,1-1, (f) M\,1-32, (g) M\,3-15, (h) M\,1-25, (i) Hen\,2-142, (j) K\,2-16, (k) NGC\,6578, (l) NGC\,6567, (m) NGC\,6629, and (n) Sa\,3-107, followed by the associated synthetic P--V diagrams from the morpho-kinematic model of all the PNe, except for PB\,6 and Sa\,3-107.
\newline
\textbf{Fig. Set~\ref{wc1:model:shape}.} \textsc{shape} mesh models for: (a) M\,3-30, (b) Hb\,4, (c) IC\,1297, (d) Pe\,1-1, (e) M\,1-32, (f) M\,3-15, (g) M\,1-25, (h) Hen\,2-142, (i) K\,2-16, (j) NGC\,6578, (k) NGC\,6567, and (l) NGC\,6629, along with archival \textit{HST} images or narrow-band H$\alpha$+[N\,{\sc ii}] images with the 3.5-m ESO NTT from \citet{Schwarz1992}.
\newline
\textbf{Fig.~\ref{wc1:figures:shape:models}.} 3D mesh models of the PNe M\,3-30, Hb\,4, IC\,1297, 
Th\,2-A, Pe\,1-1, M\,1-32, M\,3-15, M\,1-25, Hen\,2-142, K\,2-16, MGC\,6578, M\,2-42, NGC\,6567 and NGC\,6629 
in an interactive X3D file viewer.

\end{appendix}

{ \small 
\begin{center}
\textbf{ORCID iDs}
\end{center}
\vspace{-5pt}

\noindent A.~Danehkar \orcidauthor{0000-0003-4552-5997} \url{https://orcid.org/0000-0003-4552-5997}

}



\begin{thebibliography}{141}
\expandafter\ifx\csname natexlab\endcsname\relax\def\natexlab#1{#1}\fi

\bibitem[{{Acker}(1976)}]{Acker1976}
{Acker}, A. 1976, Publication de l'Observatoire de Strasbourg,
  \href{https://ui.adsabs.harvard.edu/abs/1976POStr...4....1A}{4, 1}

\bibitem[{{Acker} {et~al.}(2002){Acker}, {Gesicki}, {Grosdidier}, \&
  {Durand}}]{Acker2002}
{Acker}, A., {Gesicki}, K., {Grosdidier}, Y., \& {Durand}, S. 2002,
  {\href{https://dx.doi.org/10.1051/0004-6361:20020009}{\color{magenta}\aap}},
  \href{https://ui.adsabs.harvard.edu/abs/2002A%26A...384..620A}{384, 620}

\bibitem[{{Acker} {et~al.}(1992){Acker}, {Marcout}, {Ochsenbein}, {Stenholm},
  {Tylenda}, \& {Schohn}}]{Acker1992}
{Acker}, A., {Marcout}, J., {Ochsenbein}, F., {et~al.} 1992, {The
  Strasbourg-ESO Catalogue of Galactic Planetary Nebulae. Parts I, II.}
  (Garching: ESO)

\bibitem[{{Acker} \& {Neiner}(2003)}]{Acker2003}
{Acker}, A. \& {Neiner}, C. 2003,
  {\href{https://dx.doi.org/10.1051/0004-6361:20030391}{\color{magenta}\aap}},
  \href{https://ui.adsabs.harvard.edu/abs/2003A%26A...403..659A}{403, 659}

\bibitem[{{Akashi} {et~al.}(2015){Akashi}, {Sabach}, {Yogev}, \&
  {Soker}}]{Akashi2015}
{Akashi}, M., {Sabach}, E., {Yogev}, O., \& {Soker}, N. 2015,
  {\href{https://dx.doi.org/10.1093/mnras/stv1666}{\color{magenta}\mnras}},
  \href{https://ui.adsabs.harvard.edu/abs/2015MNRAS.453.2115A}{453, 2115}

\bibitem[{{Akashi} \& {Soker}(2016)}]{Akashi2016}
{Akashi}, M. \& {Soker}, N. 2016,
  {\href{https://dx.doi.org/10.1093/mnras/stw1683}{\color{magenta}\mnras}},
  \href{https://ui.adsabs.harvard.edu/abs/2016MNRAS.462..206A}{462, 206}

\bibitem[{{Akashi} \& {Soker}(2017)}]{Akashi2017}
{Akashi}, M. \& {Soker}, N. 2017,
  {\href{https://dx.doi.org/10.1093/mnras/stx1058}{\color{magenta}\mnras}},
  \href{https://ui.adsabs.harvard.edu/abs/2017MNRAS.469.3296A}{469, 3296}

\bibitem[{{Akras} {et~al.}(2015){Akras}, {Boumis}, {Meaburn}, {Alikakos},
  {L{\'o}pez}, \& {Gon{\c{c}}alves}}]{Akras2015}
{Akras}, S., {Boumis}, P., {Meaburn}, J., {et~al.} 2015,
  {\href{https://dx.doi.org/10.1093/mnras/stv1468}{\color{magenta}\mnras}},
  \href{https://ui.adsabs.harvard.edu/abs/2015MNRAS.452.2911A}{452, 2911}

\bibitem[{{Akras} {et~al.}(2016){Akras}, {Clyne}, {Boumis}, {Monteiro},
  {Gon{\c{c}}alves}, {Redman}, \& {Williams}}]{Akras2016}
{Akras}, S., {Clyne}, N., {Boumis}, P., {et~al.} 2016,
  {\href{https://dx.doi.org/10.1093/mnras/stw038}{\color{magenta}\mnras}},
  \href{https://ui.adsabs.harvard.edu/abs/2016MNRAS.457.3409A}{457, 3409}

\bibitem[{{Akras} \& {Gon{\c{c}}alves}(2016)}]{Akras2016a}
{Akras}, S. \& {Gon{\c{c}}alves}, D.~R. 2016,
  {\href{https://dx.doi.org/10.1093/mnras/stv2139}{\color{magenta}\mnras}},
  \href{https://ui.adsabs.harvard.edu/abs/2016MNRAS.455..930A}{455, 930}

\bibitem[{{Akras} {et~al.}(2017){Akras}, {Gon{\c{c}}alves}, \&
  {Ramos-Larios}}]{Akras2017}
{Akras}, S., {Gon{\c{c}}alves}, D.~R., \& {Ramos-Larios}, G. 2017,
  {\href{https://dx.doi.org/10.1093/mnras/stw2736}{\color{magenta}\mnras}},
  \href{https://ui.adsabs.harvard.edu/abs/2017MNRAS.465.1289A}{465, 1289}

\bibitem[{{Akras} {et~al.}(2020){Akras}, {Gon{\c{c}}alves}, {Ramos-Larios}, \&
  {Aleman}}]{Akras2020}
{Akras}, S., {Gon{\c{c}}alves}, D.~R., {Ramos-Larios}, G., \& {Aleman}, I.
  2020,
  {\href{https://dx.doi.org/10.1093/mnras/staa515}{\color{magenta}\mnras}},
  \href{https://ui.adsabs.harvard.edu/abs/2020MNRAS.493.3800A}{493, 3800}

\bibitem[{{Akras} \& {L{\'o}pez}(2012)}]{Akras2012}
{Akras}, S. \& {L{\'o}pez}, J.~A. 2012,
  {\href{https://dx.doi.org/10.1111/j.1365-2966.2012.21578.x}{\color{magenta}\mnras}},
  \href{https://ui.adsabs.harvard.edu/abs/2012MNRAS.425.2197A}{425, 2197}

\bibitem[{{Akras} \& {Steffen}(2012)}]{Akras2012a}
{Akras}, S. \& {Steffen}, W. 2012,
  {\href{https://dx.doi.org/10.1111/j.1365-2966.2012.20928.x}{\color{magenta}\mnras}},
  \href{https://ui.adsabs.harvard.edu/abs/2012MNRAS.423..925A}{423, 925}

\bibitem[{{Arnaud} {et~al.}(1996){Arnaud}, {Borkowski}, \&
  {Harrington}}]{Arnaud1996}
{Arnaud}, K., {Borkowski}, K.~J., \& {Harrington}, J.~P. 1996,
  {\href{https://dx.doi.org/10.1086/310037}{\color{magenta}\apjl}},
  \href{https://ui.adsabs.harvard.edu/abs/1996ApJ...462L..75A}{462, L75}

\bibitem[{{Balick}(1987)}]{Balick1987}
{Balick}, B. 1987,
  {\href{https://dx.doi.org/10.1086/114504}{\color{magenta}\aj}},
  \href{https://ui.adsabs.harvard.edu/abs/1987AJ.....94..671B}{94, 671}

\bibitem[{{Balick} \& {Frank}(2002)}]{Balick2002}
{Balick}, B. \& {Frank}, A. 2002,
  {\href{https://dx.doi.org/10.1146/annurev.astro.40.060401.093849}{\color{magenta}\araa}},
  \href{https://ui.adsabs.harvard.edu/abs/2002ARA%26A..40..439B}{40, 439}

\bibitem[{{Balick} {et~al.}(1994){Balick}, {Perinotto}, {Maccioni}, {Terzian},
  \& {Hajian}}]{Balick1994}
{Balick}, B., {Perinotto}, M., {Maccioni}, A., {Terzian}, Y., \& {Hajian}, A.
  1994, {\href{https://dx.doi.org/10.1086/173932}{\color{magenta}\apj}},
  \href{https://ui.adsabs.harvard.edu/abs/1994ApJ...424..800B}{424, 800}

\bibitem[{{Balick} {et~al.}(1987){Balick}, {Preston}, \& {Icke}}]{Balick1987a}
{Balick}, B., {Preston}, H.~L., \& {Icke}, V. 1987,
  {\href{https://dx.doi.org/10.1086/114595}{\color{magenta}\aj}},
  \href{https://ui.adsabs.harvard.edu/abs/1987AJ.....94.1641B}{94, 1641}

\bibitem[{{Balick} {et~al.}(1993){Balick}, {Rugers}, {Terzian}, \&
  {Chengalur}}]{Balick1993}
{Balick}, B., {Rugers}, M., {Terzian}, Y., \& {Chengalur}, J.~N. 1993,
  {\href{https://dx.doi.org/10.1086/172881}{\color{magenta}\apj}},
  \href{https://ui.adsabs.harvard.edu/abs/1993ApJ...411..778B}{411, 778}

\bibitem[{{Behr} {et~al.}(2009){Behr}, {Eschler}, {Jung}, \&
  {Z\"{o}llner}}]{Behr2009}
{Behr}, J., {Eschler}, P., {Jung}, Y., \& {Z\"{o}llner}, M. 2009, in
  Proceedings of the 14th International Conference on 3D Web Technology, Web3D
  '09 (New York, NY, USA: Association for Computing Machinery),
  \href{https://doi.org/10.1145/1559764.1559784}{127--135}

\bibitem[{{Bl\"{o}cker}(1995)}]{Bloecker1995b}
{Bl\"{o}cker}, T. 1995, \aap,
  \href{https://ui.adsabs.harvard.edu/abs/1995A%26A...299..755B}{299, 755}

\bibitem[{{Castellanos-Ram{\'\i}rez} {et~al.}(2021){Castellanos-Ram{\'\i}rez},
  {Rodr{\'\i}guez-Gonz{\'a}lez}, {Meliani}, {Rivera-Ortiz}, {Raga}, \&
  {Cant{\'o}}}]{Castellanos-Ramirez2021}
{Castellanos-Ram{\'\i}rez}, A., {Rodr{\'\i}guez-Gonz{\'a}lez}, A., {Meliani},
  Z., {et~al.} 2021,
  {\href{https://dx.doi.org/10.1093/mnras/stab2373}{\color{magenta}\mnras}},
  \href{https://ui.adsabs.harvard.edu/abs/2021MNRAS.507.4044C}{507, 4044}

\bibitem[{{Chen} {et~al.}(2016){Chen}, {Nordhaus}, {Frank}, {Blackman}, \&
  {Balick}}]{Chen2016}
{Chen}, Z., {Nordhaus}, J., {Frank}, A., {Blackman}, E.~G., \& {Balick}, B.
  2016,
  {\href{https://dx.doi.org/10.1093/mnras/stw1305}{\color{magenta}\mnras}},
  \href{https://ui.adsabs.harvard.edu/abs/2016MNRAS.460.4182C}{460, 4182}

\bibitem[{{Cignoni} {et~al.}(2008){Cignoni}, {Callieri}, {Corsini},
  {Dellepiane}, {Ganovelli}, \& {Ranzuglia}}]{Cignoni2008}
{Cignoni}, P., {Callieri}, M., {Corsini}, M., {et~al.} 2008, in Eurographics
  Italian Chapter Conference, ed. V.~{Scarano}, R.~{De~Chiara}, \& U.~{Erra},
  Vol. 2008 (Salerno: The Eurographics Association),
  \href{http://dx.doi.org/10.2312/LocalChapterEvents/ItalChap/ItalianChapConf2008/129-136}{129--136}

\bibitem[{{Clark} {et~al.}(2013){Clark}, {L{\'o}pez}, {Steffen}, \&
  {Richer}}]{Clark2013}
{Clark}, D.~M., {L{\'o}pez}, J.~A., {Steffen}, W., \& {Richer}, M.~G. 2013,
  {\href{https://dx.doi.org/10.1088/0004-6256/145/3/57}{\color{magenta}\aj}},
  \href{https://ui.adsabs.harvard.edu/abs/2013AJ....145...57C}{145, 57}

\bibitem[{{Clegg} {et~al.}(1999){Clegg}, {Miller}, {Storey}, \&
  {Kisielius}}]{Clegg1999}
{Clegg}, R.~E.~S., {Miller}, S., {Storey}, P.~J., \& {Kisielius}, R. 1999,
  {\href{https://dx.doi.org/10.1051/aas:1999178}{\color{magenta}\aaps}},
  \href{https://ui.adsabs.harvard.edu/abs/1999A%26AS..135..359C}{135, 359}

\bibitem[{{Corradi} {et~al.}(1996){Corradi}, {Manso}, {Mampaso}, \&
  {Schwarz}}]{Corradi1996}
{Corradi}, R.~L.~M., {Manso}, R., {Mampaso}, A., \& {Schwarz}, H.~E. 1996,
  \aap, \href{https://ui.adsabs.harvard.edu/abs/1996A%26A...313..913C}{313,
  913}

\bibitem[{{Corradi} \& {Schwarz}(1995)}]{Corradi1995}
{Corradi}, R.~L.~M. \& {Schwarz}, H.~E. 1995, \aap,
  \href{https://ui.adsabs.harvard.edu/abs/1995A%26A...293..871C}{293, 871}

\bibitem[{{Crowther} {et~al.}(1998){Crowther}, {De Marco}, \&
  {Barlow}}]{Crowther1998}
{Crowther}, P.~A., {De Marco}, O., \& {Barlow}, M.~J. 1998,
  {\href{https://dx.doi.org/10.1046/j.1365-8711.1998.01360.x}{\color{magenta}\mnras}},
  \href{https://ui.adsabs.harvard.edu/abs/1998MNRAS.296..367C}{296, 367}

\bibitem[{{Danehkar}(2014)}]{Danehkar2014b}
{Danehkar}, A. 2014,
  \href{https://ui.adsabs.harvard.edu/abs/2014PhDT........76D}{\href{https://ui.adsabs.harvard.edu/abs/2014PhDT........76D}{\color{magenta}{Evolution
  of Planetary Nebulae with WR-type Central Stars}}}, PhD thesis, Macquarie
  University

\bibitem[{{Danehkar}(2015)}]{Danehkar2015a}
{Danehkar}, A. 2015,
  {\href{https://dx.doi.org/10.1088/0004-637X/815/1/35}{\color{magenta}\apj}},
  \href{https://ui.adsabs.harvard.edu/abs/2015ApJ...815...35D}{815, 35}

\bibitem[{{Danehkar}(2018)}]{Danehkar2018}
{Danehkar}, A. 2018,
  {\href{https://dx.doi.org/10.1017/pasa.2018.1}{\color{magenta}\pasa}},
  \href{https://ui.adsabs.harvard.edu/abs/2018PASA...35....5D}{35, e005}

\bibitem[{{Danehkar}(2021)}]{Danehkar2021}
{Danehkar}, A. 2021,
  {\href{https://dx.doi.org/10.3847/1538-4365/ac2310}{\color{magenta}\apjs}},
  \href{https://ui.adsabs.harvard.edu/abs/2021ApJS..257...58D}{257, 58}

\bibitem[{{Danehkar}(2022)}]{Danehkar2022}
{Danehkar}, A. 2022,
  {\href{https://dx.doi.org/10.1093/mnras/stab3735}{\color{magenta}\mnras}},
  \href{https://ui.adsabs.harvard.edu/abs/2022MNRAS.511.1022D/abstract}{511, 1022}

\bibitem[{{Danehkar} {et~al.}(2018){Danehkar}, {Karovska}, {Maksym}, \&
  {Montez}}]{Danehkar2018a}
{Danehkar}, A., {Karovska}, M., {Maksym}, W.~P., \& {Montez}, Rodolfo, J. 2018,
  {\href{https://dx.doi.org/10.3847/1538-4357/aa9e8c}{\color{magenta}\apj}},
  \href{https://ui.adsabs.harvard.edu/abs/2018ApJ...852...87D}{852, 87}

\bibitem[{{Danehkar} \& {Parker}(2015)}]{Danehkar2015}
{Danehkar}, A. \& {Parker}, Q.~A. 2015,
  {\href{https://dx.doi.org/10.1093/mnrasl/slv022}{\color{magenta}\mnras}},
  \href{https://ui.adsabs.harvard.edu/abs/2015MNRAS.449L..56D}{449, L56}

\bibitem[{{Danehkar} {et~al.}(2016){Danehkar}, {Parker}, \&
  {Steffen}}]{Danehkar2016}
{Danehkar}, A., {Parker}, Q.~A., \& {Steffen}, W. 2016,
  {\href{https://dx.doi.org/10.3847/0004-6256/151/2/38}{\color{magenta}\aj}},
  \href{https://ui.adsabs.harvard.edu/abs/2016AJ....151...38D}{151, 38}

\bibitem[{{Danehkar} {et~al.}(2014){Danehkar}, {Todt}, {Ercolano}, \&
  {Kniazev}}]{Danehkar2014}
{Danehkar}, A., {Todt}, H., {Ercolano}, B., \& {Kniazev}, A.~Y. 2014,
  {\href{https://dx.doi.org/10.1093/mnras/stu203}{\color{magenta}\mnras}},
  \href{https://ui.adsabs.harvard.edu/abs/2014MNRAS.439.3605D}{439, 3605}

\bibitem[{{De Marco}(2009)}]{DeMarco2009}
{De Marco}, O. 2009,
  {\href{https://dx.doi.org/10.1086/597765}{\color{magenta}\pasp}},
  \href{https://ui.adsabs.harvard.edu/abs/2009PASP..121..316D}{121, 316}

\bibitem[{{De Marco} \& {Soker}(2011)}]{DeMarco2011}
{De Marco}, O. \& {Soker}, N. 2011,
  {\href{https://dx.doi.org/10.1086/659846}{\color{magenta}\pasp}},
  \href{https://ui.adsabs.harvard.edu/abs/2011PASP..123..402D}{123, 402}

\bibitem[{{Depew} {et~al.}(2011){Depew}, {Parker}, {Miszalski}, {De Marco},
  {Frew}, {Acker}, {Kovacevic}, \& {Sharp}}]{Depew2011}
{Depew}, K., {Parker}, Q.~A., {Miszalski}, B., {et~al.} 2011,
  {\href{https://dx.doi.org/10.1111/j.1365-2966.2011.18337.x}{\color{magenta}\mnras}},
  \href{https://ui.adsabs.harvard.edu/abs/2011MNRAS.414.2812D}{414, 2812}

\bibitem[{{Derlopa} {et~al.}(2019){Derlopa}, {Akras}, {Boumis}, \&
  {Steffen}}]{Derlopa2019}
{Derlopa}, S., {Akras}, S., {Boumis}, P., \& {Steffen}, W. 2019,
  {\href{https://dx.doi.org/10.1093/mnras/stz193}{\color{magenta}\mnras}},
  \href{https://ui.adsabs.harvard.edu/abs/2019MNRAS.484.3746D}{484, 3746}

\bibitem[{{Devine} {et~al.}(1997){Devine}, {Bally}, {Reipurth}, \&
  {Heathcote}}]{Devine1997}
{Devine}, D., {Bally}, J., {Reipurth}, B., \& {Heathcote}, S. 1997,
  {\href{https://dx.doi.org/10.1086/118629}{\color{magenta}\aj}},
  \href{https://ui.adsabs.harvard.edu/abs/1997AJ....114.2095D}{114, 2095}

\bibitem[{{Dopita} {et~al.}(2007){Dopita}, {Hart}, {McGregor}, {Oates},
  {Bloxham}, \& {Jones}}]{Dopita2007}
{Dopita}, M., {Hart}, J., {McGregor}, P., {et~al.} 2007,
  {\href{https://dx.doi.org/10.1007/s10509-007-9510-z}{\color{magenta}\apss}},
  \href{https://ui.adsabs.harvard.edu/abs/2007Ap%26SS.310..255D}{310, 255}

\bibitem[{{Dopita} {et~al.}(2010){Dopita}, {Rhee}, {Farage}, {McGregor},
  {Bloxham}, {Green}, {Roberts}, {Neilson}, {Wilson}, {Young}, {Firth},
  {Busarello}, \& {Merluzzi}}]{Dopita2010}
{Dopita}, M., {Rhee}, J., {Farage}, C., {et~al.} 2010,
  {\href{https://dx.doi.org/10.1007/s10509-010-0335-9}{\color{magenta}\apss}},
  \href{https://ui.adsabs.harvard.edu/abs/2010Ap%26SS.327..245D}{327, 245}

\bibitem[{{Dopita} {et~al.}(1985){Dopita}, {Lawrence}, {Ford}, \&
  {Webster}}]{Dopita1985}
{Dopita}, M.~A., {Lawrence}, C.~J., {Ford}, H.~C., \& {Webster}, B.~L. 1985,
  {\href{https://dx.doi.org/10.1086/163457}{\color{magenta}\apj}},
  \href{https://ui.adsabs.harvard.edu/abs/1985ApJ...296..390D}{296, 390}

\bibitem[{{Dopita} \& {Meatheringham}(1990)}]{Dopita1990}
{Dopita}, M.~A. \& {Meatheringham}, S.~J. 1990,
  {\href{https://dx.doi.org/10.1086/168900}{\color{magenta}\apj}},
  \href{https://ui.adsabs.harvard.edu/abs/1990ApJ...357..140D}{357, 140}

\bibitem[{{Dopita} \& {Meatheringham}(1991)}]{Dopita1991}
{Dopita}, M.~A. \& {Meatheringham}, S.~J. 1991,
  {\href{https://dx.doi.org/10.1086/170377}{\color{magenta}\apj}},
  \href{https://ui.adsabs.harvard.edu/abs/1991ApJ...377..480D}{377, 480}

\bibitem[{{Dopita} {et~al.}(1988){Dopita}, {Meatheringham}, {Webster}, \&
  {Ford}}]{Dopita1988}
{Dopita}, M.~A., {Meatheringham}, S.~J., {Webster}, B.~L., \& {Ford}, H.~C.
  1988, {\href{https://dx.doi.org/10.1086/166221}{\color{magenta}\apj}},
  \href{https://ui.adsabs.harvard.edu/abs/1988ApJ...327..639D}{327, 639}

\bibitem[{{Dopita} {et~al.}(1996){Dopita}, {Vassiliadis}, {Meatheringham},
  {Bohlin}, {Ford}, {Harrington}, {Wood}, {Stecher}, \& {Maran}}]{Dopita1996}
{Dopita}, M.~A., {Vassiliadis}, E., {Meatheringham}, S.~J., {et~al.} 1996,
  {\href{https://dx.doi.org/10.1086/176972}{\color{magenta}\apj}},
  \href{https://ui.adsabs.harvard.edu/abs/1996ApJ...460..320D}{460, 320}

\bibitem[{{Durand} {et~al.}(1998){Durand}, {Acker}, \& {Zijlstra}}]{Durand1998}
{Durand}, S., {Acker}, A., \& {Zijlstra}, A. 1998,
  {\href{https://dx.doi.org/10.1051/aas:1998356}{\color{magenta}\aaps}},
  \href{https://ui.adsabs.harvard.edu/abs/1998A%26AS..132...13D}{132, 13}

\bibitem[{{Fanning}(2011)}]{Fanning2011}
{Fanning}, D.~W. 2011, {Coyote's Guide to Traditional IDL Graphics} (Fort
  Collins, Colorado: Coyote Book Publishing)

\bibitem[{{Frank} \& {Blackman}(2004)}]{Frank2004}
{Frank}, A. \& {Blackman}, E.~G. 2004,
  {\href{https://dx.doi.org/10.1086/382018}{\color{magenta}\apj}},
  \href{https://ui.adsabs.harvard.edu/abs/2004ApJ...614..737F}{614, 737}

\bibitem[{{Frew} {et~al.}(2016){Frew}, {Parker}, \&
  {Boji{\v{c}}i{\'c}}}]{Frew2016}
{Frew}, D.~J., {Parker}, Q.~A., \& {Boji{\v{c}}i{\'c}}, I.~S. 2016,
  {\href{https://dx.doi.org/10.1093/mnras/stv1516}{\color{magenta}\mnras}},
  \href{https://ui.adsabs.harvard.edu/abs/2016MNRAS.455.1459F}{455, 1459}

\bibitem[{{Garc{\'{\i}}a-D{\'{\i}}az}
  {et~al.}(2009){Garc{\'{\i}}a-D{\'{\i}}az}, {Clark}, {L{\'o}pez}, {Steffen},
  \& {Richer}}]{Garcia-Diaz2009}
{Garc{\'{\i}}a-D{\'{\i}}az}, M.~T., {Clark}, D.~M., {L{\'o}pez}, J.~A.,
  {Steffen}, W., \& {Richer}, M.~G. 2009,
  {\href{https://dx.doi.org/10.1088/0004-637X/699/2/1633}{\color{magenta}\apj}},
  \href{https://ui.adsabs.harvard.edu/abs/2009ApJ...699.1633G}{699, 1633}

\bibitem[{{Garc{\'{\i}}a-Rojas} {et~al.}(2009){Garc{\'{\i}}a-Rojas},
  {Pe{\~n}a}, \& {Peimbert}}]{Garcia-Rojas2009}
{Garc{\'{\i}}a-Rojas}, J., {Pe{\~n}a}, M., \& {Peimbert}, A. 2009,
  {\href{https://dx.doi.org/10.1051/0004-6361:200811185}{\color{magenta}\aap}},
  \href{https://ui.adsabs.harvard.edu/abs/2009A%26A...496..139G}{496, 139}

\bibitem[{{Garc{\'{\i}}a-Segura}(1997)}]{Garcia-Segura1997}
{Garc{\'{\i}}a-Segura}, G. 1997,
  {\href{https://dx.doi.org/10.1086/316796}{\color{magenta}\apjl}},
  \href{https://ui.adsabs.harvard.edu/abs/1997ApJ...489L.189G}{489, L189}

\bibitem[{{Garc{\'{\i}}a-Segura} {et~al.}(1999){Garc{\'{\i}}a-Segura},
  {Langer}, {R{\'o}{\.z}yczka}, \& {Franco}}]{Garcia-Segura1999}
{Garc{\'{\i}}a-Segura}, G., {Langer}, N., {R{\'o}{\.z}yczka}, M., \& {Franco},
  J. 1999, {\href{https://dx.doi.org/10.1086/307205}{\color{magenta}\apj}},
  \href{https://ui.adsabs.harvard.edu/abs/1999ApJ...517..767G}{517, 767}

\bibitem[{{Garc{\'{\i}}a-Segura} \& {L{\'o}pez}(2000)}]{Garcia-Segura2000}
{Garc{\'{\i}}a-Segura}, G. \& {L{\'o}pez}, J.~A. 2000,
  {\href{https://dx.doi.org/10.1086/317186}{\color{magenta}\apj}},
  \href{https://ui.adsabs.harvard.edu/abs/2000ApJ...544..336G}{544, 336}

\bibitem[{{Garc{\'\i}a-Segura} {et~al.}(2001){Garc{\'\i}a-Segura}, {L{\'o}pez},
  \& {Franco}}]{GarciaSegura2001}
{Garc{\'\i}a-Segura}, G., {L{\'o}pez}, J.~A., \& {Franco}, J. 2001,
  {\href{https://dx.doi.org/10.1086/323072}{\color{magenta}\apj}},
  \href{https://ui.adsabs.harvard.edu/abs/2001ApJ...560..928G}{560, 928}

\bibitem[{{Garc{\'\i}a-Segura} {et~al.}(2018){Garc{\'\i}a-Segura}, {Ricker}, \&
  {Taam}}]{Garcia-Segura2018}
{Garc{\'\i}a-Segura}, G., {Ricker}, P.~M., \& {Taam}, R.~E. 2018,
  {\href{https://dx.doi.org/10.3847/1538-4357/aac08c}{\color{magenta}\apj}},
  \href{https://ui.adsabs.harvard.edu/abs/2018ApJ...860...19G}{860, 19}

\bibitem[{{Garc{\'\i}a-Segura} {et~al.}(2021){Garc{\'\i}a-Segura}, {Taam}, \&
  {Ricker}}]{Garcia-Segura2021}
{Garc{\'\i}a-Segura}, G., {Taam}, R.~E., \& {Ricker}, P.~M. 2021,
  {\href{https://dx.doi.org/10.3847/1538-4357/abfc4e}{\color{magenta}\apj}},
  \href{https://ui.adsabs.harvard.edu/abs/2021ApJ...914..111G}{914, 111}

\bibitem[{{Garc{\'\i}a-Segura} {et~al.}(2014){Garc{\'\i}a-Segura}, {Villaver},
  {Langer}, {Yoon}, \& {Manchado}}]{Garcia-Segura2014}
{Garc{\'\i}a-Segura}, G., {Villaver}, E., {Langer}, N., {Yoon}, S.~C., \&
  {Manchado}, A. 2014,
  {\href{https://dx.doi.org/10.1088/0004-637X/783/2/74}{\color{magenta}\apj}},
  \href{https://ui.adsabs.harvard.edu/abs/2014ApJ...783...74G}{783, 74}

\bibitem[{{Gesicki} \& {Acker}(1996)}]{Gesicki1996}
{Gesicki}, K. \& {Acker}, A. 1996,
  {\href{https://dx.doi.org/10.1007/BF00645502}{\color{magenta}\apss}},
  \href{https://ui.adsabs.harvard.edu/abs/1996Ap%26SS.238..101G}{238, 101}

\bibitem[{{Gesicki} \& {Zijlstra}(2000)}]{Gesicki2000}
{Gesicki}, K. \& {Zijlstra}, A.~A. 2000, \aap,
  \href{https://ui.adsabs.harvard.edu/abs/2000A%26A...358.1058G}{358, 1058}

\bibitem[{{Gesicki} \& {Zijlstra}(2007)}]{Gesicki2007}
{Gesicki}, K. \& {Zijlstra}, A.~A. 2007,
  {\href{https://dx.doi.org/10.1051/0004-6361:20077250}{\color{magenta}\aap}},
  \href{https://ui.adsabs.harvard.edu/abs/2007A%26A...467L..29G}{467, L29}

\bibitem[{{Gon{\c c}alves} {et~al.}(2001){Gon{\c c}alves}, {Corradi}, \&
  {Mampaso}}]{Gonccalves2001}
{Gon{\c c}alves}, D.~R., {Corradi}, R.~L.~M., \& {Mampaso}, A. 2001,
  {\href{https://dx.doi.org/10.1086/318364}{\color{magenta}\apj}},
  \href{https://ui.adsabs.harvard.edu/abs/2001ApJ...547..302G}{547, 302}

\bibitem[{{Gon{\c c}alves} {et~al.}(2009){Gon{\c c}alves}, {Mampaso},
  {Corradi}, \& {Quireza}}]{Gonccalves2009}
{Gon{\c c}alves}, D.~R., {Mampaso}, A., {Corradi}, R.~L.~M., \& {Quireza}, C.
  2009,
  {\href{https://dx.doi.org/10.1111/j.1365-2966.2009.15266.x}{\color{magenta}\mnras}},
  \href{https://ui.adsabs.harvard.edu/abs/2009MNRAS.398.2166G}{398, 2166}

\bibitem[{{G{\'o}rny} {et~al.}(2009){G{\'o}rny}, {Chiappini}, {Stasi{\'n}ska},
  \& {Cuisinier}}]{Gorny2009}
{G{\'o}rny}, S.~K., {Chiappini}, C., {Stasi{\'n}ska}, G., \& {Cuisinier}, F.
  2009,
  {\href{https://dx.doi.org/10.1051/0004-6361/200810841}{\color{magenta}\aap}},
  \href{https://ui.adsabs.harvard.edu/abs/2009A%26A...500.1089G}{500, 1089}

\bibitem[{{G{\'o}rny} {et~al.}(2004){G{\'o}rny}, {Stasi{\'n}ska}, {Escudero},
  \& {Costa}}]{Gorny2004}
{G{\'o}rny}, S.~K., {Stasi{\'n}ska}, G., {Escudero}, A.~V., \& {Costa},
  R.~D.~D. 2004,
  {\href{https://dx.doi.org/10.1051/0004-6361:20047064}{\color{magenta}\aap}},
  \href{https://ui.adsabs.harvard.edu/abs/2004A%26A...427..231G}{427, 231}

\bibitem[{{G{\'o}rny} {et~al.}(1997){G{\'o}rny}, {Stasi{\'n}ska}, \&
  {Tylenda}}]{Gorny1997}
{G{\'o}rny}, S.~K., {Stasi{\'n}ska}, G., \& {Tylenda}, R. 1997, \aap,
  \href{https://ui.adsabs.harvard.edu/abs/1997A%26A...318..256G}{318, 256}

\bibitem[{{Hambly} {et~al.}(2001){Hambly}, {MacGillivray}, {Read}, {Tritton},
  {Thomson}, {Kelly}, {Morgan}, {Smith}, {Driver}, {Williamson}, {Parker},
  {Hawkins}, {Williams}, \& {Lawrence}}]{Hambly2001}
{Hambly}, N.~C., {MacGillivray}, H.~T., {Read}, M.~A., {et~al.} 2001,
  {\href{https://dx.doi.org/10.1111/j.1365-8711.2001.04660.x}{\color{magenta}\mnras}},
  \href{https://ui.adsabs.harvard.edu/abs/2001MNRAS.326.1279H}{326, 1279}

\bibitem[{{Hegazi} {et~al.}(2020){Hegazi}, {Bear}, \& {Soker}}]{Hegazi2020}
{Hegazi}, A., {Bear}, E., \& {Soker}, N. 2020,
  {\href{https://dx.doi.org/10.1093/mnras/staa1551}{\color{magenta}\mnras}},
  \href{https://ui.adsabs.harvard.edu/abs/2020MNRAS.496..612H}{496, 612}

\bibitem[{{Huckvale} {et~al.}(2013){Huckvale}, {Prouse}, {Jones}, {Lloyd},
  {Pollacco}, {L{\'o}pez}, {O'Brien}, {Sabin}, \& {Vaytet}}]{Huckvale2013}
{Huckvale}, L., {Prouse}, B., {Jones}, D., {et~al.} 2013,
  {\href{https://dx.doi.org/10.1093/mnras/stt1109}{\color{magenta}\mnras}},
  \href{https://ui.adsabs.harvard.edu/abs/2013MNRAS.434.1505H}{434, 1505}

\bibitem[{{Jones} {et~al.}(2010){Jones}, {Lloyd}, {Santander-Garc{\'{\i}}a},
  {L{\'o}pez}, {Meaburn}, {Mitchell}, {O'Brien}, {Pollacco},
  {Rubio-D{\'{\i}}ez}, \& {Vaytet}}]{Jones2010a}
{Jones}, D., {Lloyd}, M., {Santander-Garc{\'{\i}}a}, M., {et~al.} 2010,
  {\href{https://dx.doi.org/10.1111/j.1365-2966.2010.17277.x}{\color{magenta}\mnras}},
  \href{https://ui.adsabs.harvard.edu/abs/2010MNRAS.408.2312J}{408, 2312}

\bibitem[{{Jones} {et~al.}(2012){Jones}, {Mitchell}, {Lloyd}, {Pollacco},
  {O'Brien}, {Meaburn}, \& {Vaytet}}]{Jones2012}
{Jones}, D., {Mitchell}, D.~L., {Lloyd}, M., {et~al.} 2012,
  {\href{https://dx.doi.org/10.1111/j.1365-2966.2011.20192.x}{\color{magenta}\mnras}},
  \href{https://ui.adsabs.harvard.edu/abs/2012MNRAS.420.2271J}{420, 2271}

\bibitem[{{Kahn} \& {West}(1985)}]{Kahn1985}
{Kahn}, F.~D. \& {West}, K.~A. 1985, \mnras,
  \href{https://ui.adsabs.harvard.edu/abs/1985MNRAS.212..837K}{212, 837}

\bibitem[{{Kaler} {et~al.}(1991){Kaler}, {Shaw}, {Feibelman}, \&
  {Imhoff}}]{Kaler1991}
{Kaler}, J.~B., {Shaw}, R.~A., {Feibelman}, W.~A., \& {Imhoff}, C.~L. 1991,
  {\href{https://dx.doi.org/10.1086/132796}{\color{magenta}\pasp}},
  \href{https://ui.adsabs.harvard.edu/abs/1991PASP..103...67K}{103, 67}

\bibitem[{{Kastner} {et~al.}(2008){Kastner}, {Montez}, {Balick}, \& {De
  Marco}}]{Kastner2008}
{Kastner}, J.~H., {Montez}, Jr., R., {Balick}, B., \& {De Marco}, O. 2008,
  {\href{https://dx.doi.org/10.1086/523890}{\color{magenta}\apj}},
  \href{https://ui.adsabs.harvard.edu/abs/2008ApJ...672..957K}{672, 957}

\bibitem[{{Kastner} {et~al.}(2012){Kastner}, {Montez}, {Balick}, {Frew},
  {Miszalski}, {Sahai}, {Blackman}, {Chu}, {De Marco}, {Frank}, {Guerrero},
  {Lopez}, {Rapson}, {Zijlstra}, {Behar}, {Bujarrabal}, {Corradi}, {Nordhaus},
  {Parker}, {Sandin}, {Sch{\"o}nberner}, {Soker}, {Sokoloski}, {Steffen},
  {Ueta}, \& {Villaver}}]{Kastner2012}
{Kastner}, J.~H., {Montez}, Jr., R., {Balick}, B., {et~al.} 2012,
  {\href{https://dx.doi.org/10.1088/0004-6256/144/2/58}{\color{magenta}\aj}},
  \href{https://ui.adsabs.harvard.edu/abs/2012AJ....144...58K}{144, 58}

\bibitem[{{Koesterke} \& {Hamann}(1997)}]{Koesterke1997}
{Koesterke}, L. \& {Hamann}, W.-R. 1997, \aap,
  \href{https://ui.adsabs.harvard.edu/abs/1997A%26A...320...91K}{320, 91}

\bibitem[{{Kohoutek}(1977)}]{Kohoutek1977}
{Kohoutek}, L. 1977, \aap,
  \href{https://ui.adsabs.harvard.edu/abs/1977A%26A....59..137K}{59, 137}

\bibitem[{{Kwok}(2010)}]{Kwok2010}
{Kwok}, S. 2010,
  {\href{https://dx.doi.org/10.1071/AS09027}{\color{magenta}\pasa}},
  \href{https://ui.adsabs.harvard.edu/abs/2010PASA...27..174K}{27, 174}

\bibitem[{{Kwok} {et~al.}(1978){Kwok}, {Purton}, \& {Fitzgerald}}]{Kwok1978}
{Kwok}, S., {Purton}, C.~R., \& {Fitzgerald}, P.~M. 1978,
  {\href{https://dx.doi.org/10.1086/182621}{\color{magenta}\apjl}},
  \href{https://ui.adsabs.harvard.edu/abs/1978ApJ...219L.125K}{219, L125}

\bibitem[{{Landsman}(1993)}]{Landsman1993}
{Landsman}, W.~B. 1993, in ASP Conf. Ser., Vol.~52, Astronomical Data Analysis
  Software and Systems II, ed. R.~J. {Hanisch}, R.~J.~V. {Brissenden}, \&
  J.~{Barnes} (San Francisco, CA: ASP),
  \href{https://ui.adsabs.harvard.edu/abs/1993ASPC...52..246L}{246}

\bibitem[{{Lasker} {et~al.}(2008){Lasker}, {Lattanzi}, {McLean}, {Bucciarelli},
  {Drimmel}, {Garcia}, {Greene}, {Guglielmetti}, {Hanley}, {Hawkins},
  {Laidler}, {Loomis}, {Meakes}, {Mignani}, {Morbidelli}, {Morrison},
  {Pannunzio}, {Rosenberg}, {Sarasso}, {Smart}, {Spagna}, {Sturch},
  {Volpicelli}, {White}, {Wolfe}, \& {Zacchei}}]{Lasker2008}
{Lasker}, B.~M., {Lattanzi}, M.~G., {McLean}, B.~J., {et~al.} 2008,
  {\href{https://dx.doi.org/10.1088/0004-6256/136/2/735}{\color{magenta}\aj}},
  \href{https://ui.adsabs.harvard.edu/abs/2008AJ....136..735L}{136, 735}

\bibitem[{{Lee} \& {Sahai}(2004)}]{Lee2004}
{Lee}, C.-F. \& {Sahai}, R. 2004,
  {\href{https://dx.doi.org/10.1086/381677}{\color{magenta}\apj}},
  \href{https://ui.adsabs.harvard.edu/abs/2004ApJ...606..483L}{606, 483}

\bibitem[{{Leuenhagen} \& {Hamann}(1998)}]{Leuenhagen1998}
{Leuenhagen}, U. \& {Hamann}, W.-R. 1998, \aap,
  \href{https://ui.adsabs.harvard.edu/abs/1998A%26A...330..265L}{330, 265}

\bibitem[{{Leuenhagen} {et~al.}(1996){Leuenhagen}, {Hamann}, \&
  {Jeffery}}]{Leuenhagen1996}
{Leuenhagen}, U., {Hamann}, W.-R., \& {Jeffery}, C.~S. 1996, \aap,
  \href{https://ui.adsabs.harvard.edu/abs/1996A%26A...312..167L}{312, 167}

\bibitem[{{L{\'o}pez} {et~al.}(2012){L{\'o}pez}, {Garc{\'{\i}}a-D{\'{\i}}az},
  {Steffen}, {Riesgo}, \& {Richer}}]{Lopez2012}
{L{\'o}pez}, J.~A., {Garc{\'{\i}}a-D{\'{\i}}az}, M.~T., {Steffen}, W.,
  {Riesgo}, H., \& {Richer}, M.~G. 2012,
  {\href{https://dx.doi.org/10.1088/0004-637X/750/2/131}{\color{magenta}\apj}},
  \href{https://ui.adsabs.harvard.edu/abs/2012ApJ...750..131L}{750, 131}

\bibitem[{{L{\'o}pez} {et~al.}(1997){L{\'o}pez}, {Steffen}, \&
  {Meaburn}}]{Lopez1997}
{L{\'o}pez}, J.~A., {Steffen}, W., \& {Meaburn}, J. 1997,
  {\href{https://dx.doi.org/10.1086/304472}{\color{magenta}\apj}},
  \href{https://ui.adsabs.harvard.edu/abs/1997ApJ...485..697L}{485, 697}

\bibitem[{{Markwardt}(2009)}]{Markwardt2009}
{Markwardt}, C.~B. 2009, in Astronomical Society of the Pacific Conference
  Series, Vol. 411, Astronomical Data Analysis Software and Systems XVIII, ed.
  D.~A. {Bohlender}, D.~{Durand}, \& P.~{Dowler},
  \href{https://ui.adsabs.harvard.edu/abs/2009ASPC..411..251M}{251}

\bibitem[{{Medina} {et~al.}(2006){Medina}, {Pe{\~n}a}, {Morisset}, \&
  {Stasi{\'n}ska}}]{Medina2006}
{Medina}, S., {Pe{\~n}a}, M., {Morisset}, C., \& {Stasi{\'n}ska}, G. 2006,
  \rmxaa, \href{https://ui.adsabs.harvard.edu/abs/2006RMxAA..42...53M}{42, 53}

\bibitem[{{Miszalski} {et~al.}(2009{\natexlab{a}}){Miszalski}, {Acker},
  {Moffat}, {Parker}, \& {Udalski}}]{Miszalski2009a}
{Miszalski}, B., {Acker}, A., {Moffat}, A.~F.~J., {Parker}, Q.~A., \&
  {Udalski}, A. 2009{\natexlab{a}},
  {\href{https://dx.doi.org/10.1051/0004-6361/200811380}{\color{magenta}\aap}},
  \href{https://ui.adsabs.harvard.edu/abs/2009A%26A...496..813M}{496, 813}

\bibitem[{{Miszalski} {et~al.}(2009{\natexlab{b}}){Miszalski}, {Acker},
  {Parker}, \& {Moffat}}]{Miszalski2009b}
{Miszalski}, B., {Acker}, A., {Parker}, Q.~A., \& {Moffat}, A.~F.~J.
  2009{\natexlab{b}},
  {\href{https://dx.doi.org/10.1051/0004-6361/200912176}{\color{magenta}\aap}},
  \href{https://ui.adsabs.harvard.edu/abs/2009A%26A...505..249M}{505, 249}

\bibitem[{{Mitchell} {et~al.}(2007){Mitchell}, {Pollacco}, {O'Brien}, {Bryce},
  {L{\'o}pez}, {Meaburn}, \& {Vaytet}}]{Mitchell2007}
{Mitchell}, D.~L., {Pollacco}, D., {O'Brien}, T.~J., {et~al.} 2007,
  {\href{https://dx.doi.org/10.1111/j.1365-2966.2006.11251.x}{\color{magenta}\mnras}},
  \href{https://ui.adsabs.harvard.edu/abs/2007MNRAS.374.1404M}{374, 1404}

\bibitem[{{Montez} {et~al.}(2005){Montez}, {Kastner}, {De Marco}, \&
  {Soker}}]{Montez2005}
{Montez}, Jr., R., {Kastner}, J.~H., {De Marco}, O., \& {Soker}, N. 2005,
  {\href{https://dx.doi.org/10.1086/497262}{\color{magenta}\apj}},
  \href{https://ui.adsabs.harvard.edu/abs/2005ApJ...635..381M}{635, 381}

\bibitem[{{Nordhaus} \& {Blackman}(2006)}]{Nordhaus2006}
{Nordhaus}, J. \& {Blackman}, E.~G. 2006,
  {\href{https://dx.doi.org/10.1111/j.1365-2966.2006.10625.x}{\color{magenta}\mnras}},
  \href{https://ui.adsabs.harvard.edu/abs/2006MNRAS.370.2004N}{370, 2004}

\bibitem[{{Nordhaus} {et~al.}(2007){Nordhaus}, {Blackman}, \&
  {Frank}}]{Nordhaus2007}
{Nordhaus}, J., {Blackman}, E.~G., \& {Frank}, A. 2007,
  {\href{https://dx.doi.org/10.1111/j.1365-2966.2007.11417.x}{\color{magenta}\mnras}},
  \href{https://ui.adsabs.harvard.edu/abs/2007MNRAS.376..599N}{376, 599}

\bibitem[{{Nordhaus} {et~al.}(2010){Nordhaus}, {Spiegel}, {Ibgui}, {Goodman},
  \& {Burrows}}]{Nordhaus2010}
{Nordhaus}, J., {Spiegel}, D.~S., {Ibgui}, L., {Goodman}, J., \& {Burrows}, A.
  2010,
  {\href{https://dx.doi.org/10.1111/j.1365-2966.2010.17155.x}{\color{magenta}\mnras}},
  \href{https://ui.adsabs.harvard.edu/abs/2010MNRAS.408..631N}{408, 631}

\bibitem[{{Nugis} \& {Lamers}(2000)}]{Nugis2000}
{Nugis}, T. \& {Lamers}, H.~J.~G.~L.~M. 2000, \aap,
  \href{https://ui.adsabs.harvard.edu/abs/2000A%26A...360..227N}{360, 227}

\bibitem[{{Orosz} {et~al.}(2019){Orosz}, {G{\'o}mez}, {Imai}, {Tafoya},
  {Torrelles}, {Burns}, {Frau}, {Guerrero}, {Miranda}, {Perez-Torres},
  {Ramos-Larios}, {Rizzo}, {Su{\'a}rez}, \& {Uscanga}}]{Orosz2019}
{Orosz}, G., {G{\'o}mez}, J.~F., {Imai}, H., {et~al.} 2019,
  {\href{https://dx.doi.org/10.1093/mnrasl/sly177}{\color{magenta}\mnras}},
  \href{https://ui.adsabs.harvard.edu/abs/2019MNRAS.482L..40O}{482, L40}

\bibitem[{{Paczynski}(1976)}]{Paczynski1976}
{Paczynski}, B. 1976, in IAU Symposium, Vol.~73, Structure and Evolution of
  Close Binary Systems, ed. P.~{Eggleton}, S.~{Mitton}, \& J.~{Whelan} (Dordrecht: D. Reidel),
  \href{https://ui.adsabs.harvard.edu/abs/1976IAUS...73...75P}{75}

\bibitem[{{Parker} {et~al.}(2005){Parker}, {Phillipps}, {Pierce}, \&
  et~al.}]{Parker2005}
{Parker}, Q.~A., {Phillipps}, S., {Pierce}, M., \& et~al. 2005,
  {\href{https://dx.doi.org/10.1111/j.1365-2966.2005.09350.x}{\color{magenta}\mnras}},
  \href{https://ui.adsabs.harvard.edu/abs/2005MNRAS.362..689P}{362, 689}

\bibitem[{{Pauldrach} {et~al.}(1988){Pauldrach}, {Puls}, {Kudritzki},
  {M{\'e}ndez}, \& {Heap}}]{Pauldrach1988}
{Pauldrach}, A., {Puls}, J., {Kudritzki}, R.~P., {M{\'e}ndez}, R.~H., \&
  {Heap}, S.~R. 1988, \aap,
  \href{https://ui.adsabs.harvard.edu/abs/1988A%26A...207..123P}{207, 123}

\bibitem[{{Pe{\~n}a} {et~al.}(1998){Pe{\~n}a}, {Stasi{\'n}ska}, {Esteban},
  {Koesterke}, {Medina}, \& {Kingsburgh}}]{Pena1998}
{Pe{\~n}a}, M., {Stasi{\'n}ska}, G., {Esteban}, C., {et~al.} 1998, \aap,
  \href{https://ui.adsabs.harvard.edu/abs/1998A%26A...337..866P}{337, 866}

\bibitem[{{Pe{\~n}a} {et~al.}(2001){Pe{\~n}a}, {Stasi{\'n}ska}, \&
  {Medina}}]{Pena2001}
{Pe{\~n}a}, M., {Stasi{\'n}ska}, G., \& {Medina}, S. 2001,
  {\href{https://dx.doi.org/10.1051/0004-6361:20000497}{\color{magenta}\aap}},
  \href{https://ui.adsabs.harvard.edu/abs/2001A%26A...367..983P}{367, 983}

\bibitem[{{Rapoport} {et~al.}(2021){Rapoport}, {Bear}, \&
  {Soker}}]{Rapoport2021}
{Rapoport}, I., {Bear}, E., \& {Soker}, N. 2021,
  {\href{https://dx.doi.org/10.1093/mnras/stab1774}{\color{magenta}\mnras}},
  \href{https://ui.adsabs.harvard.edu/abs/2021MNRAS.506..468R}{506, 468}

\bibitem[{{Richer} {et~al.}(2009){Richer}, {B{\'a}ez}, {L{\'o}pez}, {Riesgo},
  \& {Garc{\'{\i}}a-D{\'{\i}}az}}]{Richer2009}
{Richer}, M.~G., {B{\'a}ez}, S.-H., {L{\'o}pez}, J.~A., {Riesgo}, H., \&
  {Garc{\'{\i}}a-D{\'{\i}}az}, M.~T. 2009, \rmxaa,
  \href{https://ui.adsabs.harvard.edu/abs/2009RMxAA..45..239R}{45, 239}

\bibitem[{{Sahai} {et~al.}(2011){Sahai}, {Morris}, \& {Villar}}]{Sahai2011}
{Sahai}, R., {Morris}, M.~R., \& {Villar}, G.~G. 2011,
  {\href{https://dx.doi.org/10.1088/0004-6256/141/4/134}{\color{magenta}\aj}},
  \href{https://ui.adsabs.harvard.edu/abs/2011AJ....141..134S}{141, 134}

\bibitem[{{Sch{\"o}nberner} {et~al.}(2010){Sch{\"o}nberner}, {Jacob}, {Sandin},
  \& {Steffen}}]{Schonberner2010}
{Sch{\"o}nberner}, D., {Jacob}, R., {Sandin}, C., \& {Steffen}, M. 2010,
  {\href{https://dx.doi.org/10.1051/0004-6361/200913427}{\color{magenta}\aap}},
  \href{https://ui.adsabs.harvard.edu/abs/2010A%26A...523A..86S}{523, A86}

\bibitem[{{Sch{\"o}nberner} {et~al.}(2005b){Sch{\"o}nberner}, {Jacob}, \&
  {Steffen}}]{Schonberner2005b}
{Sch{\"o}nberner}, D., {Jacob}, R., \& {Steffen}, M. 2005b,
  {\href{https://dx.doi.org/10.1051/0004-6361:20053108}{\color{magenta}\aap}},
  \href{https://ui.adsabs.harvard.edu/abs/2005A%26A...441..573S}{441, 573}

\bibitem[{{Sch{\"o}nberner} {et~al.}(2005a){Sch{\"o}nberner}, {Jacob},
  {Steffen}, {Perinotto}, {Corradi}, \& {Acker}}]{Schonberner2005a}
{Sch{\"o}nberner}, D., {Jacob}, R., {Steffen}, M., {et~al.} 2005a,
  {\href{https://dx.doi.org/10.1051/0004-6361:20041669}{\color{magenta}\aap}},
  \href{https://ui.adsabs.harvard.edu/abs/2005A%26A...431..963S}{431, 963}

\bibitem[{{Schwarz} {et~al.}(1992){Schwarz}, {Corradi}, \&
  {Melnick}}]{Schwarz1992}
{Schwarz}, H.~E., {Corradi}, R.~L.~M., \& {Melnick}, J. 1992, \aaps,
  \href{https://ui.adsabs.harvard.edu/abs/1992A%26AS...96...23S}{96, 23}

\bibitem[{{Shaw} \& {Kaler}(1989)}]{Shaw1989}
{Shaw}, R.~A. \& {Kaler}, J.~B. 1989,
  {\href{https://dx.doi.org/10.1086/191320}{\color{magenta}\apjs}},
  \href{https://ui.adsabs.harvard.edu/abs/1989ApJS...69..495S}{69, 495}

\bibitem[{{Soker}(1990)}]{Soker1990}
{Soker}, N. 1990,
  {\href{https://dx.doi.org/10.1086/115465}{\color{magenta}\aj}},
  \href{https://ui.adsabs.harvard.edu/abs/1990AJ.....99.1869S}{99, 1869}

\bibitem[{{Soker}(2006)}]{Soker2006}
{Soker}, N. 2006,
  {\href{https://dx.doi.org/10.1086/498829}{\color{magenta}\pasp}},
  \href{https://ui.adsabs.harvard.edu/abs/2006PASP..118..260S}{118, 260}

\bibitem[{{Soker} \& {Harpaz}(1992)}]{Soker1992}
{Soker}, N. \& {Harpaz}, A. 1992,
  {\href{https://dx.doi.org/10.1086/133076}{\color{magenta}\pasp}},
  \href{https://ui.adsabs.harvard.edu/abs/1992PASP..104..923S}{104, 923}

\bibitem[{{Soker} \& {Livio}(1994)}]{Soker1994}
{Soker}, N. \& {Livio}, M. 1994,
  {\href{https://dx.doi.org/10.1086/173639}{\color{magenta}\apj}},
  \href{https://ui.adsabs.harvard.edu/abs/1994ApJ...421..219S}{421, 219}

\bibitem[{{Stanghellini} {et~al.}(1993){Stanghellini}, {Corradi}, \&
  {Schwarz}}]{Stanghellini1993}
{Stanghellini}, L., {Corradi}, R.~L.~M., \& {Schwarz}, H.~E. 1993, \aap,
  \href{https://ui.adsabs.harvard.edu/abs/1993A%26A...279..521S}{279, 521}

\bibitem[{{Stanghellini} \& {Haywood}(2010)}]{Stanghellini2010}
{Stanghellini}, L. \& {Haywood}, M. 2010,
  {\href{https://dx.doi.org/10.1088/0004-637X/714/2/1096}{\color{magenta}\apj}},
  \href{https://ui.adsabs.harvard.edu/abs/2010ApJ...714.1096S}{714, 1096}

\bibitem[{{Stanghellini} {et~al.}(2008){Stanghellini}, {Shaw}, \&
  {Villaver}}]{Stanghellini2008}
{Stanghellini}, L., {Shaw}, R.~A., \& {Villaver}, E. 2008,
  {\href{https://dx.doi.org/10.1086/592395}{\color{magenta}\apj}},
  \href{https://ui.adsabs.harvard.edu/abs/2008ApJ...689..194S}{689, 194}

\bibitem[{{Steffen} {et~al.}(2009){Steffen}, {Garc{\'{\i}}a-Segura}, \&
  {Koning}}]{Steffen2009}
{Steffen}, W., {Garc{\'{\i}}a-Segura}, G., \& {Koning}, N. 2009,
  {\href{https://dx.doi.org/10.1088/0004-637X/691/1/696}{\color{magenta}\apj}},
  \href{https://ui.adsabs.harvard.edu/abs/2009ApJ...691..696S}{691, 696}

\bibitem[{Steffen {et~al.}(2011)Steffen, Koning, Wenger, Morisset, \&
  Magnor}]{Steffen2011}
Steffen, W., Koning, N., Wenger, S., Morisset, C., \& Magnor, M. 2011,
  {\href{https://dx.doi.org/10.1109/TVCG.2010.62}{\color{magenta}IEEE Trans.
  Vis. Comput. Graphics}}, 
  \href{https://ui.adsabs.harvard.edu/abs/2011ITVCG..17..454S/abstract}{17, 454}

\bibitem[{{Steffen} \& {L{\'o}pez}(2006)}]{Steffen2006}
{Steffen}, W. \& {L{\'o}pez}, J.~A. 2006, \rmxaa,
  \href{https://ui.adsabs.harvard.edu/abs/2006RMxAA..42...99S}{42, 99}

\bibitem[{{Tajitsu} \& {Tamura}(1998)}]{Tajitsu1998}
{Tajitsu}, A. \& {Tamura}, S. 1998,
  {\href{https://dx.doi.org/10.1086/300315}{\color{magenta}\aj}},
  \href{https://ui.adsabs.harvard.edu/abs/1998AJ....115.1989T}{115, 1989}

\bibitem[{{Tylenda} {et~al.}(1993){Tylenda}, {Acker}, \&
  {Stenholm}}]{Tylenda1993}
{Tylenda}, R., {Acker}, A., \& {Stenholm}, B. 1993, \aaps,
  \href{https://ui.adsabs.harvard.edu/abs/1993A%26AS..102..595T}{102, 595}

\bibitem[{{Tylenda} {et~al.}(2003){Tylenda}, {Si{\'o}dmiak}, {G{\'o}rny},
  {Corradi}, \& {Schwarz}}]{Tylenda2003}
{Tylenda}, R., {Si{\'o}dmiak}, N., {G{\'o}rny}, S.~K., {Corradi}, R.~L.~M., \&
  {Schwarz}, H.~E. 2003,
  {\href{https://dx.doi.org/10.1051/0004-6361:20030645}{\color{magenta}\aap}},
  \href{https://ui.adsabs.harvard.edu/abs/2003A%26A...405..627T}{405, 627}

\bibitem[{{Tyndall} {et~al.}(2012){Tyndall}, {Jones}, {Lloyd}, {O'Brien}, \&
  {Pollacco}}]{Tyndall2012}
{Tyndall}, A.~A., {Jones}, D., {Lloyd}, M., {O'Brien}, T.~J., \& {Pollacco}, D.
  2012,
  {\href{https://dx.doi.org/10.1111/j.1365-2966.2012.20755.x}{\color{magenta}\mnras}},
  \href{https://ui.adsabs.harvard.edu/abs/2012MNRAS.422.1804T}{422, 1804}

\bibitem[{{van der Hucht}(2001)}]{vanderHucht2001}
{van der Hucht}, K.~A. 2001,
  {\href{https://dx.doi.org/10.1016/S1387-6473(00)00112-3}{\color{magenta}\nar}},
  \href{https://ui.adsabs.harvard.edu/abs/2001NewAR..45..135V}{45, 135}

\bibitem[{{van der Hucht} {et~al.}(1981){van der Hucht}, {Conti}, {Lundstrom},
  \& {Stenholm}}]{vanderHucht1981}
{van der Hucht}, K.~A., {Conti}, P.~S., {Lundstrom}, I., \& {Stenholm}, B.
  1981, {\href{https://dx.doi.org/10.1007/BF00173260}{\color{magenta}\ssr}},
  \href{https://ui.adsabs.harvard.edu/abs/1981SSRv...28..227V}{28, 227}

\bibitem[{{van Dokkum}(2001)}]{Dokkum2001}
{van Dokkum}, P.~G. 2001,
  {\href{https://dx.doi.org/10.1086/323894}{\color{magenta}\pasp}},
  \href{https://ui.adsabs.harvard.edu/abs/2001PASP..113.1420V}{113, 1420}

\bibitem[{{Weidmann} {et~al.}(2008){Weidmann}, {Gamen}, {D{\'{\i}}az}, \&
  {Niemela}}]{Weidmann2008}
{Weidmann}, W.~A., {Gamen}, R., {D{\'{\i}}az}, R.~J., \& {Niemela}, V.~S. 2008,
  {\href{https://dx.doi.org/10.1051/0004-6361:200809989}{\color{magenta}\aap}},
  \href{https://ui.adsabs.harvard.edu/abs/2008A%26A...488..245W}{488, 245}

\bibitem[{{Weidmann} {et~al.}(2015){Weidmann}, {M{\'e}ndez}, \&
  {Gamen}}]{Weidmann2015}
{Weidmann}, W.~A., {M{\'e}ndez}, R.~H., \& {Gamen}, R. 2015,
  {\href{https://dx.doi.org/10.1051/0004-6361/201526096}{\color{magenta}\aap}},
  \href{https://ui.adsabs.harvard.edu/abs/2015A&A...579A..86W}{579, A86}

\bibitem[{{Weidmann} {et~al.}(2016){Weidmann}, {Schmidt}, {Vena Valdarenas},
  {Ahumada}, {Volpe}, \& {Mudrik}}]{Weidmann2016}
{Weidmann}, W.~A., {Schmidt}, E.~O., {Vena Valdarenas}, R.~R., {et~al.} 2016,
  {\href{https://dx.doi.org/10.1051/0004-6361/201527199}{\color{magenta}\aap}},
  \href{https://ui.adsabs.harvard.edu/abs/2016A&A...592A.103W}{592, A103}

\bibitem[{{Weinberger}(1989)}]{Weinberger1989}
{Weinberger}, R. 1989, \aaps,
  \href{https://ui.adsabs.harvard.edu/abs/1989A%26AS...78..301W}{78, 301}

\bibitem[{{Werner} \& {Herwig}(2006)}]{Werner2006}
{Werner}, K. \& {Herwig}, F. 2006,
  {\href{https://dx.doi.org/10.1086/500443}{\color{magenta}\pasp}},
  \href{https://ui.adsabs.harvard.edu/abs/2006PASP..118..183W}{118, 183}

\bibitem[{{Zhang}(1995)}]{Zhang1995}
{Zhang}, C.~Y. 1995,
  {\href{https://dx.doi.org/10.1086/192173}{\color{magenta}\apjs}},
  \href{https://ui.adsabs.harvard.edu/abs/1995ApJS...98..659Z}{98, 659}

\bibitem[{{Zou} {et~al.}(2020){Zou}, {Frank}, {Chen}, {Reichardt}, {De Marco},
  {Blackman}, {Nordhaus}, {Balick}, {Carroll-Nellenback}, {Chamandy}, \&
  {Liu}}]{Zou2020}
{Zou}, Y., {Frank}, A., {Chen}, Z., {et~al.} 2020,
  {\href{https://dx.doi.org/10.1093/mnras/staa2145}{\color{magenta}\mnras}},
  \href{https://ui.adsabs.harvard.edu/abs/2020MNRAS.497.2855Z}{497, 2855}

\bibitem[{{Zuckerman} \& {Aller}(1986)}]{Zuckerman1986}
{Zuckerman}, B. \& {Aller}, L.~H. 1986,
  {\href{https://dx.doi.org/10.1086/163943}{\color{magenta}\apj}},
  \href{https://ui.adsabs.harvard.edu/abs/1986ApJ...301..772Z}{301, 772}

\end{thebibliography}

\newpage

\onecolumngrid

\begin{appendix}

\onecolumngrid

\setcounter{section}{1}

\vspace{20pt}

\section{Supplementary Data} 

\bigskip

\noindent The following figure sets and interactive figures  are available for the electronic edition of this article:
\medskip\newline
\textbf{Fig. Set~\ref{wc1:ifu_map}.} From left to right, the spatial distribution maps of logarithmic flux intensity, continuum, LSR velocity, and velocity dispersion of the H$\alpha$ $\lambda$6563 and [N\,{\sc ii}] $\lambda$6584 emission lines for: (a) PB\,6, (b) M\,3-30, (c) Hb\,4, (d) IC\,1297, (e) Pe\,1-1, (f) M\,1-32, (g) M\,3-15, (h) M\,1-25, (i) Hen\,2-142, (j) K\,2-16, (k) NGC\,6578, (l) NGC\,6567, (m) NGC\,6629, and (n) Sa\,3-107. Fluxes are in logarithm of $10^{-15}$~erg\,s${}^{-1}$\,cm${}^{-2}$\,spaxel${}^{-1}$ unit, continua in $10^{-15}$~erg\,s${}^{-1}$\,cm${}^{-2}$\,{\AA}$^{-1}$\,spaxel${}^{-1}$, and LSR velocities and velocity dispersion in km\,s${}^{-1}$. The white/black contour in each panel corresponds to $\sim 10$ percent of the mean surface brightness of each object in the H$\alpha$ emission (or $R$-band) retrieved from the SHS (or SSS). North is up and east is toward the left-hand side.
\medskip\newline
\textbf{Fig. Set~\ref{wc1:vel:slic}.} Velocity slices along the H$\alpha$ $\lambda$6563 and [N\,{\sc ii}] $\lambda$6584 emission-line profiles for: (a) PB\,6, (b) M\,3-30, (c) Hb\,4, (d) IC\,1297, (e) Pe\,1-1, (f) M\,1-32, (g) M\,3-15, (h) M\,1-25, (i) Hen\,2-142, (j) K\,2-16, (k) NGC\,6578, (l) NGC\,6567, (m) NGC\,6629, and (n) Sa\,3-107, followed by the associated synthetic velocity-resolved channel maps obtained for all the PNe, except for PB\,6 and Sa\,3-107, produced by the best-fitting morpho-kinematic models with the parameters given in Table~4. Each observed slice has a $\sim 21$ km\,s${}^{-1}$ width, whose central velocity is given in km\,s${}^{-1}$ unit at the top of the panel. The LSR systemic velocity ($v_{\rm sys}$) of each object is given in km\,s${}^{-1}$ unit in the right bottom corner of each observed velocity channel map. The flux color  in each observed slice is in logarithm of $10^{-15}$~erg\,s${}^{-1}$\,cm${}^{-2}$\,spaxel${}^{-1}$ unit. The gray contour in each panel corresponds to $\sim 10$ percent of the mean surface brightness of each object in the H$\alpha$ emission (or $R$-band) retrieved from the SHS (or SSS). North is up and east is toward the left-hand side.
\medskip\newline
\textbf{Fig. Set~\ref{wc1:pv:diagram}.} P--V arrays in the H$\alpha$ $\lambda$6563 and [N\,{\sc ii}] $\lambda$6584 emissions for: (a) PB\,6, (b) M\,3-30, (c) Hb\,4, (d) IC\,1297, (e) Pe\,1-1, (f) M\,1-32, (g) M\,3-15, (h) M\,1-25, (i) Hen\,2-142, (j) K\,2-16, (k) NGC\,6578, (l) NGC\,6567, (m) NGC\,6629, and (n) Sa\,3-107, followed by the associated synthetic P--V diagrams obtained from the best-fitting morpho-kinematic model of all the PNe, except for PB\,6 and Sa\,3-107, with the parameters given in Table~4. Slits oriented with the position angles (PA) along and vertical to the symmetric axis of the morpho-kinematic model passing through the central star, respectively. The velocity in each observed P--V arrays is with respect to the systemic velocity of the object, given in km\,s${}^{-1}$ unit. The angular offset at 0 arcsec is the nebular center.
\medskip\newline
\textbf{Fig. Set~\ref{wc1:model:shape}.} \textsc{shape} wireframe models of (a) M\,3-30, (b) Hb\,4, (c) IC\,1297, (d) Pe\,1-1, (e) M\,1-32, (f) M\,3-15, (g) M\,1-25, (h) Hen\,2-142, (i) K\,2-16, (j) NGC\,6578, (k) NGC\,6567, and (l) NGC\,6629, in the top view ($i=0^{\circ}$; first column), the front view ($i=90^{\circ}$; second column), and the best-fitting inclination and orientation (third column; see Figure~5 for an interactive viewer in the online journal), followed by the rendered image (fourth column) and the associated archival imaging observation (fifth column), respectively. The parameters of the best-fitting models are listed in Table~4. The archival data include the \textit{HST} images of Hb\,4, Pe\,1-1, M\,3-15, M\,1-25, Hen\,2-142, NGC\,6578, NGC\,6567 and NGC\,6629, with the instruments and filters/gratings listed in Table~2; and the narrow-band H$\alpha$+[N\,{\sc ii}] filter images of M\,3-30, IC\,1297, M\,1-32 and K\,2-16 taken with the 3.5-m ESO NTT from Schwarz et al. (1992), with the image scales shown by solid lines. North is up and east is toward the left-hand side in each archival image. 
\medskip\newline
\textbf{Fig.~5.} 3D mesh models of the PNe M\,3-30, Hb\,4, IC\,1297, 
Th\,2-A, Pe\,1-1, M\,1-32, M\,3-15, M\,1-25, Hen\,2-142, K\,2-16, MGC\,6578, M\,2-42, NGC\,6567 and NGC\,6629 
in an interactive X3D file viewer.

\end{appendix}



\newpage

\renewcommand{\arraystretch}{0.68}


\newpage
\FloatBarrier
\normalsize



\figurenum{1}
\begin{figure*}
\begin{center}
{\footnotesize (a) PB\,6 H$\alpha$ $\lambda$6563}\\ 
\includegraphics[width=6.8in]{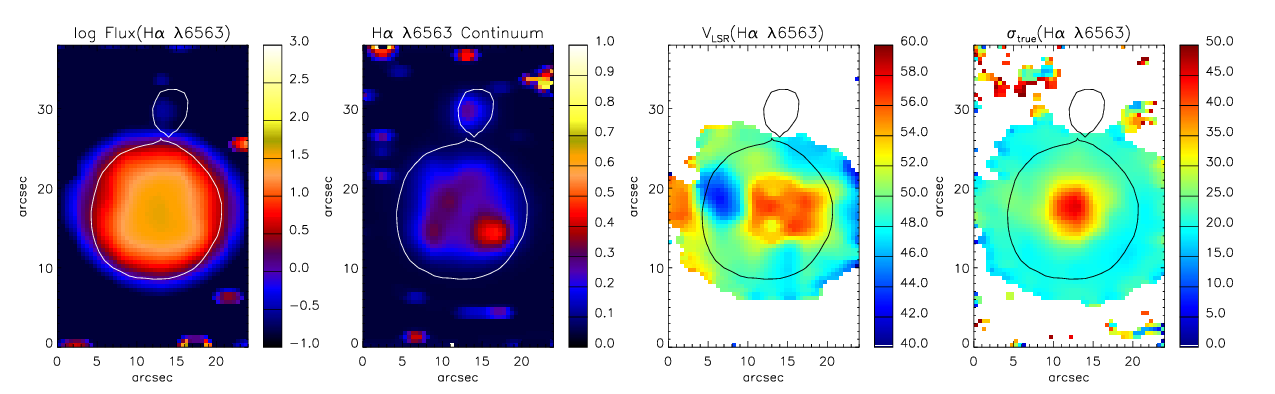}\\
{\footnotesize (a) PB\,6 [N\,{\sc ii}] $\lambda$6584}\\ 
\includegraphics[width=6.8in]{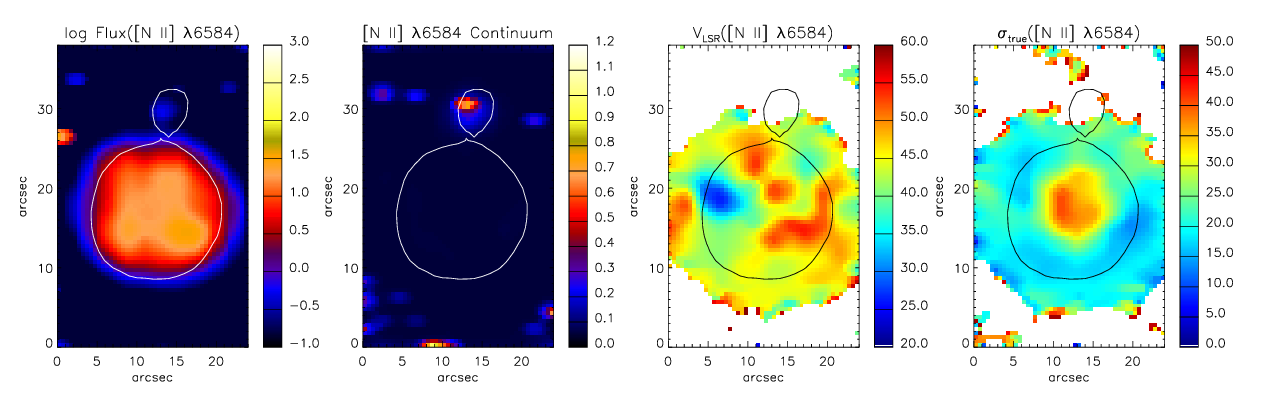}\\
{\footnotesize (b) M\,3-30 H$\alpha$ $\lambda$6563}\\ 
\includegraphics[width=6.8in]{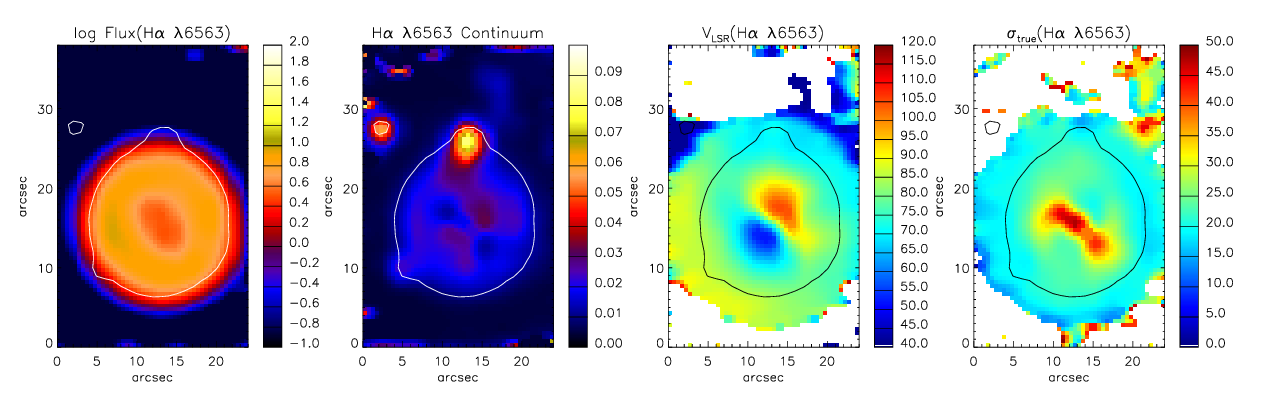}\\
{\footnotesize (b) M\,3-30 [N\,{\sc ii}] $\lambda$6584}\\ 
\includegraphics[width=6.8in]{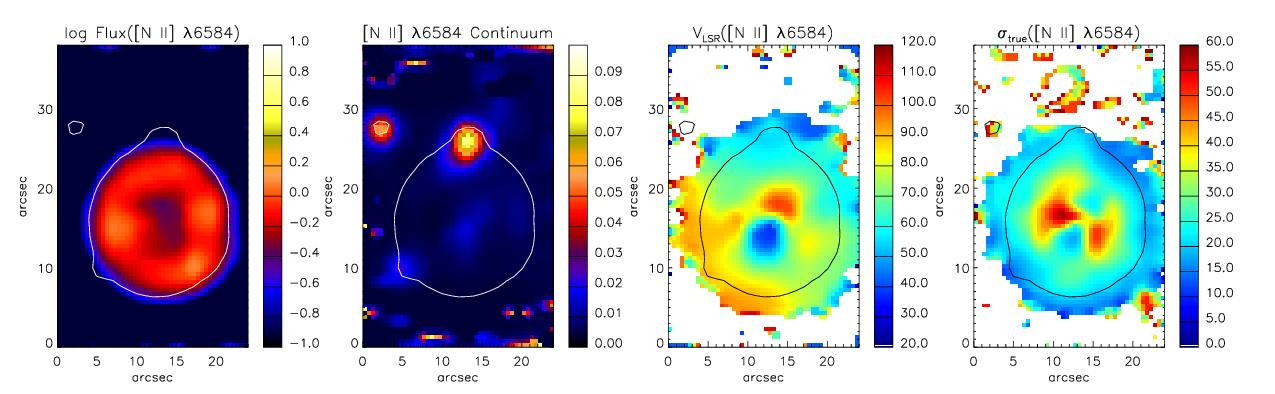}
\caption{From left to right, the spatial distribution maps of logarithmic flux intensity, continuum, LSR velocity, and velocity dispersion of the H$\alpha$ $\lambda$6563 and [N\,{\sc ii}] $\lambda$6584 emission lines for: (a) PB\,6, (b) M\,3-30, (c) Hb\,4, (d) IC\,1297, (e) Pe\,1-1, (f) M\,1-32, (g) M\,3-15, (h) M\,1-25, (i) Hen\,2-142, (j) K\,2-16, (k) NGC\,6578, (l) NGC\,6567, (m) NGC\,6629, and (n) Sa\,3-107. Fluxes are in logarithm of $10^{-15}$~erg\,s${}^{-1}$\,cm${}^{-2}$\,spaxel${}^{-1}$ unit, continua in $10^{-15}$~erg\,s${}^{-1}$\,cm${}^{-2}$\,{\AA}$^{-1}$\,spaxel${}^{-1}$, and LSR velocities and velocity dispersion in km\,s${}^{-1}$. The white/black contour in each panel corresponds to $\sim 10$ percent of the mean surface brightness of each object in the H$\alpha$ emission (or $R$-band) retrieved from the SHS (or SSS). North is up and east is toward the left-hand side.
}
\label{wc1:ifu_map}%
\end{center}
\end{figure*}

\newpage
\FloatBarrier
\normalsize

\figurenum{1}
\begin{figure*}
\begin{center}
{\footnotesize (c) Hb\,4 H$\alpha$ $\lambda$6563}\\ 
\includegraphics[width=6.8in]{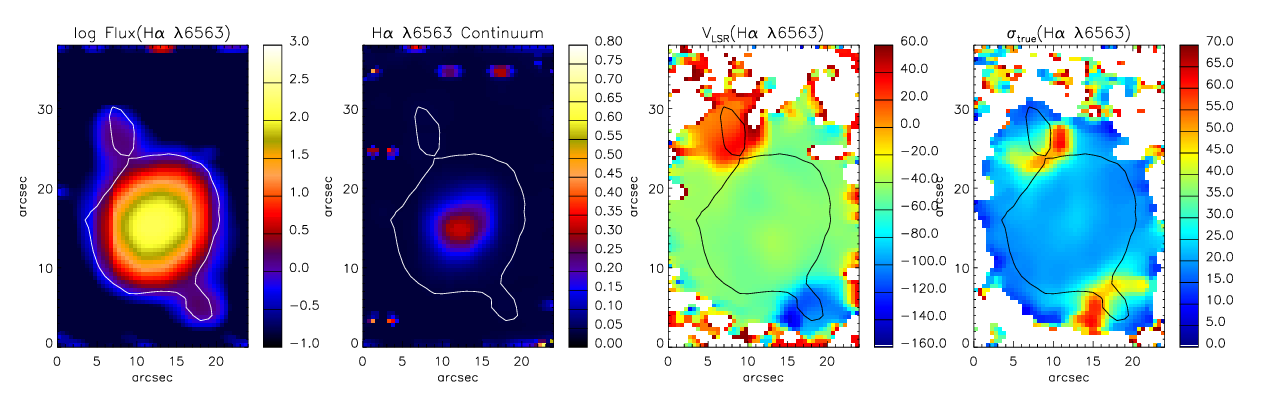}\\
{\footnotesize (c) Hb\,4 [N\,{\sc ii}] $\lambda$6584}\\ 
\includegraphics[width=6.8in]{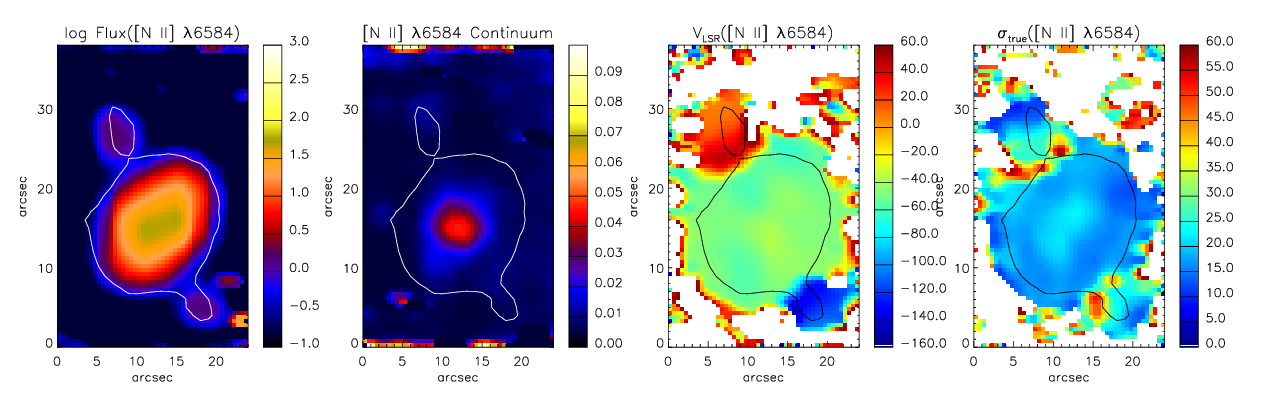}\\
{\footnotesize (d) IC\,1297 H$\alpha$ $\lambda$6563}\\ 
\includegraphics[width=6.8in]{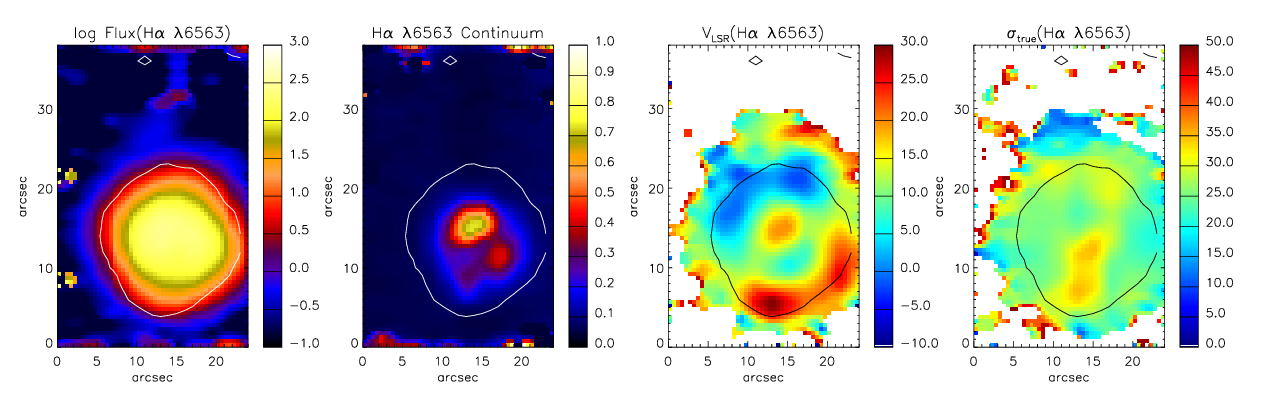}\\
{\footnotesize (d) IC\,1297 [N\,{\sc ii}] $\lambda$6584}\\ 
\includegraphics[width=6.8in]{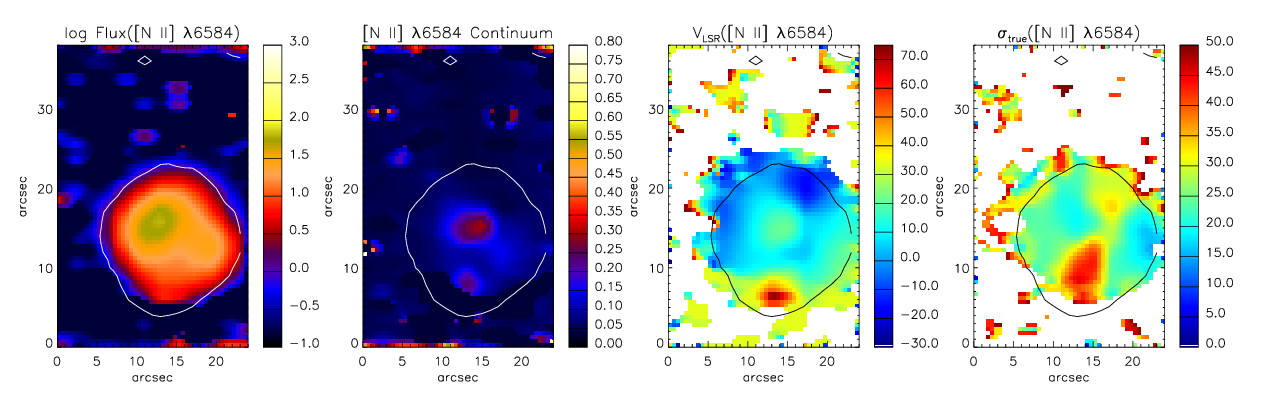}\\
\caption{\textit{-- continued}}
\end{center}
\end{figure*}

\newpage
\FloatBarrier
\normalsize

\figurenum{1}
\begin{figure*}
\begin{center}
{\footnotesize (e) Pe\,1-1 H$\alpha$ $\lambda$6563}\\ 
\includegraphics[width=6.8in]{figure1/fig1_pe1_1_6563.eps}\\
{\footnotesize (e) Pe\,1-1 [N\,{\sc ii}] $\lambda$6584}\\ 
\includegraphics[width=6.8in]{figure1/fig1_pe1_1_6584.eps}\\
{\footnotesize (f) M\,1-32 H$\alpha$ $\lambda$6563}\\ 
\includegraphics[width=6.8in]{figure1/fig1_m1_32_6563.eps}\\
{\footnotesize (f) M\,1-32 [N\,{\sc ii}] $\lambda$6584}\\ 
\includegraphics[width=6.8in]{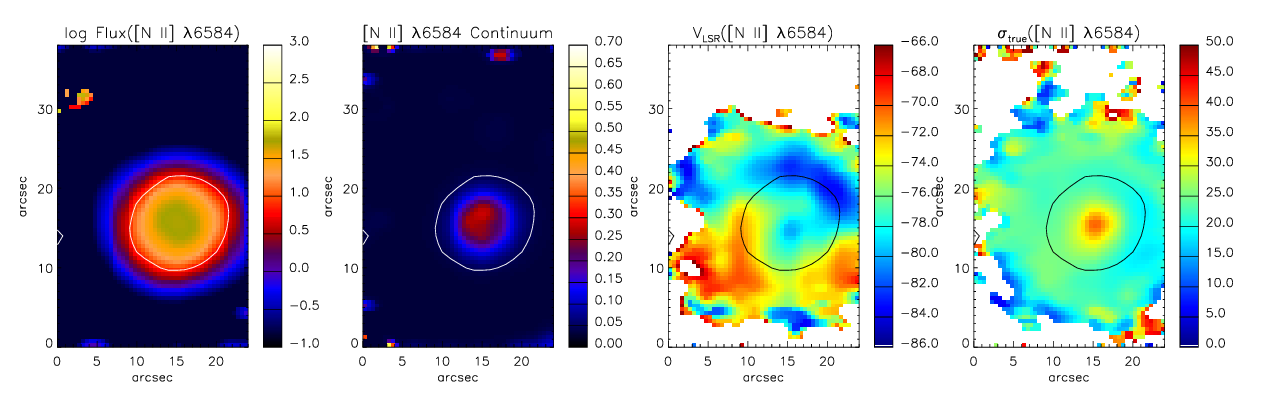}\\
\caption{\textit{-- continued}}
\end{center}
\end{figure*}

\newpage
\FloatBarrier
\normalsize

\figurenum{1}
\begin{figure*}
\begin{center}
{\footnotesize (g) M\,3-15 H$\alpha$ $\lambda$6563}\\ 
\includegraphics[width=6.8in]{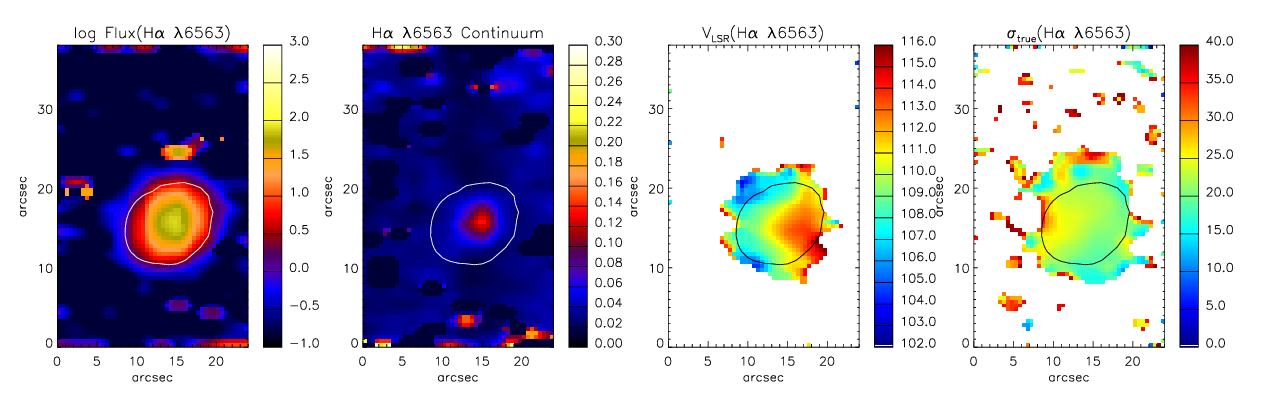}\\
{\footnotesize (g) M\,3-15 [N\,{\sc ii}] $\lambda$6584}\\ 
\includegraphics[width=6.8in]{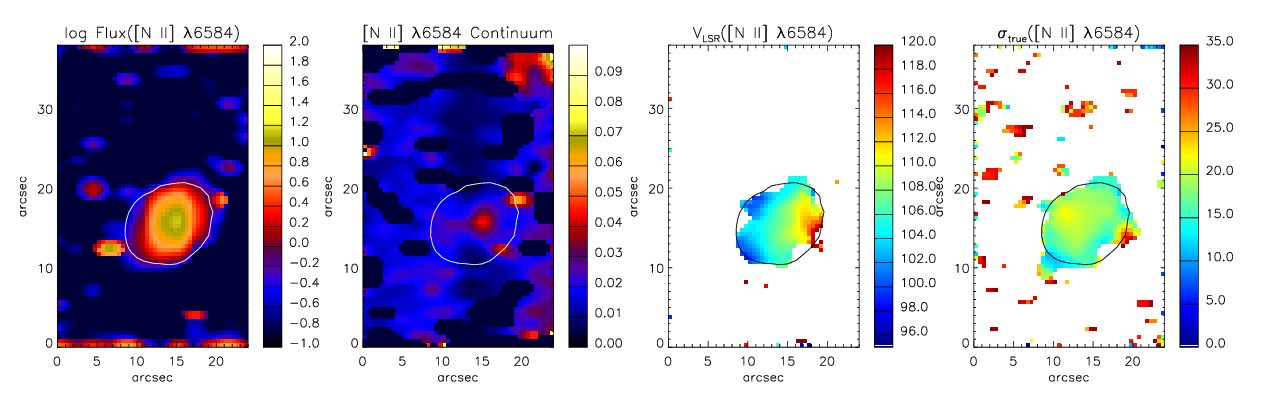}\\
{\footnotesize (h) M\,1-25 H$\alpha$ $\lambda$6563}\\ 
\includegraphics[width=6.8in]{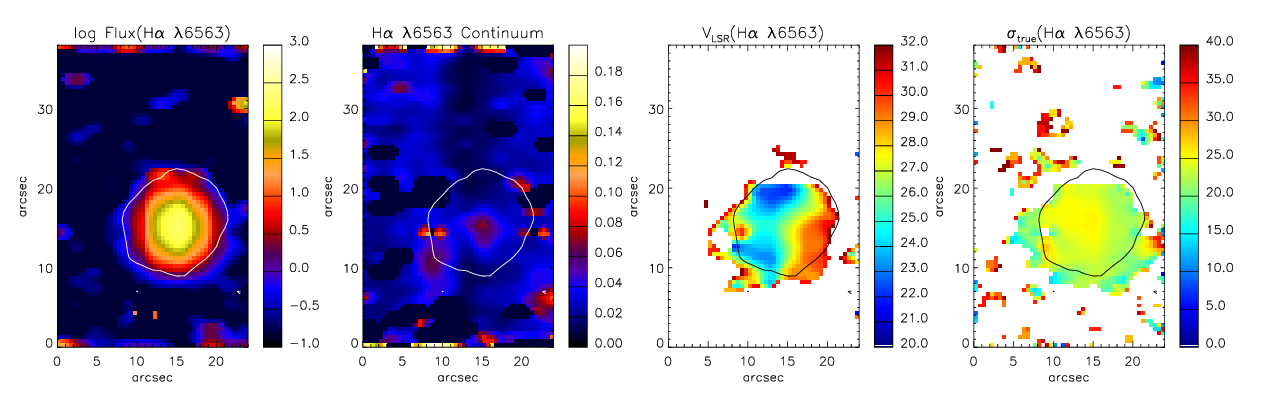}\\
{\footnotesize (h) M\,1-25 [N\,{\sc ii}] $\lambda$6584}\\ 
\includegraphics[width=6.8in]{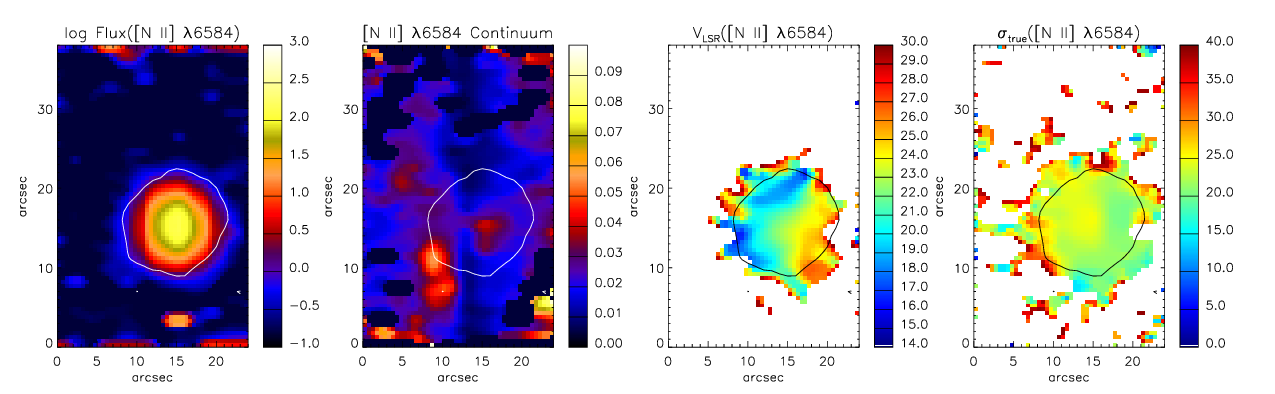}\\
\caption{\textit{-- continued}}
\end{center}
\end{figure*}

\newpage
\FloatBarrier
\normalsize

\figurenum{1}
\begin{figure*}
\begin{center}
{\footnotesize (i) Hen\,2-142 H$\alpha$ $\lambda$6563}\\ 
\includegraphics[width=6.8in]{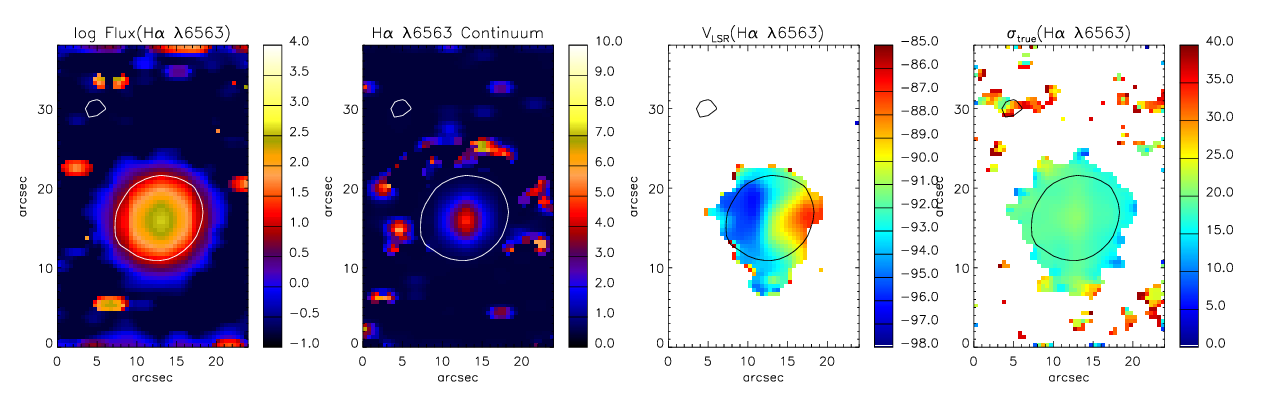}\\
{\footnotesize (i) Hen\,2-142 [N\,{\sc ii}] $\lambda$6584}\\ 
\includegraphics[width=6.8in]{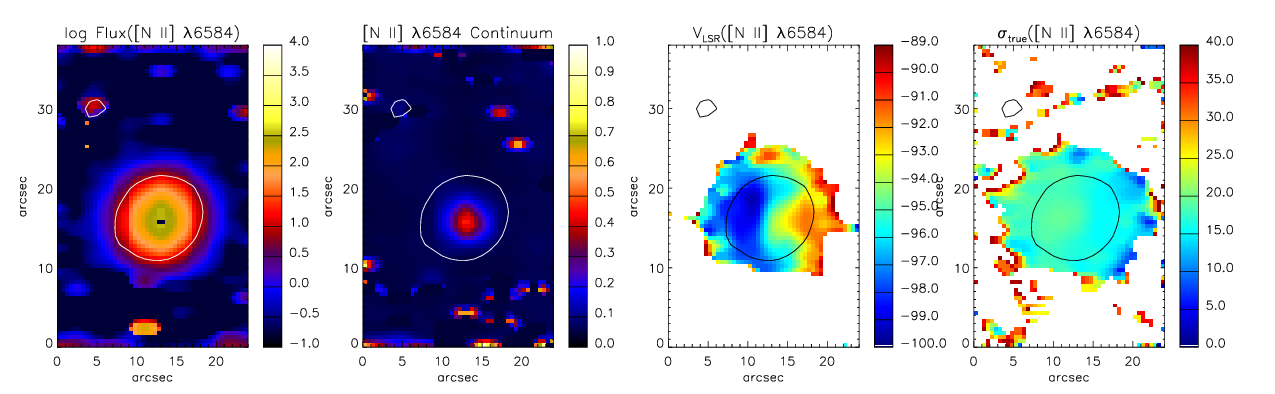}\\
{\footnotesize (j) K\,2-16 H$\alpha$ $\lambda$6563}\\ 
\includegraphics[width=6.8in]{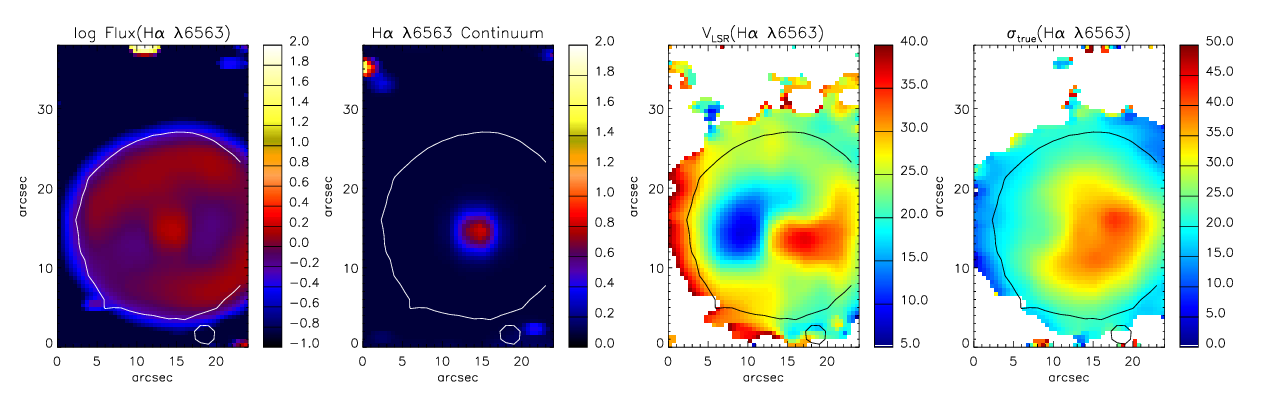}\\
{\footnotesize (j) K\,2-16 [N\,{\sc ii}] $\lambda$6584}\\ 
\includegraphics[width=6.8in]{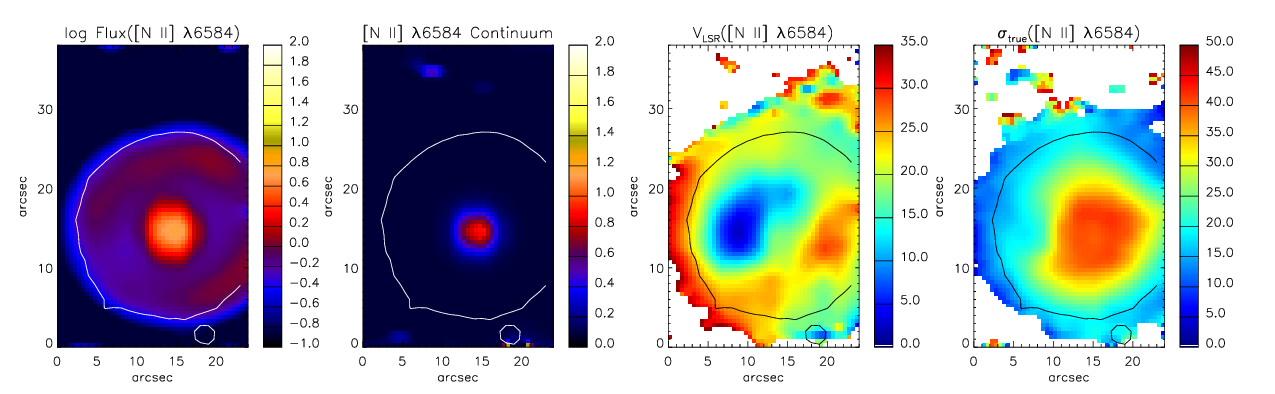}\\
\caption{\textit{-- continued}}
\end{center}
\end{figure*}

\newpage
\FloatBarrier
\normalsize

\figurenum{1}
\begin{figure*}
\begin{center}
{\footnotesize (k) NGC\,6578 H$\alpha$ $\lambda$6563}\\ 
\includegraphics[width=6.8in]{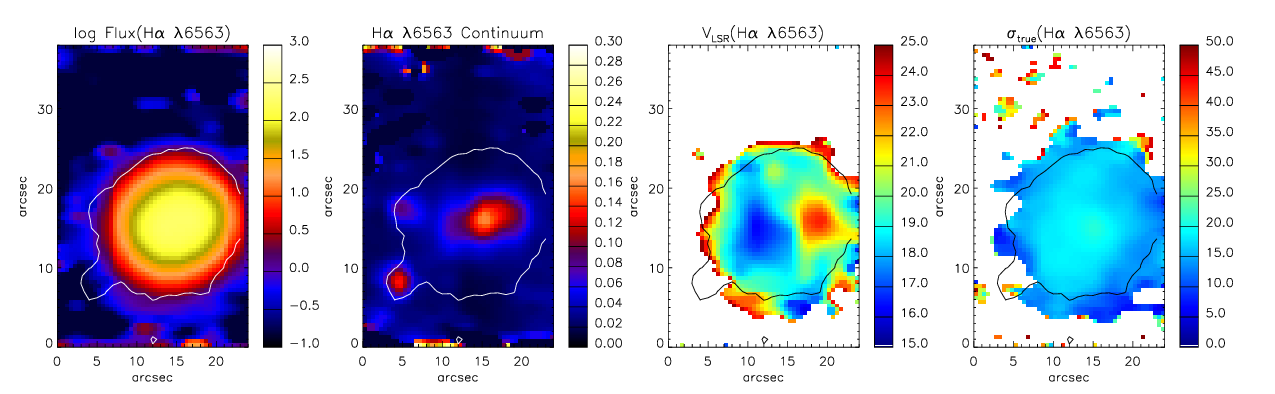}\\
{\footnotesize (k) NGC\,6578 [N\,{\sc ii}] $\lambda$6584}\\ 
\includegraphics[width=6.8in]{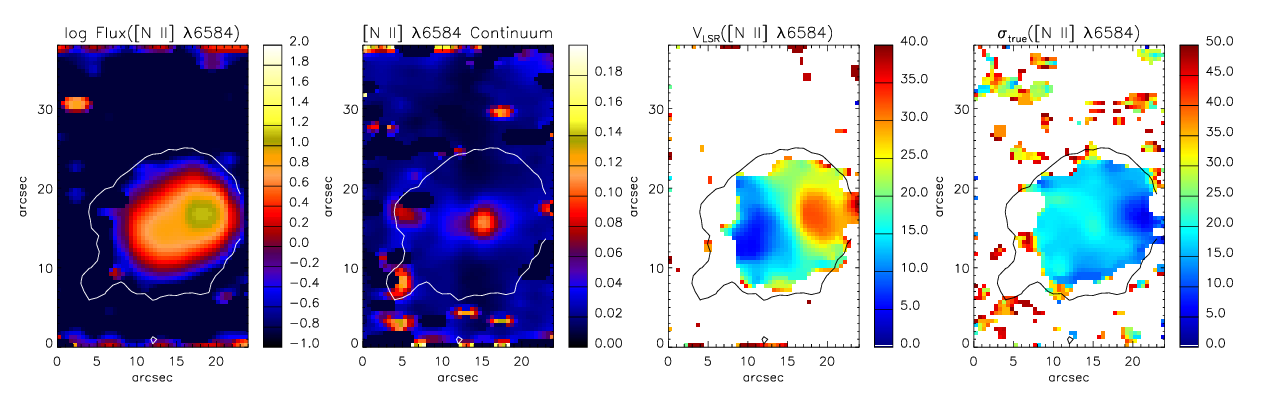}\\
{\footnotesize (l) NGC\,6567 H$\alpha$ $\lambda$6563}\\ 
\includegraphics[width=6.8in]{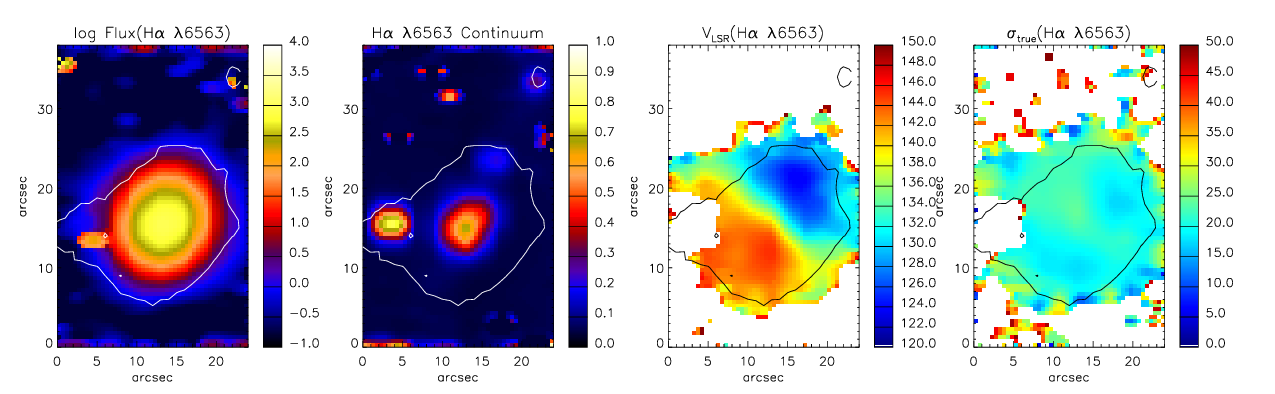}\\
{\footnotesize (l) NGC\,6567 [N\,{\sc ii}] $\lambda$6584}\\ 
\includegraphics[width=6.8in]{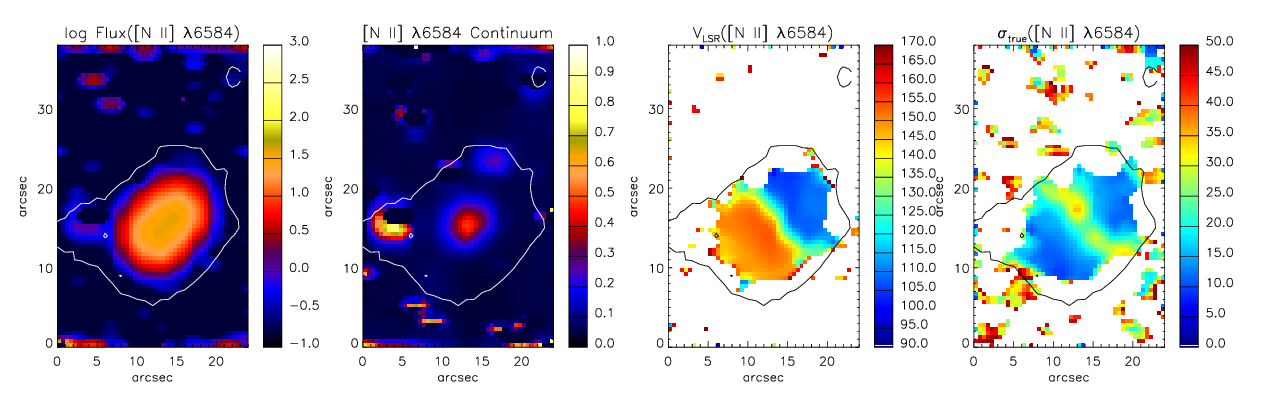}\\
\caption{\textit{-- continued}}
\end{center}
\end{figure*}

\newpage
\FloatBarrier
\normalsize

\figurenum{1}
\begin{figure*}
\begin{center}
{\footnotesize (m) NGC\,6629 H$\alpha$ $\lambda$6563}\\ 
\includegraphics[width=6.8in]{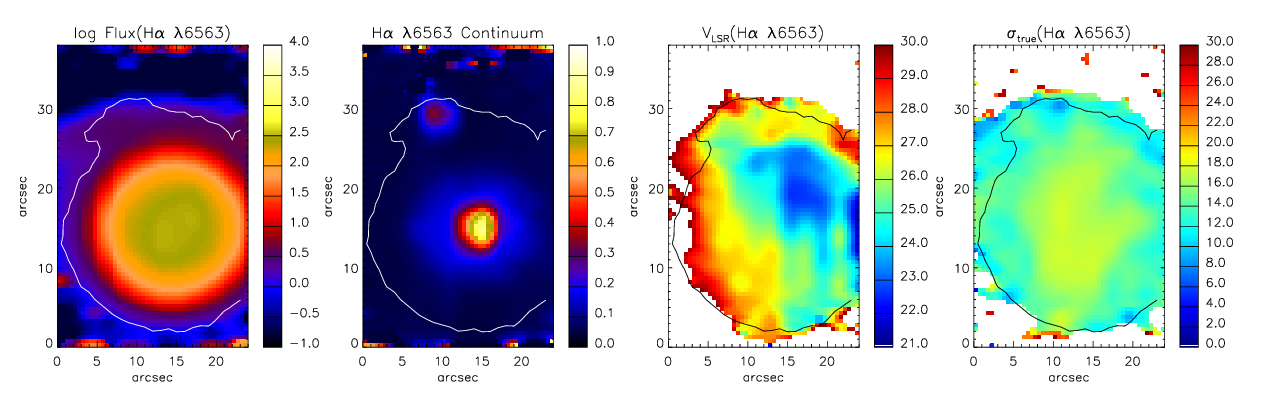}\\
{\footnotesize (m) NGC\,6629 [N\,{\sc ii}] $\lambda$6584}\\ 
\includegraphics[width=6.8in]{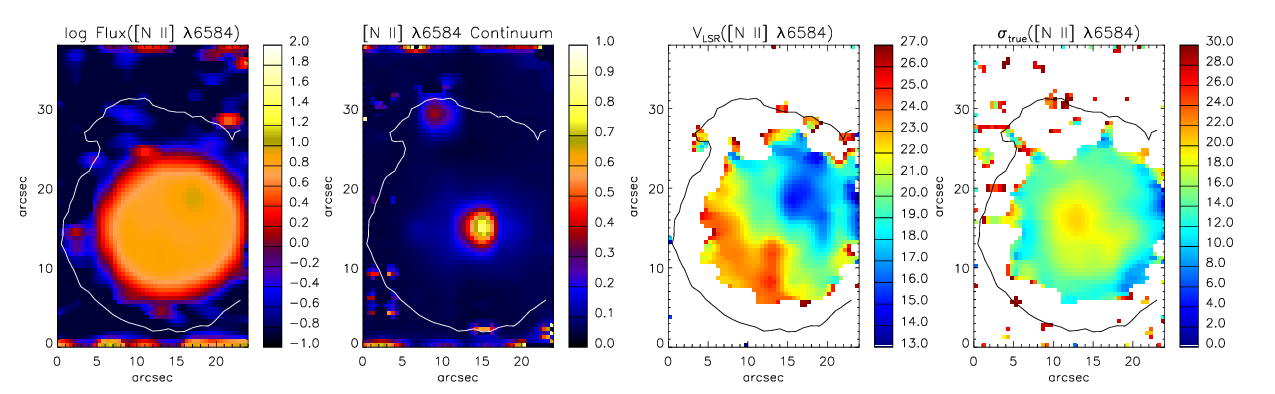}\\
{\footnotesize (n) Sa\,3-107 H$\alpha$ $\lambda$6563}\\ 
\includegraphics[width=6.8in]{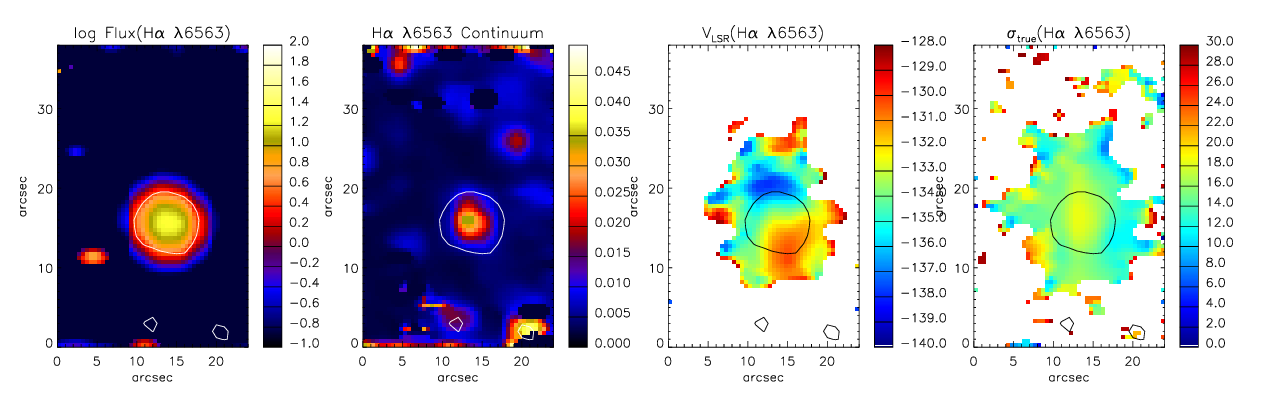}\\
{\footnotesize (n) Sa\,3-107 [N\,{\sc ii}] $\lambda$6584}\\ 
\includegraphics[width=6.8in]{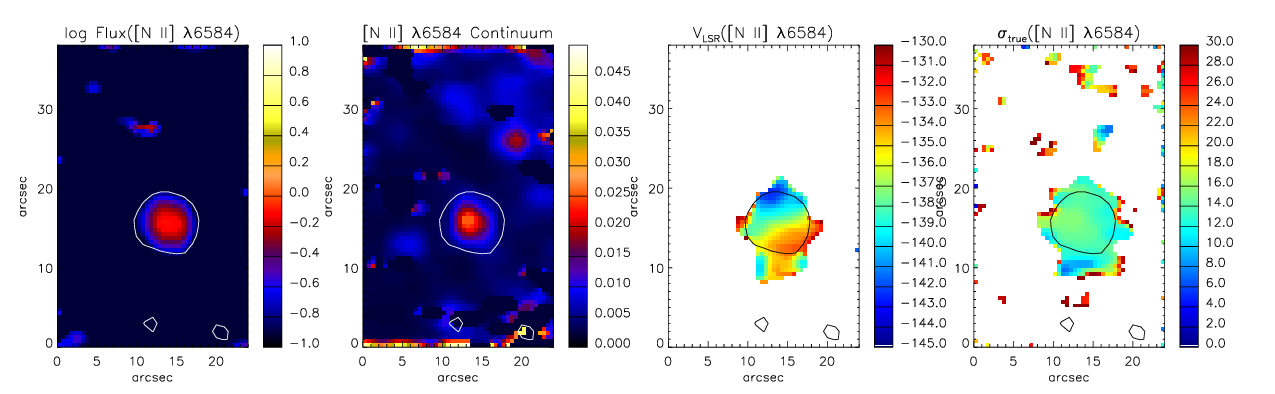}\\
\caption{\textit{-- continued}}
\end{center}
\end{figure*}

\newpage
\FloatBarrier
\normalsize

\figurenum{2}
\begin{figure*}
\begin{center}
{\footnotesize (a) PB\,6 H$\alpha$ $\lambda$6563}\\ 
\includegraphics[width=6.in]{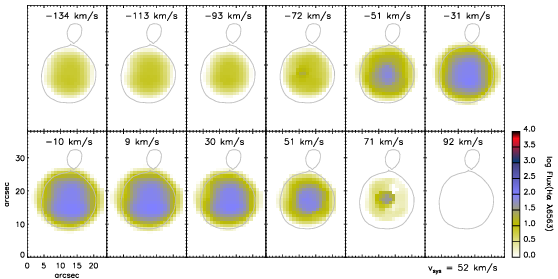}\\
{\footnotesize (a) PB\,6 [N\,{\sc ii}] $\lambda$6584}\\ 
\includegraphics[width=6.in]{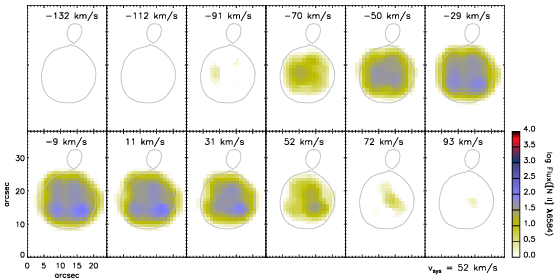}\\
\caption{Velocity slices along the H$\alpha$ $\lambda$6563 and [N\,{\sc ii}] $\lambda$6584 emission-line profiles for: (a) PB\,6, (b) M\,3-30, (c) Hb\,4, (d) IC\,1297, (e) Pe\,1-1, (f) M\,1-32, (g) M\,3-15, (h) M\,1-25, (i) Hen\,2-142, (j) K\,2-16, (k) NGC\,6578, (l) NGC\,6567, (m) NGC\,6629, and (n) Sa\,3-107, followed by the associated synthetic velocity-resolved channel maps obtained for all the PNe, except for PB\,6 and Sa\,3-107, produced by the best-fitting morpho-kinematic models with the parameters given in Table~4. Each observed slice has a $\sim 21$ km\,s${}^{-1}$ width, whose central velocity is given in km\,s${}^{-1}$ unit at the top of the panel. The LSR systemic velocity ($v_{\rm sys}$) of each object is given in km\,s${}^{-1}$ unit in the right bottom corner of each observed velocity channel map. The flux color  in each observed slice is in logarithm of $10^{-15}$~erg\,s${}^{-1}$\,cm${}^{-2}$\,spaxel${}^{-1}$ unit. The gray contour in each panel corresponds to $\sim 10$ percent of the mean surface brightness of each object in the H$\alpha$ emission (or $R$-band) retrieved from the SHS (or SSS). North is up and east is toward the left-hand side.
}
\label{wc1:vel:slic}%
\end{center}
\end{figure*}

\newpage
\FloatBarrier
\normalsize

\figurenum{2}
\begin{figure*}
\begin{center}
{\footnotesize (b) M\,3-30 H$\alpha$ $\lambda$6563}\\ 
\includegraphics[width=6.in]{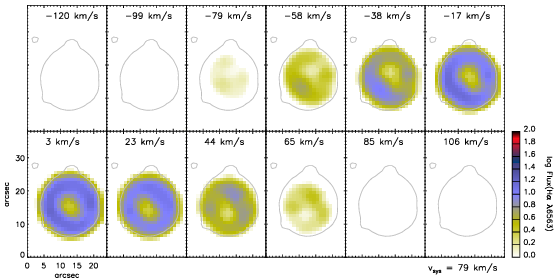}\\
{\footnotesize (b) M\,3-30 Model (H$\alpha$ $\lambda$6563)}\\ 
\includegraphics[width=5.8in]{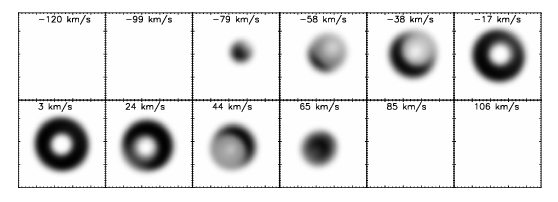}\\
\caption{\textit{-- continued}}
\end{center}
\end{figure*}

\newpage
\FloatBarrier
\normalsize

\figurenum{2}
\begin{figure*}
\begin{center}
{\footnotesize (b) M\,3-30 [N\,{\sc ii}] $\lambda$6584}\\ 
\includegraphics[width=6.in]{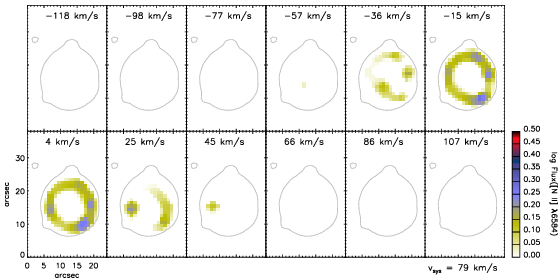}\\
{\footnotesize (b) M\,3-30 Model ([N\,{\sc ii}] $\lambda$6584)}\\ 
\includegraphics[width=5.8in]{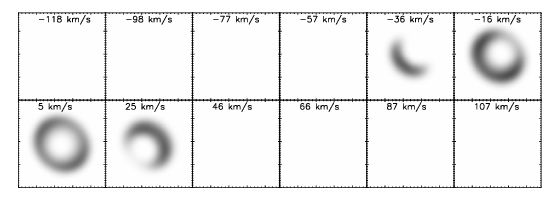}\\
\caption{\textit{-- continued}}
\end{center}
\end{figure*}

\newpage
\FloatBarrier
\normalsize

\figurenum{2}
\begin{figure*}
\begin{center}
{\footnotesize (c) Hb\,4 H$\alpha$ $\lambda$6563}\\ 
\includegraphics[width=6.in]{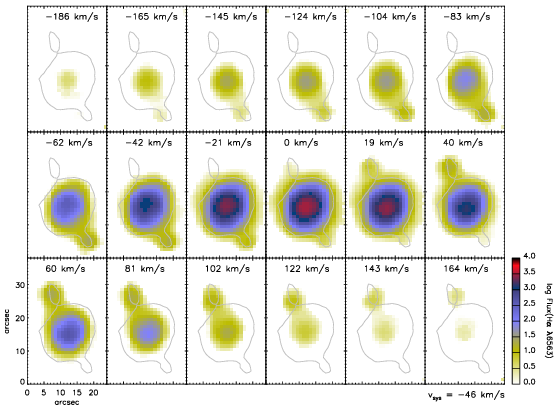}\\
{\footnotesize (c) Hb\,4 Model (H$\alpha$ $\lambda$6563)}\\ 
\includegraphics[width=5.8in]{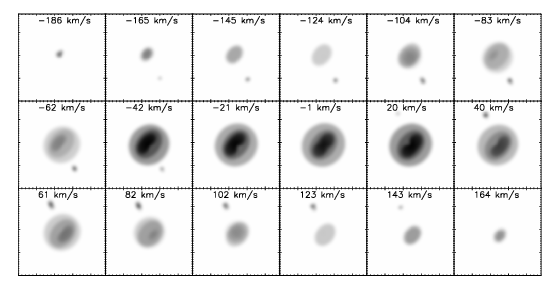}\\
\caption{\textit{-- continued}}
\end{center}
\end{figure*}

\newpage
\FloatBarrier
\normalsize

\figurenum{2}
\begin{figure*}
\begin{center}
{\footnotesize (c) Hb\,4 [N\,{\sc ii}] $\lambda$6584}\\ 
\includegraphics[width=6.in]{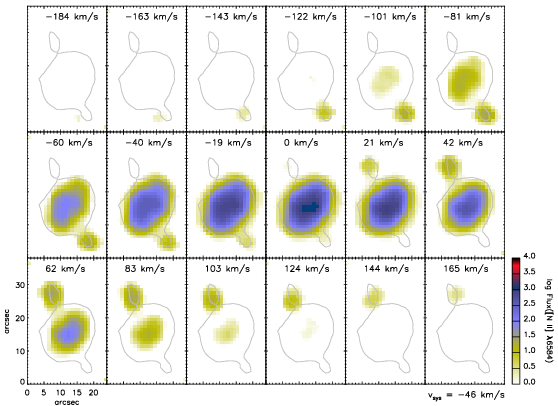}\\
{\footnotesize (c) Hb\,4 Model ([N\,{\sc ii}] $\lambda$6584)}\\ 
\includegraphics[width=5.8in]{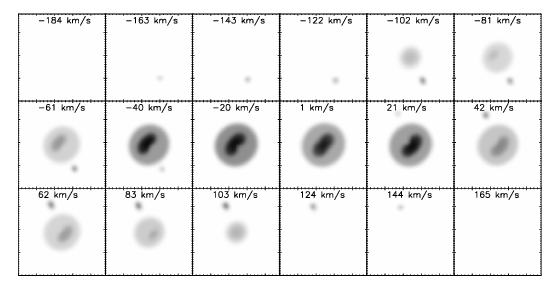}\\
\caption{\textit{-- continued}}
\end{center}
\end{figure*}

\newpage
\FloatBarrier
\normalsize

\figurenum{2}
\begin{figure*}
\begin{center}
{\footnotesize (d) IC\,1297 H$\alpha$ $\lambda$6563}\\ 
\includegraphics[width=6.in]{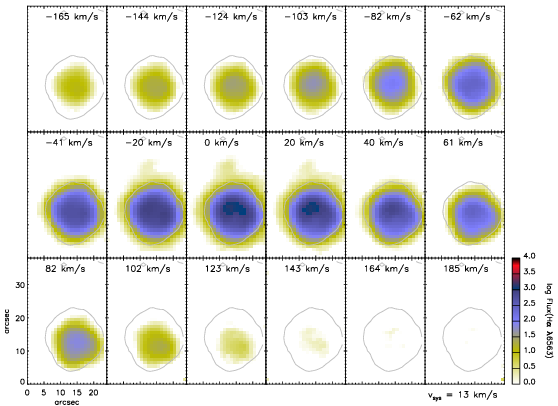}\\
{\footnotesize (d) IC\,1297 Model (H$\alpha$ $\lambda$6563)}\\ 
\includegraphics[width=5.8in]{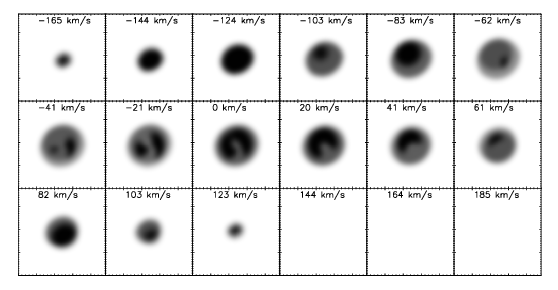}\\
\caption{\textit{-- continued}}
\end{center}
\end{figure*}

\newpage
\FloatBarrier
\normalsize

\figurenum{2}
\begin{figure*}
\begin{center}
{\footnotesize (d) IC\,1297 [N\,{\sc ii}] $\lambda$6584}\\ 
\includegraphics[width=6.in]{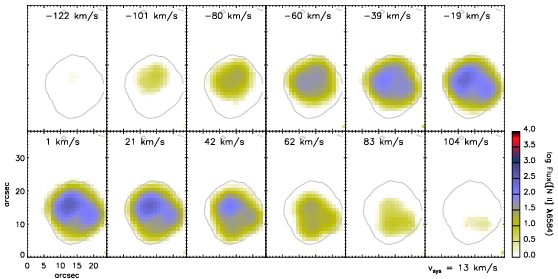}\\
{\footnotesize (d) IC\,1297 Model ([N\,{\sc ii}] $\lambda$6584)}\\ 
\includegraphics[width=5.8in]{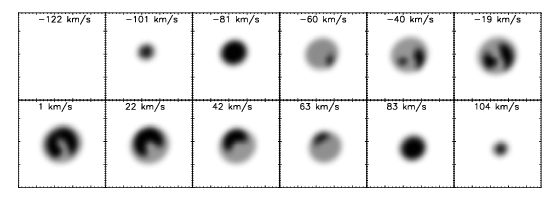}\\
\caption{\textit{-- continued}}
\end{center}
\end{figure*}

\newpage
\FloatBarrier
\normalsize

\figurenum{2}
\begin{figure*}
\begin{center}
{\footnotesize (e) Pe\,1-1 H$\alpha$ $\lambda$6563}\\ 
\includegraphics[width=6.in]{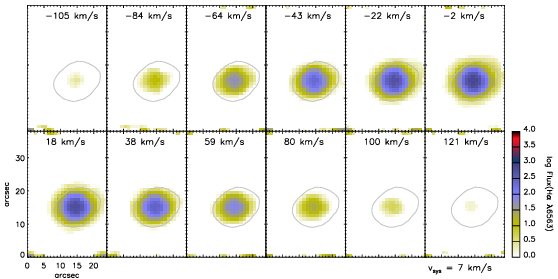}\\
{\footnotesize (e) Pe\,1-1 Model (H$\alpha$ $\lambda$6563)}\\ 
\includegraphics[width=5.8in]{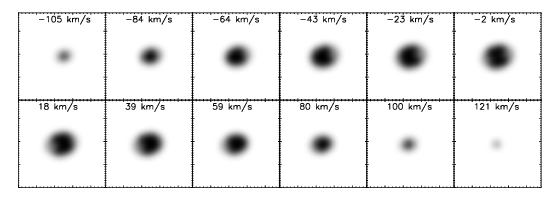}\\
\caption{\textit{-- continued}}
\end{center}
\end{figure*}

\newpage
\FloatBarrier
\normalsize

\figurenum{2}
\begin{figure*}
\begin{center}
{\footnotesize (e) Pe\,1-1 [N\,{\sc ii}] $\lambda$6584}\\ 
\includegraphics[width=6.in]{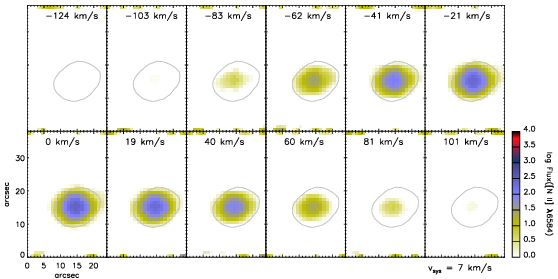}\\
{\footnotesize (e) Pe\,1-1 Model ([N\,{\sc ii}] $\lambda$6584)}\\ 
\includegraphics[width=5.8in]{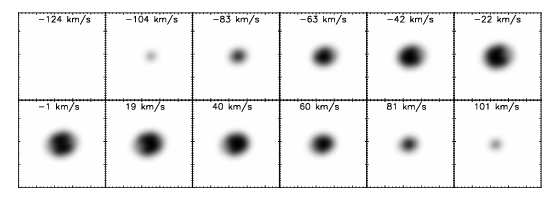}\\
\caption{\textit{-- continued}}
\end{center}
\end{figure*}

\newpage
\FloatBarrier
\normalsize

\figurenum{2}
\begin{figure*}
\begin{center}
{\footnotesize (f) M\,1-32 H$\alpha$ $\lambda$6563}\\ 
\includegraphics[width=6.in]{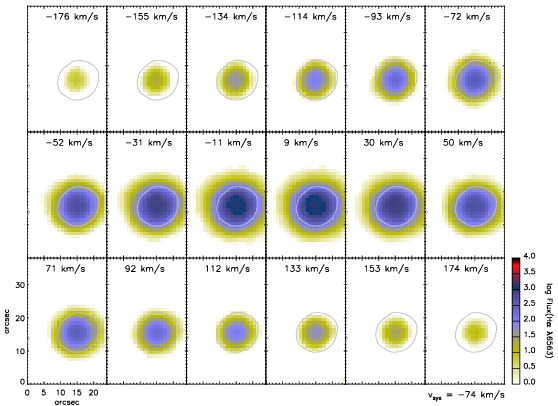}\\
{\footnotesize (f) M\,1-32 Model (H$\alpha$ $\lambda$6563)}\\ 
\includegraphics[width=5.8in]{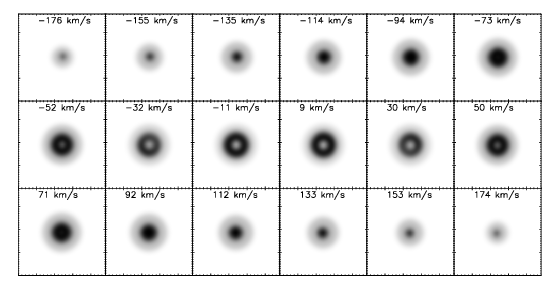}\\
\end{center}
\end{figure*}

\newpage
\FloatBarrier
\normalsize

\figurenum{2}
\begin{figure*}
\begin{center}
{\footnotesize (f) M\,1-32 [N\,{\sc ii}] $\lambda$6584}\\ 
\includegraphics[width=6.in]{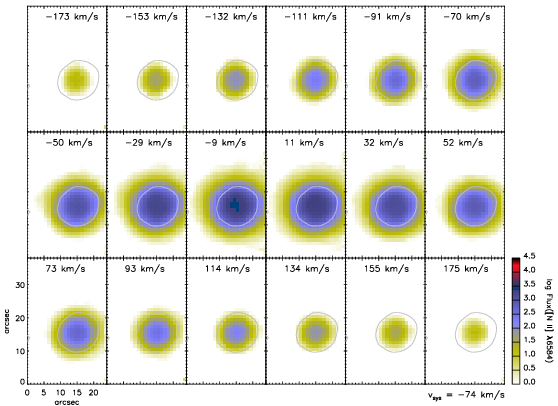}\\
{\footnotesize (f) M\,1-32 Model ([N\,{\sc ii}] $\lambda$6584)}\\ 
\includegraphics[width=5.8in]{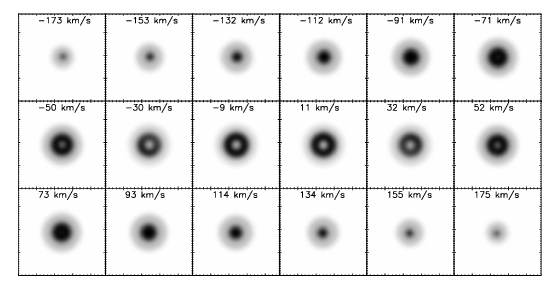}\\
\end{center}
\end{figure*}

\newpage
\FloatBarrier
\normalsize

\figurenum{2}
\begin{figure*}
\begin{center}
{\footnotesize (g) M\,3-15 H$\alpha$ $\lambda$6563}\\ 
\includegraphics[width=6.in]{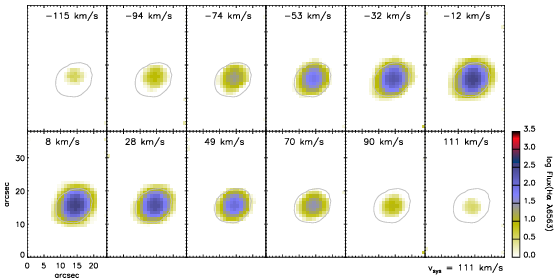}\\
{\footnotesize (g) M\,3-15 Model (H$\alpha$ $\lambda$6563)}\\ 
\includegraphics[width=5.8in]{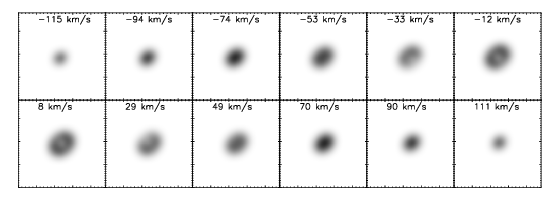}\\
\caption{\textit{-- continued}}
\end{center}
\end{figure*}

\newpage
\FloatBarrier
\normalsize

\figurenum{2}
\begin{figure*}
\begin{center}
{\footnotesize (g) M\,3-15 [N\,{\sc ii}] $\lambda$6584}\\ 
\includegraphics[width=6.in]{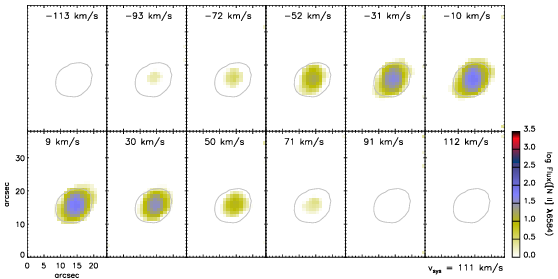}\\
{\footnotesize (g) M\,3-15 Model ([N\,{\sc ii}] $\lambda$6584)}\\ 
\includegraphics[width=5.8in]{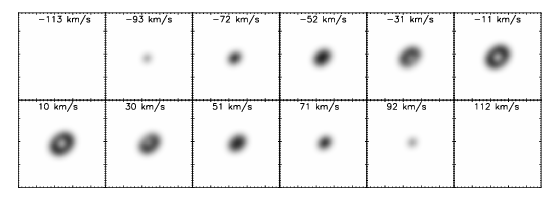}\\
\caption{\textit{-- continued}}
\end{center}
\end{figure*}

\newpage
\FloatBarrier
\normalsize

\figurenum{2}
\begin{figure*}
\begin{center}
{\footnotesize (h) M\,1-25 H$\alpha$ $\lambda$6563}\\ 
\includegraphics[width=6.in]{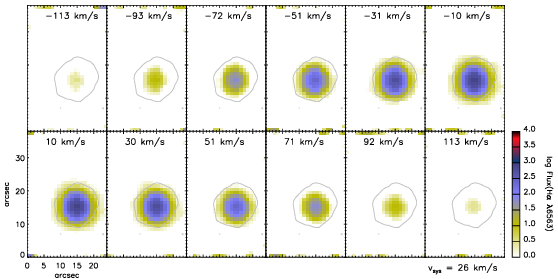}\\
{\footnotesize (h) M\,1-25 Model (H$\alpha$ $\lambda$6563)}\\ 
\includegraphics[width=5.8in]{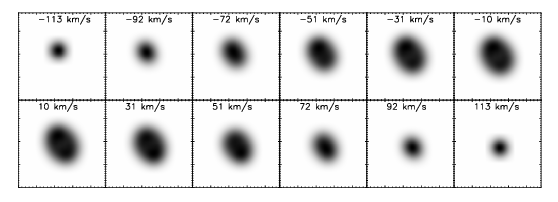}\\
\end{center}
\end{figure*}

\newpage
\FloatBarrier
\normalsize

\figurenum{2}
\begin{figure*}
\begin{center}
{\footnotesize (h) M\,1-25 [N\,{\sc ii}] $\lambda$6584}\\
\includegraphics[width=6.in]{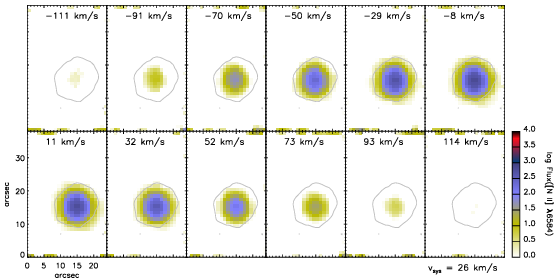}\\
{\footnotesize (h) M\,1-25 Model ([N\,{\sc ii}] $\lambda$6584)}\\ 
\includegraphics[width=5.8in]{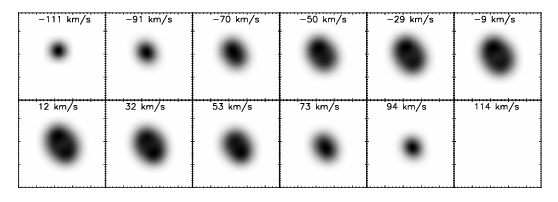}\\
\end{center}
\end{figure*}

\newpage
\FloatBarrier
\normalsize

\figurenum{2}
\begin{figure*}
\begin{center}
{\footnotesize (i) Hen\,2-142 H$\alpha$ $\lambda$6563}\\ 
\includegraphics[width=6.in]{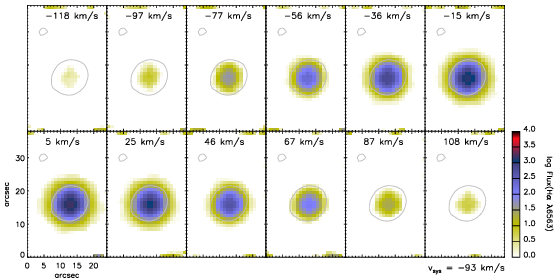}\\
{\footnotesize (i) Hen\,2-142 Model (H$\alpha$ $\lambda$6563)}\\ 
\includegraphics[width=5.8in]{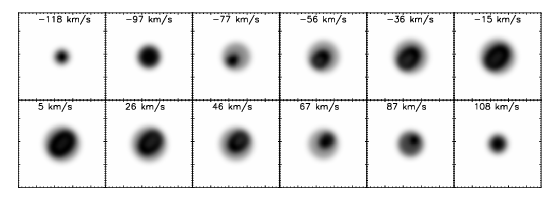}\\
\caption{\textit{-- continued}}
\end{center}
\end{figure*}

\newpage
\FloatBarrier
\normalsize

\figurenum{2}
\begin{figure*}
\begin{center}
{\footnotesize (i) Hen\,2-142 [N\,{\sc ii}] $\lambda$6584}\\ 
\includegraphics[width=6.in]{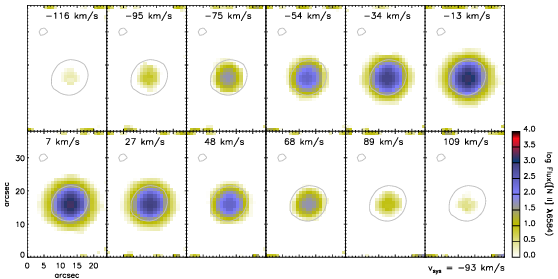}\\
{\footnotesize (i) Hen\,2-142 Model ([N\,{\sc ii}] $\lambda$6584)}\\ 
\includegraphics[width=5.8in]{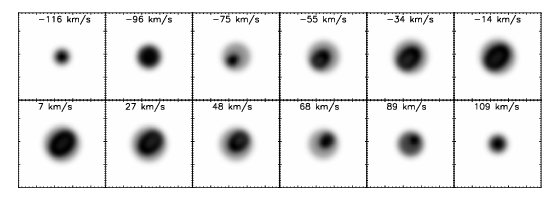}\\
\caption{\textit{-- continued}}
\end{center}
\end{figure*}

\newpage
\FloatBarrier
\normalsize

\figurenum{2}
\begin{figure*}
\begin{center}
{\footnotesize (j) K\,2-16 H$\alpha$ $\lambda$6563}\\ 
\includegraphics[width=6.in]{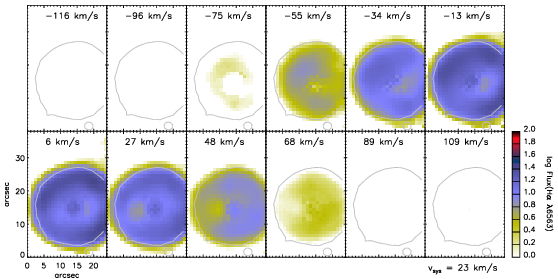}\\
{\footnotesize (j) K\,2-16 Model (H$\alpha$ $\lambda$6563)}\\ 
\includegraphics[width=5.8in]{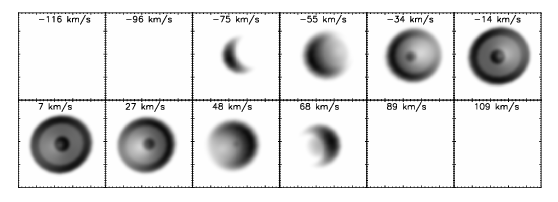}\\
\caption{\textit{-- continued}}
\end{center}
\end{figure*}

\newpage
\FloatBarrier
\normalsize

\figurenum{2}
\begin{figure*}
\begin{center}
{\footnotesize (j) K\,2-16 [N\,{\sc ii}] $\lambda$6584}\\ 
\includegraphics[width=6.in]{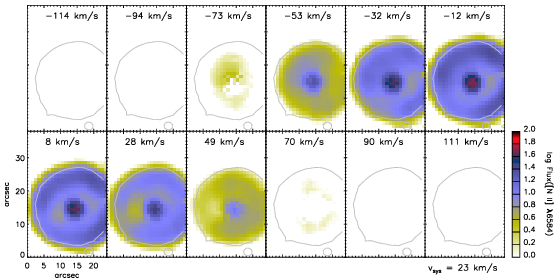}\\
{\footnotesize (j) K\,2-16 Model ([N\,{\sc ii}] $\lambda$6584)}\\ 
\includegraphics[width=5.8in]{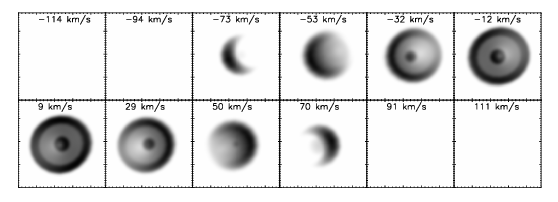}\\
\caption{\textit{-- continued}}
\end{center}
\end{figure*}

\newpage
\FloatBarrier
\normalsize

\figurenum{2}
\begin{figure*}
\begin{center}
{\footnotesize (k) NGC\,6578 H$\alpha$ $\lambda$6563}\\ 
\includegraphics[width=6.in]{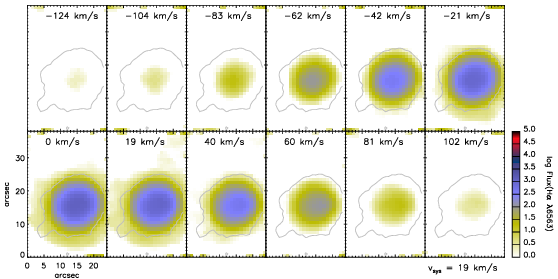}\\
{\footnotesize (k) NGC\,6578 Model (H$\alpha$ $\lambda$6563)}\\ 
\includegraphics[width=5.8in]{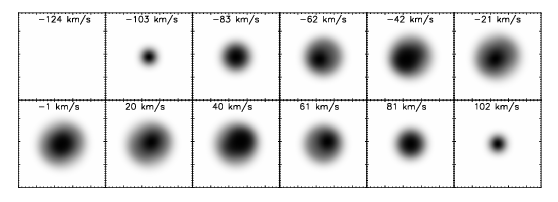}\\
\caption{\textit{-- continued}}
\end{center}
\end{figure*}

\newpage
\FloatBarrier
\normalsize

\figurenum{2}
\begin{figure*}
\begin{center}
{\footnotesize (k) NGC\,6578 [N\,{\sc ii}] $\lambda$6584}\\ 
\includegraphics[width=6.in]{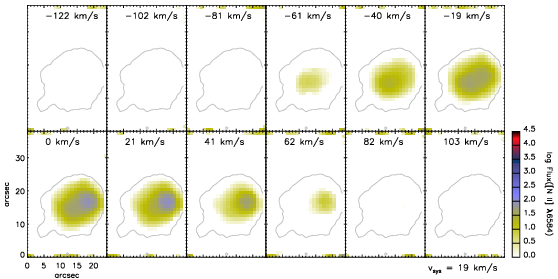}\\
{\footnotesize (k) NGC\,6578 Model ([N\,{\sc ii}] $\lambda$6584)}\\ 
\includegraphics[width=5.8in]{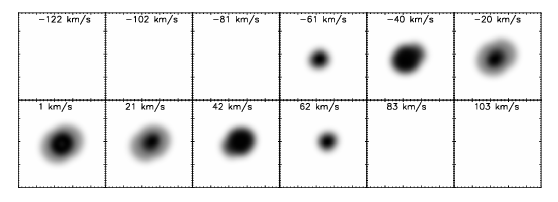}\\
\caption{\textit{-- continued}}
\end{center}
\end{figure*}

\newpage
\FloatBarrier
\normalsize

\figurenum{2}
\begin{figure*}
\begin{center}
{\footnotesize (l) NGC\,6567 H$\alpha$ $\lambda$6563}\\ 
\includegraphics[width=6.in]{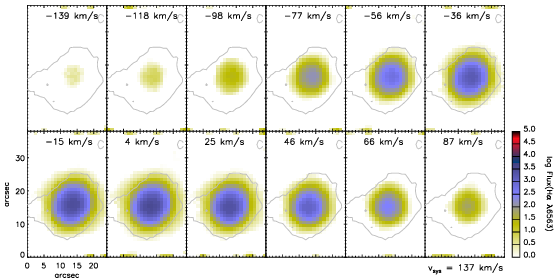}\\
{\footnotesize (l) NGC\,6567 Model (H$\alpha$ $\lambda$6563)}\\ 
\includegraphics[width=5.8in]{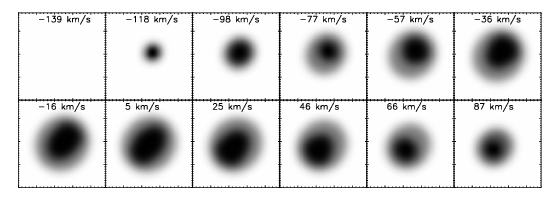}\\
\caption{\textit{-- continued}}
\end{center}
\end{figure*}

\newpage
\FloatBarrier
\normalsize

\figurenum{2}
\begin{figure*}
\begin{center}
{\footnotesize (l) NGC\,6567 [N\,{\sc ii}] $\lambda$6584}\\ 
\includegraphics[width=6.in]{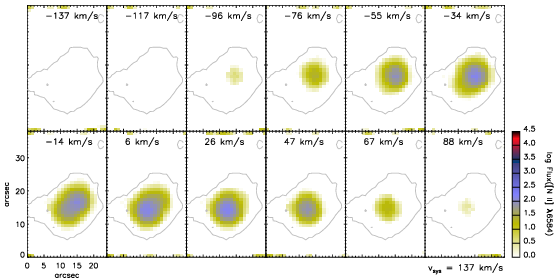}\\
{\footnotesize (l) NGC\,6567 Model ([N\,{\sc ii}] $\lambda$6584)}\\ 
\includegraphics[width=5.8in]{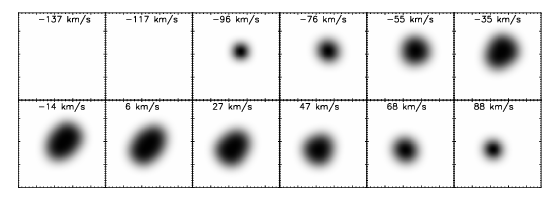}\\
\caption{\textit{-- continued}}
\end{center}
\end{figure*}

\newpage
\FloatBarrier
\normalsize

\figurenum{2}
\begin{figure*}
\begin{center}
{\footnotesize (m) NGC\,6629 H$\alpha$ $\lambda$6563}\\ 
\includegraphics[width=6.in]{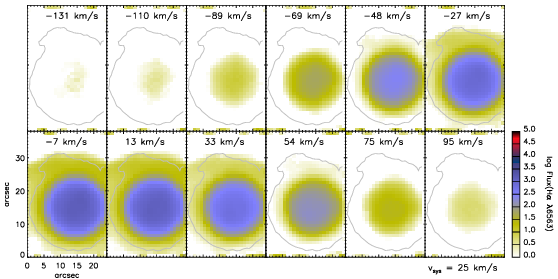}\\
{\footnotesize (m) NGC\,6629 Model (H$\alpha$ $\lambda$6563)}\\ 
\includegraphics[width=5.8in]{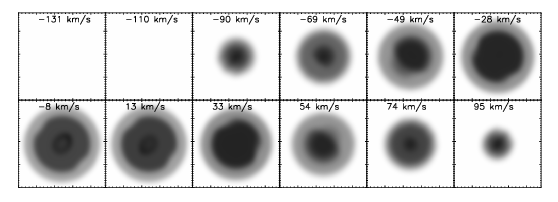}\\
\caption{\textit{-- continued}}
\end{center}
\end{figure*}

\newpage
\FloatBarrier
\normalsize

\figurenum{2}
\begin{figure*}
\begin{center}
{\footnotesize (m) NGC\,6629 [N\,{\sc ii}] $\lambda$6584}\\ 
\includegraphics[width=6.in]{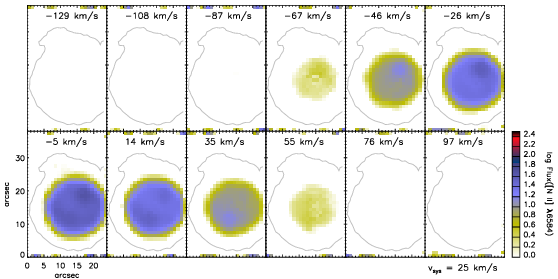}\\
{\footnotesize (m) NGC\,6629 Model ([N\,{\sc ii}] $\lambda$6584)}\\ 
\includegraphics[width=5.8in]{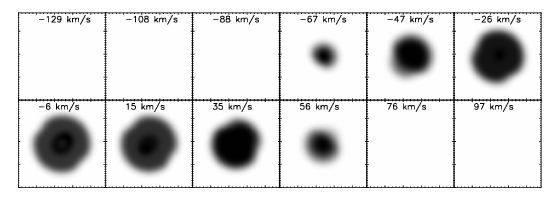}\\
\caption{\textit{-- continued}}
\end{center}
\end{figure*}

\newpage
\FloatBarrier
\normalsize

\figurenum{2}
\begin{figure*}
\begin{center}
{\footnotesize (n) Sa\,3-107 H$\alpha$ $\lambda$6563}\\ 
\includegraphics[width=6.8in]{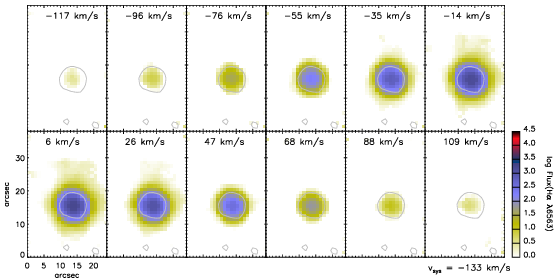}\\
{\footnotesize (n) Sa\,3-107 [N\,{\sc ii}] $\lambda$6584}\\ 
\includegraphics[width=6.8in]{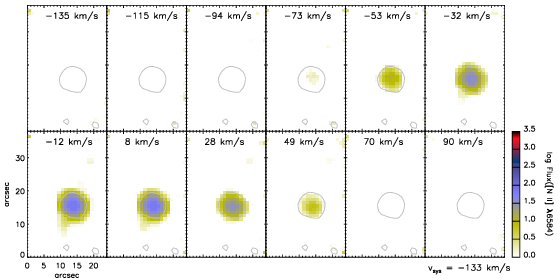}\\
\caption{\textit{-- continued}}
\end{center}
\end{figure*}

\newpage
\FloatBarrier
\normalsize

\figurenum{3}
\begin{figure*}
\begin{center}
{\footnotesize (a) PB\,6 H$\alpha$ $\lambda$6563}\\ 
\includegraphics[width=3.45in]{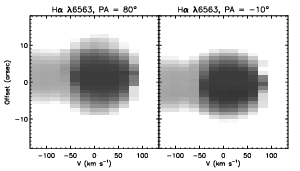}\\
{\footnotesize (a) PB\,6 [N\,{\sc ii}] $\lambda$6584}\\ 
\includegraphics[width=3.45in]{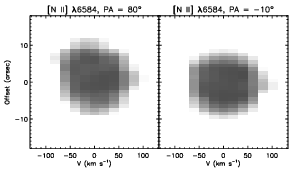}\\
\caption{P--V arrays in the H$\alpha$ $\lambda$6563 and [N\,{\sc ii}] $\lambda$6584 emissions for: (a) PB\,6, (b) M\,3-30, (c) Hb\,4, (d) IC\,1297, (e) Pe\,1-1, (f) M\,1-32, (g) M\,3-15, (h) M\,1-25, (i) Hen\,2-142, (j) K\,2-16, (k) NGC\,6578, (l) NGC\,6567, (m) NGC\,6629, and (n) Sa\,3-107, followed by the associated synthetic P--V diagrams obtained from the best-fitting morpho-kinematic model of all the PNe, except for PB\,6 and Sa\,3-107, with the parameters given in Table~4. Slits oriented with the position angles (PA) along and vertical to the symmetric axis of the morpho-kinematic model passing through the central star, respectively. The velocity in each observed P--V arrays is with respect to the systemic velocity of the object, given in km\,s${}^{-1}$ unit. The angular offset at 0 arcsec is the nebular center. }
\label{wc1:pv:diagram}%
\end{center}
\end{figure*}

\newpage
\FloatBarrier
\normalsize

\figurenum{3}
\begin{figure*}
\begin{center}
{\footnotesize (b) M\,3-30 H$\alpha$ $\lambda$6563}\\ 
\includegraphics[width=3.45in]{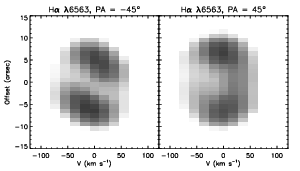}\\
{\footnotesize (b) M\,3-30 Model (H$\alpha$ $\lambda$6563)} \\ 
\includegraphics[width=3.45in]{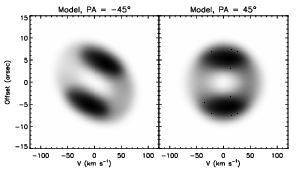}\\
{\footnotesize (b) M\,3-30 [N\,{\sc ii}] $\lambda$6584}\\ 
\includegraphics[width=3.45in]{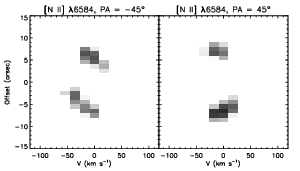}\\
{\footnotesize (b) M\,3-30 Model ([N\,{\sc ii}] $\lambda$6584)} \\ 
\includegraphics[width=3.45in]{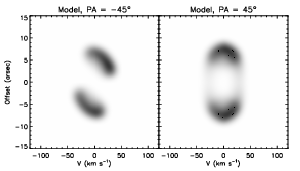}\\
\caption{\textit{-- continued}}
\end{center}
\end{figure*}

\newpage
\FloatBarrier
\normalsize

\figurenum{3}
\begin{figure*}
\begin{center}
{\footnotesize (c) Hb\,4 H$\alpha$ $\lambda$6563}\\ 
\includegraphics[width=3.45in]{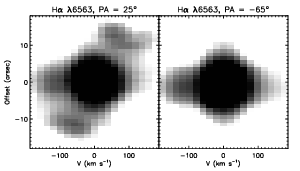}\\
{\footnotesize (c) Hb\,4 Model (H$\alpha$ $\lambda$6563)} \\ 
\includegraphics[width=3.45in]{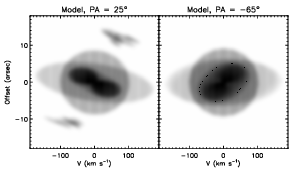}\\
{\footnotesize (c) Hb\,4 [N\,{\sc ii}] $\lambda$6584}\\ 
\includegraphics[width=3.45in]{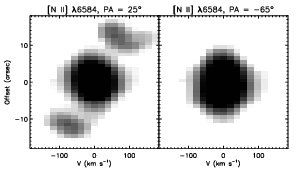}\\
{\footnotesize (c) Hb\,4 Model ([N\,{\sc ii}] $\lambda$6584)} \\ 
\includegraphics[width=3.45in]{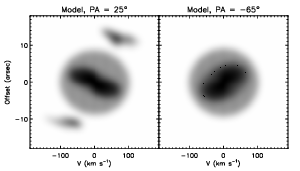}\\
\caption{\textit{-- continued}}
\end{center}
\end{figure*}

\newpage
\FloatBarrier
\normalsize

\figurenum{3}
\begin{figure*}
\begin{center}
{\footnotesize (d) IC\,1297 H$\alpha$ $\lambda$6563}\\ 
\includegraphics[width=3.45in]{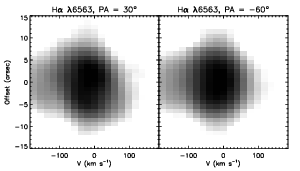}\\
{\footnotesize (d) IC\,1297 Model (H$\alpha$ $\lambda$6563)} \\ 
\includegraphics[width=3.45in]{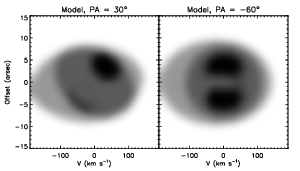}\\
{\footnotesize (d) IC\,1297 [N\,{\sc ii}] $\lambda$6584}\\ 
\includegraphics[width=3.45in]{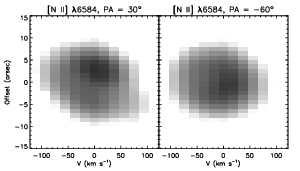}\\
{\footnotesize (d) IC\,1297 Model ([N\,{\sc ii}] $\lambda$6584)} \\ 
\includegraphics[width=3.45in]{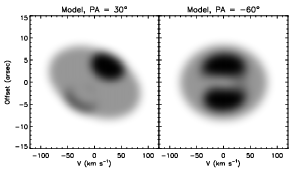}\\
\caption{\textit{-- continued}}
\end{center}
\end{figure*}

\newpage
\FloatBarrier
\normalsize

\figurenum{3}
\begin{figure*}
\begin{center}
{\footnotesize (e) Pe\,1-1 H$\alpha$ $\lambda$6563}\\ 
\includegraphics[width=3.45in]{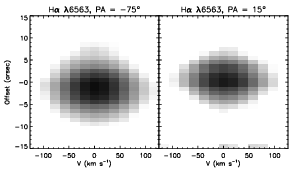}\\
{\footnotesize (e) Pe\,1-1 Model (H$\alpha$ $\lambda$6563)} \\ 
\includegraphics[width=3.45in]{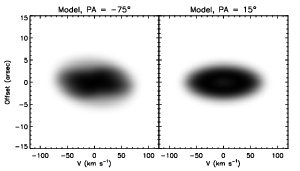}\\
{\footnotesize (e) Pe\,1-1 [N\,{\sc ii}] $\lambda$6584}\\ 
\includegraphics[width=3.45in]{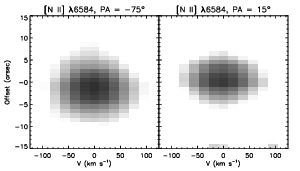}\\
{\footnotesize (e) Pe\,1-1 Model ([N\,{\sc ii}] $\lambda$6584)} \\ 
\includegraphics[width=3.45in]{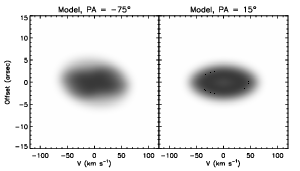}\\
\caption{\textit{-- continued}}
\end{center}
\end{figure*}

\newpage
\FloatBarrier
\normalsize

\figurenum{3}
\begin{figure*}
\begin{center}
{\footnotesize (f) M\,1-32 H$\alpha$ $\lambda$6563}\\ 
\includegraphics[width=3.45in]{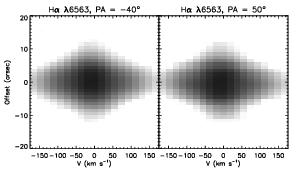}\\
{\footnotesize (f) M\,1-32 Model (H$\alpha$ $\lambda$6563)} \\ 
\includegraphics[width=3.45in]{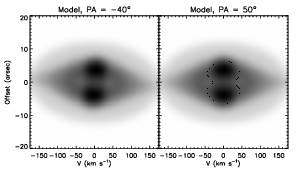}\\
{\footnotesize (f) M\,1-32 [N\,{\sc ii}] $\lambda$6584}\\ 
\includegraphics[width=3.45in]{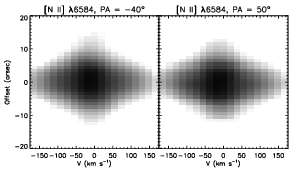}\\
{\footnotesize (f) M\,1-32 Model ([N\,{\sc ii}] $\lambda$6584)} \\ 
\includegraphics[width=3.45in]{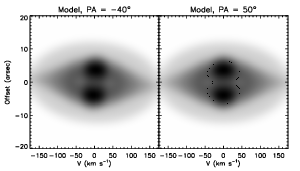}\\
\caption{\textit{-- continued}}
\end{center}
\end{figure*}

\newpage
\FloatBarrier
\normalsize

\figurenum{3}
\begin{figure*}
\begin{center}
{\footnotesize (g) M\,3-15 H$\alpha$ $\lambda$6563}\\ 
\includegraphics[width=3.45in]{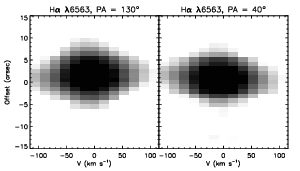}\\
{\footnotesize (g) M\,3-15 Model (H$\alpha$ $\lambda$6563)} \\ 
\includegraphics[width=3.45in]{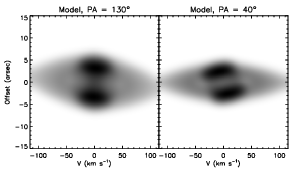}\\
{\footnotesize (g) M\,3-15 [N\,{\sc ii}] $\lambda$6584}\\ 
\includegraphics[width=3.45in]{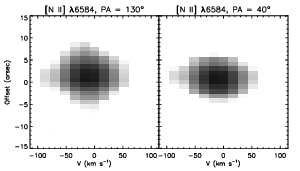}\\
{\footnotesize (g) M\,3-15 Model ([N\,{\sc ii}] $\lambda$6584)} \\ 
\includegraphics[width=3.45in]{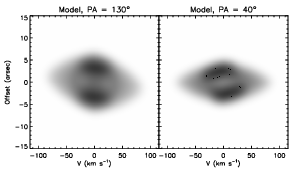}\\
\caption{\textit{-- continued}}
\end{center}
\end{figure*}

\newpage
\FloatBarrier
\normalsize

\figurenum{3}
\begin{figure*}
\begin{center}
{\footnotesize (h) M\,1-25 H$\alpha$ $\lambda$6563}\\ 
\includegraphics[width=3.45in]{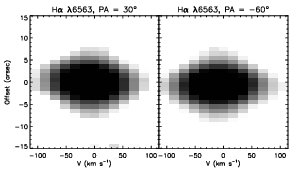}\\
{\footnotesize (h) M\,1-25 Model (H$\alpha$ $\lambda$6563)} \\ 
\includegraphics[width=3.45in]{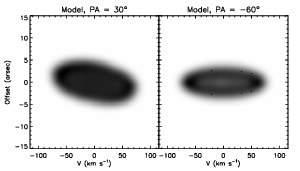}\\
{\footnotesize (h) M\,1-25 [N\,{\sc ii}] $\lambda$6584}\\ 
\includegraphics[width=3.45in]{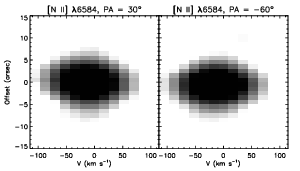}\\
{\footnotesize (h) M\,1-25 Model ([N\,{\sc ii}] $\lambda$6584)} \\ 
\includegraphics[width=3.45in]{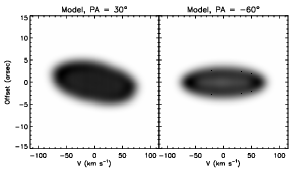}\\
\caption{\textit{-- continued}}
\end{center}
\end{figure*}

\newpage
\FloatBarrier
\normalsize

\figurenum{3}
\begin{figure*}
\begin{center}
{\footnotesize (i) Hen\,2-142 H$\alpha$ $\lambda$6563}\\ 
\includegraphics[width=3.45in]{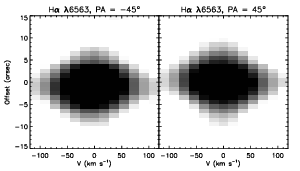}\\
{\footnotesize (i) Hen\,2-142 Model (H$\alpha$ $\lambda$6563)} \\ 
\includegraphics[width=3.45in]{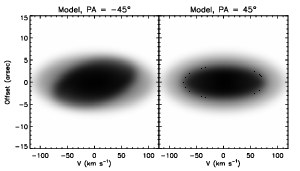}\\
{\footnotesize (i) Hen\,2-142 [N\,{\sc ii}] $\lambda$6584}\\ 
\includegraphics[width=3.45in]{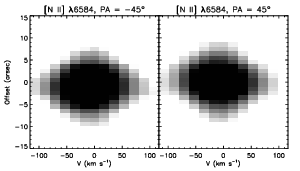}\\
{\footnotesize (i) Hen\,2-142 Model ([N\,{\sc ii}] $\lambda$6584)} \\ 
\includegraphics[width=3.45in]{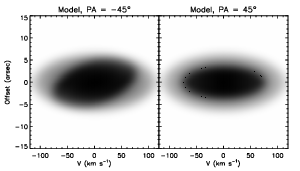}\\
\caption{\textit{-- continued}}
\end{center}
\end{figure*}

\newpage
\FloatBarrier
\normalsize

\figurenum{3}
\begin{figure*}
\begin{center}
{\footnotesize (j) K\,2-16 H$\alpha$ $\lambda$6563}\\ 
\includegraphics[width=3.45in]{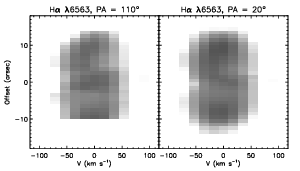}\\
{\footnotesize (j) K\,2-16 Model (H$\alpha$ $\lambda$6563)} \\ 
\includegraphics[width=3.45in]{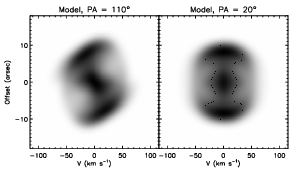}\\
{\footnotesize (j) K\,2-16 [N\,{\sc ii}] $\lambda$6584}\\ 
\includegraphics[width=3.45in]{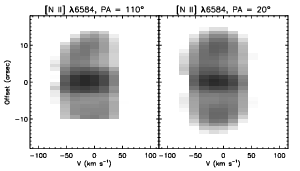}\\
{\footnotesize (j) K\,2-16 Model ([N\,{\sc ii}] $\lambda$6584)} \\ 
\includegraphics[width=3.45in]{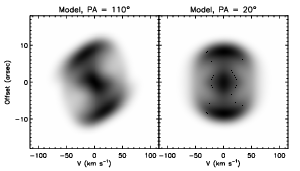}\\
\caption{\textit{-- continued}}
\end{center}
\end{figure*}

\newpage
\FloatBarrier
\normalsize

\figurenum{3}
\begin{figure*}
\begin{center}
{\footnotesize (k) NGC\,6578 H$\alpha$ $\lambda$6563}\\ 
\includegraphics[width=3.45in]{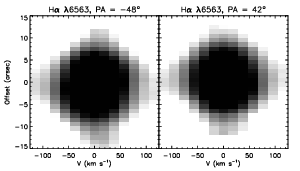}\\
{\footnotesize (k) NGC\,6578 Model (H$\alpha$ $\lambda$6563)} \\ 
\includegraphics[width=3.45in]{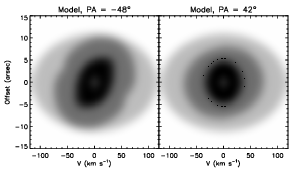}\\
{\footnotesize (k) NGC\,6578 [N\,{\sc ii}] $\lambda$6584}\\ 
\includegraphics[width=3.45in]{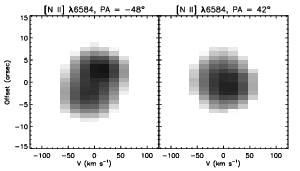}\\
{\footnotesize (k) NGC\,6578 Model ([N\,{\sc ii}] $\lambda$6584)} \\ 
\includegraphics[width=3.45in]{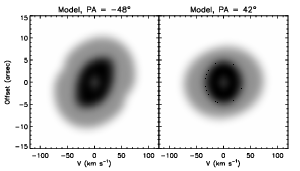}\\
\caption{\textit{-- continued}}
\end{center}
\end{figure*}

\newpage
\FloatBarrier
\normalsize

\figurenum{3}
\begin{figure*}
\begin{center}
{\footnotesize (l) NGC\,6567 H$\alpha$ $\lambda$6563}\\ 
\includegraphics[width=3.45in]{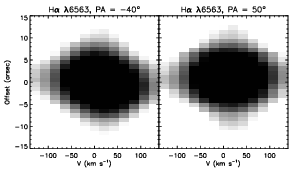}\\
{\footnotesize (l) NGC\,6567 Model (H$\alpha$ $\lambda$6563)} \\ 
\includegraphics[width=3.45in]{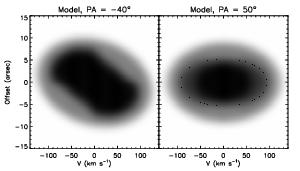}\\
{\footnotesize (l) NGC\,6567 [N\,{\sc ii}] $\lambda$6584}\\ 
\includegraphics[width=3.45in]{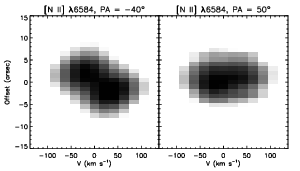}\\
{\footnotesize (l) NGC\,6567 Model ([N\,{\sc ii}] $\lambda$6584)} \\ 
\includegraphics[width=3.45in]{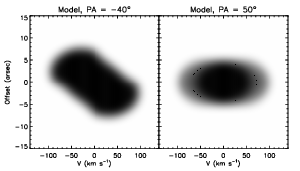}\\
\caption{\textit{-- continued}}
\end{center}
\end{figure*}

\newpage
\FloatBarrier
\normalsize

\figurenum{3}
\begin{figure*}
\begin{center}
{\footnotesize (m) NGC\,6629 H$\alpha$ $\lambda$6563}\\ 
\includegraphics[width=3.45in]{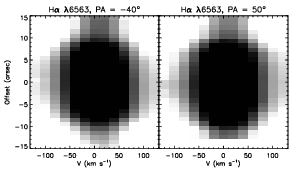}\\
{\footnotesize (m) NGC\,6629 Model (H$\alpha$ $\lambda$6563)} \\ 
\includegraphics[width=3.45in]{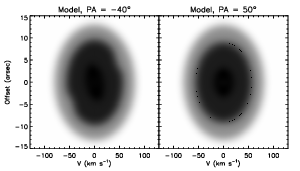}\\
{\footnotesize (m) NGC\,6629 [N\,{\sc ii}] $\lambda$6584}\\ 
\includegraphics[width=3.45in]{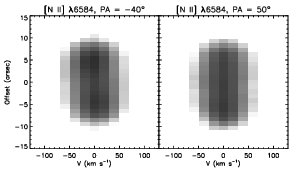}\\
{\footnotesize (m) NGC\,6629 Model ([N\,{\sc ii}] $\lambda$6584)} \\ 
\includegraphics[width=3.45in]{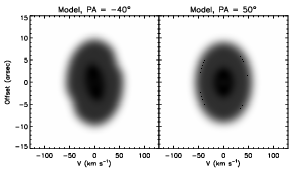}\\
\caption{\textit{-- continued}}
\end{center}
\end{figure*}

\newpage
\FloatBarrier
\normalsize

\figurenum{3}
\begin{figure*}
\begin{center}
{\footnotesize (n) Sa\,3-107 H$\alpha$ $\lambda$6563}\\ 
\includegraphics[width=3.45in]{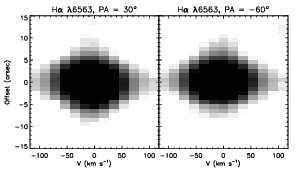}\\
{\footnotesize (n) Sa\,3-107 [N\,{\sc ii}] $\lambda$6584}\\ 
\includegraphics[width=3.45in]{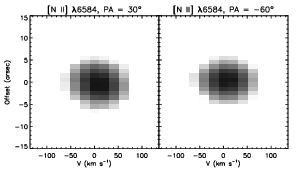}\\
\caption{\textit{-- continued}}
\end{center}
\end{figure*}

\newpage
\FloatBarrier
\normalsize

\figurenum{4}
\begin{figure*}
\begin{center}
{\footnotesize (a) M\,3-30}\\ 
\includegraphics[width=7.in]{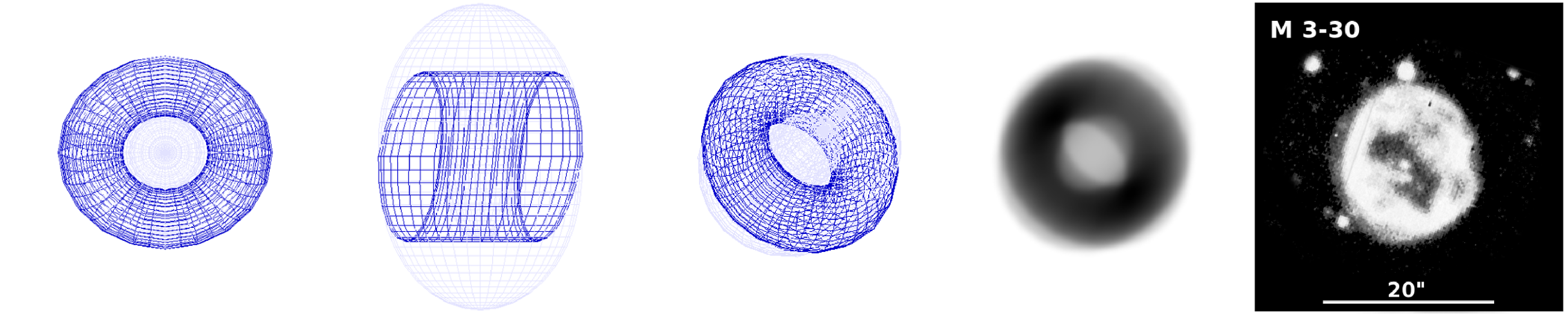}\\
{\footnotesize (b) Hb\,4}\\ 
\includegraphics[width=7.in]{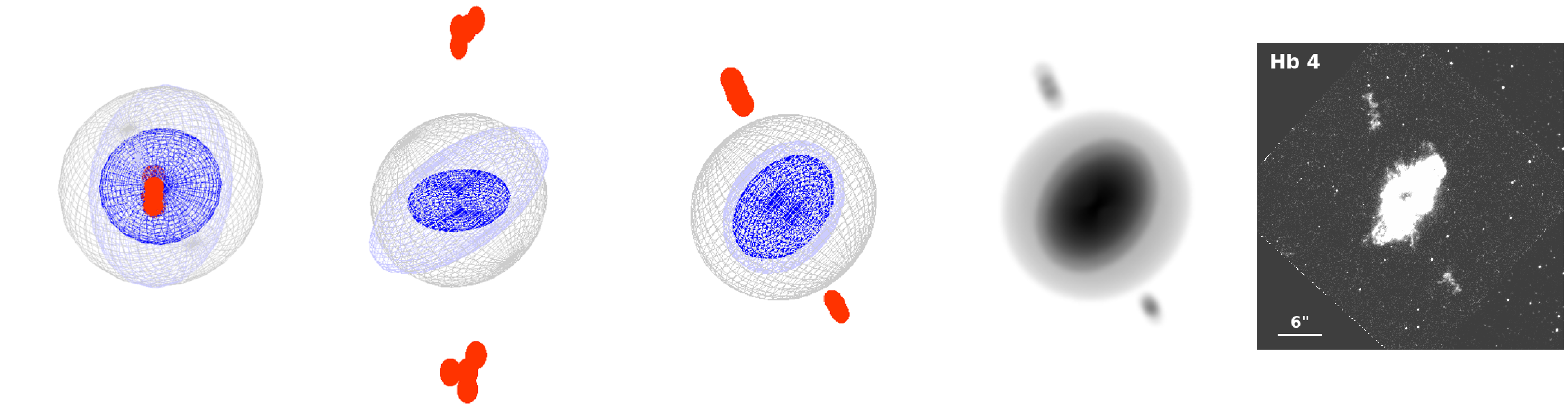}\\
{\footnotesize (c) IC\,1297}\\ 
\includegraphics[width=7.in]{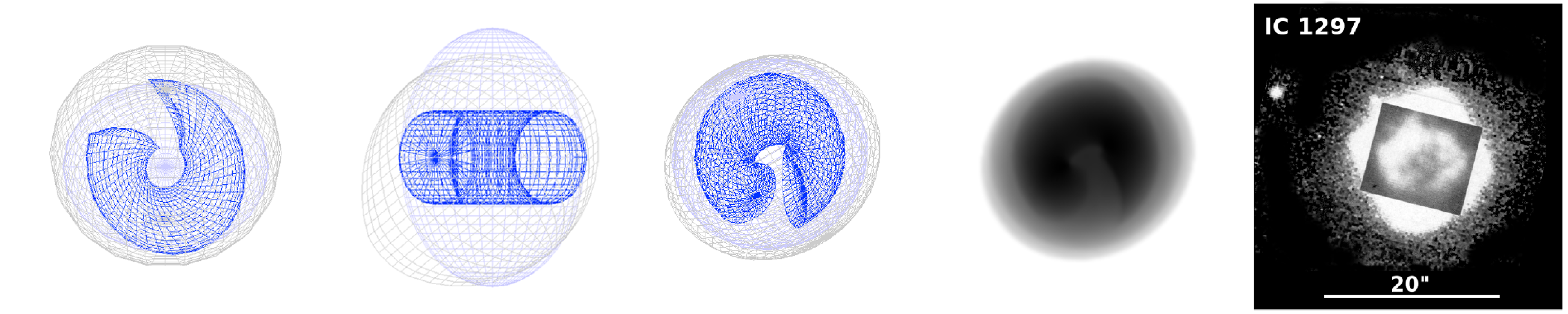}\\
{\footnotesize (d) Pe\,1-1}\\ 
\includegraphics[width=7.in]{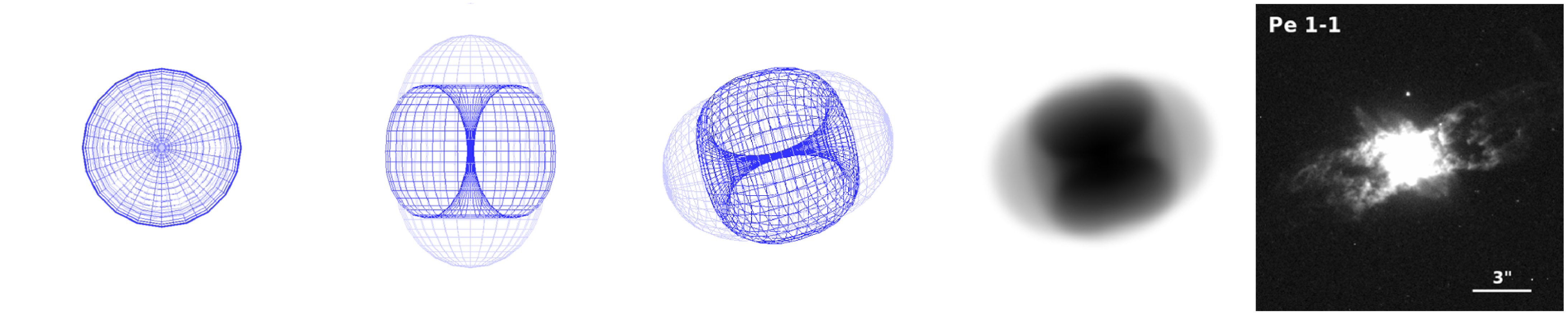}\\
\caption{\textsc{shape} wireframe models of (a) M\,3-30, (b) Hb\,4, (c) IC\,1297, (d) Pe\,1-1, (e) M\,1-32, (f) M\,3-15, (g) M\,1-25, (h) Hen\,2-142, (i) K\,2-16, (j) NGC\,6578, (k) NGC\,6567, and (l) NGC\,6629, in the top view ($i=0^{\circ}$; first column), the front view ($i=90^{\circ}$; second column), and the best-fitting inclination and orientation (third column; see Figure~5 for an interactive viewer in the online journal), followed by the rendered image (fourth column) and the associated archival imaging observation (fifth column), respectively. The parameters of the best-fitting models are listed in Table~4. The archival data include the \textit{HST} images of Hb\,4, Pe\,1-1, M\,3-15, M\,1-25, Hen\,2-142, NGC\,6578, NGC\,6567 and NGC\,6629, with the instruments and filters/gratings listed in Table~2; and the narrow-band H$\alpha$+[N\,{\sc ii}] filter images of M\,3-30, IC\,1297, M\,1-32 and K\,2-16 taken with the 3.5-m ESO NTT from Schwarz et al. (1992), with the image scales shown by solid lines. North is up and east is toward the left-hand side in each archival image. 
}
\label{wc1:model:shape}%
\end{center}
\end{figure*}

\newpage
\FloatBarrier
\normalsize

\figurenum{4}
\begin{figure*}
\begin{center}
{\footnotesize (e) M\,1-32}\\ 
\includegraphics[width=7.in]{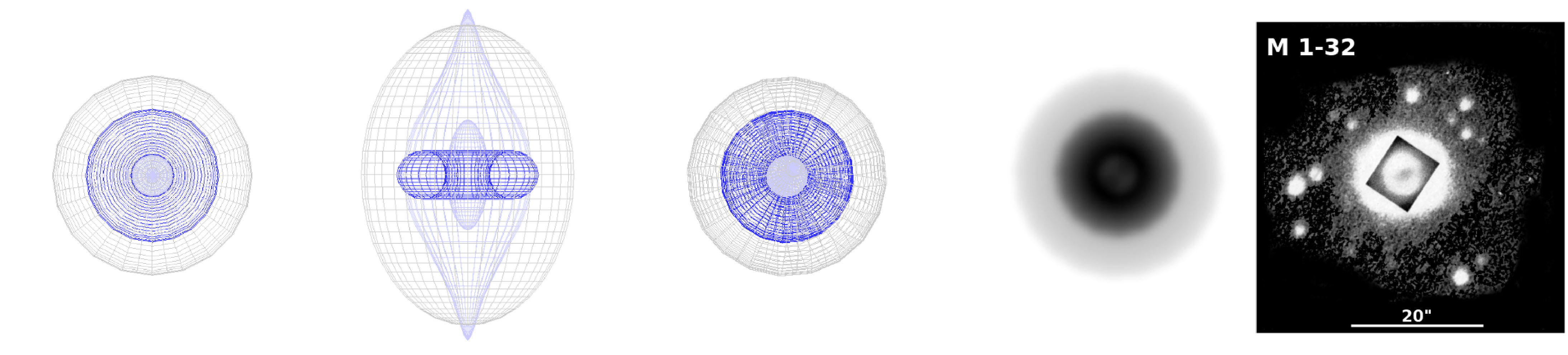}\\
{\footnotesize (f) M\,3-15}\\ 
\includegraphics[width=7.in]{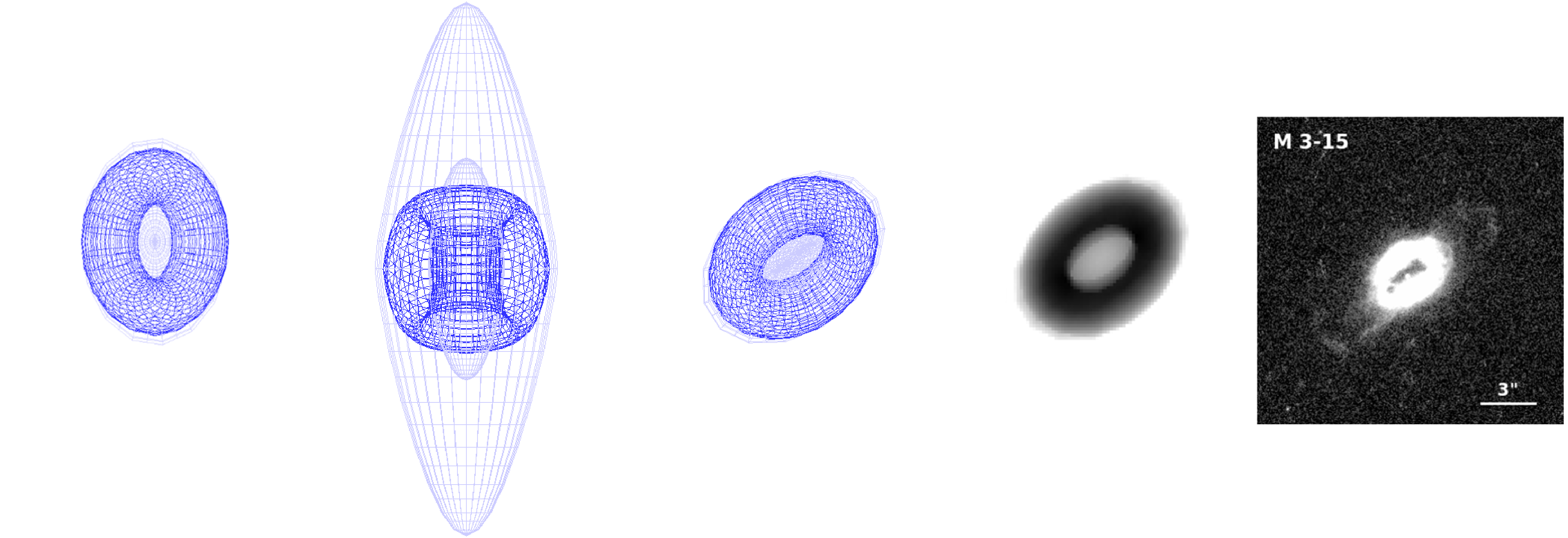}\\
{\footnotesize (g) M\,1-25}\\ 
\includegraphics[width=7.in]{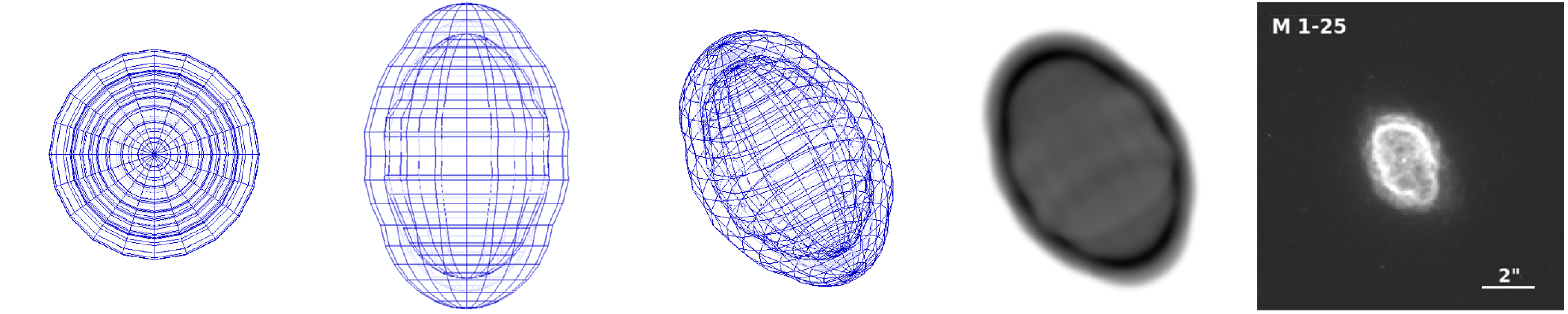}\\
\caption{\textit{-- continued}}
\end{center}
\end{figure*}

\newpage
\FloatBarrier
\normalsize

\figurenum{4}
\begin{figure*}
\begin{center}
{\footnotesize (h) Hen\,2-142}\\ 
\includegraphics[width=7.in]{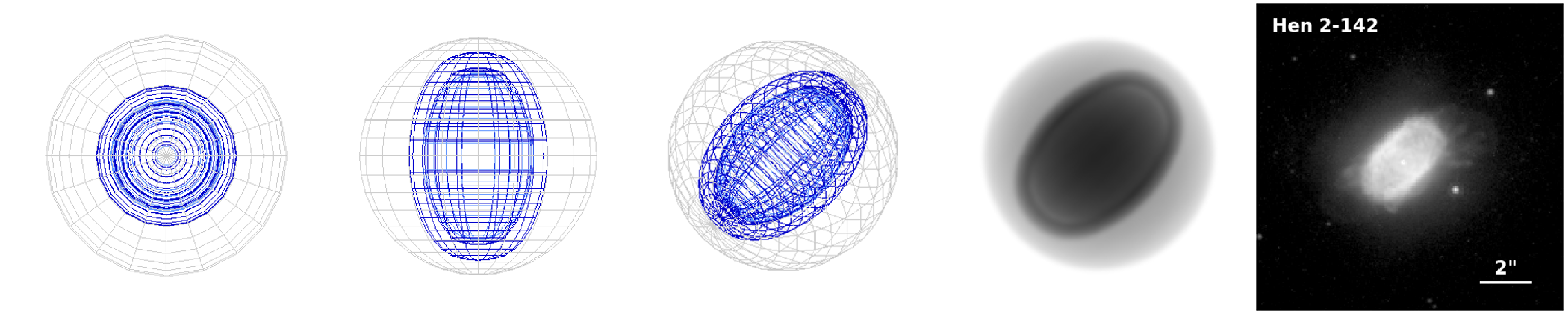}\\
{\footnotesize (i) K\,2-16}\\ 
\includegraphics[width=7.in]{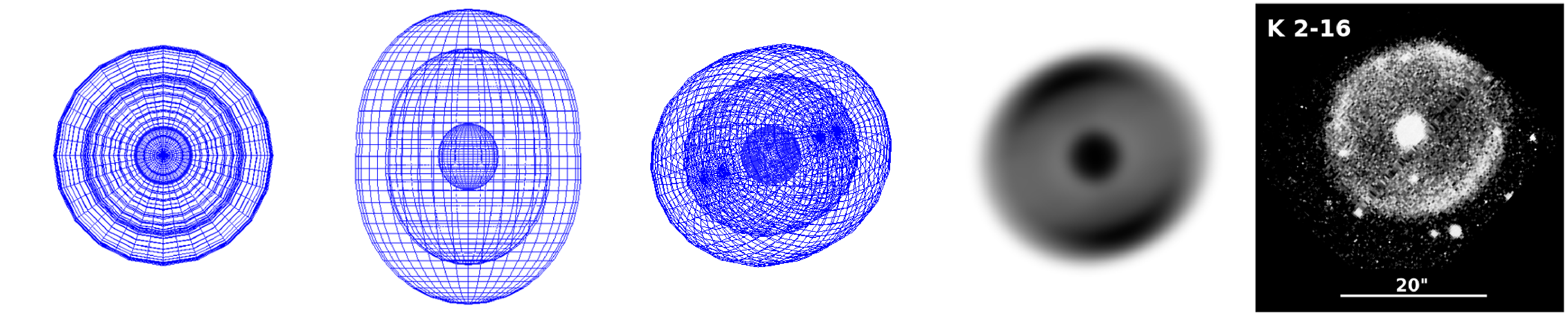}\\
{\footnotesize (j) NGC\,6578}\\ 
\includegraphics[width=7.in]{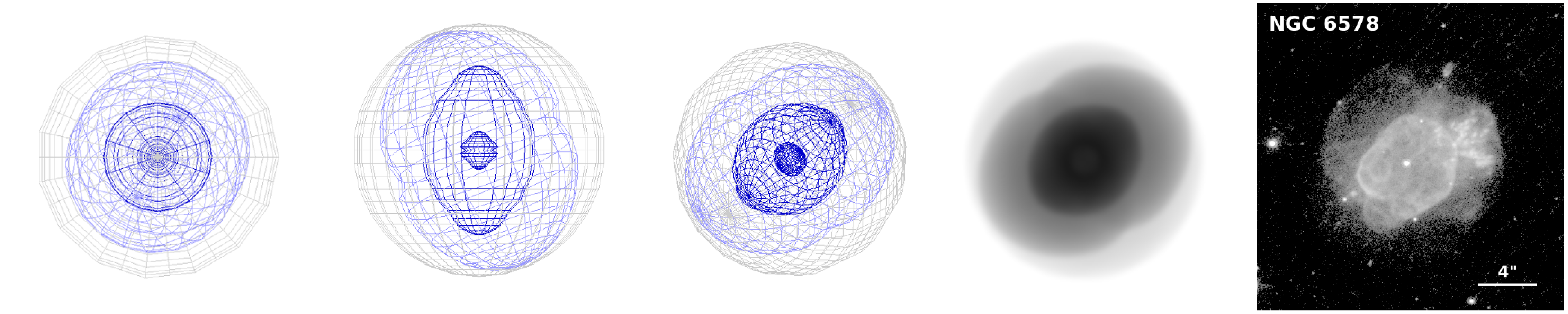}\\
{\footnotesize (k) NGC\,6567}\\ 
\includegraphics[width=7.in]{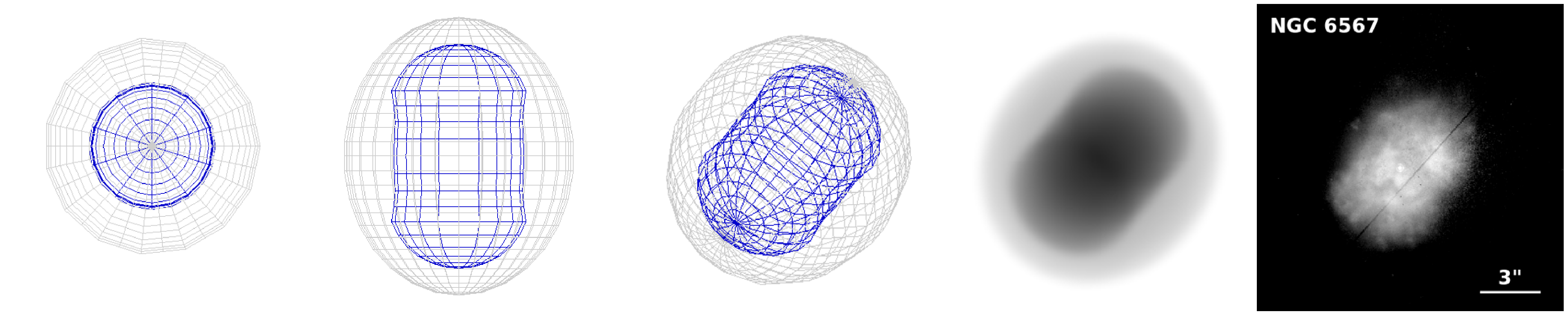}\\
{\footnotesize (l) NGC\,6629}\\ 
\includegraphics[width=7.in]{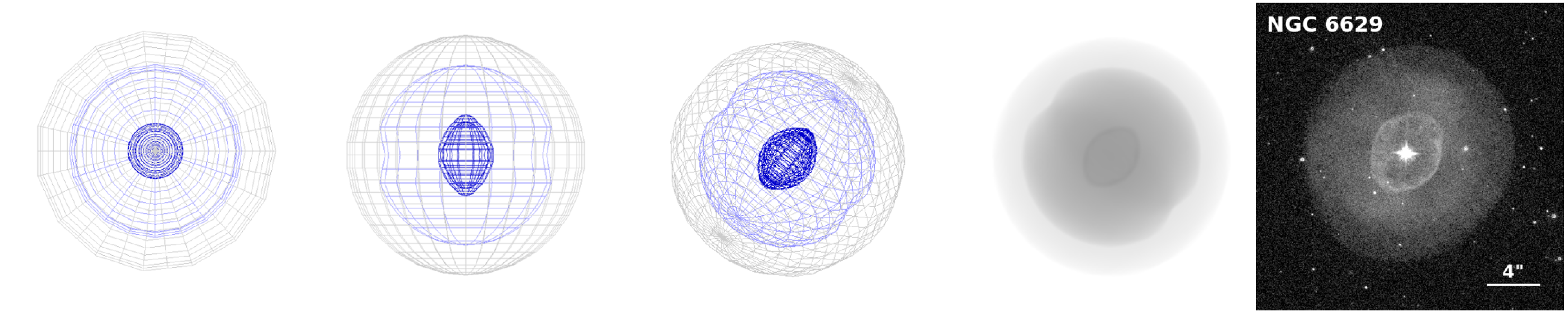}\\
\caption{\textit{-- continued}}
\end{center}
\end{figure*}

\newpage
\FloatBarrier
\normalsize


\FloatBarrier

\end{document}